\documentclass[11pt, oneside]{article}   	
\usepackage{jheppub}
\usepackage{amsmath}
\usepackage{amssymb}
\usepackage{amsfonts}
\usepackage{amsthm}
\usepackage{amsbsy}
\usepackage{array}
\usepackage{mathtools}
\usepackage[usenames,dvipsnames]{xcolor}
\usepackage{subcaption}
\usepackage{dsdshorthand}
\usepackage{graphicx}
\usepackage[vcentermath]{youngtab}
\usepackage{slashed}

\usepackage{tikz}
\usetikzlibrary{decorations.pathmorphing}
\usetikzlibrary{positioning,decorations.pathreplacing}
\usetikzlibrary{calc}
\tikzset{snake it/.style={decorate, decoration=snake}}
\usepackage{tikz-feynman}

\usepackage{ifthen}
\newboolean{mostlyminus}
\setboolean{mostlyminus}{true} 
\newcommand{\signminus}{\ifthenelse{\boolean{mostlyminus}}{-}{}}
\newcommand{\signplus}{\ifthenelse{\boolean{mostlyminus}}{}{-}}
\newcommand{\signminusplus}{\ifthenelse{\boolean{mostlyminus}}{-}{+}}
\newcommand{\signplusminus}{\ifthenelse{\boolean{mostlyminus}}{+}{-}}

\renewcommand\vol{\mathop{\mathrm{vol}}}

\newcommand{\DGLAP}{\textrm{DGLAP}}
\newcommand{\BFKL}{\textrm{BFKL}}
\newcommand\scriplus{\mathscr{I}^+}
\newcommand{\QCD}{\textrm{QCD}}
\newcommand{\symF}{K}

\newcommand*\link[1]{\hspace*{0em plus 1fill}\makebox{#1}}

\makeatletter
\def\@fpheader{\ }
\makeatother

\title{Seeing through the confinement screen: DGLAP/BFKL mixing and light-ray matching in QCD}
\author{Cyuan-Han Chang$^{1,2}$, Hao Chen$^{3}$, David Simmons-Duffin$^2$, Hua Xing Zhu$^{4,5}$}
\affiliation{
${}^1$Kadanoff Center for Theoretical Physics \& Enrico Fermi Institute, University of Chicago, Chicago,
Illinois 60637, USA \\
${}^2$Walter Burke Institute for Theoretical Physics, Caltech, Pasadena, California 91125, USA \\
${}^3$Center for Theoretical Physics - a Leinweber Institute, Massachusetts Institute of Technology, 77 Massachusetts Avenue, Cambridge, MA 02139, U.S.A\\
${}^4$School of Physics, Peking University, Beijing 100871, China\\
${}^5$Center for High Energy Physics, Peking University, Beijing 100871, China
}

\emailAdd{cchang10@uchicago.edu}
\emailAdd{hao\_chen@mit.edu}
\emailAdd{dsd@caltech.edu}
\emailAdd{zhuhx@pku.edu.cn}

\date{}
\abstract{We argue that collider observables such as hadron number flux can be matched onto a linear combination of detectors/light-ray operators in perturbative QCD. The spectrum of detectors in QCD is subtle, due to recombination between the DGLAP and BFKL trajectories. We explain how to define and renormalize these trajectories at one-loop, systematically incorporating their recombination. The leading and subleading soft gluon theorems play an important role, and our analysis suggests the presence of an infinite series of further subleading soft theorems for squared-amplitudes/form factors. Combined with our light-ray matching hypothesis, the anomalous dimensions of recombined DGLAP/BFKL detectors yield a prediction for the energy dependence of the number of particles in a jet, as well as other predictions for more general energy-weighted hadron measurements. We compare these predictions to Monte-Carlo simulations, finding good agreement.}

\preprint{MIT-CTP 5876\\
\link{CALT-TH 2025-018}}

\renewcommand\beforetochook{\pagestyle{myplain}\pagenumbering{roman}\renewcommand{\baselinestretch}{0.947}}

\begin{document}

\maketitle
\renewcommand{\baselinestretch}{1}
\pagenumbering{roman}
\setcounter{page}{2}
\newpage
\pagenumbering{arabic}
\setcounter{page}{1}

\section{Introduction}

Can we predict the number of particles in a jet analytically? The naive answer is: ``no, particle number is not an IR-safe observable in QCD." However, particle number is clearly well-defined for an experimentalist: they can count the blips in their calorimeters. We will argue that, contrary to the naive answer, particle number {\it does} correspond to a well-defined IR-safe operator in perturbative QCD, but its definition is subtle. It is a nontrivial linear combination of light-ray operators.

Exactly the same question was asked in the early days of QCD. Particle multiplicity is naturally related to the $n \to 1$ limit of the Mellin-$n$ moment of the single-particle inclusive cross section $E_p d\sigma/d^3 p$ in $e^+e^- \to \text{hadron}(p) + X$. Remarkably, it was found that the $Q^2$ evolution of average multiplicity can be calculated within perturbation theory~\cite{Bassetto:1979nt,Amati:1980ch,Furmanski:1979jx,Konishi:1979ft,Mueller:1981ex}. To leading logarithmic accuracy in QCD, the corresponding anomalous dimension governing the resummation begins at ${\cal O}(\sqrt{\alpha_s})$, in contrast to the usual perturbative expansion, which starts at ${\cal O}(\alpha_s)$. The dominant contributions at this order are captured by a ladder-diagram approximation, a consequence of soft gluon color coherence~\cite{Mueller:1981ex,Ermolaev:1981cm,Dokshitzer:1982fh,Dokshitzer:1982xr,Bassetto:1982ma,Mueller:1983cq}, which underpins the angular-ordered parton shower formalism~\cite{Marchesini:1987cf,Webber:1983if}. Since these early developments, the average particle multiplicity has been computed to higher orders in perturbation theory~\cite{Vogt:2011jv,Albino:2011cm,Kom:2012hd,Bolzoni:2012ii,Bolzoni:2013rsa}, and has found applications from extraction of the strong coupling constant~\cite{Perez-Ramos:2013eba} to quark-gluon jet discrimination~\cite{Gallicchio:2011xq}.

There is a clear physical picture for the growth of particle multiplicity in a jet. As partons propagate away from a collision, they repeatedly split. Splitting preserves energy but not particle number. After many splittings, the partons hadronize. Hadronization is intrinsically nonperturbative, but in a collision with high energy $Q\gg \Lambda_{\QCD}$, the initial splitting process is perturbative. We can thus imagine using perturbation theory to describe how much {\it more\/} splitting occurs when we change the energy of a collision $Q\to Q+ \delta Q$, and in this way predict how particle number grows with $Q$.

In this work, we revisit the calculation of particle multiplicity, and argue that the correct framework to organize this calculation, and to separate the effects of splitting and hadronization, is ``running and matching" for light-ray operators. Let $\mathbb{N}(\vec n)$ be the operator measuring hadron number flux in the direction of a unit vector $\vec n \in S^{d-2}$. This is a well-defined operator in the IR effective field theory (EFT) of hadrons --- for example, we can write it in terms of hadron creation and annihilation operators. We propose that $\mathbb{N}(\vec n)$ {\it matches\/} onto a linear combination of IR-safe asymptotic measurements in perturbative QCD. We make an educated guess that a basis for such measurements is given by light-ray operators at future null infinity, sometimes called ``detectors" \cite{Hofman:2008ar,Kravchuk:2018htv,Caron-Huot:2022eqs}. Schematically, we have
\be
\label{eq:schematicmatching}
\mathbb{N}(\vec n) &= \sum_\cD C_\cD \cD(\vec n),
\ee
where $\cD$ runs over an appropriate set of detectors in perturbative QCD, and $C_\cD$ are Wilson coefficients. Note that this matching is {\it backwards\/} compared to the usual matching for local operators in EFT: for local operators, we expand UV local operators in a series of IR local operators. For detectors, we expand IR detectors in a series of UV detectors.

The expansion (\ref{eq:schematicmatching}) is constrained by symmetries. In particular, because hadronization (however mysterious it might be) is Lorentz-invariant, the Lorentz representations of both sides must match. In the approximation where the outgoing hadrons have high energy and can be treated as massless, the particle number flux operator $\mathbb{N}(\vec n)$ has weight $2-d$ under boosts in the $\vec n$ direction.\footnote{This follows from the fact that Lorentz boosts act like dilatations on the celestial sphere, and the dimension of the celestial sphere is $d-2$. We explain how to modify this discussion to take into account hadron masses in section~\ref{sec:matching}.} Meanwhile,  the detectors of perturbative QCD come in continuous families $\cD_{J_L,i}(\vec n)$ called ``Regge trajectories," labeled by a boost weight $J_L\in \C$ and a trajectory label $i$.\footnote{Note, these are Regge trajectories of {\it operators\/}, not particles. The two are related in AdS/CFT, where Regge trajectories of bulk particles correspond to Regge trajectories of boundary operators.} (See \cite{Caron-Huot:2022eqs,Homrich:2022cfq,Henriksson:2023cnh,Brizio:2024nso,Herrmann:2024yai,Homrich:2024nwc,Balitsky:2024xvi,Ekhammar:2024neh,YZUpcoming} for recent explorations of Regge trajectories of detectors/light-ray operators in various theories.) Thus, Lorentz invariance dictates that only detectors with $J_L=2-d$ can appear on the right-hand side of (\ref{eq:schematicmatching}). Consequently, a more precise version of (\ref{eq:schematicmatching}) is
\be
\label{eq:particlenumbermatching}
\mathbb{N}(\vec n) &= \sum_i C_i(J_L,\mu) \cD_{J_L,i}(\vec n,\mu)\Big|_{J_L=2-d},
\ee
where we have allowed the Wilson coefficients and detectors to depend on a renormalization scale $\mu$.

The Wilson coefficients $C_i(J_L,\mu)$ encode the nonperturbative effects of hadronization. Meanwhile, the detectors $\cD_{J_L,i}(\vec n,\mu)$ have anomalous dimensions that encode the effects of splitting. By analyzing how matrix elements of $\cD_{J_L,i}(\vec n,\mu)$ change with scale, we can predict how matrix elements $\<\mathbb{N}(\vec n)\>_Q$ depend on $Q$. The expansion (\ref{eq:particlenumbermatching}) essentially becomes an expansion in $(\Lambda_\QCD/Q)^{\Delta_{L,i}}$, where $-\Delta_{L,i}$ is the scaling dimension of the detector $\cD_{J_L,i}$. More precisely, because scaling dimensions depend on the running coupling, the matrix elements of $\cD_{J_L,i}$ satisfy an RG equation with eigenvalues $-\Delta_{L,i}(J_L;\a_s(\mu))$.

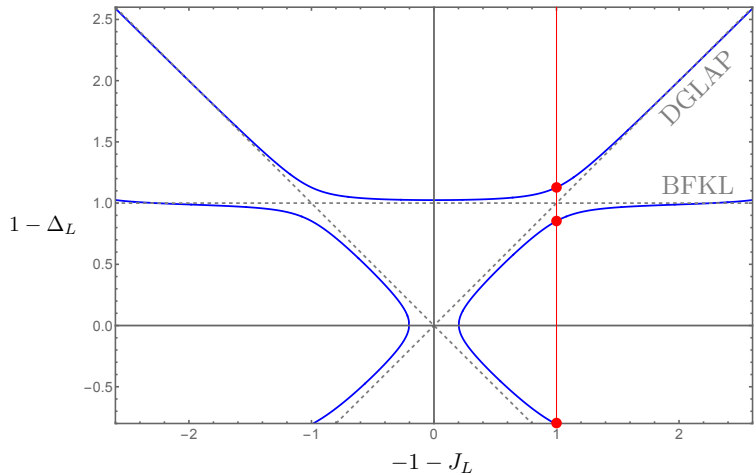
\begin{figure}
    \centering
\begin{tikzpicture}
\node at (0,0) {\includegraphics[width = 10cm]{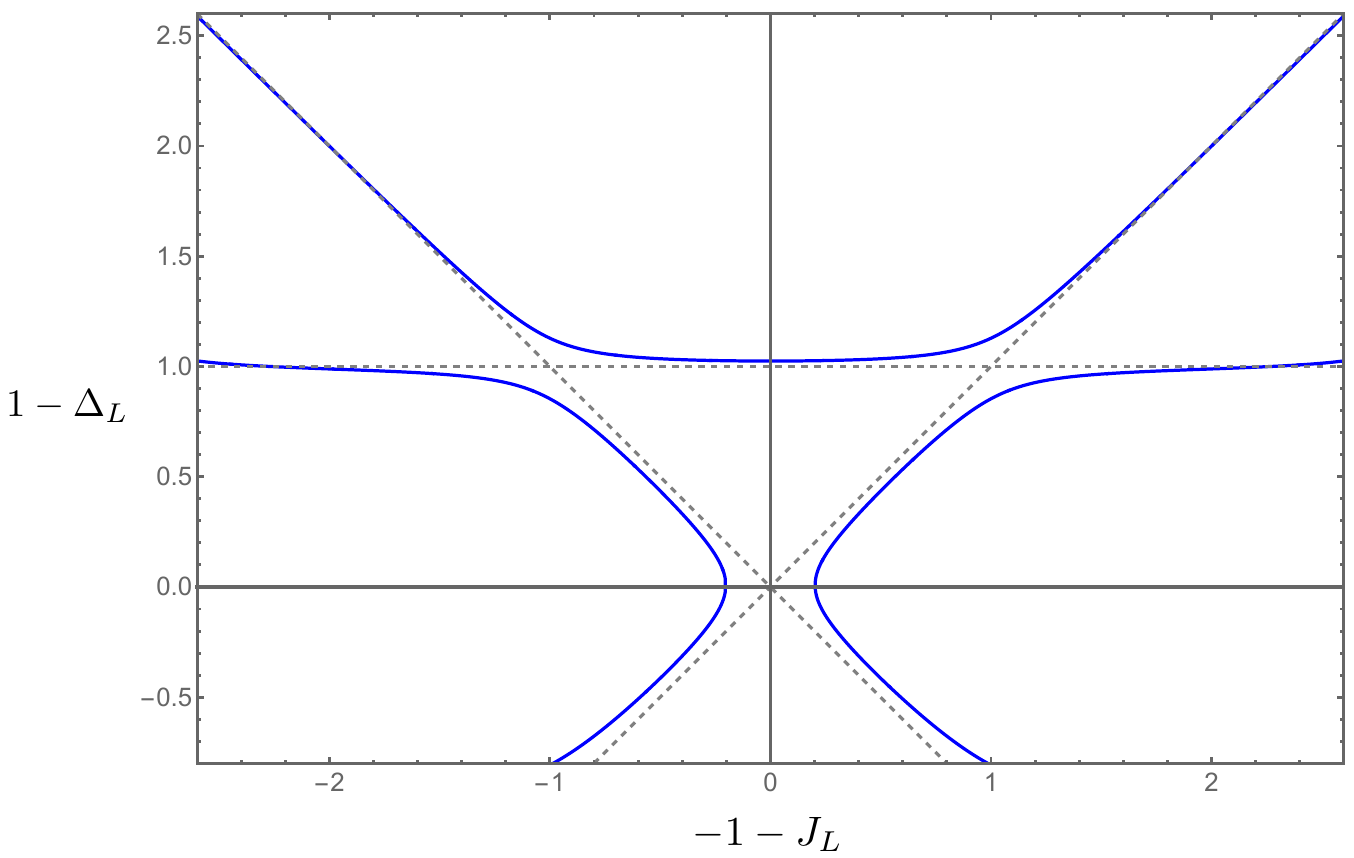}};
	\draw[red] (2.32,-2.465) -- (2.32,3.075);
	\fill[red] (2.32,-2.465) circle (2pt);
	\fill[red] (2.32,0.223) circle (2pt);
	\fill[red] (2.32,0.67) circle (2pt);
	\draw (4.2,2.0) node[rotate=45] {\small \color{gray} DGLAP}; 
	\draw (4.2,0.7) node[] {\small \color{gray} BFKL}; 
\end{tikzpicture}
    \caption{Detectors appearing in the EFT expansion (\ref{eq:particlenumbermatching}) of the particle number operator $\mathbb{N}(\vec n)$ are obtained by drawing a vertical line (red) through the Chew-Frautschi plot at $J_L=2-d$, and tabulating its intersections (red circles) with Regge trajectories. Each detector gives a contribution like $(\Lambda_\mathrm{QCD}/Q)^{\Delta_L}$ to matrix elements of the particle number operator at high energies. Thus, the highest intersection gives the leading contribution at large $Q$. For the purposes of illustration, we show part of the Chew-Frautschi plot of pure YM theory in $d=4$ at one-loop --- the full Chew-Fratuschi plot of QCD will have an infinite number of additional trajectories. Recombination of the DGLAP and BFKL trajectories (dashed gray lines) creates an enhanced correction (\ref{eq:crazysquareroot}) to $\Delta_L$ for the leading detector.}
    \label{fig:CFplot_YM_illustration}
\end{figure}

We can visualize the detectors of perturbative QCD using an (operator) Chew-Frautschi plot, which displays $1-\Delta_L$ versus $\frac{2-d}{2}-J_L$. The terms in (\ref{eq:particlenumbermatching}) come from drawing a vertical line through the Chew-Frautschi plot at $J_L=2-d$, and tabulating its intersections with different Regge trajectories. The highest intersection gives the leading contribution at large $Q$, while lower intersections give power-suppressed corrections, see figure~\ref{fig:CFplot_YM_illustration}.

One of the features of QCD that makes (\ref{eq:particlenumbermatching}) subtle is that the leading trajectory on the right-hand side of (\ref{eq:particlenumbermatching}) lives near a nontrivial {\it recombination\/} of two tree level trajectories: the Dokshitzer-Gribov-Lipatov-Altarelli-Parisi (DGLAP) trajectory and the Balitsky-Fadin-Kuraev-Lipatov (BFKL) trajectory, illustrated as dashed gray lines in figure~\ref{fig:CFplot_YM_illustration}. One of our main goals in this work is to explore the physics of these trajectories at one-loop, in particular understanding their recombination, with applications to observables like particle number.

The DGLAP trajectory, which follows a 45$^\circ$ line on the Chew-Frautschi plot at tree level, is related to analytic continuation in spin of twist-2 local operators \cite{Balitsky:1987bk,Braun:2003rp,Hofman:2008ar,Caron-Huot:2017vep,Kravchuk:2018htv}. Roughly speaking, it counts partons weighted by a power of their energy. We review DGLAP detectors in section~\ref{sec:twisttwo}. Meanwhile, the BFKL trajectory is horizontal at tree level, and measures pairwise color correlations on the celestial sphere. BFKL detectors are related by a change of conformal frame to the operators described by BFKL \cite{Lipatov:1976zz,Kuraev:1977fs,Balitsky:1978ic} that play an important role in Regge physics. We explain the detailed construction of BFKL detectors in section~\ref{sec:bfkltraj}.

The tree-level DGLAP and BFKL trajectories meet precisely at $J_L=2-d$. When the coupling is turned on, they recombine, creating a smooth nondegenerate curve on the Chew-Frautschi plot \cite{Jaroszewicz:1982gr,Lipatov:1996ts,Kotikov:2000pm,Kotikov:2002ab,Kotikov:2004er,Brower:2006ea}. In this work, we will describe this recombination of DGLAP and BFKL trajectories at the level of operators (not just curves on the Chew-Frautschi plot), using similar techniques to those used for the Wilson-Fisher theory  \cite{Caron-Huot:2022eqs} and $O(N)$ models \cite{YZUpcoming}.

We will see that certain universal divergences in DGLAP matrix elements near $J_L=2-d$ are controlled by the soft gluon theorem. Similarly, universal divergences in BFKL matrix elements are controlled by collinear splitting. Together, these effects combine to create mixing between the trajectories, as we explain for pure Yang-Mills theory in section~\ref{sec:dglapbfklmixing}, and for QCD in section~\ref{sec:QCD}. This mixing involves a nontrivial interplay of $1/\e$ divergences and rapidity divergences (which are not regularized by dimensional regularization). Following \cite{Caron-Huot:2022eqs}, we use analytic continuation in $J_L$ as a rapidity regulator. This has the advantage that Lorentz symmetry remains manifest throughout and it puts strong constraints on the pattern of operator mixing.

An important consequence of DGLAP-BFKL mixing is that the anomalous dimension at $J_L=-2$ (specializing for the moment to $d=4$) is proportional to the {\it square-root\/} of $\a_s$:
\be
\label{eq:crazysquareroot}
-\De_L(J_L=-2;\a_s) &= \sqrt{\frac{2C_A}{\pi} \a_s} + O(\a_s).
\ee
In other words, a combination of IR and collinear effects conspire to enhance the value of $-\Delta_L$ near $J_L=-2$ --- precisely the value of $J_L$ relevant for counting particle number. This is why it is essential to correctly map out the Chew-Frautschi plot of QCD to understand the $Q$-dependence of particle number.
This $\sqrt \alpha_s$ effect agrees with previous jet multiplicity calculations \cite{Bassetto:1979nt,Amati:1980ch,Furmanski:1979jx,Konishi:1979ft,Mueller:1981ex}. For a simple derivation using DGLAP evolution and certain scaling assumptions see e.g.\ \cite{Ellis:1996mzs,Larkoski:2024uoc}. However, our analysis gives a different conceptual understanding for this resummation, in particular highlighting its connection to BFKL physics.
 
 Note that there are infinite number of additional BFKL-type trajectories that have $\Delta_L=0$ at tree-level, built from higher products of color detectors at null infinity, see e.g.\ \cite{Caron-Huot:2013fea}. In $\cN=4$ SYM, some horizontal trajectories exhibit anomalous dimensions of order $\sqrt \a_s$, even for generic $J_L$ \cite{Ekhammar:2024neh}. It will be important to determine whether the recombined DGLAP-BFKL trajectory that we study has the highest value of $-\Delta_L$ at $J_L=-2$ among all of these trajectories at leading order (and hence gives the leading contribution to the running of $\mathbb{N}(\vec n)$). 

The recombination of the DGLAP and BFKL trajectories near $J_L=2-d$ is one of an infinite number of recombinations between tree-level trajectories in the Chew-Frautschi plot. In section~\ref{sec:shadowDGLAP}, we turn our attention to mixing of the DGLAP trajectory with its ``shadow," which occurs at $J_L=\frac{2-d}{2}$ (in the middle of the Chew-Frautschi plot). We find that this mixing is controlled by the subleading soft gluon theorem.

In general, we argue that each recombination of the DGLAP trajectory with other trajectories expresses a new kind of ``soft theorem" for squared amplitudes and form factors. For example, recombination of the DGLAP trajectory with another trajectory at $J_L=n$ encodes a (sub-)$^{n+2}$leading soft gluon theorem for squared form factors, where the coefficient of the soft behavior is a matrix element of an operator on the other trajectory. We only study the cases $J_L=-2,-1$ in this work, but it would be interesting to determine these operators explicitly for other values of $J_L$ as well.

Overall, we find a consistent picture for DGLAP-BFKL mixing at 1-loop in QCD, and thereby begin the process of mapping out the operator Chew-Frautschi plot in the interacting theory. But how can we test our results? First, it is useful to access more locations on the Chew-Frautschi plot than just $J_L=2-d$. An important observation is that the argument for matching of particle number flux $\mathbb{N}(\vec n)$ onto a linear combination of detectors (\ref{eq:particlenumbermatching}) generalizes to any IR measurement with definite boost weight $J_L$, as we explain in section~\ref{sec:matching}. For example, consider an operator $\mathbb{N}_{J_L}(\vec n)$ that counts hadrons weighted by powers of their energy $E^{2-d-J_L}$, again in the approximation where the hadrons are massless. (This is essentially a DGLAP-type operator in the EFT of hadrons.) Lorentz symmetry dictates that
\be
\label{eq:particlenumbermatchingagain}
\mathbb{N}_{J_L}(\vec n) &= \sum_i C_i(J_L,\mu) \cD_{J_L,i}(\vec n,\mu),
\ee
where the Wilson coefficients $C_i(J_L,\mu)$ encode the effects of hadronization.

The matching equation (\ref{eq:particlenumbermatchingagain}) is the motivation for the title of this paper: ``Seeing through the confinement screen." The idea is that confinement provides a kind of ``screen" or ``filter" that sits between our measurement devices and the hard processes that produce outgoing partons. In optics or image processing, one can undo/deconvolve a filter by diagonalizing the filter and dividing by its eigenvalues. Equation (\ref{eq:particlenumbermatchingagain}) can be understood in the same spirit. Because hadronization is Lorentz-invariant, it is ``diagonalized" by restricting to definite $J_L$. The Wilson coefficients $C_i(J_L,\mu)$ are analogous to the eigenvalues of a filter. If we divide by them, we can access the underlying detectors $\cD_{J_L,i}(\vec n,\mu)$ that directly probe partons at the scale $\mu$.

\begin{figure}
    \centering
    \includegraphics[width = 10cm]{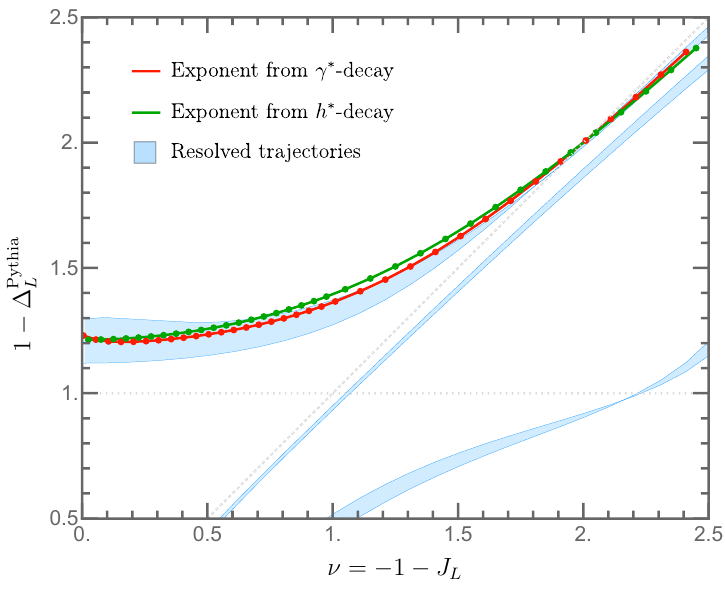}
    \caption{A plot of $1-\Delta_L$ as a function of $\nu=-1-J_L$, determined via (\ref{eq:mapoutcfplot}) for simulated $\g^*$ and $h^*$ decay in Pythia. The 1-loop analytical results for the recombined trajectories (taking into account uncertainties in the choice of matching scale) are shown in light blue.}
    \label{fig:data_intersection_comparison_intro}
\end{figure}

A prediction of (\ref{eq:particlenumbermatchingagain}) is that the leading $Q$-scaling of matrix elements of $\mathbb{N}_{J_L}(\vec n)$ should be controlled by the QCD Regge trajectory $i=i_\mathrm{min}$ with the smallest value of $\Re(\De_L)$: 
\be
\label{eq:mapoutcfplot}
\frac{d}{d\log Q} \log \<\mathbb{N}_{J_L}(\vec n)\>_Q &= -\De_{L,i_\mathrm{min}}(J_L,\a_s(Q)) + \dots.
\ee
Here, $\<\cdot\>_Q$ denotes a matrix element in a state with energy $Q$, and ``$\dots$" represents power suppressed terms in $Q$ and contributions from the $\beta$-function that are subleading in perturbation theory. Thus, by measuring the left-hand side, and letting $J_L$ vary, we can effectively map out the leading Regge trajectory on the operator Chew-Frautschi plot using experimental data. (With precise-enough data, we could map out subleading trajectories as well.) We perform this analysis in section~\ref{sec:comparisontodata} using both artificial data generated with Pythia and CMS Open Data. The results (for real $J_L$) are shown in figure~\ref{fig:data_intersection_comparison_intro}, where we compare our 1-loop prediction for $\De_{L,i_\mathrm{min}}$ to data generated with Pythia.\footnote{As we explain in section~\ref{sec:comparisontodata}, a direct comparison to LHC data requires additional work beyond the scope of this work.} We find good agreement, providing strong evidence for our conceptual framework of matching onto light-ray operators, and for our quantitative picture of the interacting Chew-Frautschi plot of QCD. 

We conclude in section~\ref{sec:discussion} with discussion and future directions.

\section{The DGLAP trajectory}
\label{sec:twisttwo}

In this section, we review DGLAP detectors in perturbative massless QCD and provide details about their renormalization. We start with the definition of DGLAP detectors in the free theory, and then discuss how to calculate their one-loop matrix elements in the interacting theory with dimensional regularization $d=4-2\epsilon$ and renormalize them in the $\overline{\mathrm{MS}}$ scheme.

\subsection{Definition of bare DGLAP detectors}

DGLAP detectors are generalizations of the energy flow operator $\cE(n)$ in perturbation theory. In the free theory, the definition of DGLAP detectors for gluons and quarks is:
\be
\cD^{\DGLAP}_{J_L,g}(z) &= \sum_{\l,c} \int_0^\oo \frac{E^{-J_L}dE}{(2\pi)^{d-1}2E}\left.\left[a^{\dag}_{\l,c}(p)a_{\l,c}(p)\right]\right|_{p=E z}\,, \label{eq:DGLAP_detector_g_def1}\\
\cD^{\DGLAP}_{J_L,q}(z) &= \sum_{s,i} \int_0^\oo \frac{E^{-J_L}dE}{(2\pi)^{d-1}2E}\left.\left[b^{\dag}_{s,i}(p)b_{s,i}(p) + d^{\dag}_{s,i}(p)d_{s,i}(p)\right]\right|_{p=E z}. \label{eq:DGLAP_detector_q_def1}
\ee
Here $\l,s$ are the helicity and spin labels, and $c,i$ are the color indices for adjoint and fundamental representations. $a_{\l,c}, b_{s,i}, d_{s,i}$ are the annihilation operators for gluons, quarks, and anti-quarks, respectively.\footnote{The commutation/anti-commutation relations are given by $[a_{\l,c}(p), a^{\dag}_{\l',c'}(p')] = (2\pi)^{d-1}2p^0\de^{(d-1)}(\vec{p}-\vec{p}')\de_{cc'}\de_{\l\l'}$, $\{b_{s,i}(p), b^{\dag}_{s',i'}(p')\} = (2\pi)^{d-1}2p^0\de^{(d-1)}(\vec{p}-\vec{p}')\de_{ii'}\de_{ss'}$, $\{d_{s,i}(p), d^{\dag}_{s',i'}(p')\} = (2\pi)^{d-1}2p^0\de^{(d-1)}(\vec{p}-\vec{p}')\de_{ii'}\de_{ss'}$.}
 The physical meaning of these detectors is that they measure gluons and quarks in the null direction $z$ and weight them by $E^{2-d-J_L}$, where $J_L\in \C$ is a complex parameter.
 
 The detectors are homogeneous functions of $z$ with degree $J_L$. This guarantees that they transform in irreducible representations of the Lorentz group with Lorentz spin $J_L$. In particular, $J_L$ is the weight under a boost in the $z$ direction.\footnote{The representation-theoretic definition of $J_L$ as a boost weight generalizes to other detectors, which can also be classified by their $J_L$. In particular, it is better to think of $J_L$ as a boost weight, and not simply as a Mellin moment, since the description as a Mellin moment is special to bare DGLAP detectors, and is not tied to a symmetry.} To obtain detectors that depend on a unit vector $\vec n$, as discussed in the introduction, one should substitute $z=(1,\vec n)$. Since the Lorentz group $\SO(d-1,1)$ is the conformal group on the celestial sphere, one can also think of detectors $\cD_{J_L}(z)$ that are homogeneous in $z$ as embedding-space primary operators with ``dimension" $-J_L$ in a fictitious $d-2$-dimensional CFT. See \cite{Caron-Huot:2022eqs} for more discussion of representations of the Lorentz group and how it relates to detectors.
  
  Once we turn on interactions, the detectors (\ref{eq:DGLAP_detector_g_def1}) and (\ref{eq:DGLAP_detector_q_def1}) are in general not infrared-collinear safe: their matrix elements have IR divergences that manifest as poles in $\e$. (Only the special combination with $J_L=1-d$ corresponding to the energy flow operator $\cE(n)= \cD^{\DGLAP}_{1-d,g}(z) + \cD^{\DGLAP}_{1-d,q}(z)$  is IR-collinear safe.) Thus, in the interacting theory \eqref{eq:DGLAP_detector_g_def1} and \eqref{eq:DGLAP_detector_q_def1} should be interpreted as bare operators requiring renormalization.

\subsection{DGLAP detectors from fields at $\scriplus$}

The bare DGLAP detectors (\ref{eq:DGLAP_detector_g_def1}) and (\ref{eq:DGLAP_detector_q_def1}) can also be viewed as non-local operators constructed by smearing fields along a light-ray at future null infinity $\scriplus$. 
Let us consider the gluon DGLAP detector as an example. To construct it, we take the gluon field strength $F_{\mu\nu}=F_{\mu\nu}^aT_{\mathrm{adj}}^a$ to future null infinity along a null direction $z$:
\be\label{eq:null_limit_def_special}
F^{(\bar z)}_\nu(\a,z) \equiv \lim_{L \to \infty}  {L^{\De_A} \over 4} \bar{z}^\mu F_{\mu \nu}(L z + \a \bar{z}/4) \,,
\ee
where $\De_A=\frac{d-2}{2}$, and $\bar{z}$ is an auxiliary null vector satisfying $z\cdot\bar{z}= \signplus 2$. We have set the position of $F_{\mu\nu}$ to be $x=Lz + \a \bar z/4$ so that its retarded time at $\scriplus$ is $\a=2x\.z$. We additionally contracted one of the indices of $F$ with $\bar z$. 

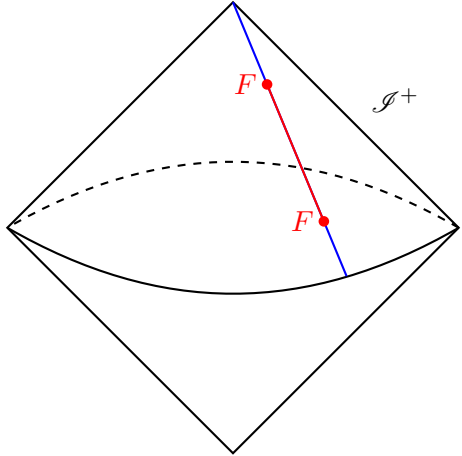
\begin{figure}[htbp]
\begin{center}
\begin{tikzpicture}
    \coordinate (V) at (0,3);
    \coordinate (A) at (1.51,-0.644);
    \coordinate (F1) at ($(V)!0.3!(A)$);
    \coordinate (F2) at ($(V)!0.8!(A)$);
    \draw[thick] (0,3) -- (3,0) -- (0,-3) -- (-3,0) -- cycle;
    \draw[thick, dashed] (3,0) to[out=150, in=30] (-3,0);
    \draw[thick] (3,0) to[out=-150, in=-30] (-3,0);
    \draw[thick,blue] (V) -- (A);
    \draw[thick,red] (F1) -- (F2);
    \fill[red] (F1) circle (2pt);
    \fill[red] (F2) circle (2pt);
    \node[red, left] at (F1) {$F$};
    \node[red, left] at (F2) {$F$};
    \node[black, right] at (1.7, 1.7) {$\scriplus$};
\end{tikzpicture}
\end{center}
\caption{The Penrose diagram of the construction of the gluon DGLAP detector from two field strength tensors $F$ at future null infinity $\scriplus$.}
\label{fig:Penrose_diagram_DGLAP}
\end{figure}

The bare gluon DGLAP detector is a smeared product of two field strengths $F$ connected by an adjoint Wilson line along $\scriplus$, see figure~\ref{fig:Penrose_diagram_DGLAP}.
Its precise definition is
\be\label{eq:DGLAP_detector_g_def}
\cD^{\DGLAP(\bar z)}_{J_L,g}(z)\equiv \frac{1}{C_{J_L}}\int d\a_1 d\a_2\ &\p{(\a_1-\a_2+i\e)^{2\De_A+J_L}+(\a_2-\a_1+i\e)^{2\De_A+J_L}}\nn \\
&\times :F_a^{(\bar z)\nu}(\a_1,z)W^{(\bar z)ab}_{\mathrm{adj}}(\a_1,\a_2)F^{(\bar z)}_{b\nu}(\a_2,z):,
\ee
where 
\be
W^{(\bar z)}_{\mathrm{adj}}(\a_1,\a_2) = \lim_{L\to\infty} P\exp\left(ig\int_{\a_1}^{\a_2} \frac{d\a}{4}\;\bar{z}\cdot A^a(L z+\a \bar{z}/4)T^a_{\mathrm{adj}}\right)\,.
\ee
The normalization constant 
\be\label{eq:CJL_definition}
C_{J_L} = 2^{2\De_A+J_L+1}e^{\frac{i\pi}{2}(2\De_A+J_L)}\sin(\pi(2\De_A+J_L))\G(2\De_A+J_L+1)
\ee
is chosen such that in the free theory, \eqref{eq:DGLAP_detector_g_def} becomes equivalent to \eqref{eq:DGLAP_detector_g_def1} using the mode expansion of the gluon field
\be\label{eq:Amu_mode_expansion}
A^{\mu}_c(x) = \sum_{\l} \int \frac{d^{d-1}\vec{p}}{(2\pi)^{d-1}2 p^0} \left[
  \varepsilon_{\lambda}^\mu(p)  a_{\l,c}(p)e^{\signminus ip\cdot x} + {\varepsilon_{\lambda}^*}^\mu(p) a^{\dag}_{\l,c}(p)e^{\signplus ip\cdot x}
\right]\,,
\ee
where $\varepsilon_{\lambda}^\mu(p)$ is a polarization vector that depends on the gauge choice. 

We provide a detailed derivation of this equivalence between~(\ref{eq:DGLAP_detector_g_def}) and~(\ref{eq:DGLAP_detector_g_def1}) in appendix \ref{app:twist-2}. Note that even though (\ref{eq:DGLAP_detector_g_def}) depends on $\bar z$, its expression in terms of creation and annihilation operators is independent of $\bar z$. Thus, we usually omit the $(\bar z)$ label.

Let us briefly comment on this way of defining detectors. We take the point of view that perturbative QCD admits some class of light-ray operators at $\scriplus$, and any finite observable at $\scriplus$ with the correct symmetries will be a linear combination of these light-ray operators. We can construct finite observables by either (1) constructing bare operators at $\scriplus$ and then renormalizing them (as we do in this work), or (2) constructing renormalized operators at some finite location, and then taking a limit as they approach $\scriplus$ (taking care to adjust the limit to obtain a finite result). These different methods might also involve different choices --- e.g.\ one could take Wilson lines to connect the field strengths $F_{\mu\nu}$ to timelike infinity instead of to each other.\footnote{In the particular case of connecting Wilson lines to future infinity, it is known that different $i\epsilon$ prescriptions for the Wilson lines can lead to time-reversal-odd effects \cite{Collins:2002kn}. Our interpretation is that such constructions yield a nontrivial contribution from time-reversal-odd light-ray operators in the limit that the operators are taken to $\scriplus$. It would be interesting to explore this in more detail.} Ultimately, any method should yield access to the same underlying light-ray operators, potentially with coefficients that depend on the method of construction. This is analogous to the idea that any  measurement at a point can be expanded in a basis of (finite) local operators.

\begin{figure}[htbp]
\begin{center}
        \subfloat[Gluon case]{
            \begin{tikzpicture}
                \begin{feynman}
                \coordinate (vertex) at (0,1.5);
                
                \draw[dashed] (0,-1) -- (0,2);
                
                \draw[gluon] (1.5,0) -- (vertex);
                \draw[gluon] (-1.5,0) -- (vertex);
                \node[below] at (1.5,0) {$a,\mu$};
                \node[below] at (-1.5,0) {$b,\nu$};
                
                \filldraw[red] (vertex) circle (3pt);
                \node[right] at (0.2, 1.8) {$\cD^{\DGLAP}_{J_L,g}(z)$};
                \draw[->] (1.5,0.5) -- (1.0, 1.0);
                \node[right] at (1.5,0.5) {$p$};
                \end{feynman}
            \end{tikzpicture}
        }
        \qquad
        \subfloat[Quark case]{
            \begin{tikzpicture}
                \begin{feynman}
                \coordinate (vertex) at (0,1.5);
                
                \draw[dashed] (0,-1) -- (0,2);
                
                \draw[fermion] (1.5,0) -- (vertex);
                \draw[fermion] (vertex) -- (-1.5,0);
                \node[below] at (1.5,0) {$j$};
                \node[below] at (-1.5,0) {$i$};
                
                \filldraw[red] (vertex) circle (3pt);
                \node[right] at (0.2, 1.8) {$\cD^{\DGLAP}_{J_L,q}(z)$};
                \draw[->] (1.5,0.5) -- (1.0, 1.0);
                \node[right] at (1.5,0.5) {$p$};
                \end{feynman}
            \end{tikzpicture}
        }
        \qquad
        \subfloat[Anti-quark case]{
            \begin{tikzpicture}
                \begin{feynman}
                \coordinate (vertex) at (0,1.5);
                
                \draw[dashed] (0,-1) -- (0,2);
                
                \draw[fermion] (-1.5,0) -- (vertex);
                \draw[fermion] (vertex) -- (1.5,0);
                \node[below] at (1.5,0) {$i$};
                \node[below] at (-1.5,0) {$j$};
                
                \filldraw[red] (vertex) circle (3pt);
                \node[right] at (0.2, 1.8) {$\cD^{\DGLAP}_{J_L,q}(z)$};
                \draw[->] (1.5,0.5) -- (1.0, 1.0);
                \node[right] at (1.5,0.5) {$p$};
                \end{feynman}
            \end{tikzpicture}
        }
\end{center}
\caption{The Feynman diagrams of the DGLAP detector for (a) gluon case, (b) quark case, and (c) anti-quark case.}
\label{fig:DGLAP_detector_Feynman}
\end{figure}
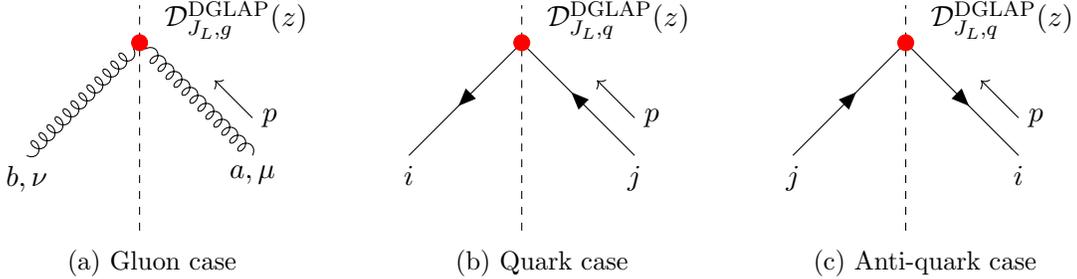

The Feynman rule for the gluon DGLAP detector can be computed from the tree-level matrix element (figure~\ref{fig:DGLAP_detector_Feynman} (a))
\be
\label{eq:dglapfeynmanrule}
\<0|A_{\nu}^b(-q) \cD^{\DGLAP}_{J_L,g}(z) A_{\mu}^a(p)|0\> = (2\pi)^d\de^{(d)}(p-q)\left[\de^{ab} \Pi_{\mu\nu}(z) V_{J_L}(z;p)\right],
\ee
where the vertex $V_{J_L}(z;p)$ is given by
\be\label{eq:VJL_definition}
V_{J_L}(z;p) = \pi  \int_0^{\oo} d\b\, \b^{-J_L-1}\de^{(d)}(p-\b z).
\ee
The tensor $\Pi^{\mu\nu}(z) = \sum_{\lambda} {\varepsilon_{\lambda}^*}^\mu(z)\varepsilon_{\lambda}^\nu(z)$ involves a sum over  polarization vectors. It is gauge-dependent because $\<0|A_{\nu}^b(-q) \cdots A_{\mu}^a(p)|0\>$ is not a gauge-invariant state. A convenient choice is lightcone gauge
\be\label{eq:Pi_lightconegauge}
\Pi^{\mu\nu}(z) = \signminus g^{\mu\nu} \signplusminus {z^\mu n^\nu + z^\nu n^\mu \over z\cdot n},
\ee
where $n^\mu$ is an arbitrary auxiliary null vector. The matrix element of DGLAP detectors inside gauge-invariant states does not depend on the choice of $n^\mu$.

Alternatively, we can work entirely in a covariant gauge, and determine the Feynman rule for $\cD^{\DGLAP}_{J_L,g}(z)$ from the definition (\ref{eq:DGLAP_detector_g_def}). In Feynman gauge we find (\ref{eq:dglapfeynmanrule}) with 
\be
\label{eq:polarizationtensorfeynmangauge}
\Pi^{\mu\nu}(z) = \signminus g^{\mu\nu} \signplusminus {z^\mu \bar z^\nu + z^\nu \bar z^\mu \over z\cdot \bar z} \qquad(\textrm{Feynman gauge}),
\ee
see appendix~\ref{app:twist-2}. Interestingly, in Feynman gauge, dependence on $\bar z$ has persisted in the Feynman rules for the DGLAP detector: $\bar z$ appears as a kind of ``effective" choice of lightcone gauge for the polarization sum. However, this $\bar z$-dependence again goes away when we consider matrix elements in gauge-invariant states.

The Feynman rules for the quark DGLAP detector can be obtained in a similar way from their tree-level matrix elements
\be
\<0|\psi_{i,\a}(-q) \cD^{\DGLAP}_{J_L,q}(z) \bar{\psi}_{j,\b}(p)|0\> = (2\pi)^d\de^{(d)}(p-q)\left[\de_{ij} \slashed{z}_{\a \b} V_{J_L}(z;p)\right], \\
\<0|\bar{\psi}_{j,\b}(-q) \cD^{\DGLAP}_{J_L,q}(z) \psi_{i,\a}(p)|0\> = (2\pi)^d\de^{(d)}(p-q)\left[\de_{ij} \slashed{z}_{\a \b} V_{J_L}(z;p)\right],
\ee
where $\a,\b$ are spinor indices for quark fields. The corresponding Feynman diagrams are shown in figure~\ref{fig:DGLAP_detector_Feynman} (b) and (c). Again, these operators can alternatively be constructed from a smeared product of quark fields connected by a Wilson line along $\scriplus$.

\subsection{Wilson line contributions at null infinity}
\label{sec:wilsonlinedetails}

One might worry that \eqref{eq:DGLAP_detector_g_def} could have terms with more than two creation/annihilation operators coming from the $gf^{abc}A^b_{\mu}A^c_{\nu}$ term in $F^a_{\mu\nu}$ and the Wilson line $W_{\mathrm{adj}}(\a_1,\a_2)$. However, such terms are suppressed in the limit $L\to\infty$, provided we use a gauge that behaves sufficiently nicely at null infinity.

For example, if we choose a fixed gauge where the $\varepsilon^\mu_\l(p)$ are nonsingular for all null $p$, then using the distributional identity
\be
\label{eq:distributionalidentity}
e^{-ip\.L z} \de(p^2) &\sim  \frac 1 2 \p{\frac{2\pi}{L\, e^{i\pi/2}}}^{\frac{d-2}{2}} \int_0^\oo d\b\, \b^{\frac{d-4}{2}} \de(p-\b z) \qquad (L\to \oo)
\ee
inside the mode expansion (\ref{eq:Amu_mode_expansion}), we find that $A^\mu(x+L z)$ scales like $L^{\frac{2-d}{2}}$ as $L\to \oo$. Consequently, the Wilson line $W^{(\bar z)ab}_{\mathrm{adj}}(\a_1,\a_2)$ takes the form $1+O(L^{\frac{2-d}{2}})$, which tends to $1$ as $L\to \oo$. Similarly, the $gf^{abc}A^b_{\mu}A^c_{\nu}$ term in $F^a_{\mu\nu}$ does not contribute to (\ref{eq:DGLAP_detector_g_def}) in the limit $L\to \oo$. This leads to the equivalence of \eqref{eq:DGLAP_detector_g_def} and  \eqref{eq:DGLAP_detector_g_def1}.

Covariant gauges cannot be interpreted in terms of a fixed choice of $\varepsilon_\lambda^\mu(p)$, but we can reach a similar conclusion in Feynman gauge as follows. Recall that the expectation value of a detector $\cD$ is computed in the Schwinger-Keldysh formalism, with two time sheets (one for the bra and one for the ket) connected by a time fold where $\cD$ lives. (See e.g.\ \cite{Caron-Huot:2022eqs} for a summary of the Feynman rules.) Propagators connecting $A_\mu(p)$ to $\cD$ are Wightman propagators. In Feynman gauge, the Wightman propagator takes the form
\be
\<0|A_\nu(x+L z) A_\mu(p)|0\> &=  2\pi (-g_{\mu\nu}) \de^+(p^2) e^{-ip\.(x+L z)},
\ee
(where $\de^+(p^2) = \de(p^2) \th(p^0)$),
which again scales as $L^{\frac{2-d}{2}}$ by the distributional identity (\ref{eq:distributionalidentity}). Consequently, the Feynman rules for the DGLAP detector in Feynman gauge are still very simple, see (\ref{eq:dglapfeynmanrule}) and (\ref{eq:polarizationtensorfeynmangauge}) above.

By contrast, in $R_\xi$-gauge for general $\xi$, the Wightman propagator has extra distributional terms supported near the null cone. These terms cause the Wightman propagator $\<0|A_\nu(x+L z) A_\mu(p)|0\>$ to scale as $L^{\frac{4-d}{2}}$ as $L\to \oo$, instead of $L^{\frac{2-d}{2}}$. If $d>4$, then we can still ignore such terms in the limit $L\to \oo$. Formally, in dimensional regularization, we can imagine analytically continuing from $d>4$, so these terms can be ignored in dim-reg for any $d\neq 4$ as well. However, with a different regulator, such terms might need to be taken into account. In this sense, general $R_\xi$-gauges do not behave as nicely at null infinity as Feynman gauge.

For future reference, we define a ``sufficiently nice" gauge as one where the position-space Wightman propagator falls off in the same way at large distance as the mode expansion (\ref{eq:Amu_mode_expansion}) does when $\varepsilon_\l^\mu(p)$ is a smooth function of $p$ on the null cone. In particular, this implies that the propagator falls off as $L^{\frac{2-d}{2}}$ at null infinity, and as $L^{2-d}$ at spatial or timelike infinity. Feynman gauge is ``sufficiently nice" according to our definition, but $R_{\xi\neq 1}$ gauge is not.

\subsection{Tree level matrix elements of DGLAP detectors}

The matrix elements of DGLAP detectors can be computed as generalized weighted cross sections
\be
\<\Psi|\cD^{\DGLAP}_{J_L,g}(z)|\Phi\> &= \sum_{n} \sum_{X_n}\frac{1}{\symF_{X_n}} \int d\mathrm{LIPS}_n\, \cM_{\Phi\to X_n} \cM_{\Psi\to X_n}^* \sum_{\substack{i\in X_n,\\ i = g}} E_i^{2-d-J_L} \delta^{(d-2)}(\hat{p}_i-\hat{z}) ,\\
\<\Psi|\cD^{\DGLAP}_{J_L,q}(z)|\Phi\> &= \sum_{n} \sum_{X_n} \frac{1}{\symF_{X_n}} \int d\mathrm{LIPS}_n\, \cM_{\Phi\to X_n} \cM_{\Psi\to X_n}^* \sum_{\substack{i\in X_n,\\ i = q, \bar{q}}} E_i^{2-d-J_L} \delta^{(d-2)}(\hat{p}_i-\hat{z}) ,
\ee
where $\cM_{\Phi\to X_n}$ is the amplitude of producing an $n$-particle state $X_n$ from the initial state $\Phi$ with total momentum $p$ and $\symF_{X_n}$ is the corresponding symmetry factor accounting for permutations of identical particles. $p_i$ is the momentum of each particle in the final state $X_n$, $\hat{p}_i, \hat{z}$ are unit vectors of the corresponding null vectors, and $d\mathrm{LIPS}_n$ is the Lorentz invariant phase space measure for $n$-particle states
\be
d\mathrm{LIPS}_n =  \left[\prod_{i=1}^n {d^{d-1}\vec{p}_i \over (2\pi)^{d-1} 2E_i}\right] (2\pi)^d \de^{(d)}(p-\sum_{i=1}^n p_i)
=  \left[\prod_{i=1}^n {d^{d}p_i \, \de^+(p_i^2)\over (2\pi)^{d-1}}\right] (2\pi)^d \de^{(d)}(p-\sum_{i=1}^n p_i)\,,
\ee
in which we define the final-state on-shell constraint $\de^+(p_i^2) = \de(p_i^2) \theta(p_i^0)$.

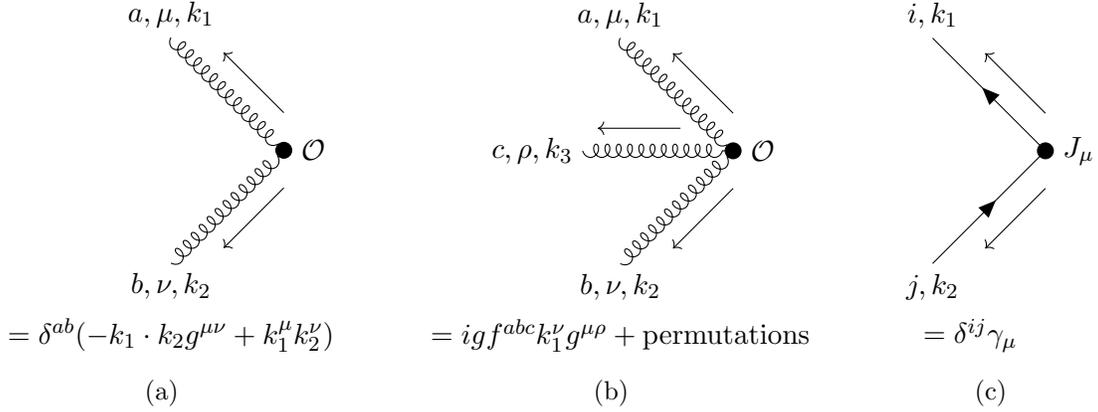
\begin{figure}[htbp]
    \centering
    \subfloat[]{\label{fig:diagrama}
        \begin{tikzpicture}
            \begin{feynman}
                \coordinate (v) at (1.5,0);
                
                \coordinate (a) at (0,1.5);
                \coordinate (b) at (0,-1.5);
                \node[above] at (a) {$a,\mu,k_1$};
                \node[below] at (b) {$b,\nu,k_2$};
                
                \draw[gluon] (a) -- (v);
                \draw[gluon] (b) -- (v);
                \draw[->] (1.5,0.5) -- (0.7, 1.3);
                \draw[->] (1.5, -0.5) -- (0.7, -1.3);
                
                \filldraw[black] (v) circle (3pt);
                \node[right] at (1.6,0) {$\cO$};
                \node[below] at (0,-2.1) {$=\de^{ab}(-k_1\cdot k_2 g^{\mu\nu} + k_1^\mu k_2^\nu)$};
            \end{feynman}
        \end{tikzpicture}
    }
    \qquad
    \subfloat[]{\label{fig:diagramb}
        \begin{tikzpicture}
            \begin{feynman}
                \coordinate (v) at (1.5,0);
                
                \coordinate (a) at (0,1.5);
                \coordinate (b) at (0,-1.5);
                \coordinate (c) at (-0.5,0);
                \node[above] at (a) {$a,\mu,k_1$};
                \node[below] at (b) {$b,\nu,k_2$};
                \node[left] at (c) {$c,\rho,k_3$};
                
                \draw[gluon] (a) -- (v);
                \draw[gluon] (b) -- (v);
                \draw[gluon] (c) -- (v);
                \draw[->] (1.5,0.5) -- (0.7, 1.3);
                \draw[->] (1.5, -0.5) -- (0.7, -1.3);
                \draw[->] (0.8, 0.3) -- (-0.3, 0.3);

                \filldraw[black] (v) circle (3pt);
                \node[right] at (1.6,0) {$\cO$};
                \node[below] at (0,-2.1) {$=\signplus i g f^{abc} k_1^\nu g^{\mu\rho} +\text{permutations}$};
            \end{feynman}
        \end{tikzpicture}
    }
    \qquad
    \subfloat[]{\label{fig:diagramc}
        \begin{tikzpicture}
            \begin{feynman}
                \coordinate (v) at (1.5,0);
                
                \coordinate (a) at (0,1.5);
                \coordinate (b) at (0,-1.5);
                \node[above] at (a) {$i,k_1$};
                \node[below] at (b) {$j,k_2$};
                
                \draw[fermion] (v) -- (a);
                \draw[fermion] (b) -- (v);
                \draw[->] (1.5,0.5) -- (0.7, 1.3);
                \draw[->] (1.5, -0.5) -- (0.7, -1.3);
                
                \filldraw[black] (v) circle (3pt);
                \node[right] at (1.6,0) {$J_\mu$};
                \node[below] at (0.5,-2.1) {$=\de^{ij} \gamma_\mu$};
            \end{feynman}
        \end{tikzpicture}
    }
    \caption{Feynman diagrams and corresponding Feynman rules for the composite operators $\cO$ and $J_\mu$, with the convention of all-outgoing momenta. Diagrams (\ref{fig:diagrama}) and (\ref{fig:diagramb}) represent the $g^0$ and $g^1$ contributions to $\cO$, respectively, while the $g^2$ contribution is not shown. In diagram (\ref{fig:diagramb}), permutations refer to cyclic permutations of the triplets $\{a,\mu,k_1\}$, $\{b,\nu,k_2\}$, and $\{c,\rho,k_3\}$. Diagram (\ref{fig:diagramc}) shows the Feynman rule for the current operator $J_\mu$.}
    \label{fig:source_operators}
\end{figure}

Let us compute some explicit examples of the tree-level matrix elements for the DGLAP detectors. The simplest gauge-invariant source operators that create particle excitations are $\cO = {1 \over 4 N_c} \textrm{Tr}(F_{\mu\nu}F^{\mu\nu})$ and $J_\mu = \sum_{i}\bar{\psi}_i\gamma_\mu\psi_i$. The Feynman rules for these composite operators that we will need are shown in figure~\ref{fig:source_operators}. Their tree-level two-particle form factors are given by 
\be
& \cF_{g,\text{tree}}^{a_1 a_2}(k_1,\l_1;k_2,\l_2) = \<g_1 g_2|\cO|0\> = \de^{a_1 a_2}[(k_1\cdot \varepsilon_{\l_2}^*)(k_2\cdot \varepsilon_{\l_1}^*) - (k_1\cdot k_2)(\varepsilon_{\l_1}^* \cdot \varepsilon_{\l_2}^*)]\,,\label{eq:cF_g_tree}\\
& \cF_{q,\text{tree}}^{i_1 i_2, \mu}(k_1,s_1;k_2,s_2) =\<q_1\bar{q}_2|J^\mu|0\> = \de^{i_1 i_2} \bar{u}_{s_1}(k_1)\gamma^\mu v_{s_2}(k_2),\label{eq:cF_q_tree}
\ee
where $\varepsilon_{\l_n}$ is the polarization vector of the gluon with momentum $k_n$, and $u_{s_n}(k_n)$ and $v_{s_n}(k_n)$ are spinors for the quark and anti-quark with momentum $k_n$ and spin $s_n$.

\begin{figure}[htbp]
    \centering
    \subfloat[]{
        \begin{tikzpicture}
            \begin{feynman}
                \coordinate (O1) at (1.5,0);
                \coordinate (O2) at (-1.5,0);
                \coordinate (V1) at (0,1.5);
                \coordinate (V2) at (0,-1.5);
                
                \draw[dashed] (0,-1.8) -- (0,1.8);
                
                \draw[gluon] (O1) -- (V1);
                \draw[gluon] (O1) -- (V2);
                \draw[gluon] (O2) -- (V1);
                \draw[gluon] (O2) -- (V2);
                
                \filldraw[red] (V1) circle (3pt);
                \filldraw[black] (O1) circle (3pt);
                \filldraw[black] (O2) circle (3pt);
                \node[above] at (V1) {$\cD_{J_L,g}^{\DGLAP}(z)$};
                \node[right] at (O1) {$\cO$};
                \node[left] at (O2) {$\cO$};
            \end{feynman}
        \end{tikzpicture}
    }
    \qquad
    \subfloat[]{
        \begin{tikzpicture}
            \begin{feynman}
                \coordinate (O1) at (1.5,0);
                \coordinate (O2) at (-1.5,0);
                \coordinate (V1) at (0,1.5);
                \coordinate (V2) at (0,-1.5);
                
                \draw[dashed] (0,-1.8) -- (0,1.8);
                
                \draw[fermion] (O1) -- (V1);
                \draw[fermion] (V2) -- (O1);
                \draw[fermion] (O2) -- (V2);
                \draw[fermion] (V1) -- (O2);
                
                \filldraw[red] (V1) circle (3pt);
                \filldraw[black] (O1) circle (3pt);
                \filldraw[black] (O2) circle (3pt);
                \node[above] at (V1) {$\cD_{J_L,q}^{\DGLAP}(z)$};
                \node[right] at (O1) {$J_\mu$};
                \node[left] at (O2) {$J^\mu$};
                \node[right] at (2.1,0) {$+$};
                \coordinate (J1) at (6.3,0);
                \coordinate (J2) at (3.3,0);
                \coordinate (W2) at (4.8,1.5);
                \coordinate (W1) at (4.8,-1.5);
                
                \draw[dashed] (4.8,-1.8) -- (4.8,1.8);
                
                \draw[fermion] (J1) -- (W1);
                \draw[fermion] (W2) -- (J1);
                \draw[fermion] (J2) -- (W2);
                \draw[fermion] (W1) -- (J2);
                
                \filldraw[red] (W2) circle (3pt);
                \filldraw[black] (J1) circle (3pt);
                \filldraw[black] (J2) circle (3pt);
                \node[above] at (W2) {$\cD_{J_L,q}^{\DGLAP}(z)$};
                \node[right] at (J1) {$J_\mu$};
                \node[left] at (J2) {$J^\mu$};
            \end{feynman}
        \end{tikzpicture}
    }
    \caption{Feynman diagrams for the tree-level matrix elements of the DGLAP detectors.}
    \label{fig:tree_level_diagrams}
\end{figure}
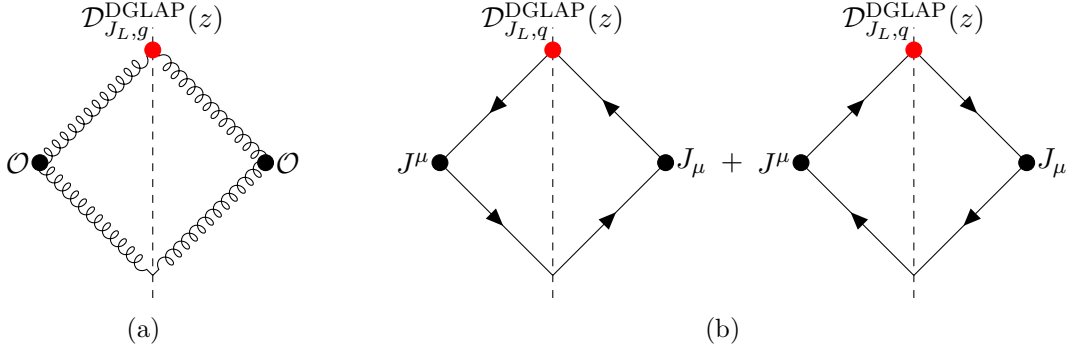

To study matrix elements of detectors $\cD$ in states created by local operators, we adopt a shorthand notation where we strip off the overall momentum-conserving $\de$-function:
\be
\label{eq:stripoffdeltaconvention}
\<0|\cO(-p')\cD\cO(p)|0\> &= (2\pi)^d \de(p-p') \<0|\cO(-p)\cD\cO(p)|0\>.
\ee
With gauge-invariant source operators $\cO$ and $J_\mu$, we can construct the following non-vanishing tree-level matrix elements for DGLAP detectors inside scalar sources from \eqref{eq:cF_g_tree} and \eqref{eq:cF_q_tree}: 
\be
\<0|\cO(-p) \cD_{J_L,g}^{\DGLAP}(z) \cO(p)|0\>^{\text{tree}} &=  \int_0^\infty \!  {E^{-J_L} dE\over (2\pi)^{d-1}2E}
\left[2\pi \de^+(k_2^2) \,  \cI_{g,\text{tree}}(k_1,k_2)\right]\Big|_{\substack{\hspace{-1em} k_1\to E z\\ k_2 \to p-E z }}\,, \\
\<0|J^\mu(-p) \cD_{J_L,q}^{\DGLAP}(z) J_\mu(p)|0\>^{\text{tree}} &= \int_0^\infty \! {E^{-J_L} dE\over (2\pi)^{d-1}2E}
\left[2\pi \de^+(k_2^2) \,  \cI_{q,\text{tree}}(k_1,k_2)+(k_1\leftrightarrow k_2)\right]\Big|_{\substack{\hspace{-1em} k_1\to E z\\ k_2 \to p-E z }}\,,
\label{eq:dglaptreelevel}
\ee  
where the integrands $\cI_{g,\text{tree}}(k_1,k_2)$ and $\cI_{q,\text{tree}}(k_1,k_2)$ are the squares of the tree-level form factors, summed over all possible color indices and polarization states
\be
\cI_{g,\text{tree}}(k_1,k_2) &= \sum_{a_1,a_2, \l_1,\l_2} \left| \cF_{g,\text{tree}}^{a_1 a_2}(k_1,\l_1;k_2,\l_2) \right|^2= (d-2)(N_c^2-1)(k_1\cdot k_2)^2\,,\label{eq:form_factor_sq_O_to_gg}\\
\cI_{q,\text{tree}}(k_1,k_2) &= g_{\mu\nu}\sum_{i_1,i_2, s_1,s_2} \left(\cF_{q,\text{tree}}^{i_1 i_2,\mu}(k_1,s_1;k_2,s_2)\right)^* \cF_{q,\text{tree}}^{i_1 i_2,\nu}(k_1,s_1;k_2,s_2)=4(d-2)N_c (k_1\cdot k_2)\,.
\ee
The corresponding Feynman diagrams are shown in figure~\ref{fig:tree_level_diagrams}. For the gluon case, the measurement of $\cD_{J_L,g}^{\DGLAP}(z)$ on two legs cancels the symmetry factor from the phase space of identical particles; while for the quark case, detecting a quark or anti-quark are two different processes, though these two diagrams give the same result due to charge conjugation symmetry.

Note that we have chosen to contract the $\mu$ indices of the current $J_\mu$ in (\ref{eq:dglaptreelevel}), which means we are forming a rotationally-invariant density matrix instead of a pure state. One could alternatively leave these indices un-contracted, for example to study detectors with nonzero transverse spin \cite{Chang:2020qpj,Chen:2022jhb,Chang:2022ryc}. 

The results for the two matrix elements above are
\be
\<\cD_{J_L,g}^{\DGLAP}(z)\>_{\cO(p)}^{\text{tree}}  &=  {d-2 \over 2^{d+1}\pi^{d-2}} (N_c^2-1) (\signplus 2 z\cdot p)^{J_L}(\signplus p^2)^{1-J_L}\,,\label{eq:DDGLAP_tree_matrixelement_gluon}\\
\<\cD_{J_L,q}^{\DGLAP}(z)\>_{J(p)}^{\text{tree}}  &=  {d-2 \over 2^{d-3}\pi^{d-2}} N_c (\signplus 2 z\cdot p)^{J_L}(\signplus p^2)^{-J_L}\,,\label{eq:DDGLAP_tree_matrixelement_quark}
\ee
where we have additionally introduced the shorthand notation 
\be\label{eq:short-hand_notation}
\<\cD(z)\>_{\cO(p)}  \equiv \<0|\cO(-p)\cD(z) \cO(p)|0\>\,,\quad
\<\cD(z)\>_{J(p)}  \equiv \signminus \<0|J^\mu(-p) \cD(z) J_\mu(p)|0\>\,.
\ee

\subsection{One-loop matrix elements of DGLAP detectors}\label{sec:DGLAP_1loop}

Now let us calculate some one-loop matrix elements of DGLAP detectors and renormalize their IR divergences. From this data, we will be able to read off the familiar leading-order (LO) DGLAP anomalous dimensions~\cite{Gross:1973zrg,Gross:1973ju}. We will also need these results later when we study mixing of DGLAP operators with other Regge trajectories.

Matrix elements of DGLAP operators are the moments of hard functions that are widely used in QCD factorizations, such as semi-inclusive electron-positron annihilation~\cite{Rijken:1996ns,Mitov:2006wy,He:2025hin} and collinear factorization of energy correlators~\cite{Dixon:2019uzg,Chen:2020vvp}. Similar discussions about the RG structure of QCD DGLAP detectors and the connection to hard functions can be found in~\cite{Chen:2023zzh}.

At one loop, due to quark-gluon mixing, we must compute matrix elements of $\cD^{\DGLAP}_{J_L,g}$ and $\cD^{\DGLAP}_{J_L,q}$ in states created by both $J_\mu(p)$ and $\cO(p)$ to extract anomalous dimensions. For generic $J_L$, the $\epsilon$-poles have the form
\be
\< \cD^{\DGLAP}_{J_L,q}(z)\>_{[J]_R(p)}^{\text{1-loop}} &= {\alpha_s \over 4\pi} \frac{\hat{\gamma}^{(0)}_{qq}(J_L)}{\epsilon} \< \cD^{\DGLAP}_{J_L,q}(z)\>_{J(p)}^{\text{tree}} + \cO(\epsilon^0)\,,\\
\< \cD^{\DGLAP}_{J_L,q}(z)\>_{[\cO]_R(p)}^{\text{1-loop}} &= {\alpha_s \over 4\pi} \frac{\hat{\gamma}^{(0)}_{qg}(J_L)}{\epsilon} \< \cD^{\DGLAP}_{J_L,g}(z)\>_{\cO(p)}^{\text{tree}} + \cO(\epsilon^0)\,,\\
\< \cD^{\DGLAP}_{J_L,g}(z)\>_{[J]_R(p)}^{\text{1-loop}} &= {\alpha_s \over 4\pi} \frac{\hat{\gamma}^{(0)}_{gq}(J_L)}{\epsilon} \< \cD^{\DGLAP}_{J_L,q}(z)\>_{J(p)}^{\text{tree}} + \cO(\epsilon^0)\,,\\
\< \cD^{\DGLAP}_{J_L,g}(z)\>_{[\cO]_R(p)}^{\text{1-loop}} &={\alpha_s \over 4\pi} \frac{\hat{\gamma}^{(0)}_{gg}(J_L)}{\epsilon} \< \cD^{\DGLAP}_{J_L,g}(z)\>_{\cO(p)}^{\text{tree}} + \cO(\epsilon^0)\,. \label{eq:1-loop-div-gg}
\ee
where $[\cO]_R$, $[J]_R$ denote the renormalized operators for $\cO, J$. 

In the next few subsections, we present a detailed computation of $\< \cD^{\DGLAP}_{J_L,g}(z)\>_{[\cO]_R(p)}^{\text{1-loop}}$ and explain how to extract $\hat{\gamma}^{(0)}_{gg}(J_L)$. The other cases are left to appendix~\ref{app:QCD_1loop}. The one-loop computation consists of contributions from real emission, virtual corrections, wavefunction renormalization counterterms, and local composite operator renormalization counterterms. 

\subsubsection{One-loop computation details}\label{sec:1loop_details}
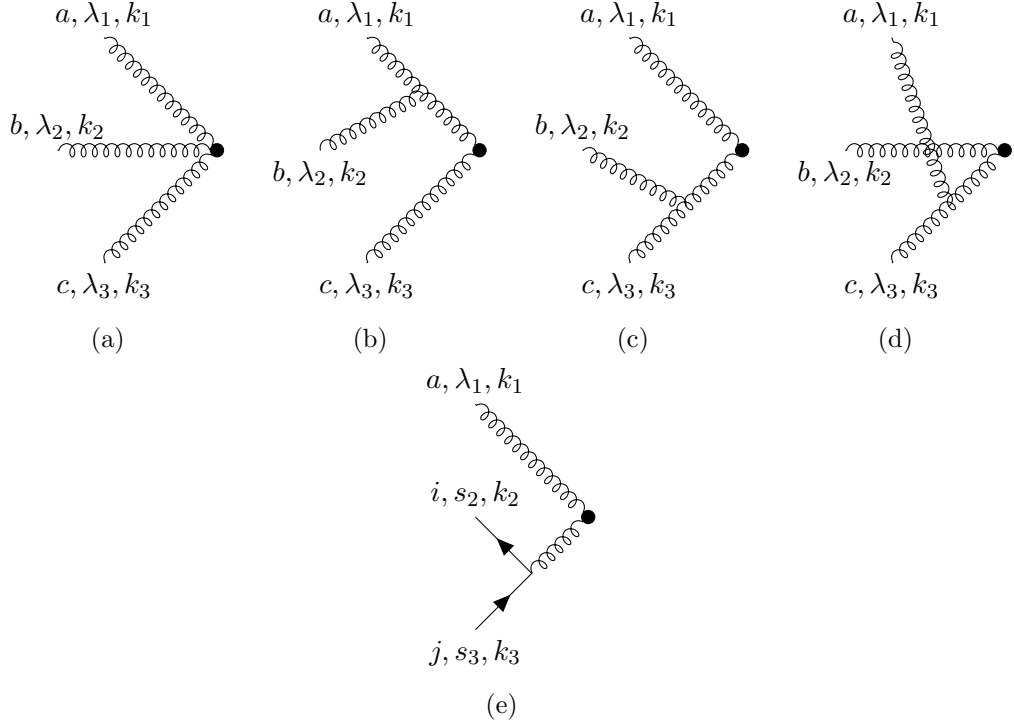
\begin{figure}[ht]
    \centering
    
    \subfloat[]{%
        \begin{tikzpicture}
            \node[draw, black, fill=black, circle, inner sep=0pt, minimum size=5pt] (a) at (0, 0){};

            \begin{feynman}
                \vertex[label={[label distance=0.0cm]90:{\(a,\l_1,k_1\)}}] (b) at (-1.5, 1.5);
                \vertex[label={[label distance=0.0cm]-90:{\(c,\l_3,k_3\)}}] (c) at (-1.5, -1.5);
                \vertex[label={[label distance=0.0cm]90:{\(b,\l_2,k_2\)}}] (d) at (-{sqrt(2)*1.5}, 0);

                \diagram* {
                    (a) -- [gluon] (b),
                    (a) -- [gluon] (c),
                    (a) -- [gluon] (d),
                };
            \end{feynman}
        \end{tikzpicture}
    }\quad
    \subfloat[]{%
        \begin{tikzpicture}
            \node[draw, black, fill=black, circle, inner sep=0pt, minimum size=5pt] (a) at (0, 0){};

            \begin{feynman}
                \vertex[label={[label distance=0.0cm]90:{\(a,\l_1,k_1\)}}] (b) at (-1.5, 1.5);
                \vertex[label={[label distance=0.0cm]-90:{\(c,\l_3,k_3\)}}] (c) at (-1.5, -1.5);
                \vertex[label={[label distance=0.0cm]-90:{\(b,\l_2,k_2\)}}] (d) at (-{sqrt(2)*1.5}, 0);
                \vertex (e) at (-0.75, 0.75);

                \diagram* {
                    (a) -- [gluon] (b),
                    (a) -- [gluon] (c),
                    (e) -- [gluon] (d),
                };
            \end{feynman}
        \end{tikzpicture}
    }\quad
    \subfloat[]{%
        \begin{tikzpicture}
            \node[draw, black, fill=black, circle, inner sep=0pt, minimum size=5pt] (a) at (0, 0){};

            \begin{feynman}
                \vertex[label={[label distance=0.0cm]90:{\(a,\l_1,k_1\)}}] (b) at (-1.5, 1.5);
                \vertex[label={[label distance=0.0cm]-90:{\(c,\l_3,k_3\)}}] (c) at (-1.5, -1.5);
                \vertex[label={[label distance=0.0cm]90:{\(b,\l_2,k_2\)}}] (d) at (-{sqrt(2)*1.5}, 0);
                \vertex (e) at (-0.75, -0.75);

                \diagram* {
                    (a) -- [gluon] (b),
                    (a) -- [gluon] (c),
                    (e) -- [gluon] (d),
                };
            \end{feynman}
        \end{tikzpicture}
    }\quad
    \subfloat[]{%
        \begin{tikzpicture}
            \node[draw, black, fill=black, circle, inner sep=0pt, minimum size=5pt] (a) at (0, 0){};

            \begin{feynman}
                \vertex[label={[label distance=0.0cm]90:{\(a,\l_1,k_1\)}}] (b) at (-1.5, 1.5);
                \vertex[label={[label distance=0.0cm]-90:{\(c,\l_3,k_3\)}}] (c) at (-1.5, -1.5);
                \vertex[label={[label distance=0.0cm]-90:{\(b,\l_2,k_2\)}}] (d) at (-{sqrt(2)*1.5}, 0);
                \vertex (e) at (-0.75, -0.75);

                \diagram* {
                    (a) -- [gluon] (d),
                    (a) -- [gluon] (c),
                    (e) -- [gluon] (b),
                };
            \end{feynman}
        \end{tikzpicture}
    }\quad
    \subfloat[]{%
        \begin{tikzpicture}
            \node[draw, black, fill=black, circle, inner sep=0pt, minimum size=5pt] (a) at (0, 0){};

            \begin{feynman}
                \vertex[label={[label distance=0.0cm]90:{\(a,\l_1,k_1\)}}] (b) at (-1.5, 1.5);
                \vertex[label={[label distance=0.0cm]-90:{\(j,s_3,k_3\)}}] (c) at (-1.5, -1.5);
                \vertex[label={[label distance=0.0cm]90:{\(i,s_2,k_2\)}}] (d) at (-1.5, 0);
                \vertex (e) at (-0.75, -0.75);

                \diagram* {
                    (a) -- [gluon] (b),
                    (a) -- [gluon] (e),
                    (c) -- [fermion] (e) -- [fermion] (d),
                };
            \end{feynman}
        \end{tikzpicture}
    }
    
    \caption{Real emission diagrams at one loop for $\cO$. (a)-(d) are Feynman diagrams that produce 3 gluon final states, while (e) corresponds to the production of one gluon and a pair of quark and anti-quark.}
    \label{fig:1loop_real_emission_O}
\end{figure}

\noindent\textbf{Real Emission}\\
The integrand for the one-loop real emission contribution can be obtained from the tree-level 3-particle form factors $\<g_1 g_2 g_3|\cO|0\>_{\text{tree}}$ and $\<g_1 q_2 \bar{q}_3|\cO|0\>_{\text{tree}}$:
 \be
 \cI^{\cO, \text{tree}}_{ggg}(k_1,k_2,k_3) &= \sum_{\text{color}} \sum_{\text{polarization}}\left|\<g_1 g_2 g_3|\cO|0\>_{\text{tree}}\right|^2\,,\\
 \cI^{\cO,\text{tree}}_{gq\bar{q}}(k_1,k_2,k_3) &= \sum_{\text{color}} \sum_{\text{polarization}}\left|\<g_1 q_2 \bar{q}_3|\cO|0\>_{\text{tree}}\right|^2\,.
\ee
The Feynman diagrams for the tree-level 3-particle form factors are shown in figure~\ref{fig:1loop_real_emission_O}. The expressions for $\cI^{\cO,\text{tree}}_{ggg}$ and $\cI^{\cO,\text{tree}}_{gq\bar{q}}$ are 
\be
&\cI^{\cO,\text{tree}}_{ggg}(k_1, k_2, k_3) \nn\\
& = (d-2) g^2 \tilde{\mu}^{2\epsilon} C_A(N_c^2-1)
\Big[{k_1\. k_2\,k_1\.(k_2+k_3) \, k_1\.(k_2+3k_3) \over (k_2\. k_3)(k_3\. k_1)}
+{8d-20\over d-2}{(k_1\. k_2)^2\over k_2\. k_3} +\text{perm.} \Big],\label{eq:form_factor_sq_O_to_ggg}\\
&\cI^{\cO,\text{tree}}_{gq\bar{q}}(k_1, k_2, k_3) = {(d-2)g^2 \tilde{\mu}^{2\epsilon}n_f T_F C_F \over k_2\. k_3}
\Big[(k_1\. (k_2+k_3))^2 -{4 \over d-2} (k_1\.k_2) (k_1\. k_3) \Big]\,,
\ee
where $n_f$ is the number of quarks, $C_F, C_A$ are the Casimir invariants of the fundamental and adjoint representations of $SU(N_c)$, $T_F$ is the trace normalization factor for the fundamental representation $\mathrm{tr}(t^a t^b)=T_F \delta^{ab}$, and $\text{``perm.''}$ in $\cI^{\cO,\text{tree}}_{ggg}$ stands for permutations of $k_1, k_2, k_3$. From these, we can calculate the real emission contribution $\< \cD^{\DGLAP}_{J_L,g}(z)\>_{[\cO]_R(p)}^{\text{1-loop,R}}$ from phase space integration 
\be
\< \cD^{\DGLAP}_{J_L,g}(z)\>_{[\cO]_R(p)}^{\text{1-loop,R}} &= \int_0^\oo { E^{-J_L} dE \over (2\pi)^{d-1}2E} \int \left[\prod_{n=2}^3 {d^d k_n\, \de^+(k_n^2)\over (2\pi)^{d-1}}\right] (2\pi)^d \de^{(d)}(p-E z-k_2-k_3)\nn\\
&\hspace{2cm}\times \left[\frac{1}{2} \cI^{\cO,\text{tree}}_{ggg}(k_1, k_2, k_3)+\cI^{\cO,\text{tree}}_{gq\bar{q}}(k_1, k_2, k_3)\right]\Big|_{k_1\to E z}\,, \\
&= g^2 \tilde{\mu}^{2\epsilon} (N_c^2 -1) (\signplus 2 z\. p)^{J_L} (\signplus p^2)^{-J_L+{d\over 2}-1}\times C^{\text{1-loop,R}}_{g,\cO}(J_L, d)\,.
\ee
where $\frac{1}{2}$ in front of $\cI^{\cO,\text{tree}}_{ggg}$ accounts for the symmetry factor arising from the presence of two identical gluons in the final state that are not measured by the detector. The coefficient $C^{\text{1-loop,R}}_{g,\cO}(J_L, d)$ is 
\be
&C^{\text{1-loop,R}}_{g,\cO}(J_L, d)= {\Gamma(\frac{d}{2}-2) \over 2^{3d} \pi^{\frac{3d-5}{2}} \Gamma({d+1\over 2})\Gamma(\frac{d}{2}-J_L)}\Bigg[
4 n_f T_F (d-2)^2\Gamma(2-J_L)  \nn\\  
&\hspace{1.3 cm} + C_A {\Gamma(-2-J_L)\over d-4}\Big(
    3 (d-2)^2 (3 d-4) J_L^4
    -2 \left(10 d^4-115 d^3+448 d^2-708 d+368\right) J_L^3     \nn\\
&\hspace{4cm}
    +\left(14 d^5-234 d^4+1459 d^3-4256 d^2+5724 d-2704\right) J_L^2 \nn\\
&\hspace{4cm}
    -2 \left(2 d^6-45 d^5+391 d^4-1691 d^3+3856 d^2-4356 d+1840\right) J_L \nn\\
&\hspace{4cm}
        +(d-4) (d-2) (d-1) \left(d^4-11d^3+64 d^2-188 d+208\right)
    \Big)
\Bigg]\,,
\ee
which has poles at $\epsilon = 0$ 
\be
&C^{\text{1-loop,R}}_{g,\cO}(J_L, d)= {C_A \over 64\pi^4 \epsilon^2} + {1 \over 64\pi^4 \epsilon}\Big[-{n_f T_F \over 3 } + C_A\Big( \psi(-J_L-1) + {1\over J_L+2} - {2\over J_L+1}+{1\over J_L} \nn\\
 &\hspace{6cm}
 -{1 \over J_L-1}-{1\over 12} + 4\log(2)+3\log(\pi)\Big)\Big]+ O(\epsilon^0)\,, 
 \ee
and $J_L = -2 + \mathbb{N}$
 \be
& C^{\text{1-loop,R}}_{g,\cO}(J_L, d) ={C_A \over J_L+2} { (-1+\epsilon)\Gamma(-\epsilon)\over 4^{3-2\epsilon} \pi^{4-3\epsilon}\Gamma(1-2\epsilon)} + O((J_L+2)^0) \label{eq:DGLAP_CR_JLminus2pole}\,,\\
& C^{\text{1-loop,R}}_{g,\cO}(J_L, d) ={C_A \over J_L+1} { (-1+\epsilon)^2\Gamma(-\epsilon)\over 4^{3-2\epsilon} \pi^{4-3\epsilon} \Gamma(1-2\epsilon)} + O((J_L+1)^0)\,,\,\dots\,. \label{eq:DGLAP_CR_JLminus1pole}
\ee

\begin{figure}[ht]
    \centering
    
    \subfloat[]{%
        \begin{tikzpicture}
            \node[draw, black, fill=black, circle, inner sep=0pt, minimum size=5pt] (a) at (0, 0){};

            \begin{feynman}
                \vertex[label={[label distance=0.0cm]90:{\(a,\l_1,k_1\)}}] (b) at (-1.5, 1.5);
                \vertex[label={[label distance=0.0cm]-90:{\(b,\l_2,k_2\)}}] (c) at (-1.5, -1.5);
                \vertex (i1) at (-0.75, 0.75);
                \vertex (i2) at (-0.75, -0.75);

                \diagram* {
                    (a) -- [gluon] (b),
                    (a) -- [gluon] (c),
                    (i1) -- [gluon, half right, looseness=1.1] (i2),
                };
            \end{feynman}
        \end{tikzpicture}
    }\qquad
    \subfloat[]{%
        \begin{tikzpicture}
            \node[draw, black, fill=black, circle, inner sep=0pt, minimum size=5pt] (a) at (0, 0){};
            \node[left] at (-1.8,0) {$\frac{1}{2}$};
            \begin{feynman}
                \vertex[label={[label distance=0.0cm]90:{\(a,\l_1,k_1\)}}] (b) at (-1.5, 1.5);
                \vertex[label={[label distance=0.0cm]-90:{\(b,\l_2,k_2\)}}] (c) at (-1.5, -1.5);
                \vertex (e) at (-0.9, 0.9);

                \diagram* {
                    (a) -- [gluon] (b),
                    (a) -- [gluon] (c),
                    (a) -- [gluon, half left, looseness=1.1] (e),
                };
            \end{feynman}
        \end{tikzpicture}
    }\qquad
    \subfloat[]{%
        \begin{tikzpicture}
            \node[draw, black, fill=black, circle, inner sep=0pt, minimum size=5pt] (a) at (0, 0){};
            \node[left] at (-1.8,0) {$\frac{1}{2}$};
            \begin{feynman}
                \vertex[label={[label distance=0.0cm]90:{\(a,\l_1,k_1\)}}] (b) at (-1.5, 1.5);
                \vertex[label={[label distance=0.0cm]-90:{\(b,\l_2,k_2\)}}] (c) at (-1.5, -1.5);
                \vertex (e) at (-0.9, -0.9);

                \diagram* {
                    (a) -- [gluon] (b),
                    (a) -- [gluon] (c),
                    (a) -- [gluon, half right, looseness=1.1] (e),
                };
            \end{feynman}
        \end{tikzpicture}
    }\qquad
    \subfloat[]{%
        \begin{tikzpicture}
            \node[draw, black, fill=black, circle, inner sep=0pt, minimum size=5pt] (a) at (1.5, 0){};
            \node[left] at (-1.8,0) {$\frac{1}{2}$};
            \begin{feynman}
                \vertex[label={[label distance=0.0cm]90:{\(a,\l_1,k_1\)}}] (b) at (-1.5, 1.5);
                \vertex[label={[label distance=0.0cm]-90:{\(b,\l_2,k_2\)}}] (c) at (-1.5, -1.5);
                \vertex (d) at (0, 0);

                \diagram* {
                    (a) -- [gluon, half right, looseness=1.1] (d),
                    (a) -- [gluon, half left, looseness=1.1] (d),
                    (d) -- [gluon] (b),
                    (d) -- [gluon] (c),
                };
            \end{feynman}
        \end{tikzpicture}
    }
    
    \caption{One-loop virtual diagrams for $\< \cD^{\DGLAP}_{J_L,g}(z)\>_{[\cO]_R(p)}^{\text{1-loop}}$. Diagrams (b)-(d) contain symmetry factor ${1\over 2}$ due to the symmetry of internal loop.}
    \label{fig:1loop_virtual_O}
\end{figure}
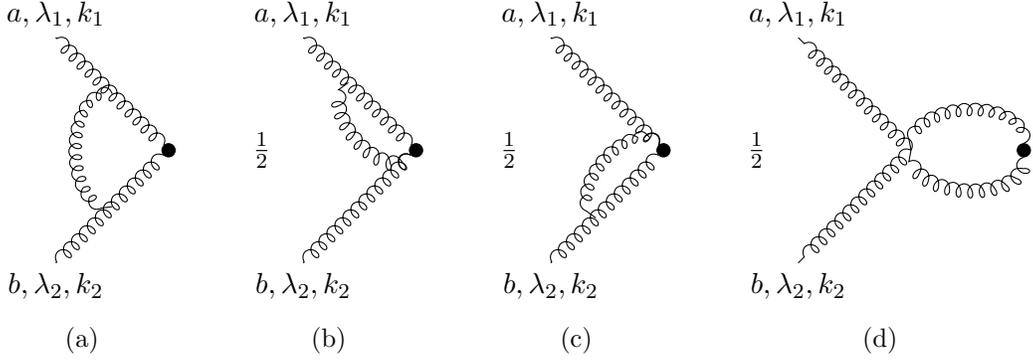

\noindent\textbf{Virtual Corrections}\\
The one-loop virtual corrections can be calculated from the one-loop two-particle form factor $\cF_{gg, \text{1-loop}}^{a_1 a_2}$. The corresponding Feynman diagrams are shown in figure~\ref{fig:1loop_virtual_O}. The one-loop form factor is proportional to the tree-level form factor $\cF_{g,\text{tree}}^{a_1 a_2}(k_1,\l_1; k_2, \l_2)$:
\be
{\cF_{gg, \text{1-loop}}^{a_1 a_2}(k_1,\l_1; k_2, \l_2)  \over \cF_{g,\text{tree}}^{a_1 a_2}(k_1,\l_1; k_2, \l_2)}  =g^2 \tilde{\mu}^{2\epsilon} C_A
{\Gamma(\frac{d}{2}-2)(d^3-16d^2+68d-88)\over 2^{d+1}\pi^{\frac{d}{2}-1}\sin(\frac{d}{2}\pi)\Gamma(d-1)} (\signminus 2 k_1\.k_2-i0)^{\frac{d}{2}-2}\,,
\ee
which is independent of colors and polarizations. This simplifies the computation of the virtual contribution to $\< \cD^{\DGLAP}_{J_L,g}(z)\>_{[\cO]_R(p)}^{\text{1-loop}}$ because momentum conservation implies $2k_1\. k_2 = p^2$ and the phase space integration is same as the tree-level case:
\be
\< \cD^{\DGLAP}_{J_L,g}(z)\>_{[\cO]_R(p)}^{\text{1-loop,V}} 
&= \left[{\cF_{gg, \text{1-loop}}^{a_1 a_2}(k_1,\l_1; k_2, \l_2)  \over \cF_{g,\text{tree}}^{a_1 a_2}(k_1,\l_1; k_2, \l_2)} + \text{c.c.}\right] \< \cD^{\DGLAP}_{J_L,g}(z)\>_{\cO(p)}^{\text{tree}} \nn\\
& = g^2 \tilde{\mu}^{2\epsilon} (N_c^2-1) 
 (\signplus 2 z\cdot p)^{J_L}(\signplus p^2)^{-J_L+\frac{d}{2}-1} \times C^{\text{1-loop,V}}_{g,\cO}(d)\,.
\ee
where the coefficient $C^{\text{1-loop,V}}_{g,\cO}(d)$ is
\be
C^{\text{1-loop,V}}_{g,\cO}(d) &= C_A {\Gamma(\frac{d}{2}-2)(d^3-16d^2+68d-88)\over 2^{2d+1}\pi^{\frac{3d}{2}-3}\tan(\frac{d}{2}\pi)\Gamma(d-2)}\\
& = -{C_A \over 64\pi^4 \epsilon^2} + {C_A \over 64 \pi^4 \epsilon} (1+\gamma_E-4\log(2)-3\log(\pi)) + O(\epsilon^0)\,. \label{eq:CV_g_1loop}
\ee
We observe that the double pole in $\epsilon$ present in $C^{\text{1-loop,V}}_{g,\cO}(d)$ precisely cancels the corresponding double pole in $\epsilon$ appearing in $C^{\text{1-loop,R}}_{g,\cO}(J_L,d)$. This cancellation is an important consistency check on our calculation.

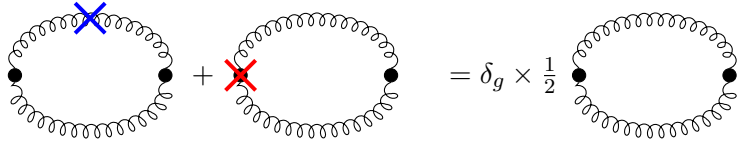
\begin{figure}
\centering
\begin{tikzpicture}
    \node[draw, black, fill=black, circle, inner sep=0pt, minimum size=5pt] (a1) at (1.0, 0){};
    \node[draw, black, fill=black, circle, inner sep=0pt, minimum size=5pt] (b1) at (-1.0, 0){};
    \node[draw, black, fill=black, circle, inner sep=0pt, minimum size=5pt] (a2) at (4.0, 0){};
    \node[draw, black, fill=black, circle, inner sep=0pt, minimum size=5pt] (b2) at (2.0, 0){};
    \node[draw, black, fill=black, circle, inner sep=0pt, minimum size=5pt] (a3) at (8.5, 0){};
    \node[draw, black, fill=black, circle, inner sep=0pt, minimum size=5pt] (b3) at (6.5, 0){};
    \node[] at (1.5,0){$+$};
    \node[] at (5.5,0){$= \de_g \times \frac{1}{2}$};
    \begin{feynman}
        \diagram* {
            (a1) -- [gluon, half right, looseness=1.1] (b1);
            (a1) -- [gluon, half left, looseness=1.1] (b1);
            (a2) -- [gluon, half right, looseness=1.1] (b2);
            (a2) -- [gluon, half left, looseness=1.1] (b2);
            (a3) -- [gluon, half right, looseness=1.1] (b3);
            (a3) -- [gluon, half left, looseness=1.1] (b3);

        };
    \end{feynman}
    \coordinate (c) at (0, 0.76);
    \coordinate (d) at (2, 0);
    \coordinate (v1) at (1,1);
    \coordinate (v2) at (1,-1); 
    \def\tempsize{0.2};
    \draw[blue, line width=1.5pt] ($(c)+\tempsize*(v1)$) -- ($(c)-\tempsize*(v1)$);
    \draw[blue, line width=1.5pt] ($(c)+\tempsize*(v2)$) -- ($(c)-\tempsize*(v2)$);
    \draw[red, line width=1.5pt] ($(d)+\tempsize*(v1)$) -- ($(d)-\tempsize*(v1)$);
    \draw[red, line width=1.5pt] ($(d)+\tempsize*(v2)$) -- ($(d)-\tempsize*(v2)$);
\end{tikzpicture}
\caption{Feynman diagrams for the counterterm contributions in the local correlation function $\<[\cO]_R(-p) [\cO]_R(p) \>$. The blue cross represents the wavefunction renormalization counterterm for gluon, while the red cross represents the composite operator counterterm for $\cO$. On the right side, the factor of $\frac{1}{2}$ is the symmetry factor for the tree-level correlation function $\<\cO(-p) \cO(p) \>^{\text{tree}}$.}
\label{fig:local_counterterm}
\end{figure}

\noindent\textbf{Counterterms}\\
The contribution of wavefunction renormalization and local composite operator counterterms to $\< \cD^{\DGLAP}_{J_L,g}(z)\>_{[\cO]_R(p)}^{\text{1-loop}}$ is also proportional to the tree-level matrix element
\be
\< \cD^{\DGLAP}_{J_L,g}(z)\>_{[\cO]_R(p)}^{\text{1-loop,C}} = \de_g \< \cD^{\DGLAP}_{J_L,g}(z)\>_{\cO(p)}^{\text{tree}}\,,
\ee
where $\de_g$ can be extracted from the counterterm contributions to the local correlation function $\<[\cO]_R(-p) [\cO]_R(p) \>$. The corresponding Feynman diagrams are shown in figure~\ref{fig:local_counterterm}, from which we can infer the relation
\be\label{eq:delta_g}
\de_g = -2\de_A + 2\de_\cO = -{g^2\over 24\pi^2 \epsilon}(11C_A - 4n_f T_F)\,, 
\ee
where we have used the one-loop wavefunction renormalization counterterm coefficient $\de_A$ and one-loop composite operator counterterm coefficient $\de_\cO$ in the $\overline{\mathrm{MS}}$ scheme
\be
\de_A = {g^2\over (4\pi)^2 \epsilon}\left(\frac{5}{3}C_A -\frac{4}{3} n_f T_F\right)\,,\qquad
\de_\cO = {g^2\over (4\pi)^2 \epsilon}\left(-2C_A\right)\,.  
\ee

\subsubsection{Pole structures and DGLAP detector renormalization}

From the calculations above, we observe that the matrix element
\be
\<\cD^{\DGLAP}_{J_L,g}(z)\>_{[\cO]_R(p)}^{\text{1-loop}} = \<\cD^{\DGLAP}_{J_L,g}(z)\>_{[\cO]_R(p)}^{\text{1-loop,R}}+\<\cD^{\DGLAP}_{J_L,g}(z)\>_{[\cO]_R(p)}^{\text{1-loop,V}}+\<\cD^{\DGLAP}_{J_L,g}(z)\>_{[\cO]_R(p)}^{\text{1-loop,C}}\,,
\ee 
exhibits poles in both $J_L$ and $\epsilon$.
$J_L$-poles arise exclusively from the real emission contributions, with the leading pole occurring at $J_L = -2$
\be
\<\cD^{\DGLAP}_{J_L,g}(z) \>_{[\cO]_R(p)}^{\text{1-loop}} =
{g^2 \tilde{\mu}^{2\epsilon} (N_c^2 -1)C_A \over J_L+2} { (-1+\epsilon)\Gamma(-\epsilon)\over 4^{3-2\epsilon} \pi^{4-3\epsilon}\Gamma(1-2\epsilon)}  { (\signplus p^2)^{3-\epsilon} \over (\signplus 2 z\. p)^{2}} + O((J_L+2)^0)\,.
\ee
We discuss this pole in more detail in the next section.

The $\epsilon$-pole of the one-loop matrix element $\<\cD^{\DGLAP}_{J_L,g}(z) \>_{[\cO]_R(p)}^{\text{1-loop}}$ is 
\be
&\<\cD^{\DGLAP}_{J_L,g}(z) \>_{[\cO]_R(p)}^{\text{1-loop}} \nn\\
&={g^2 (N_c^2 -1) \over 256 \pi^4 \epsilon} 
{(\signplus 2 z\. p)^{J_L} \over (\signplus p^2)^{J_L-1}}
\Big[4 C_A\Big( \psi(-J_L)+\gamma_E - {1 \over (J_L+2)(J_L+1)}
 -{1 \over J_L(J_L-1)} \Big)-\beta_0\Big]+ O(\epsilon^0)\,, 
\ee
where $\beta_0 = {11 \over 3}C_A-{4\over 3}n_f T_F$ is the one-loop beta function constant in QCD. Comparing with \eqref{eq:1-loop-div-gg}, we extract the coefficient $\hat{\gamma}^{(0)}_{gg}$
\be \label{eq:DGLAP_gamma_gg}
\hat{\gamma}^{(0)}_{gg}(J_L) = 
\Big[4 C_A\Big( \psi(-J_L)+\gamma_E - {1 \over (J_L+2)(J_L+1)}
 -{1 \over J_L(J_L-1)} \Big)-\beta_0\Big]\,.
\ee

Similarly, we can obtain the other coefficients
\be
&\hat{\gamma}^{(0)}_{qq}(J_L) = C_F\Big[
   4(\psi(-J_L)+\gamma_E )- {2 \over J_L(J_L+1)} -3 \Big]\,,\\
& \hat{\gamma}^{(0)}_{gq}(J_L) = C_F
    {2(2+J_L+J_L^2)\over J_L(J_L+1)(J_L+2)}\,,\\ 
& \hat{\gamma}^{(0)}_{qg}(J_L) = 2n_f T_F {2(2+J_L+J_L^2)\over (J_L-1)J_L(J_L+1)}\,.
\ee
These correspond to the well-known LO anomalous dimensions for twist-2 local operators in QCD. They are obtained through the tree-level reciprocity relation $J_L = -1 - J$, which connects the spin $J_L$ of the bare light-ray operator to the spin $J$ of the corresponding bare local operator in $d=4$ dimensions. 

To subtract IR divergences for generic $J_L$, we define renormalized DGLAP detectors in the $\overline{\mathrm{MS}}$ scheme
\be
\label{eq:dglabrenormalized}
[\vec{\cD}^{\DGLAP}_{J_L}]_R(z;\mu) = \left[\cZ_{J_L}^{\DGLAP}(\alpha_s(\mu))\right]^{-1} \vec{\cD}^{\DGLAP}_{J_L}(z)\,,
\ee
where $\vec{\cD}_{J_L}^{\DGLAP}$ denotes the DGLAP detector multiplet $(\cD^{\DGLAP}_{J_L,q},\cD^{\DGLAP}_{J_L,g})^T$, $\mu$ is the renormalization scale, and $\cZ_{J_L}^{\DGLAP}(\alpha_s(\mu))$ is a renormalization matrix that only contains poles in $\epsilon$
\be
\left[\cZ_{J_L}^{\DGLAP}(\alpha_s(\mu))\right]_{ij} = \delta_{ij} + {1\over \epsilon} {\alpha_s(\mu)\over 4\pi} 
\begin{pmatrix}
    \hat{\gamma}^{(0)}_{qq}(J_L) & \hat{\gamma}^{(0)}_{qg}(J_L) \\
    \hat{\gamma}^{(0)}_{gq}(J_L) & \hat{\gamma}^{(0)}_{gg}(J_L)
\end{pmatrix}_{ij} + O(\alpha_s^2)\,.
\ee
Renormalized DGLAP detectors satisfy the renormalization group equation
\be
\mu\frac{d}{d\mu}[\vec{\cD}_{J_L}^{\DGLAP}]_R(z;\mu) = \gamma_{J_L}^{\DGLAP}(\alpha_s(\mu))[\vec{\cD}_{J_L}^{\DGLAP}]_R(z;\mu)\,,
\ee
where $\gamma_{J_L}^{\DGLAP}(\alpha_s)$ is the anomalous dimension matrix of DGLAP detectors, commonly referred to as the timelike anomalous dimension matrix
\be
\gamma^T(J=-1-J_L,\alpha_s)=\gamma_{J_L}^{\DGLAP}(\alpha_s) &= - \left[\cZ_{J_L}^{\DGLAP}(\alpha_s(\mu))\right]^{-1} {d \over d\ln \mu} \cZ_{J_L}^{\DGLAP}(\alpha_s(\mu)) \nn\\
&={\alpha_s(\mu)\over 2\pi} 
\begin{pmatrix}
    \hat{\gamma}^{(0)}_{qq}(J_L) & \hat{\gamma}^{(0)}_{qg}(J_L) \\
    \hat{\gamma}^{(0)}_{gq}(J_L) & \hat{\gamma}^{(0)}_{gg}(J_L)
\end{pmatrix} + O(\alpha_s^2)\,,
\ee
where we have used the beta function $\mu {d \over d\mu}\alpha_s = -2\epsilon \alpha_s + O(\alpha_s^2)$.

At LO, $\gamma^{T}(J,\alpha_s)$ is the same as the spacelike anomalous dimension matrix $\gamma^S(J,\alpha_s)$ for twist-2 local operators, known as Gribov-Lipatov reciprocity relation~\cite{Gribov:1972ri}. At higher loop orders, $\gamma^T$ and $\gamma^S$ are no longer identical, but they are related through a more general reciprocity relation~\cite{Basso:2006nk,Dokshitzer:2006nm}:
\be\label{eq:reciprocity}
\gamma^S(J,\alpha_s) = \gamma^T(J+\gamma^S(J,\alpha_s),\alpha_s)\,,
\qquad 
\gamma^T(J,\alpha_s) = \gamma^S(J-\gamma^T(J,\alpha_s),\alpha_s)\,.
\ee
In the presence of quark/gluon mixing, this relation should be interpreted in terms of matrix eigenvalues, which has been explicitly checked in QCD to three loops~\cite{Chen:2020uvt}. In CFTs, reciprocity can be understood as the statement that the same underlying light-ray operators can be defined using different conformal frames \cite{Caron-Huot:2022eqs}.

Importantly, this process of removing $1/\epsilon$ poles  {\it does not\/} also remove poles in $J_L$. Poles in $J_L$ are related (by a change of conformal frame) to rapidity divergences for light-ray operators in the middle of Minkowski space, which are well known to not be regulated by dim-reg. Consequently, $[\vec{\cD}^{\DGLAP}_{J_L}]_R$ defined in (\ref{eq:dglabrenormalized}) is not actually a well-defined operator for all values of $J_L$. The poles in $J_L$ signal recombination of the DGLAP trajectory with another trajectory that we describe in the next section. Removing poles in $J_L$ will require us to resolve this mixing of trajectories. To do so, we will need to better understand the physics associated with poles in $J_L$.

\clearpage
\section{The BFKL trajectory}
\label{sec:bfkltraj}

\subsection{$J_L$-pole of DGLAP detectors from the leading soft theorem}\label{sec:JLpole_DGLAP}

To introduce the BFKL trajectory, let us start first with the familiar DGLAP trajectory and examine more closely what happens near poles in $J_L$. We will find that a BFKL-type operator encodes the physics near the pole at $J_L=-2$ in a natural way.

In our computations in section~\ref{sec:1loop_details} and appendix~\ref{app:QCD_1loop}, we found that real emission contributions to one-loop matrix elements of $\cD^{\DGLAP}_{J_L,g}$ exhibit poles at $J_L=-2+\mathbb{N}$. (The poles for the quark case are at $J_L=-1+\mathbb{N}$.) From the definition of the DGLAP detectors \eqref{eq:DGLAP_detector_g_def1} and \eqref{eq:DGLAP_detector_q_def1}, we see that these poles must arise from the soft limit $E \to 0$, since the energy integrals are bounded above by the finite-energy source operators.

The soft behavior of amplitudes and form factors in gauge theories is governed by the Weinberg soft theorem~\cite{Weinberg:1965nx}. In this section, we show how the soft theorem predicts the structure of the leading pole at $J_L=-2$. We will see that contributions from the soft theorem can be packaged together into the expectation value of a special ``BFKL" light-ray operator. The BFKL trajectory is distinct from the DGLAP trajectory at tree level, but at loop level they mix with each other. The soft theorem is the first step in understanding this mixing.

At tree level, the emission of a soft gluon does not alter the spin or momentum of the source particle, but it does change its color. Consider a tree-level $n$-point form factor defined by
\be
\<\{f_1,p_1,a_1,s_1\}\cdots \{f_n,p_n,a_n,s_n\}|\cO(p)|0\> &=
(2\pi)^d \de(p-\sum_i p_i)\cF^{a_1,\dots, a_n; s_1,\dots, s_n}_{f_1,\dots,f_n}(p_1, \dots, p_n;p),
\ee
where $f_i = q, \bar{q},g$ denotes the type of the $i$-th parton, and $a_i,s_i$ are color and helicity labels, respectively. To conveniently encode different color structures, we follow~\cite{Catani:1999ss} and define a color-helicity basis $\left\{|\{a_1, s_1\}, \dots, \{a_n, s_n\}\>\right\}$, such that $\cF^{a_1,\dots, a_n; s_1,\dots, s_n}_{f_1,\dots,f_n}$ are the components of a state $|\cF_{f_1,\dots,f_n}\>$ in color-helicity space:
\be
\cF^{a_1,\dots, a_n; s_1,\dots, s_n}_{f_1,\dots,f_n}(p_1, \dots, p_n;p) &= \<\{a_1,s_1\},\dots, \{a_n,s_n\}|\cF_{f_1,\dots, f_n}(p_1,\dots,p_n;p)\>\,.
\ee

In color space, the charge $T^a_i$ acting on the $i$-th parton has matrix elements
\be
\<c_1,\dots, c_m |T^a_i| b_1, \dots b_m \> = \delta_{c_1 b_1} \dots T^a_{c_i b_i} \dots\delta_{c_m b_m}\,,
\ee
where $T^a_{c_ib_i}$ is the color representation matrix for the $i$-th parton, given by
\be
g:\quad& T^a_{cb} = i f^{abc}\,,\\
q:\quad& T^a_{ij} = t^a_{ij} \,,\\
\bar{q}:\quad& T^a_{ij} = -t^a_{ji}\,
\ee
for gluons, quarks, and antiquarks, respectively.

Physical form factors satisfy the gauge invariance condition
\be
\label{eq:color_conservation}
\sum_{i=1}^n T_i^a |\cF_{f_1\dots f_n}(p_1,\dots, p_n;p)\> = 0\,.
\ee
Furthermore, $T_i^c T_i^c = C_i$ is the Casimir operator acting on the $i$-th parton, which is given by $C_i = C_A = N_c$ if $i$ is a gluon, and $C_i = C_F = (N_c^2 - 1)/(2N_c)$ if $i$ is a quark or antiquark.

\begin{figure}
    \centering
    \begin{tikzpicture}[scale=0.7,
        blob/.style={draw, circle, fill=gray, minimum size=1.2cm}
    ]
    \def\tempvalue{1.8};
    \coordinate (O) at (0,0);
    \coordinate (v1) at (1,-\tempvalue);
    \coordinate (v2) at (1,\tempvalue);
    \coordinate (v3) at (-1,\tempvalue);
    \coordinate (v4) at (-1,-\tempvalue);
    \coordinate (v5) at (-2,0);
    \coordinate (v6) at (2,0);
    \coordinate (A) at (0,-1.6);
    \coordinate (G) at (1.8,1.2);
    
    \begin{feynman}
    \draw[ultra thick, green] (v1) -- (O);
    \draw[ultra thick, blue] (v2) -- (O);
    \draw[ultra thick, orange] (v3) -- (O);
    \draw[ultra thick, violet] (v4) -- (O);
    \draw[ultra thick, magenta] (v5) -- (O);
    \draw[ultra thick, brown] (v6) -- (O);
    \node[scale=1.2] at (A) {$\cdots$};
    \draw[gluon,red, thick] (O) -- (G);
    \node[scale=1.2,red,right] at (G) {$a,p_s$};
    \node[blob] (blob1) at (O) {};

    \node[scale=1.5] at (2.8,0) {$\simeq$};
    \node[scale=1.5] at (3.9,-0.29) {$\displaystyle\sum_i$};

    \coordinate (O2) at (6.5,0);
    \draw[ultra thick, green] ($(v1) + (O2)$) -- (O2);
    \draw[ultra thick, blue] ($(v2) + (O2)$) -- (O2);
    \draw[ultra thick, orange] ($(v3) + (O2)$) -- (O2);
    \draw[ultra thick, violet] ($(v4) + (O2)$) -- (O2);
    \draw[ultra thick, magenta] ($(v5) + (O2)$) -- (O2);
    \draw[ultra thick, brown] ($(v6) + (O2)$) -- (O2);
    \node[scale=1.2] at ($(A)+(O2)$) {$\cdots$};
    \node[scale=1.2,above] at ($(v2) + (O2)$) {$i$};
    \draw[ultra thick, cyan] ($0.6*(v2) + (O2)$) -- (O2);
    
    \draw[gluon,red,thick] ($(O2)+0.6*(v2)$) -- ($(G)+(O2)$);
    \node[scale=1.2,red,right] at ($(G)+(O2)$) {$a,p_s$};
    \node[blob] (blob2) at (O2) {};
    
    \node[scale=1.5] at (9.3,0) {$\simeq$};
    \node[scale=1.5] at (10.4,-0.29) {$\displaystyle\sum_i$};
    \node[scale=1.3] at (13.0,-0.05) {$\displaystyle g \, {{\color{red}\varepsilon_s}\. p_i \over {\color{red} p_s}\. p_i}\, {\color{red}T_i^a}$};

    \coordinate (O3) at (17,0);
    \draw[ultra thick, green] ($(v1) + (O3)$) -- (O3);
    \draw[ultra thick, cyan] ($(v2) + (O3)$) -- (O3);
    \draw[ultra thick, orange] ($(v3) + (O3)$) -- (O3);
    \draw[ultra thick, violet] ($(v4) + (O3)$) -- (O3);
    \draw[ultra thick, magenta] ($(v5) + (O3)$) -- (O3);
    \draw[ultra thick, brown] ($(v6) + (O3)$) -- (O3);
    \node[scale=1.2] at ($(A)+(O3)$) {$\cdots$};
    \node[scale=1.2,above] at ($(v2) + (O3)$) {$i$};
    \node[blob] (blob3) at (O3) {};

    \end{feynman}
    \end{tikzpicture}
    \caption{Illustration of tree-level soft gluon theorem. The red line represents the soft gluon with a color index $a$. $\varepsilon_s$ and $p_s$ are the polarization vector and momentum of the soft gluon. $T^a_i$ is the color charge that acts on the $i$-th particle.}
    \label{fig:amp_soft_thm}
\end{figure}
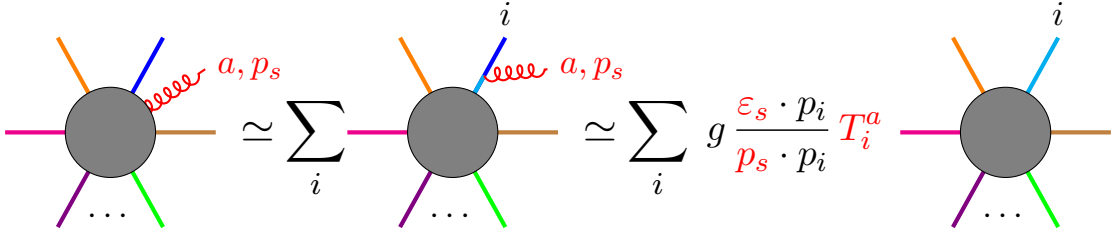

Let us now state the soft gluon theorem for an $(n+1)$-point form factor given by $\cF_{g,f_1,\dots, f_n}(p_s, p_1,\dots, p_n;p)$, where $p_s$ is the momentum of the soft gluon. As illustrated in figure~\ref{fig:amp_soft_thm}, the leading term in the limit $p_s\to 0$  factorizes into a soft factor and an $n$-point form factor~\cite{Bassetto:1983mvz}
\be\label{eq:soft_theorem}
\<\{a,\varepsilon_s\}|\cF_{g,f_1, \dots, f_n}(p_s,p_1,\dots, p_n;p)\> \simeq \sum_{i=1}^n g \tilde{\mu}^{\epsilon} {\varepsilon_s\. p_i \over p_s\. p_i} T^a_i |\cF_{f_1, \dots, f_n}(p_1,\dots, p_n;p)\>\,,
\ee
where $\<\{a,\varepsilon_s\}|$ projects onto the color index $a$ of the soft gluon and its polarization vector $\varepsilon_s$.

\begin{figure}
    \centering
    \begin{tikzpicture}[scale=0.65,
        blob/.style={draw, circle, fill=gray, minimum size=0.7cm}
    ]
    \def\tempvalue{1.8};
    \coordinate (O) at (0,0);
    \coordinate (v1) at (\tempvalue,1);
    \coordinate (v2) at (\tempvalue,-1);
    \coordinate (v3) at (1,-\tempvalue);
    \coordinate (v4) at (1,\tempvalue);
    \coordinate (A) at (0,-1.6);
    \coordinate (G) at (2,0);
    
    \begin{feynman}
    \draw[ultra thick, green!70!black] (v1) -- (O);
    \draw[ultra thick, blue] (v2) -- (O);
    \draw[ultra thick, orange] (v3) -- (O);
    \draw[ultra thick, violet] (v4) -- (O);
    \draw[gluon,red, thick] (O) -- (G);
    \node[blob] (blob1) at (O) {};
    \draw[thick] (2.2,-2) -- (2.2,2);
    \draw[thick] (-1.2,-2) -- (-1.2,2);
    \node[] at (2.5,2) {$2$};

    \node[scale=1.2] at (3.1,0) {$\simeq$};
    \node[scale=1.2] at (4.2,-0.36) {$\displaystyle\sum_{i\neq j}$};

    \coordinate (O2) at (5.8,0);
    \coordinate (O22) at (9.6,0);
    \def\tempshift{1.8};
    \coordinate (V1) at (\tempshift,0.7);
    \coordinate (V2) at (\tempshift,-0.7);
    \coordinate (V3) at (\tempshift,-\tempvalue);
    \coordinate (V4) at (\tempshift,\tempvalue);
    \draw[thick, dashed] ($0.5*(O2)+0.5*(O22)+(0,-2.2)$)--($0.5*(O2)+0.5*(O22)+(0,2.2)$);
    \draw[gluon,red,thick] ($(O2)+0.6*(V2)$) -- ($-0.6*(V2)+(O22)$);
    \draw[ultra thick, green!70!black] ($(V1) + (O2)$) -- (O2);
    \draw[ultra thick, blue] ($(V2) + (O2)$) -- (O2);
    \draw[ultra thick, orange] ($(V3) + (O2)$) -- (O2);
    \draw[ultra thick, violet] ($(V4) + (O2)$) -- (O2);
    \node[scale=0.9, blue] at ($(V2) + (O2) + (0.3,-0.3)$) {$i$};
    \draw[ultra thick, cyan] ($0.6*(V2) + (O2)$) -- (O2);
    \draw[ultra thick, green!70!black] ($-1.0*(V2) + (O22)$) -- (O22);
    \draw[ultra thick, blue] ($-1.0*(V1) + (O22)$) -- (O22);
    \draw[ultra thick, orange] ($-1.0*(V4) + (O22)$) -- (O22);
    \draw[ultra thick, violet] ($-1.0*(V3) + (O22)$) -- (O22);
    \node[scale=0.9, green!70!black] at ($-1.0*(V2) + (O22) + (-0.3,0.3)$) {$j$};
    \draw[ultra thick, green] ($ (O22)-0.6*(V2)$) -- (O22);

    \node[blob] (blob2) at (O2) {};
    \node[blob] (blob2) at (O22) {};
    
    \node[scale=1.2] at (11.0,0) {$\simeq$};
    \node[scale=1.2] at (12.0,-0.36) {$\displaystyle\sum_{i\neq j}$};
    \node[scale=1.1] at (14.5,-0.05) {$\displaystyle g^2\tilde{\mu}^{2\epsilon} \cS_{ij}({\color{red}p_s})$};

    \coordinate (O3) at (17.3,0);
    \coordinate (O32) at (21.1,0);
    \draw[thick, dashed] ($0.5*(O3)+0.5*(O32)+(0,-2.2)$)--($0.5*(O3)+0.5*(O32)+(0,2.2)$);
    \draw[ultra thick, green!70!black] ($(V1) + (O3)$) -- (O3);
    \draw[ultra thick, cyan] ($(V2) + (O3)$) -- (O3);
    \draw[ultra thick, orange] ($(V3) + (O3)$) -- (O3);
    \draw[ultra thick, violet] ($(V4) + (O3)$) -- (O3);
    \draw[ultra thick, green] ($-1.0*(V2) + (O32)$) -- (O32);
    \draw[ultra thick, blue] ($-1.0*(V1) + (O32)$) -- (O32);
    \draw[ultra thick, orange] ($-1.0*(V4) + (O32)$) -- (O32);
    \draw[ultra thick, violet] ($-1.0*(V3) + (O32)$) -- (O32);
    \node[blob] (blob3) at (O3) {};
    \node[blob] (blob3) at (O32) {};
    \fill[red] ($0.5*(O3)+0.5*(O32)-0.5*(V1)+0.5*(V2)$) circle (4pt);
    \fill[red] ($0.5*(O3)+0.5*(O32)+0.5*(V1)-0.5*(V2)$) circle (4pt);
    \node[scale=0.9, red] at ($(V2) + (O3) + (-0.25,0.4)$) {$T_i^a$};
    \node[scale=0.9, red] at ($-1.0*(V2) + (O32) + (0.3,-0.44)$) {$T^a_j$};
    \node[scale=0.9, blue] at ($(V2) + (O3) + (0.3,-0.3)$) {$i$};
    \node[scale=0.9, green!70!black] at ($-1.0*(V2) + (O32) + (-0.3,0.3)$) {$j$};

    \end{feynman}
    \end{tikzpicture}
    \caption{Illustration of the tree-level soft theorem at the cross-section level. The emission of a soft gluon induces a color rotation on the $i$-th and $j$-th partons. The red dots represent the color interference effects that arise after the soft gluon is removed.}
    \label{fig:form_factor_sq_soft}
\end{figure}
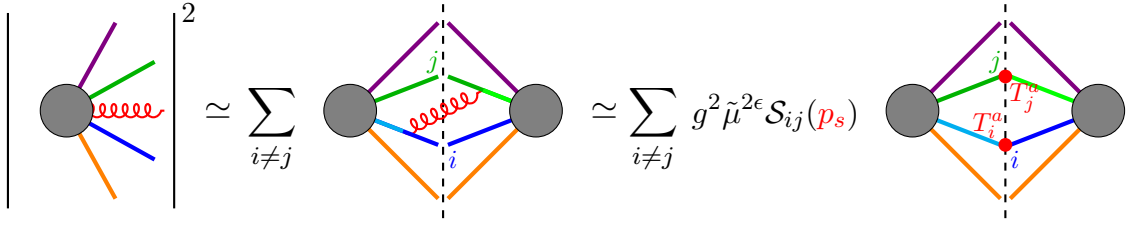

Squaring and summing over color and polarizations, we obtain a soft gluon theorem for cross-sections:
\be
&\<\cF_{g,f_1, \dots, f_n}(p_s,p_1,\dots, p_n;p)|\cF_{g,f_1, \dots, f_n}(p_s,p_1,\dots, p_n;p)\> \nn\\
&\qquad \simeq \signminus g^2\tilde{\mu}^{2\epsilon} \sum_{i\neq j}  \cS_{ij}(p_s)
\<\cF_{f_1, \dots, f_n}(p_1,\dots, p_n;p)|T^a_i T^a_j |\cF_{f_1, \dots, f_n}(p_1,\dots, p_n;p)\>\,, \label{eq:M2_soft_thm}
\ee
where $T^a_i T^a_j$ is the color charge correlation due to the soft gluon emission, and $\cS_{ij}(p_s)$ is the  soft kernel, given by
\be
\cS_{ij}(p_s) = {p_i\. p_j \over (p_s\. p_i)(p_s\. p_j)} \,.
\ee
See figure~\ref{fig:form_factor_sq_soft} for an illustration.  The result (\ref{eq:M2_soft_thm}) can be explicitly checked in the examples \eqref{eq:form_factor_sq_O_to_ggg} and \eqref{eq:form_factor_sq_O_to_gg} by taking the $k_3\to 0$ limit
\be 
\lim_{k_3\to 0} \cI_{ggg}^{\cO,\text{tree}}(k_1,k_2,k_3) \simeq 2 g^2\tilde{\mu}^{2\epsilon} C_A {k_1\. k_2\over (k_1\. k_3)(k_2\. k_3)} \cI_{g,\text{tree}}(k_1,k_2)\,.
\ee

The soft kernel $\cS_{ij}$ accounts for the leading $J_L$ pole in the matrix elements of $\cD_{J_L,g}^{\DGLAP}(z)$. To see this, we parametrize the momenta as  $p_s^\mu = E z^\mu$ and $p_{i}^\mu = E_{i} z_{i}^\mu$ in the soft kernel
\be
\cS_{ij}(p_s) = {1\over E^2}{z_i\. z_j\over (z\. z_i)(z\. z_j)}\,.
\ee
The soft gluon energy dependence scales as $1/E^2$ and leads to a pole at $J_L = -2$
\be
\int_0^{\de} {E^{-J_L} dE \over (2\pi)^{d-1}2E} {f(E)\over E^2} = -{1\over J_L+2} {f(E=0)\over 2^d \pi^{d-1}} + \text{regular terms}\,,
\ee
where, in this toy integral, $\frac{f(E)}{E^2}$ plays the role of the squared form factors, $f(E)$ is a finite smooth function, and $\de$ is an effective cutoff due the boundedness of gluon energy in matrix elements.

More concretely, consider the tree-level pure gluon form factor for simplicity
\be
|\cF_n(k_1,\dots,k_n;p)\> &= |\cF_{g\cdots g}(k_1,\dots,k_n;p)\>.
\ee
The measurement of the gluon DGLAP detector $\cD^{\DGLAP}_{J_L,g}(z)$ on the $(n+1)$-point tree-level form factor $\cF_{n+1}$ is
\be
&\<\cD_{J_L,g}^{\DGLAP}(z)\>^{\cF_{n+1}}_{\cO(p)}\nn\\
&={1\over n!}\int {E^{-J_L} dE \over (2\pi)^{d-1}2E} \int\left[\prod_{i=1}^n {d^{d-1}\vec{k}_i \over (2\pi)^{d-1}2E_i}\right] (2\pi)^d \de^{(d)}(p-E z-\sum_{i=1}^n k_i) |\cF_{n+1}(k_1,\dots, k_n,E z;p)|^2 \nn \\
&= {1\over J_L+2} {g^2 \tilde{\mu}^{2\epsilon} \over 2^d \pi^{d-1}}  {1\over n!}\int d\mathrm{LIPS}_n \sum_{i\neq j} {z_i\. z_j\over (z\. z_i)(z\. z_j)} \<\cF_n(k_1,\dots, k_n;p)|T^a_i T^a_j |\cF_n(k_1,\dots, k_n;p)\> + \cdots\,,\label{eq:YM_DGLAP_generic_pole_term}
\ee
where the dimensionless null vector $z_i$ is proportional to the momentum $p_i$, and in the third line, we have omitted the regular terms in the limit $J_L \to -2$. Since the last line corresponds to a pole of a matrix element of a light-ray operator, we anticipate that the residue can be interpreted similarly as a light-ray matrix element $\<\mathbb{O}(z)\>$, now evaluated on the form factor with one fewer gluon and with $J_L=-2$:
\be
\<\cD_{J_L,g}^{\DGLAP}(z)\>^{\cF_{n+1}}_{\cO(p)} \sim {\<\mathbb{O}_{J_L=-2}(z)\>^{\cF_{n}}_{\cO(p)}\over J_L+2}\,.
\ee

We recognize the kernel ${z_i\. z_j\over (z\. z_i)(z\. z_j)}$ in \eqref{eq:YM_DGLAP_generic_pole_term} as a special case of the horizontal trajectory kernel discussed in \cite{Caron-Huot:2022eqs}, which suggests that the light-ray operator $\mathbb{O}_{J_L=-2}$ has a similar form. To see this, we can use permutation symmetry of the gluon form factor
\be
&\<\cD_{J_L,g}^{\DGLAP}(z)\>^{\cF_{n+1}}_{\cO(p)}= {1\over J_L+2} {g^2 \tilde{\mu}^{2\epsilon} \over 2^d \pi^{d-1}} \int {d^{d-1}\vec{k}_1 \over (2\pi)^{d-1}2E_1} {d^{d-1}\vec{k}_2 \over (2\pi)^{d-1}2E_2} {z_1\. z_2\over (z\. z_1)(z\. z_2)} \nn\\
&\quad \times {1\over (n-2)!}\int\left[\prod_{i=3}^n {d^{d-1}\vec{k}_i \over (2\pi)^{d-1}2E_i}\right] (2\pi)^d \de^{(d)}(p-\sum_{i=1}^n k_i)  \<\cF_n(k_i;p)|T^a_1 T^a_2 |\cF_n(k_i;p)\> + \cdots\,,\label{eq:YM_DGLAP_generic_pole_term_v2}
\ee
where the second line can be identified as a two color interference measurement on the form factor $\cF_n$
\be
&\sum_{\l_i,b_i,c_i} (if^{ab_1 c_1})(if^{ab_2 c_2}) \<a_{\l_1, c_1}^\dagger(k_1)a_{\l_2, c_2}^\dagger(k_2) a_{\l_1, b_1}(k_1) a_{\l_2, b_2}(k_2)\>_{\cO(p)}^{\cF_{n}} \nn\\
=& {1\over (n-2)!}\int\left[\prod_{i=3}^n {d^{d-1}\vec{k}_i \over (2\pi)^{d-1}2E_i}\right] (2\pi)^d \de^{(d)}(p-\sum_{i=1}^n k_i)  \<\cF_n(k_i;p)|T^a_1 T^a_2 |\cF_n(k_i;p)\>\,.
\ee
The phase space integrations of $k_1, k_2$ do not have any extra energy weighting and thus essentially count particle number. To transform \eqref{eq:YM_DGLAP_generic_pole_term_v2} into the form of a light-ray operator matrix element, we introduce a color interference number detector for gluons
\be\label{eq:Ng_definition}
\cN_g^a(z) = i f^{abc} \sum_{\l}\int_0^\oo {E^{d-2}dE \over (2\pi)^{d-1}2E} \left[a_{\l, c}^\dagger(p)a_{\l, b}(p)\right]\Big|_{p=E z}\,.
\ee
(The counterpart for quarks will be discussed in section~\ref{sec:QCD}.) By splitting the phase space integral $d\vec{k}_i$ in \eqref{eq:YM_DGLAP_generic_pole_term_v2} into integration with respect to a unit vector $\hat{z}_i$ and energy $E_i$, i.e.\ ${d^{d-1}\vec{k}_i} = E_i^{d-2} dE_i d^{d-2}\hat{z}_i$, we find that the new light-ray operator $\mathbb{O}_{J_L=-2}(z)$ has the form
\be
\mathbb{O}_{J_L=-2}(z) \propto \int d^{d-2}\hat{z}_1 d^{d-2}\hat{z}_2 {z_1\. z_2\over (z\. z_1)(z\. z_2)} :\cN_{g}^a(z_1) \cN_{g}^a(z_2):\,. 
\label{eq:firstbfkloperator}
\ee
In the next subsection, we will present the generalization of this operator to generic $J_L$.

\subsection{BFKL detectors}

The operator $\mathbb{O}_{J_L=-2}(z)$ defined in (\ref{eq:firstbfkloperator}) is part of a larger family. We will first see how to generalize $\mathbb{O}_{J_L=-2}(z)$ to a family that depends on three continuous parameters $J_{L1}, J_{L2}, J_L\in \C$. A 1-parameter subset of this family is the BFKL trajectory that will be our focus. The remaining parameters will ultimately not be needed for this work, but it is instructive to introduce them anyway.

We start by introducing a generalized color detector $\cD^c_{J_L,g}$:
\be
\cD^c_{J_L,g}(z)=  \sum_{\l,a,b}if^{abc} \int \frac{E^{-J_L}dE}{(2\pi)^{d-1}2E}\left.\left[a^{\dag}_{\l,b}(p)a_{\l,a}(p)\right]\right|_{p=E z}.
\ee
The $\cN^c_g$ detector defined in \eqref{eq:Ng_definition} is the special case $J_L=2-d$. The Feynman rule for a color detector can be read off from the tree-level matrix element
\be\label{eq:color_vertex}
\<0|A_{\nu}^b(-q) \cD^{c}_{J_L,g}(z) A_{\mu}^a(p)|0\> = (2\pi)^d\de^{(d)}(p-q)\left[if^{abc}\Pi_{\mu\nu}(z)V_{J_L}(z;p)\right],
\ee
where $V_{J_L}(z;p)$ is given in \eqref{eq:VJL_definition}.

 A single color detector is not gauge-invariant. To build a gauge-invariant operator, we need at least two of them, so let us consider the product $:\!\cD^c_{J_{L1},g}(z_1)\cD^c_{J_{L2},g}(z_2)\!:$. However, this object does not transform irreducibly under the Lorentz group. Since we would like detectors with definite Lorentz spin, we should decompose it into Lorentz irreps. 

To find the decomposition, we can view the Lorentz group $\SO(d-1,1)$ as the conformal group for the celestial sphere $S^{d-2}$. The null vectors $z_1,z_2$ become embedding space coordinates, and we can think of $:\!\cD^c_{J_{L1},g}(z_1)\cD^c_{J_{L2},g}(z_2)\!:$ as the product of two scalar primary operators with scaling dimensions $-J_{L1}$ and $-J_{L2}$ in a fictitious $\textrm{CFT}$  on $S^{d-2}$. We obtain an irrep with definite $J_L$ by integrating against a standard conformal three-point structure with ``shadow" quantum numbers $\tl{J}_{Li}=2-d-J_{Li}$:
\be\label{eq:GeneralizedBFKL_definition}
&\cD_{J_{L1},J_{L2};J_L,g}(z) \equiv \frac{\G(d-2+J_L)}{\G(\tfrac{d-2+J_L+J_{L1}-J_{L2}}{2})\G(\tfrac{d-2+J_L-J_{L1}+J_{L2}}{2})} \nn \\
&\qquad \times \int D^{d-2}z_1 D^{d-2}z_2 \<\cP_{-\tl J_{L1}}(z_1)\cP_{-\tl J_{L2}}(z_2)\cP_{-J_L}(z)\>:\!\cD^c_{J_{L1},g}(z_1)\cD^c_{J_{L2},g}(z_2)\!:
\ee
The measure $D^{d-2}z$ is the standard $\SO(d-1,1)$-invariant measure on the projective null cone in the embedding space \cite{Simmons-Duffin:2012juh}
\be
D^{d-2}z=\frac{2d^dz \de(z^2)\th(z^0)}{\vol \SO(1,1)}.
\ee
The conformal three-point structure is\footnote{More generally, we could consider three-point structures that carry nontrivial ``transverse spin" on the celestial sphere, see e.g.\ \cite{Chang:2020qpj}. However, we focus on the case of zero transverse spin in this work.}
\be
\<\cP_{\de_1}(z_1)\cP_{\de_2}(z_2)\cP_{\de}(z)\> = \frac{1}{(2z_1\.z)^{\frac{\de_1+\de-\de_2}{2}}(2z_2\.z)^{\frac{\de_2+\de-\de_1}{2}}(2z_1\.z_2)^{\frac{\de_1+\de_2-\de}{2}}}.
\ee
We have chosen the prefactor in (\ref{eq:GeneralizedBFKL_definition}) so that $\cD_{J_{L1},J_{L2};J_L,g}$ satisfies
\be
\hat{\bS}_J[\cD_{J_{L1},J_{L2};J_L,g}] = \cD_{J_{L1},J_{L2};2-d-J_L,g}, 
\ee
where $\hat{\bS}_J$ is the normalized version of the spin shadow transform that satisfies $\hat{\bS}_J^2=1$ (defined later in \eqref{eq:SJhat_definition}).

The object $\cD_{J_{L1},J_{L2};J_L,g}(z)$ has definite Lorentz spin $J_L$. Its (bare) scaling dimension is simply the sum of the scaling dimensions of the two color detectors:
\be
\De_{L} = J_{L1}+J_{L2} +2(d-2).
\ee
Because $\De_L$ is constant as a function of $J_L$, this family of operators traces out a horizontal trajectory on the Chew-Frautschi plot as $J_L$ varies. The vertical position of the trajectory depends on the value of $J_{L1} + J_{L2}$.

We have thus defined a family of detectors $\cD_{J_{L1},J_{L2};J_L,g}$ parametrized by three continuous variables, $J_{L1},J_{L2}$, and $J_L$. However, it will turn out that only the members of this family with $J_{L1}=J_{L2}=2-d$, i.e.\ operators built from products of $\cN_g$ defined in \eqref{eq:Ng_definition}, will mix with the DGLAP trajectory. Explicitly, they are given by
\be\label{eq:DBFKL_definition}
\cD^{\BFKL}_{J_L,g}(z) &\equiv \cD_{2-d,2-d;J_L,g}(z) \nn \\
&= \frac{\G(d-2+J_L)}{\G(\tfrac{d-2+J_L}{2})^2} \int D^{d-2}z_1 D^{d-2}z_2 \p{\frac{2z_1\.z_2}{(2z_1\.z)(2z_2\.z)}}^{-\frac{J_L}{2}}:\!\cN^c_{g}(z_1)\cN^c_{g}(z_2)\!:.
\ee
We call $\cD^{\BFKL}_{J_L,g}$ the ``BFKL trajectory" for reasons that will become clear shortly. Note that $\cD^{\BFKL}_{J_L,g}$ has scaling dimension $\De_L=0$.

Since $\De_L$ of $\cD_{J_{L1},J_{L2};J_L,g}$ only depends on $J_{L1}+J_{L2}$, we could consider horizontal trajectories with other values of $J_{L1}, J_{L2}$ such that $J_{L1}+J_{L2}=2(2-d)$, and they will be degenerate with $\cD^{\BFKL}_{J_L,g}$ at tree level. Naively, one might expect BFKL detectors to mix with these other operators as well. However, we find that at one-loop $\cD^{\BFKL}_{J_L,g}$ mixes only with itself (and $\cD^{\DGLAP}_{J_L,g}$ near $J_L\sim -2$), but not with $\cD_{J_{L1},J_{L2};J_L,g}$ for other values of $J_{L1}, J_{L2}$. Overall, we will find a self-consistent picture of how intersections are resolved at one loop using only the DGLAP and BFKL trajectories. We comment about higher loops and other horizontal trajectories in section~\ref{sec:discussion}.

Including all the prefactors, the soft theorem discussed in section \ref{sec:JLpole_DGLAP} implies that the pole at $J_L=-2$ in the expectation value of $\cD^{\DGLAP}_{J_L,g}$ is given by
\be\label{eq:DGLAP_JLpole_vs_BFKL}
\<\cD^{\DGLAP}_{J_L,g}(z)\>^{\text{1-loop}} = \frac{\a_s\mu^{2\e}}{4\pi}\frac{\cR_1(\e)}{J_L+2} \<\cD^{\BFKL}_{J_L,g}(z)\>^{\text{tree}} + O((J_L+2)^0),
\ee
where 
\be\label{eq:R1_expr}
\cR_1(\e) = \frac{2^{1+2\e}(\tilde{\mu}/\mu)^{2\e}\G(-\e)^2}{\pi^{1-2\e}\G(-2\e)}.
\ee

\subsection{BFKL detectors from fields at $\scriplus$}

Schematically, a color detector $\cD_{J_L,g}^c(z)$ is an integral of a product of field strengths $F^\nu_a(\a,z)$ along $\scriplus$, similar to the definition of $\cD^{\DGLAP}_{J_L,g}$ in (\ref{eq:DGLAP_detector_g_def}), but projecting onto the color adjoint instead of the color singlet. The BFKL detector is then a product of two color detectors with color indices contracted. This suffices to define the operator in a ``sufficiently nice" gauge like Feynman gauge, where as discussed in section~\ref{sec:wilsonlinedetails}, Wilson lines along null infinity are trivial and we can safely contract color indices at different points along null infinity.

However, it is instructive to write the BFKL detector in a manifestly gauge-invariant way. To do so, we must connect up the $F$'s with adjoint Wilson lines.  One approach is to connect each $F$ by an adjoint Wilson line to spatial infinity (where the retarded time is $\a=-\oo$), and then connect up the lines at spatial infinity. More concretely, we can connect the $F$'s by adjoint Wilson lines to spacelike points at large distance, connect the points together, and then take the limit as they go to infinity.

For example, let
\be
B^{(\bar z, L)}_{b,\nu}(\a;z) &= \frac{L^{\De_A}}{4}\bar z^\mu F_{\mu\nu,a}(\a\bar z/4 + L z)W_\mathrm{adj}^{ab}(\a\bar z/4 + L z, L(z - \bar z))
\ee
be a field strength connected to an adjoint Wilson line that stretches along the $\bar z$ direction to the point $L(z-\bar z)$. We use notation where $W_\mathrm{adj}^{ab}(x_1,x_2)$ denotes an adjoint Wilson line along the line segment from $x_1$ to $x_2$.  Let us build a finite version of a color detector from a pair of these:
\be
C^{(\bar z,L)}_{c}(\a_1,\a_2;z) &= if^{ab}{}_c B^{(\bar z, L)}_{a,\nu}(\a_1;z)B^{(\bar z, L)\nu}_{b}(\a_2;z).
\ee
This object transforms like a color adjoint located at the spacelike point $L(z-\bar z)$. We can finally form a gauge-invariant operator by connecting two such $C$'s with an adjoint Wilson line along a spacelike segment:
\be
D^{(\bar z,L)}(\a_1,\a_2;z_1;\a_3,\a_4;z_2) &= C^{(\bar z,L)}_{c}(\a_1,\a_2;z_1)W_\mathrm{adj}^{cd}(L(z_1-\bar z), L(z_2-\bar z)) C^{(\bar z,L)}_{d}(\a_3,\a_4;z_2).
\ee

We define the product of two color detectors $\cD^c_{J_{L1},g}(z_1)\cD^c_{J_{L2},g}(z_2)$ (that is the building block for the BFKL detector) by taking the limit $L\to \oo$ and integrating against appropriate kernels in $\a_1,\a_2$ and $\a_3,\a_4$:
\be
\label{eq:longwindedobject}
``\cD^c_{J_{L1},g}(z_1)\cD^c_{J_{L2},g}(z_2)" &\equiv \frac{1}{C_{J_{L1}}}\int d\a_1 d\a_2\p{(\a_1-\a_2+i\e)^{2\De_A+J_{L1}}-(\a_2-\a_1+i\e)^{2\De_A+J_{L1}}} \nn\\
&\quad \frac{1}{C_{J_{L2}}}\int d\a_3 d\a_4\p{(\a_3-\a_4+i\e)^{2\De_A+J_{L2}}-(\a_3-\a_4+i\e)^{2\De_A+J_{L2}}} \nn\\
&\quad
\lim_{L\to \oo} D^{(\bar z,L)}(\a_1,\a_2;z_1;\a_3,\a_4;z_2).
\ee
Here the constant $C_{J_L}$ is given in \eqref{eq:CJL_definition}. Note that unlike the DGLAP case \eqref{eq:DGLAP_detector_g_def}, the kernels are antisymmetric in $\a_1\leftrightarrow \a_2$ and $\a_3\leftrightarrow \a_4$ due to the $f^{abc}$ color structure. 
The object (\ref{eq:longwindedobject}) should then be used on the right-hand side of (\ref{eq:GeneralizedBFKL_definition}) as usual. 

It is straightforward to see that the Feynman rules for the right-hand side of (\ref{eq:longwindedobject}) agree with the Feynman rules for $\cD^c_{J_{L1},g}(z_1)\cD^c_{J_{L2},g}(z_2)$ in a ``sufficiently nice" gauge. In such a gauge, the contribution from Wilson lines along null infinity vanishes because $A_\mu$ falls off like $L^{\frac{2-d}{2}}$ there. Furthermore, the contribution from the spacelike portion of the Wilson line vanishes because $A_\mu$ falls off even faster there --- like $L^{2-d}$, and this compensates for an extra factor of $L$ coming from the length of the Wilson line in the spacelike region. Overall, the Wilson line contributions are trivial in the $L\to \oo$ limit.\footnote{Relatedly, we could have chosen to join up the $F$'s with Wilson lines in a different way: for example, by connecting them via Wilson lines to future infinity. Different choices should lead to the same operator because the Wilson lines do not contribute to Feynman rules at null infinity (in a sufficiently nice gauge).}

In the presence of interactions, the operator $D^{(\bar z,L)}$ (for finite $L$) has divergences coming from the null segments, and also from the cusps at the ends of $W_\mathrm{adj}^{cd}(L(z_1-\bar z), L(z_2-\bar z))$. To obtain the BFKL operator, we can either (1) treat $D^{(\bar z,L)}$ as a bare operator, take the limit $L\to \oo$, and then renormalize it, or (2) renormalize the operator at finite $L$, and then take the limit $L\to \oo$, multiplying by an appropriate power of $L$ to obtain a finite result. Renormalizing the operator at finite $L$ requires additionally introducing a rapidity regulator to deal with divergences associates to the null segments and the cusps.

Products of parallel null Wilson lines in the interior of Minkowski space are associated to BFKL physics and high energy scattering in the Regge limit \cite{Mueller:1994jq,Balitsky:1995ub,Kovchegov:1999yj,Jalilian-Marian:1996mkd,Jalilian-Marian:1997jhx,Iancu:2001ad,Caron-Huot:2013fea}. In a CFT, such null lines are related by a simple change of conformal frame to null lines along $\scriplus$, see e.g.\ \cite{Caron-Huot:2013fea}. This justifies our use of the label ``BFKL" for the operator $\cD^\mathrm{BFKL}_{J_L,g}$.

\subsection{Tree level matrix elements of BFKL detectors}

\begin{figure}[htbp]
    \centering
        \begin{tikzpicture}
            \begin{feynman}
                \coordinate (O1) at (1.5,0);
                \coordinate (O2) at (-1.5,0);
                \coordinate (V1) at (0,1.5);
                \coordinate (V2) at (0,-1.5);
                \coordinate (V0) at (-2,0);
                
                \draw[dashed] (0,-1.8) -- (0,1.8);
                
                \draw[gluon] (O1) -- (V1);
                \draw[gluon] (O1) -- (V2);
                \draw[gluon] (O2) -- (V1);
                \draw[gluon] (O2) -- (V2);
                
                \filldraw[blue] (V1) circle (3pt);
                \filldraw[blue] (V2) circle (3pt);
                \filldraw[black] (O1) circle (3pt);
                \filldraw[black] (O2) circle (3pt);
                \node[above] at (V1) {$\cN_g^{c}(z_1)$};
                \node[below] at (V2) {$\cN_g^{c}(z_2)$};
                \node[right] at (O1) {$\cO$};
                \node[left] at (O2) {$\cO$};
                
                \node[left] at (V0) {$\displaystyle \int D^{d-2}z_1D^{d-2}z_2\p{\frac{2z_1\.z_2}{(2z\.z_1)(2z\.z_2)}}^{-\frac{J_L}{2}}$};
                
            \end{feynman}
        \end{tikzpicture}
    \caption{Feynman diagrams for the tree-level matrix elements of the BFKL detectors in pure Yang-Mills theory.}
    \label{fig:tree_level_diagrams_BFKL}
\end{figure}

In section \ref{sec:DGLAP_1loop}, we computed $\<\cD^{\DGLAP}_{J_L,g}(z)\>^{\text{1-loop}}$ in the state created by the gluon source operator $\cO=\frac{1}{4N_c}\Tr(F_{\mu\nu}F^{\mu\nu})$. Now, let us also compute the tree-level matrix element of $\cD^{\BFKL}_{J_L,g}$ in the same state. (We will compute the one-loop matrix element in section \ref{sec:BFKL_1loop}.) The Feynman diagrams needed for this calculation are identical to the ones considered for the DGLAP detector. However, the color structure will be different, and we have a more complicated phase space integral due to the nontrivial kernel in \eqref{eq:DBFKL_definition}. 

The diagram for the tree-level matrix element is shown in figure \ref{fig:tree_level_diagrams_BFKL}. By the color detector vertex \eqref{eq:color_vertex} and definition \eqref{eq:DBFKL_definition}, the tree-level matrix element of $\cD^{\BFKL}_{J_L,g}$ is given by
\be\label{eq:DBFKL_tree_gluon_expr0}
\<\cD^{\BFKL}_{J_L,g}(z)\>^{\text{tree}}_{\cO(p)} =& \frac{\G(J_L+d-2)}{\G(\tfrac{J_L+d-2}{2})^2}\int D^{d-2}z_1 D^{d-2}z_2 \p{\frac{2z_1\.z_2}{(2z\.z_1)(2z\.z_2)}}^{-\frac{J_L}{2}} \nn \\
&\int \frac{E_1^{d-2}dE_1}{(2\pi)^{d-1}2E_1}\frac{E_2^{d-2}dE_2}{(2\pi)^{d-1}2E_2} (2\pi)^d\de^{(d)}(p-E_1 z_1-E_2 z_2) \cI'_{\text{tree}}(E_1 z_1,E_2 z_2).
\ee
where we have used the short-hand notation \eqref{eq:short-hand_notation}, and $\cI'_{\text{tree}}(k_1,k_2)$ is similar to the square of the tree-level form factor $\cF^{a_1a_2}_{g,\text{tree}}$ given in \eqref{eq:cF_g_tree}, but with a different color sum,
\be
\cI'_{\text{tree}}(k_1,k_2)=\sum_{a_i,b_i,c,\l_i}(if^{a_1b_1c})(if^{a_2b_2c})\cF^{a_1a_2}_{g,\text{tree}}(k_i,\l_i)\p{\cF^{b_1b_2}_{g,\text{tree}}(k_i,\l_i)}^*.
\ee

We can use the momentum-conserving delta function in the integrand of \eqref{eq:DBFKL_tree_gluon_expr0} to fix $z_2$ and $\b_2$. Then, the integral becomes
\be
\<\cD^{\BFKL}_{J_L,g}(z)\>^{\text{tree}}_{\cO(p)} =& \frac{2\pi^2}{(2\pi)^d}\frac{\G(J_L+d-2)}{\G(\tfrac{J_L+d-2}{2})^2} \int_0^{\oo}d E_1\, E_1^{d-3}\int D^{d-2}z_1\, \de((p-E_1 z_1)^2) \nn \\
&\p{\frac{2z_1\.p}{(2z\.z_1)(2z\.(p-E_1 z_1))}}^{-\frac{J_L}{2}} \cI'_{\text{tree}}(E_1 z_1,p-E_1 z_1).
\ee
The remaining delta function also fixes $E_1=\frac{p^2}{2p\.z_1}$. Finally, evaluating the $z_1$-integral, we obtain
\be\label{eq:DBFKL_tree_matrixelement}
\<\cD^{\BFKL}_{J_L,g}(z)\>^{\text{tree}}_{\cO(p)} = -\frac{(d-2)C_A(N_c^2-1)}{2^{d+1}\pi^{\frac{d}{2}-1}\G(\tfrac{d-2}{2})}  (\signplus 2 z\cdot p)^{J_L}(\signplus p^2)^{\frac{d-J_L}{2}}.
\ee
As a consistency check, one can verify that the above result is indeed consistent with \eqref{eq:DGLAP_CR_JLminus2pole} and \eqref{eq:DGLAP_JLpole_vs_BFKL}.

Since color detectors $\cN^c(z_i)$ are the constituents of the BFKL detector, rather than generic $\cD^c_{J_{Li}}(z_i)$'s, we can recombine the celestial sphere integration measure $D^{d-2}z_i$ and energy integration in $\cN^c$ into a phase space integration measure
\be\label{eq:numberdetector_measure_identity}
2\pi \de(k^2)\th(k^0) = \pi \int D^{d-2}z \int_0^{\oo} d\b\, \b^{d-3}\de^{(d)}(k-\b z).
\ee
As a consequence, a matrix element of the BFKL detector can be expressed as
\be
&\<\cD^{\BFKL}_{J_L}(z)\> =\label{eq:BFKL_generic_mel}\\
&\hspace{0.8cm} \frac{\G(d-2+J_L)}{\G(\tfrac{d-2+J_L}{2})^2}\sum_n \sum_{X_n} \frac{1}{\symF_{X_n}} \int d\mathrm{LIPS}_n \sum_{\substack{i,j\in X_n\\ i\neq j}} \left({2 p_i\. p_j \over (2z\. p_i)(2z\. p_j)}\right)^{-\frac{J_L}{2}} \<\cM_{X_n}|T^c_i T^c_j|\cM_{X_n}\>\,, \nn
\ee
where we follow the notation in \eqref{eq:M2_soft_thm}. Here, $\cM_{X_n}$ is the amplitude for producing a given $n$-particle final state $X_n$, and $K_{X_n}$ is the corresponding symmetry factor. This definition will actually apply beyond pure Yang-Mills theory if we interpret $T_i^c, T_j^c$ as generators for appropriate color representations. We will include fundamental quarks in more detail in section~\ref{sec:QCD}.

\section{DGLAP/BFKL mixing}
\label{sec:dglapbfklmixing}

\subsection{Structure of DGLAP/BFKL mixing}\label{sec:mixing_structure}
\label{sec:structureofmixing}

\begin{figure}[htbp]
\begin{center}
\begin{tikzpicture}
    \draw[dashed] (-4,0) -- (4,0);
    \draw[dashed] (-4,-4) -- (4,4);
    
    \node[circle, fill=red, inner sep=0pt, minimum size=8pt] at (0,0) (dot0) {};
    \node[red, below] at (1.5,-0.3) {$\mathcal{D}_{2-d}^{\text{DGLAP}} = \mathcal{D}_{2-d}^{\text{BFKL}}$};
    
    \node[circle, fill=blue, inner sep=0pt, minimum size=8pt] at (2,0) (dot1) {};
    \node[circle, fill=blue, inner sep=0pt, minimum size=8pt] at (2,2) (dot2) {};
    \node[blue, right] at (2.5,1) {$\mathcal{D}_{J_L}^{\text{DGLAP}} \sim \frac{g^2\tilde{\mu}^{2\e}}{J_L+2}\mathcal{D}_{-2}^{\text{BFKL}}$};
    
    \draw[->, blue, dashed, line width=1.5pt] (dot2) -- (dot1);
    
    \node[circle, fill=green!70!black, inner sep=0pt, minimum size=8pt] at (-2,0) (dot3) {};
    \node[circle, fill=green!70!black, inner sep=0pt, minimum size=8pt] at (-2,-2) (dot4) {};
    \draw[->, green!70!black, dashed, line width=1.5pt] (dot3) -- (dot4);
     \node[green!70!black, left] at (-2.5,-1) {$\mathcal{D}_{J_L}^{\text{BFKL}} \sim \frac{g^2\tilde{\mu}^{2\e}}{J_L-6+2d}\mathcal{D}_{6-2d}^{\text{DGLAP}}$};
     
\end{tikzpicture}
\caption{The structure of mixing between DGLAP and BFKL operators near $J_L=2-d$. At tree-level, these operators become equal (up to a normalization) at $J_L=2-d$ (red dot). One-loop matrix elements of the DGLAP trajectory have a pole at $J_L=-2$ proportional to the BFKL trajectory (blue dots). Furthermore, one-loop matrix elements of the BFKL trajectory have a pole at $J_L=6-2d$ proportional to the DGLAP trajectory (green dots). This pattern leads to recombination between the renormalized trajectories. Both the blue and green arrows descend by the dimensionality of the coupling constant $-2\e$, as required by dimensional analysis. The arrows are vertical, as required by Lorentz symmetry.}
\label{modified}
\end{center}
\end{figure}
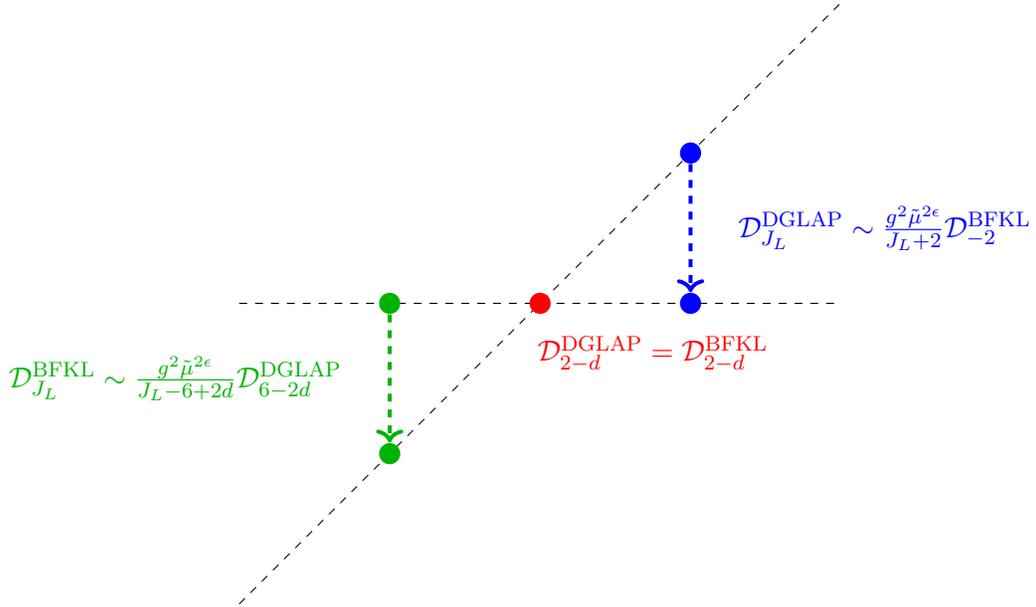

The work \cite{Caron-Huot:2022eqs} gave an example of mixing between detector trajectories at one loop in the Wilson-Fisher theory. The mixing between the DGLAP and BFKL trajectories follows a similar pattern. There are three distinguished values of $J_L$ near the tree-level intersection point, see figure~\ref{modified}. Different physical effects play a role at each value, and together these effects lead to one-loop mixing.

\begin{enumerate}
\item The first important effect is that $\cD^{\DGLAP}_{J_L,g}$ and $\cD^{\BFKL}_{J_L,g}$ become equal as operators at the tree-level intersection point $J_L=2-d=-2+2\e$ (red in figure~\ref{modified}). Relatedly, if we introduce a nondegenerate basis of operators at $J_L=2-d$, it transforms as a log multiplet under dilatations {\it at tree-level}. Mixing resolves this log multiplet into a non-log multiplet. We explain how the DGLAP and BFKL operators become equal at $J_L=2-d$ in section~\ref{eq:equalityatintersection}.

\item The second important effect is that the tree-level BFKL trajectory appears as a pole in 1-loop matrix elements of the DGLAP trajectory. Because the coupling constant $g^2 \tilde\mu^{2\e}$ has mass dimension $2\e$, by dimensional analysis this can only occur if $\De_L^{\BFKL,\mathrm{tree}}(J_L) = \De_L^{\DGLAP,\mathrm{tree}}(J_L) + 2\e$, which fixes $J_L=-2$ (blue in figure~\ref{modified}). We have already seen in section \ref{sec:JLpole_DGLAP} that this pole indeed does appear, as a consequence of the soft gluon theorem.

\item The third important effect is that the tree-level DLGAP trajectory appears as a pole in 1-loop matrix elements of the BFKL trajectory. Again, dimensional analysis dictates that this can only occur at the special value $J_L=6-2d=-2+4\e$ (green in figure~\ref{modified}). Recall that the DGLAP detector measures energy localized at a single point on the celestial sphere, while in general the BFKL detector measures delocalized color information. The pole at $J_L=6-2d=-2+4\e$ comes from two color detectors $\cN_g^c$ colliding, where they combine to measure color-singlet information in a single direction. This is related to the physics of collinear divergences, as we explain in detail in section~\ref{sec:collinear}.

\end{enumerate}

In section~\ref{sec:oneloopmixingputtingeverythingtogether}, we show how to these effects combine to control the 1-loop renormalization of the DGLAP and BFKL trajectories. Computing the 1-loop dilatation operator, we find nondegenerate eigenvalues describing recombined trajectories.

\subsubsection{Equality of bare DGLAP and BFKL detectors at $J_L=2-d$}
\label{eq:equalityatintersection}

Let us start by showing that the DGLAP detector and the BFKL detector become identical at $J_L=2-d$. First, recall that in in the definition of $\cD^{\BFKL}_{J_L,g}$ \eqref{eq:DBFKL_definition}, we integrate $:\!\cN^c_g(z_1)\cN^c_g(z_2)\!:$ against a kernel
\be
\p{\frac{2z_1\.z_2}{(2z_1\.z)(2z_2\.z)}}^{-\frac{J_L}{2}}.
\ee
It turns out that this kernel has a pole exactly at $J_L=2-d$, and the residue is a sum of delta functions that localize either $z_1$ or $z_2$ to $z$. More precisely, we show in appendix \ref{app:distributions} that
\be\label{eq:distribution_identity_JLminus2}
&\int D^{d-2}z_1 D^{d-2}z_2\ \p{\frac{2z_1\.z_2}{(2z_1\.z)(2z_2\.z)}}^{-\frac{J_L}{2}}f(z_1,z_2) \nn \\
&\sim \frac{\vol(S^{d-3})}{J_L+d-2} \p{\int D^{d-2}z' f(z,z')+ \int D^{d-2}z' f(z',z)},
\ee
where $f(z_1,z_2)$ is any test function that is smooth at $z_1=z$ and $z_2=z$.

Moreover, the prefactor in the definition of $\cD^{\BFKL}_{J_L,g}$ \eqref{eq:DBFKL_definition} has a zero at $J_L=2-d$, which cancels the pole in (\ref{eq:distribution_identity_JLminus2}). Thus, we find
\be\label{eq:BFKL_JL_2minusd}
&\cD^{\BFKL}_{2-d,g}(z)= \frac{\vol(S^{d-3})}{2} :\!\cN^c_{g}(z)\, G^c\!:,
\ee
where $G^c$ is the color charge operator
\be
G^c \equiv \int D^{d-2} z\, \cN^c_g(z).
\ee
Since $G^c$ acts as zero on gauge invariant states, it follows that the un-normal-ordered product of $\cN^c$ and $G^c$ vanishes (acting on gauge-invariant states):
\be
\cN^c_{g}(z) G^c &= 0.
\ee
On the other hand, using the commutation relations for creation and annihilation operators, we find that
\be
:\!\cN^c_{g}(z)\, G^c\!: &= \cN^c_{g}(z) G^c - C_A \cD^\mathrm{DGLAP}_{2-d,g}(z) =  - C_A \cD^\mathrm{DGLAP}_{2-d,g}(z).
\ee
We conclude that
\be
\label{eq:equivalencedglapbfkl}
\cD^{\BFKL}_{2-d,g}(z) &= - \frac{\vol S^{d-3}}{2} C_A \cD^\mathrm{DGLAP}_{2-d,g}(z).
\ee

We can also see this equivalence more explicitly by acting with (\ref{eq:BFKL_JL_2minusd}) on an $n$-gluon form factor  $|\cF_n(k_i,\l_i;p)\>$ for some gauge invariant operator.
The matrix element of $\cD^{\BFKL}_{J_L,g}$ at $J_L=2-d$ in the $n$-gluon state is
\be
\<\cD^{\BFKL}_{2-d,g}(z)\>_{\cF_n} =&\frac{\vol(S^{d-3})}{2}\frac{1}{(n-2)!}\int D^{d-2}z'  \int \frac{d^dk_1}{(2\pi)^d}\frac{d^dk_2}{(2\pi)^d}\left[\prod_{i=3}^n \frac{d^dk_i\de^{+}(k_i^2)}{(2\pi)^{d-1}}\right] \nn \\
&\times (2\pi)^d\de^{(d)}(p-\sum_i k_i)\<\cF_n|T^c_1 T^c_2 |\cF_n\> \nn \\
&\times\pi^2 \int_0^{\oo} d\b_1\, \b_1^{d-3}  \de^{(d)}(k_1-\b_1 z) \int_0^{\oo} d\b_2\, \b_2^{d-3}\de^{(d)}(k_2-\b_2 z').
\ee

Using the gauge invariance condition \eqref{eq:color_conservation} and the fact that $\cF_n$ is symmetric under $(p_i,a_i) \leftrightarrow (p_j,a_j)$, we have
\be
\<\cF_n|T^c_1 T^c_2 |\cF_n\> = \frac{1}{n-1} \sum_{i=2}^{n} \<\cF_n|T^c_1 T^c_i |\cF_n\> = -\frac{1}{n-1}\<\cF_n|T^c_1 T^c_1 |\cF_n\> = -\frac{C_A}{n-1}\<\cF_n|\cF_n\>.
\ee
Furthermore, using the identity \eqref{eq:numberdetector_measure_identity}, we can write the matrix element as
\be
\<\cD^{\BFKL}_{2-d,g}(z)\>_{\cF_n} =&-\frac{\vol(S^{d-3})}{2}C_A\Bigg(\frac{\pi}{(n-1)!} \int \frac{d^dk_1}{(2\pi)^d}\left[\prod_{i=2}^n \frac{d^dk_i\de^{+}(k_i^2)}{(2\pi)^{d-1}}\right](2\pi)^d\de^{(d)}(p-\sum_i k_i)  \nn \\
&\qquad\qquad\qquad\qquad \times \<\cF_n|\cF_n\>\int_0^{\oo} d\b_1\, \b_1^{d-3}  \de^{(d)}(k_1-\b_1 z) \Bigg)\nn \\
=& -\frac{\vol(S^{d-3})}{2}C_A \<\cD^{\DGLAP}_{2-d,g}(z)\>_{\cF_n}.
\ee
Since this holds for any form-factor $\cF_n$, we recover (\ref{eq:equivalencedglapbfkl}).

\subsubsection{$J_L$-pole of the BFKL detector from collinear divergences}
\label{sec:collinear}

In the above analysis of the BFKL detector, we have focused on the $z_1\sim z$ and $z_2\sim z$ regions, which lead to the $J_L$-pole at $J_L=2-d$. In fact, there are other poles in $J_L$ which come from the region where both $z_1$ and $z_2$ approach $z$. Consider the matrix element of $\cD^{\BFKL}_{J_L,g}$,
\be
\<\cD^{\BFKL}_{J_L,g}\>= \frac{\G(d-2+J_L)}{\G(\tfrac{d-2+J_L}{2})^2} \int D^{d-2}z_1 D^{d-2}z_2 \p{\frac{2z_1\.z_2}{(2z_1\.z)(2z_2\.z)}}^{-\frac{J_L}{2}}\<:\!\cN^c_{g}(z_1)\cN^c_{g}(z_2)\!:\>.
\ee
Note that in the $z_1\sim z_2\sim z$ limit, the matrix element of $\cN^c_g(z_1)\cN^c_g(z_2)$ can have collinear divergences that change the behavior of the integrand near the limit. Therefore, unlike the pole at $J_L=2-d$ which is a fact about the kernel itself and should be true for all states, the value of $J_L$ for which the integral localizes to $z_1\sim z_2\sim z$ actually depends on the details of the matrix element.

The collinear divergences can be parametrized by a small angle $\th$ defined by $z_1\.z_2=1-\cos\th$. As we will explain shortly, for the matrix elements we consider in this work, the collinear divergences obey
\be\label{eq:NN_collinear_behavior}
\<:\!\cN^c_{g}(z_1)\cN^c_{g}(z_2)\!:\> \sim \frac{C_{\text{coll}}}{\th^2} + \ldots,\qquad \th \to 0.
\ee
This behavior of the collinear divergences implies that the integral has a pole at $J_L=6-2d$ given by (see appendix \ref{app:distributions} for derivation)
\be\label{eq:BFKL_int_JLpole_collinear_maintext}
&\int D^{d-2}z_1 D^{d-2}z_2\ \p{\frac{2z_1\.z_2}{(2z_1\.z)(2z_2\.z)}}^{-\frac{J_L}{2}}\<:\!\cN^c_{g}(z_1)\cN^c_{g}(z_2)\!:\> \nn \\
&\sim \frac{\vol(S^{d-3})\vol(S^{d-4})}{J_L-6+2d}\frac{\pi^{\frac{3}{2}}\G(2-\tfrac{d}{2})\G(\tfrac{3d}{2}-5)}{\G(\tfrac{5-d}{2})\G(\tfrac{d}{2}-1)\G(d-3)}C_{\text{coll}}.
\ee

Now the remaining task is to understand the leading behavior $C_{\text{coll}}/\theta^2$ in the matrix element $\<:\!\!\cN_g^c(z_1) \cN_g^c(z_2)\!\!:\>$ when the two color detectors approach each other. The $1/\theta^2$ divergence is a typical collinear behavior in massless gauge theories, described by the splitting functions~\cite{Altarelli:1977zs}. A nice way of packaging the splitting function calculation is to view this collinear divergence as the leading term in the light-ray OPE between the two color detectors. In particular, we can write the tree-level OPE formula
\be\label{eq:LROPE_color_detectors}
\cN^c_{g}(z_1)\cN^c_{g}(z_2) ={g^2\tilde{\mu}^{2\epsilon}  \over (4\pi)^2} \hat C_{\textrm{color-color-DGLAP}} {1\over 2 z_1\. z_2}\cD^{\DGLAP}_{6-2d,g}(z_2) + \ldots,
\ee
where $\ldots$ are the subleading terms in the collinear limit $z_1\to z_2$.  Comparing \eqref{eq:NN_collinear_behavior} and \eqref{eq:LROPE_color_detectors}, we find the coefficient $C_{\text{coll}}$ is proportional to a matrix element of DGLAP detector
\be
C_{\text{coll}} = {g^2\tilde{\mu}^{2\epsilon}  \over (4\pi)^2} \hat C_{\textrm{color-color-DGLAP}} \<\cD^{\DGLAP}_{6-2d,g}(z)\>\,.
\ee
The OPE coefficient $\hat C_{\textrm{color-color-DGLAP}}$ can be computed using the gluon splitting function, with appropriate change of the color factor $C_A \to -{1\over 2}C_A^2$ coming from the effect of color detectors. The details of this computation can be found in \cite{Chen:2021gdk}, where we can modify the second equation in (5.29) to extract the OPE coefficient
\be\label{eq:C_colorcolorDGLAP}
\hat C_{\textrm{color-color-DGLAP}} &=  {1\over (2\pi)^{d-4}}\int_{c-i\oo}^{c+i\oo} {dJ\over 2\pi i} \left[-{1\over \pi} {\Gamma(J_1-J)\Gamma(J_2)\over \Gamma(J_1+J_2-J)}\hat{\gamma}^{(0)}_{gg}(J_L)\right]\Bigg|_{\substack{J_1=J_2=d-3,\; J_L=-1-J,\\ \hspace{-4em}C_A\to -\frac{1}{2}C_A^2}}\nn\\
&= \frac{3 (8-3 d)  \Gamma (d-2)}{2^{3 d-10} \pi^{d-\frac{7}{2}} (d-4)\Gamma \left(d-\frac{3}{2}\right)} C_A^2\,.
\ee
Here, the expression for $\hat{\gamma}_{gg}$ is given in \eqref{eq:DGLAP_gamma_gg} and the contour of the $J$-integral in the first line follows the standard choice for Barnes integrals --- lying to the R.H.S. of all the poles of $\hat{\gamma}_{gg}$ and to the L.H.S. of all the poles of $\Gamma(d-3-J)$.
The prefactor $(2\pi)^{4-d}$ and the condition $J_1=J_2=d-3$ convert the $d=4$ result in \cite{Chen:2021gdk} to generic spacetime dimension $d$.
Combining \eqref{eq:BFKL_int_JLpole_collinear_maintext}, \eqref{eq:LROPE_color_detectors}, and \eqref{eq:C_colorcolorDGLAP}, we have
\be\label{eq:BFKL_JLpole_vs_DGLAP}
\<\cD^{\BFKL}_{J_L,g}(z)\> \sim \frac{1}{J_L-6+2d}\frac{g^2\tilde{\mu}^{2\e}}{(4\pi)^2}  \frac{9\pi C_A^2(d-3)(8-3d)\G(\tfrac{3d}{2}-6)}{2^{d-4}(2d-5)\G(\tfrac{d-2}{2})\G(2d-6)} \<\cD^{\DGLAP}_{6-2d,g}(z)\>.
\ee

\subsection{One-loop matrix elements of BFKL detectors}\label{sec:BFKL_1loop}

In this subsection, we consider the one-loop matrix element of the BFKL detector in the state created by the gauge-invariant source operator $\cO=\frac{1}{4N_c}\Tr(F_{\mu\nu}F^{\mu\nu})$. As discussed in section \ref{sec:1loop_details}, there are contributions from real emission, virtual corrections, and the counterterms. In particular, the one-loop form factor from the virtual corrections and counterterms are proportional to the tree-level form factor. Therefore, their contributions can be computed using the results in section \ref{sec:1loop_details} and the tree-level matrix element \eqref{eq:DBFKL_tree_matrixelement}. Using the expressions of $C^{\text{1-loop,V}}_{g,\cO}(d)$ and $\de_g$ given by \eqref{eq:CV_g_1loop} and \eqref{eq:delta_g}, we find the virtual correction contribution is given by
\be\label{eq:BFKL_1loop_virtual}
&\< \cD^{\BFKL}_{J_L,g}(z)\>_{[\cO]_R(p)}^{\text{1-loop,V}}  \nn \\
& = -g^2 \tilde{\mu}^{2\epsilon} \frac{\pi^{\frac{d}{2}-1}C_A(N_c^2-1)}{\G(\tfrac{d-2}{2})} C^{\text{1-loop,V}}_{g,\cO}(d)\times (\signplus 2 z\cdot p)^{J_L}(\signplus p^2)^{d-2-\frac{J_L}{2}}\, \nn \\
&=g^2 \tilde{\mu}^{2\epsilon}C_A(N_c^2-1)(\signplus 2 z\cdot p)^{J_L}(\signplus p^2)^{d-2-\frac{J_L}{2}} \p{\frac{C_A}{64\pi^3\e^2} - \frac{C_A(1+2\g_E-2\log(4\pi))}{64\pi^3\e} +O(\e^0)}  \,,
\ee
and the counterterm contribution is
\be\label{eq:BFKL_1loop_counterterm}
\< \cD^{\BFKL}_{J_L,g}(z)\>_{[\cO]_R(p)}^{\text{1-loop,C}} &= \de_g \< \cD^{\BFKL}_{J_L,g}(z)\>_{\cO(p)}^{\text{tree}}\, \nn \\
&=g^2 \tilde{\mu}^{2\epsilon}C_A(N_c^2-1)(\signplus 2 z\cdot p)^{J_L}(\signplus p^2)^{d-2-\frac{J_L}{2}}  \p{\frac{11C_A}{384\pi^3\e}+O(\e^0)}.
\ee

The real emission contribution is given by
\be\label{eq:BFKL_1loop_real_eq0}
&\< \cD^{\BFKL}_{J_L,g}(z)\>_{[\cO]_R(p)}^{\text{1-loop,R}} \nn \\
&= \frac{\G(J_L+d-2)}{\G(\tfrac{J_L+d-2}{2})^2}\int D^{d-2}z_1 D^{d-2}z_2 \p{\frac{2z_1\.z_2}{(2z\.z_1)(2z\.z_2)}}^{-\frac{J_L}{2}} \nn \\
&\int \frac{E_1^{d-2}dE_1}{(2\pi)^{d-1}2E_1}\frac{E_2^{d-2}dE_2}{(2\pi)^{d-1}2E_2}\frac{d^dk_3\de^+(k_3^2)}{(2\pi)^{d-1}} (2\pi)^d\de^{(d)}(p-E_1 z_1-E_2 z_2-k_3) \cI^{\prime\cO,\text{tree}}_{ggg}(E_1 z_1,E_2 z_2,k_3),
\ee
where $\cI^{\prime\cO,\text{tree}}_{ggg}(k_1,k_2,k_3)$ can be obtained from the three-particle form-factor $\cF^{a_1a_2a_3}_{ggg,\text{tree}}$ depicted in figure~\ref{fig:1loop_real_emission_O},
\be\label{eq:Iprime_ggg_def}
\cI^{\prime\cO,\text{tree}}_{ggg}(k_1,k_2,k_3) = \sum_{a_i,b_i,c,\l_i}\de^{a_3b_3}(if^{a_1b_1c})(if^{a_2b_2c})\cF^{a_1a_2a_3}_{ggg,\text{tree}}(k_i,\l_i)\p{\cF^{b_1b_2b_3}_{ggg,\text{tree}}(k_i,\l_i)}^*.
\ee

Note that the third line of \eqref{eq:BFKL_1loop_real_eq0} is the one-loop matrix element of $\cN_g^c(z_1)\cN_g^c(z_2)$, and so it is a homogeneous function in $z_1,z_2$ with homogeneity $2-d$. By dimensional analysis, it can be written as
\be\label{eq:Fggg_definition}
&\int \frac{E_1^{d-2}dE_1}{(2\pi)^{d-1}2E_1}\frac{E_2^{d-2}dE_2}{(2\pi)^{d-1}2E_2}\frac{d^dk_3\de^+(k_3^2)}{(2\pi)^{d-1}} (2\pi)^d\de^{(d)}(p-E_1 z_1-E_2 z_2-k_3) \cI^{\prime\cO,\text{tree}}_{ggg}(E_1 z_1,E_2 z_2,k_3) \nn \\
&= (p^2)^{2d-4}(2p\.z_1)^{2-d}(2p\.z_2)^{2-d}F_{ggg}(\z),
\ee
where $F_{ggg}(\z)$ is a nontrivial function of the cross-ratio
\be\label{eq:zeta_def}
\z=\frac{p^2(2z_1\.z_2)}{(2p\.z_1)(2p\.z_2)},
\ee
which takes value between $0$ and $1$. $\z=0$ corresponds to the collinear limit, where $z_1,z_2$ approach each other, and $\z=1$ corresponds to the back-to-back limit. We compute $F_{ggg}(\z)$ explicitly in appendix~\ref{app:BFKL_1loop}, and its expression is given by \eqref{eq:Fggg_expr}.

We can now write the real-emission contribution as
\be\label{eq:BFKL_1loop_real_eq01}
&\< \cD^{\BFKL}_{J_L,g}(z)\>_{[\cO]_R(p)}^{\text{1-loop,R}} \nn \\
&= \frac{\G(J_L+d-2)}{\G(\tfrac{J_L+d-2}{2})^2}(p^2)^{2d-4}\int D^{d-2}z_1 D^{d-2}z_2 \frac{(2z_1\.z_2)^{-\frac{J_L}{2}}(2p\.z_1)^{2-d}(2p\.z_2)^{2-d}}{(2z\.z_1)^{-\frac{J_L}{2}}(2z\.z_2)^{-\frac{J_L}{2}}}F_{ggg}(\z).
\ee
Again, by dimensional analysis, the right-hand side has to be proportional to $(2p\.z)^{J_L}(p^2)^{d-2-\frac{J_L}{2}}$ with a coefficient that is a nontrivial function of $J_L$ and $d$. We will derive in appendix~\ref{app:BFKL_1loop} that the coefficient can be computed by an integral over the cross-ratio $\z$. At the end, we have
\be\label{eq:BFKL_1loop_real_eq1}
&\< \cD^{\BFKL}_{J_L,g}(z)\>_{[\cO]_R(p)}^{\text{1-loop,R}} \nn \\
&=(2p\.z)^{J_L}(p^2)^{d-2-\frac{J_L}{2}} \frac{\pi^{d-2}}{\G(\tfrac{d-2}{2})^2} \nn \\
&\qquad \times\int_0^1 d\z\, \z^{\frac{d-4-J_L}{2}}(1-\z)^{\frac{d-4}{2}}{}_2F_1\p{-\frac{J_L}{2},-\frac{J_L}{2},\frac{d-2}{2},1-\z}F_{ggg}(\z).
\ee

Our goal is to find the divergences of \eqref{eq:BFKL_1loop_real_eq1}, which include poles in $\e$ and $J_L$. Let us first consider the $\e$-poles. First, the function $F_{ggg}(\z)$ itself contains $\e$-pole, which comes from the soft behavior $E_1\to 0$ or $E_2\to 0$ in \eqref{eq:Fggg_definition}. Since it comes from the soft limit, this pole also has a universal description from the leading soft theorem, similar to the argument in section \ref{sec:JLpole_DGLAP}. We give more details in appendix~\ref{app:poles_BFKL_JL1JL2}. Another place where $\e$-pole can arise is the back-to-back region $\z\to 1$ in the $\z$-integral. More precisely, in the back-to-back limit $F_{ggg}(\z)$ contains $\frac{\log(1-\z)}{1-\z}$ and $\frac{1}{1-\z}$, and after integrating them against $(1-\z)^{\frac{d-4}{2}}$, we get poles at $d=4$.

Using \eqref{eq:Fggg_expr}, we find that the relevant terms of $F_{ggg}(\z)$ that will give rise to $\e$-poles $\< \cD^{\BFKL}_{J_L,g}(z)\>_{[\cO]_R(p)}^{\text{1-loop,R}}$ are given by
\be
&F_{ggg}(\z) \nn \\
&= \frac{C_A^2(N_c^2-1)g^2\tilde{\mu}^{2\e}}{64\pi^5}\Bigg(\frac{1}{(4-d)\z} + \frac{(d-2)\big(-\frac{2}{d-4} + \frac{2}{d-3} - \frac{4}{d-2}+2 \psi(d) + 2\g_E+\log(1-\z)\big)}{4^{d-3}\pi^{2d-8}(1-\z)}\Bigg) \nn \\
&+\ldots,
\ee
where $\ldots$ are terms that are finite at $d=4$ and less divergent than $\frac{1}{1-\z}$. Plugging this into \eqref{eq:BFKL_1loop_real_eq1} and taking the $\e\to 0$ limit, we obtain
\be\label{eq:BFKL_1loop_real_epspole}
&\< \cD^{\BFKL}_{J_L,g}(z)\>_{[\cO]_R(p)}^{\text{1-loop,R}} \nn \\
&=g^2 \tilde{\mu}^{2\e}C_A(N_c^2-1)(2p\.z)^{J_L}(p^2)^{d-2-\frac{J_L}{2}}\nn \\
&\qquad \times \p{-\frac{C_A}{64\pi^3\e^2}-\frac{C_A(5-6\g_E+12\log(4\pi)+3\psi(\tfrac{2+J_L}{2})+3\psi(-\tfrac{J_L}{2}))}{384\pi^3\e}+O(\e^0)}.
\ee

Finally, we can obtain 1-loop matrix element by combining the virtual correction \eqref{eq:BFKL_1loop_virtual}, counterterm contribution \eqref{eq:BFKL_1loop_counterterm}, and the real eimssion \eqref{eq:BFKL_1loop_real_epspole}. Again, we see that the double poles $\frac{1}{\e^2}$ exactly cancel, and the result for the $\e$-pole is
\be
&\< \cD^{\BFKL}_{J_L,g}(z)\>_{[\cO]_R(p)}^{\text{1-loop}} \nn \\
&= g^2 \tilde{\mu}^{2\e}C_A(N_c^2-1)(2p\.z)^{J_L}(p^2)^{d-2-\frac{J_L}{2}} \p{-\frac{C_A(2\g_E+\psi(\tfrac{2+J_L}{2})+\psi(-\tfrac{J_L}{2}))}{128\pi^3\e}+ O(\e^0)}.
\ee
Combining this with the tree-level result \eqref{eq:DBFKL_tree_matrixelement}, we can rewrite the one-loop matrix element as
\be
\< \cD^{\BFKL}_{J_L,g}(z)\>_{[\cO]_R(p)}^{\text{1-loop}} = \frac{\a_s}{4\pi}\frac{\g_{\BFKL}(J_L)}{\e}\<\cD^{\BFKL}_{J_L,g}(z)\>_{\cO(p)}^{\text{tree}} + O(\e^0).
\ee
$\g_{\BFKL}(J_L)$ is the detector anomalous dimension,
\be\label{eq:gamma_BFKL}
\g_{\BFKL}(J_L)= 2C_A(2\g_E+\psi(\tfrac{2+J_L}{2})+\psi(-\tfrac{J_L}{2})),
\ee
which exactly agrees with the well-known BFKL eigenvalue \cite{Balitsky:1978ic}.

Now, let us study the $J_L$-pole of the one-loop matrix element \eqref{eq:BFKL_1loop_real_eq1}. As we argue in section \ref{sec:mixing_structure}, due to the collinear divergences, $\<\cD^{\BFKL}_{J_L,g}\>^{\text{1-loop}}$ has a pole at $J_L=6-2d$ proportional to the tree-level matrix element of $\cD^{\DGLAP}_{J_L,g}$. We can see this explicitly by studying the $\z \sim 0$ region of the integral in \eqref{eq:BFKL_1loop_real_eq1}. Focusing on the $\z\to 0$ limit, we have
\be
&\< \cD^{\BFKL}_{J_L,g}(z)\>_{[\cO]_R(p)}^{\text{1-loop,R}} \nn \\
&=(2p\.z)^{J_L}(p^2)^{d-2-\frac{J_L}{2}} \p{-\frac{3 C_A^2 (N_c^2-1)g^2 \tilde{\mu}^{2\e} (3 d-8)   \G\left(\tfrac{d-1}{2}\right) \G(d-1) \G (-\tfrac{d}{2}-J_L+1)}{2^{d+3}\pi^{d-\frac{1}{2}}(d-4) \G(2d-4) \G(-\frac{J_L}{2})^2}} \nn \\
&\qquad \times \int_0^1 d\z\, \z^{\frac{J_L}{2}+d-4}\p{1+O(\z)}.
\ee
As expected, the $\z^{\frac{J_L}{2}+d-4}$ factor leads to a pole $J_L=6-2d$. By comparing with the DGLAP tree-level matrix element \eqref{eq:DDGLAP_tree_matrixelement_gluon}, we find
\be
\< \cD^{\BFKL}_{J_L,g}(z)\>_{[\cO]_R(p)}^{\text{1-loop,R}} = \frac{\a_s \mu^{2\e}}{4\pi}\frac{\cR_2(\e)}{J_L+2-4\e}\<\cD^{\DGLAP}_{J_L,g}(z)\>^{\text{tree}}_{\cO(p)} + O((J_L+2-4\e)^0),
\ee
where we have set $d=4-2\e$. The residue $\cR_2(\e)$ is given by
\be\label{eq:R2_expr}
\cR_2(\e) = -\frac{9\pi C_A^2(1-2\e)(2-3\e)(\tilde{\mu}/\mu)^{2\e}\G(-3\e)}{2^{-1-2\e}(3-4\e)\G(2-4\e)\G(1-\e)}.
\ee
Again, one can check that this result agrees with the previous prediction \eqref{eq:BFKL_JLpole_vs_DGLAP}.

\subsection{One-loop dilatation operator and recombined trajectories}
\label{sec:oneloopmixingputtingeverythingtogether}

\subsubsection{Bare detectors and their matrix elements}\label{sec:matrixelements_conditions}
We are now ready to combine all the above results and resolve the mixing between the DGLAP and BFKL trajectories at one loop, following the approach in \cite{Caron-Huot:2022eqs}. Our goal is to define renormalized DGLAP and BFKL detectors whose 1-loop matrix elements do not have any $\e$-poles and $J_L$-poles near $J_L\sim -2$. Let us first reproduce here all the divergences of the bare detectors we have computed in previous sections. At $\e\to 0$, generic $J_L$, we have (for brevity we suppress the subscript label for the source operator)
\be
\< \cD^{\DGLAP}_{J_L,g}(z)\>^{\text{1-loop}} &=\frac{\a_s}{4\pi} \frac{\hat{\g}^{(0)}_{gg}(J_L)}{\e} \< \cD^{\DGLAP}_{J_L,g}(z)\>^{\text{tree}} + \cO(\e^0) \label{eq:DGLAP_epspole_mixing}\,,  \\
\< \cD^{\BFKL}_{J_L,g}(z)\>^{\text{1-loop}} &= \frac{\a_s}{4\pi}\frac{\g_{\BFKL}(J_L)}{\e}\<\cD^{\BFKL}_{J_L,g}(z)\>^{\text{tree}} + O(\e^0). \label{eq:BFKL_epspole_mixing}
\ee
On the other hand, for generic $\e$, there are $J_L$-poles given by
\be
\<\cD^{\DGLAP}_{J_L,g}(z)\>^{\text{1-loop}} &= \frac{\a_s\mu^{2\e}}{4\pi}\frac{\cR_1(\e)}{J_L+2} \<\cD^{\BFKL}_{J_L,g}(z)\>^{\text{tree}} + O((J_L+2)^0)\label{eq:DGLAP_JLpole_mixing} \, , \\
\< \cD^{\BFKL}_{J_L,g}(z)\>^{\text{1-loop}} &= \frac{\a_s \mu^{2\e}}{4\pi}\frac{\cR_2(\e)}{J_L+2-4\e}\<\cD^{\DGLAP}_{J_L,g}(z)\>^{\text{tree}} + O((J_L+2-4\e)^0). \label{eq:BFKL_JLpole_mixing}
\ee
The expressions of $\hat{\g}^{(0)}_{gg}(J_L)$, $\g_{\BFKL}(J_L)$, $\cR_1(\e)$, and $\cR_2(\e)$ are given by \eqref{eq:DGLAP_gamma_gg}, \eqref{eq:gamma_BFKL}, \eqref{eq:R1_expr}, and \eqref{eq:R2_expr}, respectively. However, to keep the discussion below more general (so that it can also apply to other intersections discussed later in section \ref{sec:shadowDGLAP}), we will not plug in their explicit expressions until the end of this section.

In section \ref{sec:mixing_structure}, we have shown that the bare detectors $\cD^{\BFKL}_{J_L}$ and $-\frac{C_A\pi^{1-\e}}{\G(1-\e)}\cD^{\DGLAP}_{J_L}$ become identical at $J_L=-2+2\e$. Therefore, to define a basis that is non-degenerate at $J_L=-2+2\e$, we should consider
\be\label{eq:regular_basis_DGLAPBFKL}
\mathbb{D}_{J_L} = U_1 \begin{pmatrix} \mu^{J_L+2-2\e}\cD^{\DGLAP}_{J_L,g} \\ \cD^{\BFKL}_{J_L,g} \end{pmatrix}, \quad U_1= \begin{pmatrix} -\frac{C_A\pi^{1-\e}}{\G(1-\e)} & 0 \\ \frac{\frac{C_A\pi^{1-\e}}{\G(1-\e)}}{J_L+2-2\e} & \frac{1}{J_L+2-2\e} \end{pmatrix},
\ee
where we have introduced an explicit scale $\mu$ to make the multiplet dimensionless. Thanks to \eqref{eq:equivalencedglapbfkl}, this multiplet is regular for all values of $J_L$ near the intersection, including $J_L=-2+2\e$. 

One can easily check that $\mathbb{D}_{J_L}$ has finite tree-level matrix element at $J_L=-2+2\e$. Moreover, the fact that the two bare operators become exactly the same at $J_L=-2+2\e$ implies that the divergences of their one-loop matrix elements should satisfy nontrivial relations. To find the relation, we can first write
\be
\hat{\g}^{(0)}_{gg}(J_L) &= \frac{A_1}{J_L+2} + \g_1(J_L)\,, \\
\g_{\BFKL}(J_L) &= \frac{A_2}{J_L+2} + \g_2(J_L)\,, \\
\cR_1(\e) &= \p{-\frac{C_A\pi^{1-\e}}{\G(1-\e)}}^{-1}\p{\frac{A_1}{\e} + \tl \cR_1(\e)}\,, \\
\cR_2(\e) &= \p{-\frac{C_A\pi^{1-\e}}{\G(1-\e)}}\p{\frac{A_2}{\e} + \tl \cR_2(\e)}\,,
\ee
where $A_1=4C_A, A_2=-4C_A$. $\g_{1,2}(J_L)$ are regular functions of $J_L$, and $\tl \cR_{1,2}(\e)$ are regular functions of $\e$. First, one can notice that the $J_L$-pole of $\hat{\g}^{(0)}_{gg}$ at $J_L=-2$ has the same residue of the $\e$-pole of $\p{-\frac{C_A\pi^{1-\e}}{\G(1-\e)}}\cR_1(\e)$ at $\e=0$. This ensures that \eqref{eq:DGLAP_epspole_mixing} and \eqref{eq:DGLAP_JLpole_mixing} are consistent with each other at $\e\to 0, J_L\to -2$ (after using \eqref{eq:equivalencedglapbfkl}). Similarly, the $J_L$-pole of $\g_{\BFKL}(J_L)$ and the $\e$-pole of $\p{-\frac{C_A\pi^{1-\e}}{\G(1-\e)}}^{-1}\cR_2(\e)$ have the same residues.

Combining \eqref{eq:DGLAP_epspole_mixing}, \eqref{eq:BFKL_epspole_mixing}, \eqref{eq:DGLAP_JLpole_mixing}, and \eqref{eq:BFKL_JLpole_mixing}, we can write down all the divergences of $\<\cD^{\DGLAP}_{-2+2\e,g}\>^{\text{1-loop}}$ and $\<\cD^{\BFKL}_{-2+2\e,g}\>^{\text{1-loop}}$. We have
\be
&\p{-\frac{C_A\pi^{1-\e}}{\G(1-\e)}}\<\cD^{\DGLAP}_{-2+2\e,g}\>^{\text{1-loop}} -\<\cD^{\BFKL}_{-2+2\e,g}\>^{\text{1-loop}} \nn \\
&= \frac{\a_s}{4\pi}\p{\frac{A_1+A_2}{2\e^2} +\frac{2\g_1(-2)-2\g_2(-2)+\tl \cR_1(0)+\tl\cR_2(0)}{2\e}}\<\cD^{\BFKL}_{-2+2\e}\>^{\text{tree}} \nn \\
&\quad+ (\text{regular at $\e\to 0$})\,.
\ee
For this to be consistent with \eqref{eq:equivalencedglapbfkl}, we must have
\be\label{eq:degenerate_condition}
A_1+A_2=0,\qquad 2\g_1(-2)-2\g_2(-2)+\tl \cR_1(0)+\tl\cR_2(0) =0.
\ee
One can plug in the expressions of $A_{1,2}$, $\g_{1,2}$, and $\tl \cR_{1,2}$ given previously and verify the above relations. More generally, for any two trajectories that intersect and exhibit the mixing structure described in section \ref{sec:mixing_structure}, there should be corresponding relations satisfied by the divergences of their one-loop matrix elements. As we will see later, the condition is essential for ensuring that the anomalous dimension matrix is finite.

\subsubsection{Renormalized detectors}
The renormalized operator multiplet is defined by
\be
\left[\mathbb{D}_{J_L} \right]_R \equiv \cZ_{J_L}^{-1}\mathbb{D}_{J_L}.
\ee
The renormalization factor $\cZ_{J_L}$ should be chosen such that the one-loop matrix elements of $\left[\mathbb{D}_{J_L} \right]_R$ are finite. Since the basis \eqref{eq:regular_basis_DGLAPBFKL} is regular at all $J_L$ near the intersection, $\cZ_{J_L}$ should only contain divergences at $\e \to 0$, $J_L\to 2$, and $J_L \to -2+4\e$, and have no additional poles. Indeed, we find that the renormalization factor $\cZ_{J_L}$ is given by\footnote{Here, we choose a renormalization scheme of the ``minimal-subtraction" type, where the numerator of the $\frac{1}{\e(J_L+2)}$ term is a constant, the numerator of $\frac{1}{J_L+2}$ is independent of $J_L$, and the numerator of $\frac{1}{\e}$ is independent of $\e$.}
\be\label{eq:DGLAPBFKL_DivMat}
&\cZ_{J_L} = 1 + \frac{\a_s}{4\pi}M^{\DGLAP/\BFKL}_{J_L}\, , \nn \\
&M^{\DGLAP/\BFKL}_{J_L}\nn \\
&= \begin{pmatrix} \frac{4C_A}{\e(J_L+2)} + \frac{\tl R_1(\e)}{J_L+2} + \frac{\g_1(J_L)}{\e} && -\frac{2(4C_A+\e \tl R_1(\e))}{J_L+2} \\ -\frac{8C_A}{\e(J_L+2)(J_L+2-4\e)} + \frac{\tl \cR_1(\e)}{2\e(J_L+2)} + \frac{\tl \cR_2(\e)}{2\e(J_L+2-4\e)} + \frac{\g_{21}(J_L)}{\e} && -\frac{4C_A}{\e(J_L+2)} - \frac{\tl \cR_1(\e)}{J_L+2} + \frac{\g_2(J_L)}{\e} \end{pmatrix},
\ee
where we have plugged in $A_1=-A_2=4C_A$, and
\be
\g_{21}(J_L) = \frac{\g_2(J_L)-\g_2(-2)-\g_1(J_L)+\g_1(-2)}{J_L+2}.
\ee

Next, we would like to compute how the dilatation operator acts on the renormalized detectors. The dilatation operator is given by
\be
D=D_{\text{eng}}-\frac{\ptl}{\ptl \log \mu},
\ee
where $D_{\text{eng}}$ counts the engineering mass dimension. For the bare multiplet $\mathbb{D}_{J_L}$, we have
\be
D\mathbb{D}_{J_L}  =\mathscr{D}_0 \mathbb{D}_{J_L},\qquad \mathscr{D}_0 =  \begin{pmatrix} -2+2\e-J_L && 0 \\ 1 && 0\end{pmatrix}.
\ee
The matrix $\mathscr{D}_0$ takes a non-trivial Jordan form, and consequently the bare operators transform like a log-multiplet under dilatation.

For the renormalized multiplet, we have
\be
D\left[\mathbb{D}_{J_L}\right]_R = \mathscr{D}\left[\mathbb{D}_{J_L}\right]_R.
\ee
The dilatation matrix $\mathscr{D}$ is determined by the general formula
\be
\mathscr{D} &= \cZ_{J_L}^{-1}\p{\mathscr{D}_0+\b(g)\frac{\ptl}{\ptl g}}\cZ_{J_L}, \label{eq:Dilatation_oneloop_eq}
\ee
which in this case gives
\be
\mathscr{D} &= \mathscr{D}_0 + \frac{\a_s}{2\pi}\begin{pmatrix} -\g_1(J_L) && 4C_A + \e \tl \cR_1(\e) \\ \frac{(4+2J_L-8\e)\g_{21}(J_L)+2\g_1(J_L)-2\g_2(J_L)+\tl \cR_1(\e)+\tl \cR_2(\e)}{4\e}&& -\g_2(J_L) \end{pmatrix} + O(\a_s^2).
\ee
We expect this matrix to be regular at $\e\to 0$ and $J_L\to -2$. However, the off-diagonal element appears to have a $\frac{1}{\e}$ pole. It turns out that the condition that $\cD^{\DGLAP}_{J_L,g}$ and $\cD^{\BFKL}_{J_L,g}$ become identical at $J_L=2-d$ ensures that this divergence will vanish. In particular, imposing \eqref{eq:degenerate_condition} and taking the $\e\to 0$ limit, we obtain
\be
\left.\mathscr{D}\right|_{\e \to 0} = \mathscr{D}_0 + \frac{\a_s}{2\pi}\begin{pmatrix} -\g_1(J_L) && 4C_A \\ \frac{1}{4}\p{-8\g_{21}(J_L)+\tl \cR'_1(0) + \tl \cR'_2(0)} && -\g_2(J_L) \end{pmatrix} + O(\a_s^2).
\ee

\begin{figure}
    \centering
    \includegraphics[width = 10cm]{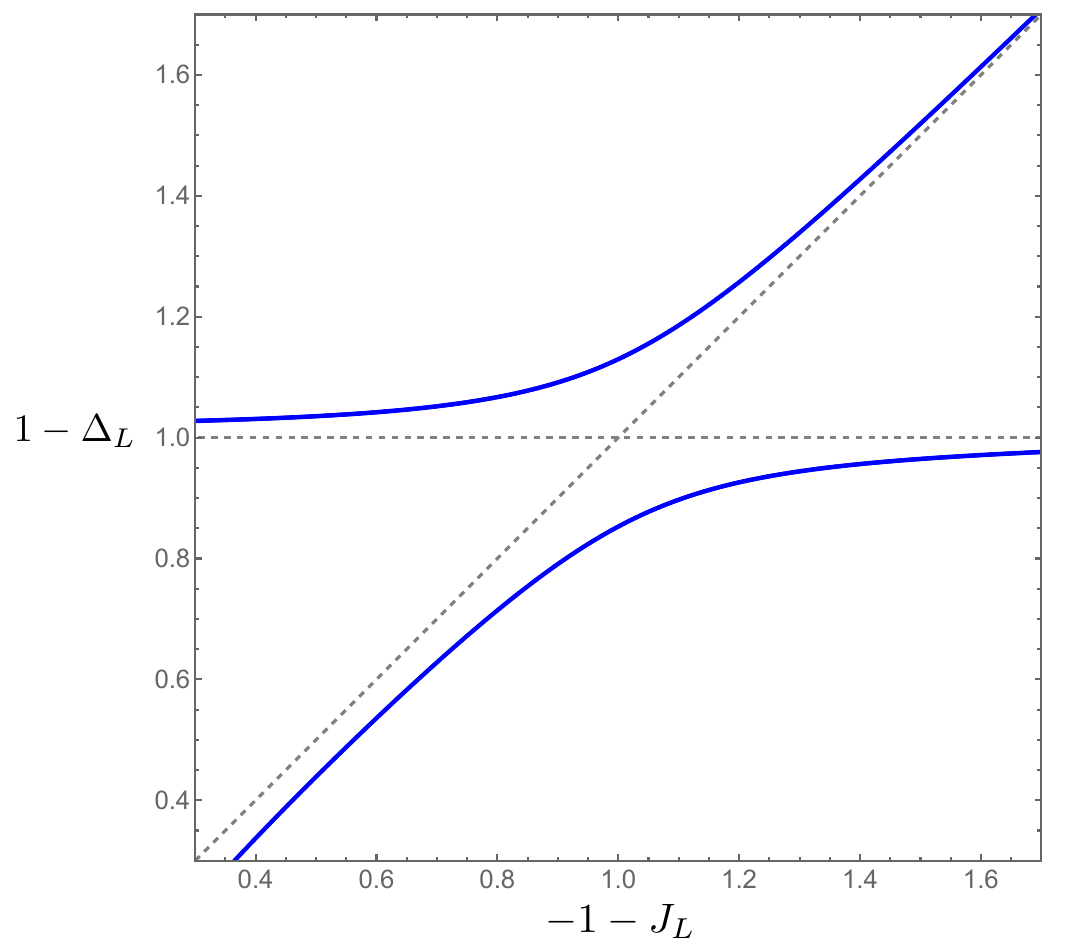}
    \caption{The renormalized Regge trajectories (blue) at one-loop near the DGLAP/BFKL intersection for $\a_s=0.01$. The gray dashed lines are the free theory trajectories.}
    \label{fig:DGLAP_BFKL}
\end{figure}

One-loop corrections have turned the tree-level dilatation matrix $\mathscr{D}_0$ into a diagonalizable matrix $\mathscr{D}$. We can determine the recombined trajectories by solving the characteristic equation,
\be\label{eq:CharEq}
\text{det}(\mathscr{D} + \De_L) = 0.
\ee
In figure \ref{fig:DGLAP_BFKL}, we plot the resulting trajectories near the intersection point $J_L=-2$ for $\a_s=0.01$. In particular, at $J_L=-2$ we find
\be\label{eq:DeltaL_DGLAPBFKL_atJLminus2}
\p{\De_{L}(J_L=-2)}_{\pm} = \pm \sqrt{\frac{2C_A}{\pi}\a_s} + \frac{11C_A}{12\pi}\a_s + O(\a_s^{3/2}),
\ee
Note that the dilataion matrix at $O(\a_s)$ predicts both the $O(\a_s^{1/2})$ and the $O(\a_s^1)$ terms in its eigenvalues at $J_L=-2$. In other words, $O(\a_s^2)$ corrections to the dilatation matrix only lead to $O(\a_s^{3/2})$ corrections to the eigenvalues at $J_L=-2$. The two corresponding left-eigenvectors of $\mathscr{D}$ are given by
\be
v_{\pm} = \left(1 \pm \frac{11}{24}\sqrt{\frac{2C_A}{\pi}\a_s}+O(\a_s),\mp \sqrt{\frac{2C_A}{\pi}\a_s} +O(\a_s^{\frac{3}{2}}) \right).
\ee
Here, we observe the same phenomenon as in the Wilson-Fisher theory \cite{Caron-Huot:2022eqs}: the renormalized trajectories exhibit $O(\sqrt{\a_s})$ splitting at the intersection point. The leading term in \eqref{eq:DeltaL_DGLAPBFKL_atJLminus2} agrees with known results from resummation of the timelike anomalous dimension \cite{Ellis:1996mzs, Ioffe:2010zz}. From the detector perspective, the fact that the detectors transform like a log-multiplet at tree-level is crucial for producing this $O(\sqrt{\a_s})$ splitting.

\subsection{All-orders predictions for poles in anomalous dimensions}

If the anomalous dimension of a trajectory has divergences at a certain point, it signals the mixing with another trajectory. In fact, resolving the recombination of two trajectories can let us predict the singularities of their anomalous dimensions near the intersection point at all loop order. This was studied in \cite{Jaroszewicz:1982gr,Lipatov:1996ts,Fadin:1998py,Kotikov:2000pm,Kotikov:2002ab,Kotikov:2007cy} in the context of DGLAP/BFKL mixing in QCD and $\cN=4$ SYM, and more recently explained in \cite{Caron-Huot:2022eqs} in terms of recombination of Regge trajectories. The idea is that away from the intersection point $J_L=-2$, the solutions to the characteristic equation \eqref{eq:CharEq} should correspond to the DGLAP and BFKL trajectories. In particular, $\De_L=J_L+2+\g^T(-1-J_L,\a_s)$ is a solution to the characteristic equation for generic $J_L$, where $\g^T$ is the timelike anomalous dimension. On the other hand, we also expect that the characteristic equation is finite at $J_L=-2$ if all the divergences can be resolved in the same way as we described above. This leads to nontrivial constraints on the poles of $\g^T(J=-1-J_L)$ at $J=1$. Using the expression of the characteristic equation at $1$-loop and assuming the higher-loop corrections are finite at $J_L=-2$,\footnote{At higher loop order, more trajectories can participate in the mixing with DGLAP and BFKL. As long as their divergences can also be resolved and the characteristic equation remains finite at $J_L=-2$, our result for the pole structures of $\g^T$ and $\g^S$ should be valid.} we can determine the leading and subleading poles $\frac{\a_s^n}{(J-1)^{2n-1}}, \frac{\a_s^n}{(J-1)^{2n-2}}$ of $\g^T(J)$ for all $n$:
\be\label{eq:timelike_poles}
\g^T(J,\a_s) &= \a _s\left(-\frac{2 C_A}{\pi  (J-1)}+\frac{11 C_A}{6 \pi }+\ldots\right)+\a _s^2 \left(\frac{4 C_A^2}{\pi ^2 (J-1)^3}-\frac{11 C_A^2}{3 \pi ^2 (J-1)^2}+\ldots\right) \nn \\
&+\a _s^3\left(-\frac{16 C_A^3}{\pi ^3 (J-1)^5}+\frac{22 C_A^3}{\pi ^3 (J-1)^4}+\ldots\right)  + \a _s^4\left(\frac{80 C_A^4}{\pi ^4 (J-1)^7}-\frac{440 C_A^4}{3 \pi ^4 (J-1)^6}+\ldots\right)+\ldots.
\ee
One can check that the first three loop orders agree with the known exact expressions \cite{Stratmann:1996hn,Moch:2007tx,Chen:2020uvt}. In QCD, the leading poles at each order $\frac{\a_s}{J-1}, \frac{\a_s^2}{(J-1)^3},\frac{\a_s^3}{(J-1)^5},\ldots$ are the same as in pure YM, while the subleading poles $\frac{\a_s}{(J-1)^0}, \frac{\a_s^2}{(J-1)^2},\frac{\a_s^3}{(J-1)^4},\ldots$ are different. The leading poles have also been obtained by mapping the BFKL physics to soft fragmentation on the celestial sphere~\cite{Neill:2020bwv}.

By replacing $\De_L\to 1-J, J_L\to 1-\De$ and plugging $\De=J+2+\g^S(J,\a_s)$ in the characteristic equation, we can also determine the leading poles of the spacelike anomalous dimension. The result is
\be\label{eq:spacelike_poles}
\g^S(J,\a_s)&=\a_s\p{-\frac{2C_A}{\pi(J-1)}+\ldots} + \a_s^2 \p{\frac{0}{(J-1)^2} + \ldots} \nn \\
&+ \a_s^3 \p{\frac{0}{(J-1)^3} + \ldots} + \a_s^4\p{-\frac{4C_A^4\z(3)}{\pi^4(J-1)^4} + \ldots} + \ldots,
\ee
which agrees with the result for QCD and $\cN=4$ SYM \cite{Fadin:1998py,Kotikov:2007cy}.

Finally, one can check that the poles of $\g^T$ in \eqref{eq:timelike_poles} and the poles of $\g^S$ in \eqref{eq:spacelike_poles} are consistent with the reciprocity relation \eqref{eq:reciprocity}. In fact, the relation leads to nontrivial relations between the poles of  $\g^T$ and poles of $\g^S$, and hence we can use it to predict one from the other. For example, the leading poles $\frac{\a_s^n}{(J-1)^{2n-1}}$ of $\g^T$ for all $n$ are fixed by the 1-loop leading pole $\frac{\a_s}{J-1}$ of $\g^S$. More generally, using reciprocity, one can determine the $\frac{\a_s^n}{(J-1)^{2n-1-k}}$ poles of $\g^T(J)$ for all $n$ if the $\frac{\a_s^m}{(J-1)^{m-p}}$ poles of $\g^S(J)$ are known for $m=1,2,\ldots,k+1$ and $p=0,1,\ldots,k+1-m$.

\section{Intersection of the DGLAP trajectory with its shadow}\label{sec:shadowDGLAP}

In figure \ref{fig:CFplot_YM_illustration}, we see that the tree-level DGLAP trajectory intersects its shadow trajectory near $J_L=-1$, mimicking the scenario described in~\cite{Caron-Huot:2022eqs} for the Wilson-Fisher theory. We also observe a corresponding pole at $J_L=-1$ in the DGLAP matrix element $\<\cD^{\DGLAP}_{J_L,g}\>$ in \eqref{eq:DGLAP_CR_JLminus1pole}. These phenomena are hallmarks of recombination of the DGLAP trajectory with its shadow. In this section, we explain how this recombination is related to the universal tree-level subleading soft behavior of amplitudes and form factors.

\subsection{The subleading soft theorem}

In perturbative gauge theory, the tree-level soft behavior of a gauge boson is universal at subleading power, a result known as the Low-Burnett-Kroll (LBK) theorem~\cite{Low:1958sn,Burnett:1967km,Bern:2014vva,Larkoski:2014bxa}.\footnote{Over the past decade, the leading and subleading soft theorems have been related to asymptotic symmetries associated to the null boundaries of the Minkowski space~\cite{Strominger:2017zoo}. In gravity, the soft behavior of graviton can be extended to higher subleading powers.} The LBK theorem generalizes the leading soft theorem \eqref{eq:soft_theorem} to include an extra term:
\be\label{eq:sub_soft_theorem}
&\<\{a,\varepsilon_s\}|\cF_{g,f_1, \dots, f_n}(p_s,p_1,\dots, p_n;p)\>  \simeq g \tilde{\mu}^{\epsilon}\left[ S^{(0)} + S^{(1)}
  +\cdots \right]|\cF_{f_1, \dots, f_n}(p_1,\dots, p_n;p)\>\,,
\ee
where $S^{(0)}$ and $S^{(1)}$ are respectively the leading and subleading soft operators 
\be
S^{(0)} = \sum_{i=1}^n  T^a_i  {\varepsilon_s\. p_i \over p_s\. p_i}\,,\qquad 
S^{(1)} =-i \sum_{i=1}^n T_i^a {\varepsilon_{s\mu} p_{s\nu}  \over p_s\cdot p_i}J_i^{\mu\nu}\,.
\ee
In the definition of $S^{(1)}$, $J_i^{\mu\nu}$ is the angular momentum generator acting on the $i$-th particle 
\be
J_i^{\mu\nu} = i \left(p_i^\mu \partial_{p_i}^\nu -p_i^\nu \partial_{p_i}^\mu \right)+ \Sigma^{\mu\nu}_i\,,
\ee
where $\Sigma^{\mu\nu}_i$ is the spin generator for the $i$-th particle.

The tree-level subleading soft theorem holds for amplitudes and form factors at both the function and distribution levels --- i.e., irrespective of whether we include the momentum-conserving delta function (see, e.g., ~\cite{Broedel:2014fsa}). It turns out that the distributional form is more appropriate for our purposes, since the Lorentz-invariant phase space integral decouples into separate integrals for each particle, and this will simplify our analysis.\footnote{There are also subtleties in the non-distributional version of the subleading soft theorem. To formulate the action of $S^{(1)}$ in (\ref{eq:sub_soft_theorem}), one must choose a sufficiently smooth extension of $\cF$ off of the momentum-conserving locus $p=\sum_i p_i$, see e.g.\ \cite{Cachazo:2014fwa}.}

To this end, let us introduce notation for a form-factor with the momentum conserving $\de$-function restored:
\be
|\boldsymbol{\cF}_{f_1,\dots f_n}(p_1,\dots , p_n;p)\> =   |\cF_{f_1,\dots f_n}(p_1,\dots , p_n)\> \times (2\pi)^d \delta\left(p-{\textstyle\sum}_i p_i\right)\,.
\ee
The distributional form of the subleading soft theorem looks the same as before
\be\label{eq:sub_soft_theorem_dist}
\<\{a,\varepsilon_s\}|\boldsymbol{\cF}_{g,f_1,\dots f_n}(p_s,p_1,\dots , p_n;p)\>  \simeq g \tilde{\mu}^{\epsilon}\left[ S^{(0)} + S^{(1)}
 +\cdots \right]|\boldsymbol{\cF}_{f_1,\dots f_n}(p_1,\dots , p_n;p)\>\,,
\ee
with the same definitions of $S^{(0)}$ and $S^{(1)}$. (The fact that (\ref{eq:sub_soft_theorem_dist}) looks the same as (\ref{eq:sub_soft_theorem}) is a consequence of the specific form of $S^{(0)}$ and $S^{(1)}$ \cite{Broedel:2014fsa}.)
Here, both sides should be interpreted as distributions to be integrated against smooth test functions of independent on-shell momenta $p_1,\dots,p_n$, and off-shell $p$. 

An inner product between tree-level form factors, after summing over soft gluon polarizations, has the following universal terms in the soft gluon limit
\be
&\<\boldsymbol{\cF}_{g,f_1, \dots, f_n}(p_s,p_1,\dots, p_n;p')|\boldsymbol{\cF}_{g,f_1, \dots, f_n}(p_s,p_1,\dots, p_n;p)\> \nn\\
=& -g^2\tilde{\mu}^{2\epsilon}\sum_{i,j}\cS_{ij}(p_s)\<\boldsymbol{\cF}_{f_1,\dots, f_n}(p_1,\dots p_n;p')|T_i^a T_j^a|\boldsymbol{\cF}_{f_1,\dots, f_n}(p_1,\dots, p_n;p)\>\nn\\
&+ g^2\tilde{\mu}^{2\epsilon} \sum_{\varepsilon_s}\left[\<S^{(0)}\boldsymbol{\cF}'_{f_1,\dots, f_n}|S^{(1)}\boldsymbol{\cF}_{f_1,\dots, f_n}\> + \<S^{(1)}\boldsymbol{\cF}'_{f_1,\dots, f_n}|S^{(0)}\boldsymbol{\cF}_{f_1,\dots, f_n}\>\right] + \dots\,, \label{eq:sub_soft_theorem_M2}
\ee
where we use the notation $S^{(i)}|\boldsymbol{\cF}\> = |S^{(i)}\boldsymbol{\cF}\>$. Here $\boldsymbol{\cF}'$ indicates a form factor with total momentum $p'$. The first term in (\ref{eq:sub_soft_theorem_M2}) is the square of the leading soft theorem computed in \eqref{eq:M2_soft_thm}, and the second term is the interference between leading and subleading soft contributions. 

Let us focus on the second term, and integrate over the non-soft momenta:
\be
&g^2\tilde{\mu}^{2\epsilon}  \frac{1}{\symF} \int d\mathrm{LIPS}'_n\sum_{\varepsilon_s}\left[\<S^{(0)}\boldsymbol{\cF}'_{f_1,\dots, f_n}|S^{(1)}\boldsymbol{\cF}_{f_1,\dots, f_n}\> + \<S^{(1)}\boldsymbol{\cF}'_{f_1,\dots, f_n}|S^{(0)}\boldsymbol{\cF}_{f_1,\dots, f_n}\>\right] \nn\\
&= g^2\tilde{\mu}^{2\epsilon} \frac{1}{\symF} \int d\mathrm{LIPS}'_n\sum_{\varepsilon_s}\left[\<\boldsymbol{\cF}'_{f_1,\dots, f_n}|S^{(0)\dag} S^{(1)} + S^{(1)\dag}S^{(0)}|\boldsymbol{\cF}_{f_1,\dots, f_n}\>\right].
\label{eq:nowintegratingovernonsoft}
\ee
Here, $K$ is a symmetry factor accounting for permutations of identical partons and $d\mathrm{LIPS}'_n$ denotes the Lorentz-invariant phase space without the momentum conservation constraint (since that is included in $|\boldsymbol{\cF}\>$ itself): 
\be
d\mathrm{LIPS}^\prime_n = \prod_{i=1}^n \frac{d^{d-1}\vec{p}_i}{(2\pi)^{d-1}2E_i} = \prod_{i=1}^n \frac{d^d p_i \delta^+(p_i)}{(2\pi)^{d-1}}\,,
\ee
In (\ref{eq:nowintegratingovernonsoft}), $S^{(k)\dag}$ denotes the adjoint of $S^{(k)}$ with respect to the inner product on final state particles (involving a sum over color indices, a sum over helicities, and integration with respect to $d\mathrm{LIPS}'_n$).

These adjoints are simple to calculate. The key observation is that the inner product on final state particles factorizes into a tensor product of inner products for each individual particle. (This is a benefit of using the distributional form of $|\boldsymbol{\cF}\>$). Furthermore, the individual inner products are Lorentz-invariant. Hence, each angular momentum generator $J_i^{\mu\nu}$ is self-adjoint:
\be
(J_i^{\mu\nu})^\dag = J_i^{\mu\nu}.
\ee
We can then calculate
\be
{S^{(1)}}^\dagger &= i \sum_{i=1}^n T_i^a J_i^{\mu\nu} {\varepsilon_{s\mu}^* p_{s_\nu}\over p_s\cdot p_i}
= i \sum_{i=1}^n T_i^a  {\varepsilon_{s\mu}^* p_{s_\nu}\over p_s\cdot p_i} J_i^{\mu\nu}  + i \sum_{i=1}^n T_i^a \varepsilon_{s\mu}^* p_{s_\nu} \left[J_i^{\mu\nu}, {1 \over p_s\cdot p_i}\right] \nonumber\\
&= i \sum_{i=1}^n T_i^a  {\varepsilon_{s\mu}^* p_{s_\nu}\over p_s\cdot p_i} J_i^{\mu\nu} + \sum_{i=1}^n T_i^a \varepsilon_{s\mu}^* p_{s_\nu} {p_i^{[\mu} p_s^{\nu]}\over (p_s\cdot p_i)^2}
= -S^{(1)}\big|_{\varepsilon_s \to \varepsilon_s^*}\,,
\ee
where we have used the on-shell conditions for the soft gluon: $p_s^2 = \varepsilon_s \cdot p_s = 0$. In other words, $S^{(1)}$ is skew-adjoint  (up to $\varepsilon_s\to \varepsilon_s^*$). Meanwhile, the leading soft operator $S^{(0)}$ is self-adjoint (up to $\varepsilon_s\to \varepsilon_s^*$)
\be
{S^{(0)}}^\dagger = S^{(0)}\big|_{\varepsilon_s \to \varepsilon_s^*}\,.
\ee
As a consequence, the operator in the last line of \eqref{eq:sub_soft_theorem_M2} simplifies to a commutator
\be\label{eq:S0S1_commutator}
\sum_{\varepsilon_s} {S^{(1)}}^\dagger S^{(0)}+{S^{(0)}}^\dagger S^{(1)} = \sum_{\varepsilon_s} [{S^{(0)}}^\dagger,S^{(1)}]\,,
\ee

 Using the commutation relation $[T^a_i, T^a_j]=0$ and polarization sum \eqref{eq:Pi_lightconegauge} for the soft gluon, we can further simplify (\ref{eq:S0S1_commutator}), giving
\be\label{eq:S0S1_commutator_simplified}
\sum_{\varepsilon_s} [{S^{(0)}}^\dagger,S^{(1)}] &= \sum_{i,j=1}^n \sum_{\varepsilon_s} T_i^a T_j^a {\varepsilon_{s\mu} p_{s\nu}  \over p_s\cdot p_j} \left[ {\varepsilon^*_s\cdot p_i \over p_s\cdot p_i}, -i  J_j^{\mu\nu}\right]= \sum_{i,j=1}^n \sum_{\varepsilon_s} T_i^a T_j^a \de_{ij} \frac{g_{\mu\nu}\varepsilon_s^{\mu}\varepsilon_s^{*\nu}}{p_s\cdot p_i} \nn\\
& = -(d-2)\sum_{i=1}^n  \frac{C_i}{p_s\cdot p_i} \,.
\ee
An important feature is that the commutator involving $J^{\mu\nu}_j$ vanishes when the particle labels are not the same $i\neq j$. This is the key point that leads to the appearance of the shadow DGLAP operator. To better appreciate this point, we can compare to the leading soft contribution at cross section level, where the soft gluon connects two different energetic particles, leading to the BFKL kernel. By contrast, the subleading soft contribution effectively arises from the emission from a single energetic particle, as shown by the typical scalar propagator $\frac{1}{p_s\cdot p_i}$ that produces a soft gluon and an on-shell energetic particle.  

\subsection{Pole of the DGLAP detector at $J_L=-1$}

Now let us discuss the relation between the $J_L=-1$ pole of DGLAP detector and the subleading soft theorem. A matrix element of $\cD^{\DGLAP}_{J_L,g}(z)$ can be written as the moment of the gluon energy distribution $f_g(E)$ in the final state
\be
\<\cD^{\DGLAP}_{J_L,g}(z)\> = \int {E^{-J_L} dE \over (2\pi)^{d-1}2E} f_g(E)\,.
\ee
The contribution from an $(n+1)$-point form factor $\cF_{n+1}$ to $f_g(E)$ is
\be
(2\pi)^d \de(p-p') f_g(E) &= \frac{1}{\symF}\int d\mathrm{LIPS}'_n \<\boldsymbol{\cF}_{n+1}(Ez,p_1,\dots, p_n;p')|\boldsymbol{\cF}_{n+1}(Ez,p_1,\dots, p_n;p)\>.
\ee

A pole in $\<\cD^{\DGLAP}_{J_L,g}\>$ at $J_L=-1$ comes from the $1/E$ term in the Laurent expansion $f_g(E)$:
\be
\label{eq:easyintegral}
\int {E^{-J_L} dE \over (2\pi)^{d-1}2E} \p{\frac{f_g^{(2)}}{E^2} + \frac{f_g^{(1)}}{E} + \dots} &= -\frac{1}{J_L+1}\frac{f_g^{(1)}}{2(2\pi)^{d-1}} + O((J_L+1)^0).
\ee
Furthermore, this term is predicted by the subleading soft theorem using (\ref{eq:nowintegratingovernonsoft}), (\ref{eq:S0S1_commutator}), and (\ref{eq:S0S1_commutator_simplified}):
\be
&(2\pi)^d \de(p-p') \frac{f_g^{(1)}}{E} \nn\\
&= -(d-2)g^2\tilde{\mu}^{2\epsilon} \frac{1}{\symF} \int d\mathrm{LIPS}'_n\sum_{i=1}^n \frac{C_i}{Ez\.p_i} \sum_{\varepsilon_s}\<\boldsymbol{\cF}'(p_1,\dots,p_n;p')|\boldsymbol{\cF}(p_1,\dots,p_n;p)\> \nn\\
&= (2\pi)^d \de(p-p') \p{-(d-2)\frac{g^2\tilde{\mu}^{2\epsilon}}{E} \frac{1}{\symF} \int d\mathrm{LIPS}_n\sum_{i=1}^n \frac{C_i}{z\.p_i} \sum_{\varepsilon_s}|\cF(p_1,\dots,p_n;p)\>|^2},
\label{eq:contributiontoffromsubleadingsoft}
\ee
where in the last line we converted from $|\boldsymbol{\cF}\>$ to $|\cF\>$, pulling out an overall momentum-conserving $\de$-function. Note that the $d\mathrm{LIPS}_n$ measure in the final line includes the conventional factor $(2\pi)^d\de(p-\sum_i p_i)$.

We would like to interpret (\ref{eq:contributiontoffromsubleadingsoft}) as the expectation value of an operator. Recall that the spin-shadow transform is defined by  \cite{Kravchuk:2018htv}
\be
\mathbf{S}_J[\cD_{J_L}](z) = \int D^{d-2}z'(2z\.z')^{2-d-J_L}\cD_{J_L}(z').
\ee
Under $\mathbf{S}_J$, the Lorentz spin transforms as $J_L \to 2-d-J_L$. It is convenient to define a normalized version of the spin shadow transform
\be\label{eq:SJhat_definition}
\hat{\mathbf{S}}_J = \frac{\G(-J_L)}{\pi^{\frac{d-2}{2}}\G(\tfrac{2-d}{2}-J_L)}\mathbf{S}_J,
\ee
so that it satisfies $\hat{\mathbf{S}}_J^2=1$. Our definition of the shadow DGLAP trajectory is
\be
\tl{\cD}_{J_L}(z) \equiv \hat{\mathbf{S}}_J[\cD_{2-d-J_L}](z).
\ee
Note that this definition implies $ \hat{\mathbf{S}}_J[\tl{\cD}_{2-d-J_L}]=\cD_{J_L}$ due to the property $\hat{\mathbf{S}}_J^2=1$. The shadow detector $\tl{\cD}_{J_L}$ has spin $J_L$, and its scaling dimension is the same as that of $\cD_{2-d-J_L}$.

One can show that the kernels $1/z\.p_i$ appearing in (\ref{eq:contributiontoffromsubleadingsoft}) are shadow transforms of DGLAP vertices:
\be
2\pi \de^+(k^2) \frac{1}{z\.k} &= 2\mathbf{S}_J[V_{3-d}](z;k) = \frac{2\pi^{1-\e}\G(-\e)}{\G(1-2\e)} \hat{\mathbf{S}}_J[V_{3-d}](z;k),
\ee
where
$V_{J_L}(z;k)$ is the DGLAP vertex defined in (\ref{eq:VJL_definition}). On the right-hand side, $\hat{\mathbf{S}}_J[V_{3-d}]$ is the vertex for the shadow trajectory $\tl \cD_{J_L=-1}$, at $J_L=-1$. It follows from (\ref{eq:easyintegral}) and (\ref{eq:contributiontoffromsubleadingsoft}) that, in the case where all particles are gluons so that $C_i=C_A$, we have
\be\label{eq:DGLAPg_JLminus1_pole}
\<\cD^{\DGLAP}_{J_L,g}(z)\>^{\text{1-loop}} = \frac{\a_s\mu^{2\e}}{4\pi}\frac{C_A \cS_1(\e)}{J_L+1}\<\tl{\cD}^{\DGLAP}_{J_L,g}(z)\>^{\text{tree}} + O((J_L+1)^0),
\ee
where
\be\label{eq:S1_expr}
\cS_1(\e) = \frac{4^{1+\e} \pi^{\e}(1-\e)(\tilde{\mu}/\mu)^{2\e}\G(-\e)}{\G(1-2\e)}.
\ee

Applying the spin shadow $\hat{\mathbf{S}}_J$ to \eqref{eq:DGLAPg_JLminus1_pole}, we see conversely that $\<\tl{\cD}^{\DGLAP}_{J_L,g}\>^{\text{1-loop}}$ has a pole at $J_L=-1+2\e$ proportional to $\<\cD^{\DGLAP}_{J_L,g}\>^{\text{tree}}$,
\be\label{eq:DGLAPShadowg_JLminus1_pole}
\<\tl{\cD}^{\DGLAP}_{J_L,g}(z)\>^{\text{1-loop}} = -\frac{\a_s\mu^{2\e}}{4\pi}\frac{C_A \cS_1(\e)}{J_L+1-2\e}\<\cD^{\DGLAP}_{J_L,g}(z)\>^{\text{tree}} + O((J_L+1-2\e)^0).
\ee
Finally, note that when $J_L=\frac{2-d}{2}$, the normalized spin shadow $\hat {\mathbf{S}}_J$ acts as the identity, so that 
\be\label{eq:Shadowsymmetricpoint_identity}
\tl{\cD}_{\frac{2-d}{2}} = \cD_{\frac{2-d}{2}}.
\ee

\subsection{Chew-Frautschi plot of pure YM at one-loop}\label{sec:CFplot_YM}

We are now ready to resolve the mixing between the DGLAP trajectory with its shadow trajectory. The identity \eqref{eq:Shadowsymmetricpoint_identity} and the two $J_L$-poles \eqref{eq:DGLAPg_JLminus1_pole}, \eqref{eq:DGLAPShadowg_JLminus1_pole} are exactly analogous to the three important effects leading to the DGLAP/BFKL mixing described in section \ref{sec:mixing_structure}. For the DGLAP/shadow DGLAP mixing, the physical origin of these three effects are different. $\tl{\cD}_{\frac{2-d}{2}} = \cD_{\frac{2-d}{2}}$ is an identity for the spin shadow transform itself, and the two $J_L$-poles are both due to the subleading soft theorem. Nonetheless, the renormalization approach for resolving the mixing will be the same. 

We can consider the following basis:
\be\label{eq:regular_basis_DGLAPwithShadow}
U_2\begin{pmatrix} \mu^{J_L+2-2\e}\cD^{\DGLAP}_{J_L,g} \\ \mu^{-J_L}\tl{\cD}^{\DGLAP}_{J_L,g} \end{pmatrix}, \quad U_2 = \begin{pmatrix}1 & 0 \\ -\frac{1}{J_L+1-\e} & \frac{1}{J_L+1-\e} \end{pmatrix}
\ee
which is nondegenerate near the tree-level intersection $J_L\sim -1$. One can check that this basis has finite tree-level and one-loop matrix elements at $J_L=-1+\e$. As discussed in section \ref{sec:matrixelements_conditions}, this is a nontrivial consistency check on our results for the $\e$-poles and $J_L$-poles near the intersection. Moreover, the renormalization matrix that removes all the divergences is given by
\be\label{eq:DGLAPwithShadow_DivMat}
&\cZ_{J_L} = 1 + \frac{\a_s}{4\pi}M^{\DGLAP/\text{ShadowDGLAP}}_{J_L}\, , \nn \\
&M^{\DGLAP/\text{ShadowDGLAP}}_{J_L}\nn \\
&= \begin{pmatrix} -\frac{4C_A}{\e(J_L+1)} + \frac{C_A\tl \cS_1(\e)}{J_L+1} + \frac{\g_3(J_L)}{\e} && -\frac{(-4C_A+\e C_A\tl \cS_1(\e))}{J_L+1} \\ \frac{8C_A}{\e(J_L+1)(J_L+1-2\e)} + \frac{C_A\tl \cS_1(\e)}{\e(J_L+1)} - \frac{C_A\tl \cS_1(\e)}{\e(J_L+1-2\e)} + \frac{\frac{\g_{3}(-2-J_L)-\g_3(J_L)}{J_L+1}}{\e} && \frac{4C_A}{\e(J_L+1)} - \frac{C_A\tl \cS_1(\e)}{J_L+1} + \frac{\g_3(-2-J_L)}{\e} \end{pmatrix},
\ee
where
\be
\g_3(J_L) &= \hat\g^{(0)}_{gg}(J_L) + \frac{4C_A}{J_L+1}, \label{eq:gammagg_atJLminus1} \\
\tl{\cS}_1(\e)&= \cS_1(\e)+\frac{4}{\e}.
\ee

So far, we have resolved the DGLAP/BFKL intersection at $J_L\sim -2$ and the DGLAP/shadow DGLAP intersection at $J_L\sim -1$. In fact, we have also resolved the intersection between the shadow DGLAP and the BFKL trajectories at $J_L\sim 0$, because it is just the spin shadow of the DGLAP/BFKL case. It turns out that we can combine these results and obtain the renormalized Regge trajectories for all $J_L$ from $0$ to $-2$. To do this, we first need to find a basis that is nondegenerate near all the intersections. Let us define
\be\label{eq:regular_basis_pureYM}
\mathbb{D}_{J_L} &= U_{\text{YM}}\begin{pmatrix} \vspace{1mm} \mu^{J_L+2-2\e}\cD^{\DGLAP}_{J_L,g} \\ \vspace{1mm} \mu^{-J_L}\tl{\cD}^{\DGLAP}_{J_L,g} \\ \cD^{\BFKL}_{J_L,g} \end{pmatrix}, \nn \\
U_{\text{YM}}&=\begin{pmatrix} \vspace{1mm} 1 & 1 & 0 \\ \vspace{1mm} -\frac{1}{J_L+1-\e} & \frac{1}{J_L+1-\e} & 0 \\ \frac{C_A\frac{\pi^{1-\e}}{\G(1-\e)}}{(-2+2\e-J_L)(-2+2\e)} & \frac{C_A\frac{\pi^{1-\e}}{\G(1-\e)}}{J_L(-2+2\e)} & \frac{1}{J_L(-2+2\e-J_L)} \end{pmatrix}.
\ee
Due to \eqref{eq:equivalencedglapbfkl} and \eqref{eq:Shadowsymmetricpoint_identity}, the matrix elements of $\mathbb{D}_{J_L}$ are indeed finite and linearly independent at $J_L\sim -2, -1, 0$. The multiplet is also defined such that it satisfies
\be\label{eq:basis_SJ_identity}
\hat{\mathbf{S}}_J[\mathbb{D}_{J_L}] = \mathbb{D}_{-2+2\e-J_L}.
\ee

The renormalized detectors are defined by
\be
[\mathbb{D}_{J_L}]_R = \cZ_{J_L}^{-1}\mathbb{D}_{J_L},
\ee
and our goal is to find the matrix $\cZ_{J_L}$ such that the renormalized detectors have finite matrix elements for all $J_L\sim -2,-1,0$ and $\e=0$. We find that the simplest way to obtain $\cZ_{J_L}$ is to take the previous results which resolve the three intersections separately, and then patch them together. For example, let us first consider the DGLAP/BFKL intersection. The renormalization matrix that removes the divergences near this intersection is given by \eqref{eq:DGLAPBFKL_DivMat}. The idea is to take this matrix and then rewrite it in the new basis \eqref{eq:regular_basis_pureYM}, and the resulting matrix should remove all the divergences in $J_L$ at $J_L=-2,-2+4\e$. However, when performing the change of basis, we could also produce additional divergences near $J_L\sim 0,-1$. These divergences should be removed since we only want this matrix to have divergences at $J_L=-2,-2+4\e$. We can remove them by imposing the following replacement rules:
\be\label{eq:JLsubtraction_rules}
\frac{F(J_L)}{(J_L+2)(J_L+2-4\e)} &\to \frac{F(-2)}{(J_L+2)(-4\e)} + \frac{F(-2+4\e)}{(4\e)(J_L+2-4\e)}
\nn\\
\frac{F(J_L)}{J_L+2} &\to \frac{F(-2)}{J_L+2},\nn\\
 \frac{F(J_L)}{J_L+2-4\e} &\to \frac{F(-2+4\e)}{J_L+2-4\e}.
\ee
At the end, the matrix is given by
\be
M^{(1)}_{J_L}=\left.U_{\text{YM}}\, \begin{pmatrix}1 & 0  \\ 0 & 0 \\ 0 & 1 \end{pmatrix} U_1^{-1}\, M^{\DGLAP/\BFKL}_{J_L}\, U_1\,\begin{pmatrix}1 & 0 & 0 \\ 0 & 0 & 1 \end{pmatrix}  U_{\text{YM}}^{-1}\right|_{\eqref{eq:JLsubtraction_rules}},
\ee
where $M^{\DGLAP/\BFKL}_{J_L}$, $U_1$, and $U_{\text{YM}}$ are given in \eqref{eq:DGLAPBFKL_DivMat}, \eqref{eq:regular_basis_DGLAPBFKL}, and \eqref{eq:regular_basis_pureYM}, respectively.

Similarly, the matrix that removes all the divergences in $J_L$ at $J_L=-1,-1+2\e$ can be written as
\be
M^{(2)}_{J_L}=\left.U_{\text{YM}}\, \begin{pmatrix}1 & 0  \\ 0 & 1 \\ 0 & 0 \end{pmatrix} U_2^{-1}\, M^{\DGLAP/\text{ShadowDGLAP}}_{J_L}\, U_2\,\begin{pmatrix}1 & 0 & 0 \\ 0 & 1 & 0 \end{pmatrix}  U_{\text{YM}}^{-1}\right|_{\eqref{eq:JLsubtraction_rules_2}},
\ee
where $M^{\DGLAP/\text{ShadowDGLAP}}_{J_L}$ and $U_2$ are given in \eqref{eq:DGLAPwithShadow_DivMat} and \eqref{eq:regular_basis_DGLAPwithShadow}, and the replacement rules are
\be\label{eq:JLsubtraction_rules_2}
\frac{F(J_L)}{(J_L+1)(J_L+1-2\e)} &\to \frac{F(-1)}{(J_L+1)(-2\e)}+ \frac{F(-1+2\e)}{(2\e)(J_L+1-2\e)}, \nn\\
\frac{F(J_L)}{J_L+1} &\to \frac{F(-1)}{J_L+1},\nn\\
\frac{F(J_L)}{J_L+1-2\e} &\to \frac{F(-1+2\e)}{J_L+1-2\e}.
\ee
Lastly, the matrix that removes the divergences near the shadow DGLAP/BFKL intersection is simply the spin shadow of the matrix for the DGLAP/BFKL intersection. In other words, $M^{(1)}_{-2+2\e-J_L}$ removes all the $J_L$-divergences at $J_L=0,4\e$.

Combining $M^{(1)}_{J_L}, M^{(1)}_{-2+2\e-J_L}$, and $M^{(2)}_{J_L}$, we can remove all the $J_L$-poles near the three intersections. There will still be remaining $\e$-poles, and they can be removed by another matrix $M^{(\e)}_{J_L}$ which have no divergences in $J_L$ near $J_L\sim -2,-1,0$. The matrix can be determined by
\be\label{eq:Divmat_eps_div_eq}
\<\mathbb{D}_{J_L}\>^{\text{1-loop}} - \frac{\a_s}{4\pi}\p{M^{(1)}_{J_L} + M^{(1)}_{-2+2\e-J_L} + M^{(2)}_{J_L}}\<\mathbb{D}_{J_L}\>^{\text{tree}} = \frac{\a_s}{4\pi}M^{(\e)}_{J_L}\<\mathbb{D}_{J_L}\>^{\text{tree}} + O(\e^0).
\ee
For $M^{(\e)}_{J_L}$, we use a slightly different scheme such that it is invariant under spin shadow $J_L\to -2+2\e-J_L$.\footnote{If we use minimal subtraction scheme, $M^{(\e)}_{J_L}$ will be invariant under $J_L \to -2 -J_L$. Therefore, we need to add some finite terms to restore exact shadow symmetry. In practice, we first use minimal subtraction scheme to find $M^{(\e)}_{J_L}$, and then replace it with $\frac{M^{(\e)}_{J_L}+M^{(\e)}_{-2+2\e-J_L}}{2}$.} This is to ensure that $\cZ_{J_L}=\cZ_{-2+2\e-J_L}$, and the renormalized detectors also satisfy \eqref{eq:basis_SJ_identity}.

Our final result for the renormalization factor is
\be\label{eq:ZJL_YM_result}
\cZ_{J_L} = 1+\frac{\a_s}{4\pi} \p{M^{(1)}_{J_L} + M^{(1)}_{-2+2\e-J_L} + M^{(2)}_{J_L} +M^{(\e)}_{J_L}}.
\ee
This matrix removes all the $\e$-divergences and $J_L$-divergences in the one-loop matrix elements of the bare detectors $\mathbb{D}_{J_L}$ near the DGLAP/BFKL, DGLAP/shadow DGLAP, and shadow DGLAP/BFKL intersections.

Careful readers might notice there is actually a pole we have not yet addressed -- the pole of $\hat \g_{gg}(J_L)$ at $J_L=0$, which will show up as a $\frac{1}{\e J_L}$ pole in $\<\cD^{\DGLAP}_{J_L,g}\>^{\text{1-loop}}$ near $J_L\sim 0$. Indeed, this divergence is not removed by the renormalization matrix given above. It comes from an intersection between the DGLAP trajectory and another trajectory at $\De_L=2, J_L=0$, which is below the three intersections we have considered on the Chew-Frautschi plot. Similarly, the shadow DGLAP trajectory should intersect with another trajectory at $\De_L=2, J_L=-2$. In order to obtain a dilatation matrix that is finite at $J_L=-2$ and $J_L=0$, we also have to resolve these two intersections. 

Currently, we do not understand what these additional trajectories that mix with DGLAP at $J_L=0$ and shadow DGLAP $J_L=-2$ are, and solving this problem is beyond the scope of this paper. However, we also expect that, once we have resolved all the intersections and found the renormalized trajectories, the behavior of the trajectories near the three intersections we consider above should not be sensitive to the details of the additional unknown trajectories. Therefore, for the purpose of this work, we introduce a ``fake trajectory" which is a horizontal trajectory at $J=-1$, or equivalently $\De_L$=2.\footnote{We thank Petr Kravchuk for suggesting this idea.} This trajectory intersects with DGLAP at $J_L=0$ and with shadow DGLAP at $J_L=-2$, and we will design its matrix elements such that we can use the same method to remove the divergences in the one-loop matrix elements of the DGLAP and shadow DGLAP detectors near those intersections. Together with \eqref{eq:ZJL_YM_result}, this allows us to construct a renormalization matrix that removes all the divergences in $\e$ and $J_L$ near $J_L\sim -2,-1,0$. We give more details on how we construct this trajectory in appendix \ref{app:fake_trajectory}.

From the renormalization matrix, we can obtain the one-loop contribution to the dilatation operator. At tree-level, the dilatation operator acts on \eqref{eq:regular_basis_pureYM} as (ignoring the fake trajectory)
\be
\mathscr{D}_0 = \begin{pmatrix} -1+\e & (1+J_L-\e)^2 & 0 \\ 1 & -1+\e & 0 \\ \frac{C_A \pi^{1-\e}}{2(-1+\e)\G(1-\e)} & 0 & 0 \end{pmatrix}.
\ee 
At one-loop, the matrix can be obtained using \eqref{eq:Dilatation_oneloop_eq}. The solution to the characteristic equation of the matrix then gives the values of $\De_L$ of the renormalized Regge trajectories as a function of $J_L$.

We give the expressions of the renormalization factor $\cZ_{J_L}-1$ and the one-loop dilatation matrix in 4d $\left.\mathscr{D}\right|_{\e\to 0}$ in the ancillary file $\tt{RenormalizationMatrices.m}$. Due to the issue of fake trajectory discussed above, the expression of the renormalization matrix is not exactly the same as the one given in \eqref{eq:ZJL_YM_result}, and similarly for the dilatation matrix. See appendix \ref{app:anc} for more details.

\begin{figure}
    \centering
    \includegraphics[width = 10cm]{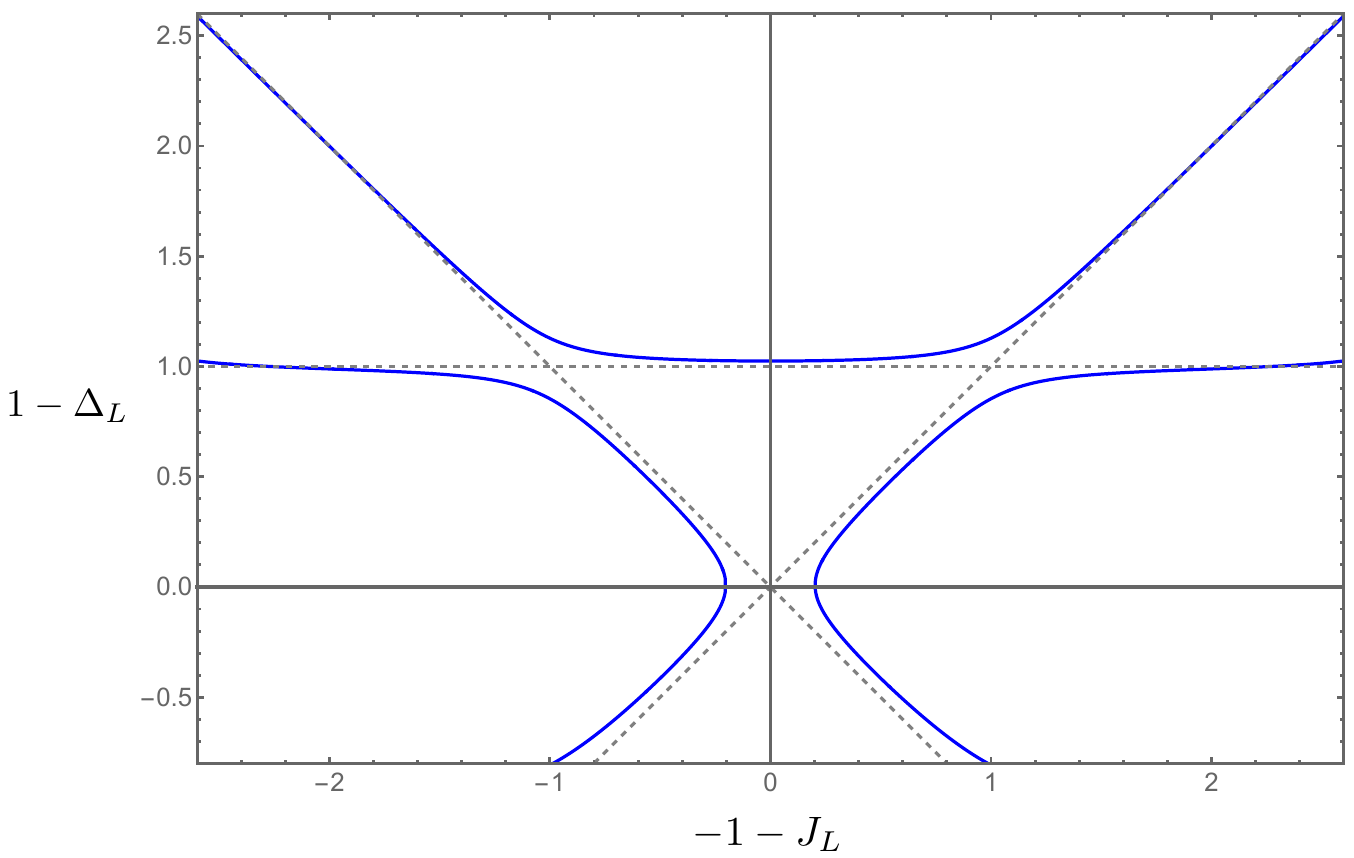}
    \caption{Chew-Frautschi plot of pure YM theory at one-loop for $\a_s=0.01$, where we have only plotted the points with real $\De_L$. Gray dashed lines are the free theory trajectories.}
    \label{fig:CFplot_YM}
\end{figure}

The renormalized Regge trajectories are plotted in figure \ref{fig:CFplot_YM}, where we only show the points with real $\De_L$. We also only plot in the region that is close to the three intersections and far enough from the other intersections which we do not consider (including the intersections with fake trajectory). Interestingly, near the intersection at $J_L\sim -1$ (which is the origin of the plot), the solutions of $\De_L$ are complex, and hence the trajectories move off to the complex plane. This phenomenon is related to the sign of the residue of the DGLAP anomalous dimension near this intersection. In particular, we find that the discriminant of the characteristic polynomial at $J_L=-1$ is given by
\be
\left.\text{Discr} \p{\text{Det}(\mathscr{D}+\De_L)}\right|_{J_L=-1}= \frac{16\a_s}{\pi}(-4C_A) + O(\a_s^2),
\ee
and therefore there are two complex solutions. The $-4C_A$ factor above is from the residue of the $J_L$-pole of $\hat \g_{gg}$ at $J_L=-1$ (see \eqref{eq:gammagg_atJLminus1}). 

One can also determine the size of the anomalous dimension near the DGLAP/shadow DGLAP intersection. Due to the orientation of the trajectories, it is better to switch to the $(\De,J)$ coordinates using $\De=1-J_L, J=1-\De_L$, and then compute the value of $\De$ at $J=0$. We find that the result also has a square-root enhancement, and it is given by
\be
\De(J=0) = 2 \pm 2\sqrt{\frac{\a_sC_A}{\pi}} + O(\a_s^{3/2}).
\ee

\section{Trajectory intersections in QCD}\label{sec:QCD}

In this section, we generalize the previous discussion on BFKL/DGLAP mixing from Yang-Mills theory to QCD. Although we primarily focus on calculations involving fundamental quarks, the analysis can easily be extended to other matter fields, including scalars and/or fermions in arbitrary color representations.

\subsection{The BFKL detector in QCD}

To build a version of the BFKL detector in QCD, we could start with gluon color detectors $\cN_g^c(z)$  defined in \eqref{eq:Ng_definition} or analogous quark color detectors $\cN_q^c(z)$, and then combine products of them to make $gg$, $qg$, or $qq$ analogs of the BFKL detector. However, it turns out that DGLAP trajectories only mix (at one loop) with BFKL operators built from the ``total" color detector
\be
\cN^c(z) = \cN^c_{g}(z) + \cN_{q}^c(z).
\ee
This is a consequence of the universality of soft gluon emission, shown in \eqref{eq:soft_theorem} and \eqref{eq:M2_soft_thm}. The soft factors depend solely on the color representation of the parent particles and do not depend on their flavors or spins.
Indeed, we will find a consistent story for BFKL/DGLAP mixing at one loop using only the total color detector $\cN^c$ and not $\cN^c_{g}$ or $\cN_{q}^c$ individually.

Individual $gg$, $gq$, or $qq$ BFKL-type operators also exhibit unfamiliar properties under renormalization. They have $1/\e^2$ divergences at $1$-loop that do not cancel between real emission and virtual diagrams. (By contrast, $1/\e^2$ divergences cancel when we use $\cN^c$.) Consequently, renormalization for these operators must work in a different way. We leave detailed investigation of these operators to future work.

For now, we focus on BFKL operators built from the total color detector $\cN^c(z)$. The 
quark component $\cN_{q}^c(z)$ incorporates all possible flavor combinations of fundamental quarks and anti-quarks\footnote{The generic construction for particles in color representation $\mathbf{R}$ is 
\be
\cN_{\mathbf{R}}^c(z) = \int_0^\oo \frac{E^{d-2}dE}{(2\pi)^{d-1}2E} \left[a_{\mathbf{R}}^\dagger(p) T^c_{\mathbf{R}} a_{\mathbf{R}}(p)\right]\Big|_{p^\mu=E z^\mu} \,,\nn
\ee
where we suppress the spin-index summation and implicitly contract color indices.}
\be
\cN_{q}^c(z) = \sum_{I=1}^{n_f} \sum_{s,i,j}\int_0^\oo \frac{E^{d-2}dE}{(2\pi)^{d-1}2E} \left[t_{ij}^c b^\dagger_{s,i,I}(p) b_{s,j,I}(p) - t_{ji}^c d^\dagger_{s,i,I}(p) d_{s,j,I}(p)\right]\Bigg|_{p^\mu=E z^\mu} \,,
\ee
where $I$ is a flavor index, $s$ a spin index, and $i,j$ are fundamental color indices. While we typically suppress the flavor index for simplicity, we retain it here to emphasize the importance of summing over all flavor contributions. Using $\cN^c(z)$ as the building block, we follow \eqref{eq:DBFKL_definition} to construct the BFKL detector in QCD:
\be\label{eq:DBFKL_def_qcd}
\cD^{\BFKL}_{J_L}(z) =  
\frac{\G(d-2+J_L)}{\G(\tfrac{d-2+J_L}{2})^2} \int D^{d-2}z_1 D^{d-2}z_2 \p{\frac{2z_1\.z_2}{(2z_1\.z)(2z_2\.z)}}^{-\frac{J_L}{2}}:\!\cN^c(z_1)\cN^c(z_2)\!:.
\ee
The matrix element of this operator is given by \eqref{eq:BFKL_generic_mel}, with $T_i^c$ being the color generator for the representation of the $i$-th particle.

In pure Yang-Mills theory, $\cD^{\DGLAP}_{J_L,g}$ is the only DGLAP detector and thus directly mixes with BFKL detector near $J_L=-2$. In contrast, in QCD, $\cD^{\DGLAP}_{J_L,g}$ mixes with $\cD^{\DGLAP}_{J_L,q}$ at all possible values of $J_L$ (see section~\ref{sec:DGLAP_1loop}). Based on the discussion of tree-level degeneracy at $J_L=2-d$ in section~\ref{eq:equalityatintersection}, we expect only one special linear combination can mix with the BFKL detector \eqref{eq:DBFKL_def_qcd}.  Inserting \eqref{eq:BFKL_JL_2minusd} into a generic matrix element corresponds to the reduction of \eqref{eq:BFKL_generic_mel} at $J_L=2-d$
\be
\<\cD^{\BFKL}_{2-d}(z)\> &= \frac{\vol(S^{d-3})}{2} \sum_n \sum_{X_n} \frac{1}{\symF_{X_n}}\sum_{\substack{1\leq i,j\leq n\\ i\neq j}} \int {d^d p_i \over (2\pi)^d} \left[\prod_{l\neq  i} {d^d p_l \delta^+(p_l^2)\over (2\pi)^{d-1}}\right] (2\pi)^d \delta(p-\sum_{l=1}^n p_l)\nn\\
&\hspace{4cm}\times \pi \int_0^\oo  \b^{d-3}d\b\,  \de^{(d)}(p_i-\b z)\<\cF_{X_n}|T^c_i T^c_j|\cF_{X_n}\>\,.
\ee
We can apply the color conservation identity $\sum_{j\neq i}T^c_j|\cF_{X_n}\> = - T_i^c|\cF_{X_n}\>$ to simplify the integrand
\be
\sum_{j\neq i}\<\cF_{X_n}|T^c_i T^c_j|\cF_{X_n}\> = - \<\cF_{X_n}|T^c_i T^c_i|\cF_{X_n}\> = - C_i \<\cF_{X_n}|\cF_{X_n}\>\,.
\ee
Now we can identify the matrix element $\<\cD^{\BFKL}_{2-d}(z)\>$ as a linear combination of DGLAP detectors
\be
\label{eq:weseeaspeciallinearcombination}
\<\cD^{\BFKL}_{2-d}(z)\> =-\frac{\vol(S^{d-3})}{2}\left[ C_A \<\cD^{\DGLAP}_{2-d,g}(z)\> + C_F\<\cD^{\DGLAP}_{2-d,q}(z)\>\right]\,.
\ee

The linear combination (\ref{eq:weseeaspeciallinearcombination}) can also be obtained from the divergence of the 1-loop DGLAP anomalous dimension near $J_L=2$:
\be
\hat{\gamma}^{(0)}_{ij}(J_L) = \begin{pmatrix}
    \hat{\gamma}^{(0)}_{qq}(J_L) & \hat{\gamma}^{(0)}_{qg}(J_L)\\
    \hat{\gamma}^{(0)}_{gq}(J_L) & \hat{\gamma}^{(0)}_{gg}(J_L)
\end{pmatrix}_{ij} 
= \frac{4}{J_L+2} \begin{pmatrix}
    0 & 0 \\
    C_F & C_A 
\end{pmatrix}_{ij} + O((J_L+2)^0)\,,
\ee
where we choose the operator basis to be $(\cD^{\DGLAP}_{J_L,q}, \cD^{\DGLAP}_{J_L,g})^T$. We see that only one eigenvalue of the DGLAP anomalous dimension matrix diverges at $J_L = -2$. The corresponding left eigenvector is $(C_F, C_A)$, which corresponds to the combination $C_F \cD^{\DGLAP}_{J_L,q} + C_A \cD^{\DGLAP}_{J_L,g}$. As a result, while DGLAP/BFKL mixing naively occurs between three trajectories, it effectively reduces to a two-trajectory problem  at LO near $J_L=-2$. 

\subsection{One-loop divergences of DGLAP and BFKL detectors in QCD}\label{subsec:QCD-divergences}

To resolve mixing of different trajectories, we need divergences of their matrix elements, both as a function of $\e$ and $J_L$. We have already seen examples of calculating these divergences in sections~\ref{sec:DGLAP_1loop}, \ref{sec:mixing_structure} and \ref{sec:BFKL_1loop} for pure Yang-Mills theory. There are no essential differences in the QCD case, beyond simply including diagrams with quarks. We leave the details of our calculations to appendix~\ref{app:QCD_1loop}.

To disentangle quark/gluon mixing, we use both the electromagnetic current $J_\mu$ and Yang-Mills Lagrangian density operator $\cO = {1 \over 4 N_c} \textrm{Tr}(F_{\mu\nu}F^{\mu\nu})$ as source operators. The non-vanishing tree-level matrix elements for DGLAP detectors are given in \eqref{eq:DDGLAP_tree_matrixelement_gluon} and \eqref{eq:DDGLAP_tree_matrixelement_quark}, while the BFKL cases are
\be
\<\cD^{\BFKL}_{J_L}(z)\>^{\text{tree}}_{\cO(p)} = -\frac{(d-2)C_A(N_c^2-1)}{2^{d+1}\pi^{\frac{d}{2}-1}\G(\tfrac{d-2}{2})}  (\signplus 2 z\cdot p)^{J_L}(\signplus p^2)^{\frac{d-J_L}{2}}\,,\\
\<\cD^{\BFKL}_{J_L}(z)\>^{\text{tree}}_{J(p)} = -\frac{(d-2)C_F N_c}{2^{d-3}\pi^{\frac{d}{2}-1}\G(\tfrac{d-2}{2})}  (\signplus 2 z\cdot p)^{J_L}(\signplus p^2)^{\frac{d-J_L-2}{2}}\,.
\ee

The $1/\epsilon$ poles for DGLAP detectors are listed in section~\ref{sec:DGLAP_1loop}. The $J_L=-2$ pole for gluon DGLAP detectors is given in \eqref{eq:DGLAP_JLpole_vs_BFKL}. It is independent of the source operator because it comes from the leading soft gluon theorem. In the same way, the quark DGLAP detector does not exhibit a pole at $J_L=-2$ because the soft limit of quark matrix elements is less divergent than those for gluons. The leading $J_L$ pole of $\cD^{\DGLAP}_{J_L,q}(z)$ occurs at $J_L=-1$, as we discuss in the next subsection.

To highlight the significance of the special combination $\cN=\cN_q+\cN_g$ in the BFKL detector \eqref{eq:DBFKL_def_qcd}, let us temporarily decompose the BFKL detector into three components
\be\label{eq:QCD_BFKL_decomposition}
\cD^{\BFKL}_{J_L}(z) = \cD^{\BFKL}_{J_L,gg}(z)+ 2\cD^{\BFKL}_{J_L,qg}(z) + \cD^{\BFKL}_{J_L,qq}(z)\,,
\ee
where $\cD^{\BFKL}_{J_L,ab}(z)$ is defined as
\be
\cD^{\BFKL}_{J_L,ab}(z) =  
\frac{\G(d-2+J_L)}{\G(\tfrac{d-2+J_L}{2})^2} \int D^{d-2}z_1 D^{d-2}z_2 \p{\frac{2z_1\.z_2}{(2z_1\.z)(2z_2\.z)}}^{-\frac{J_L}{2}}\cN^c_a(z_1)\cN^c_b(z_2).
\ee

We list the one-loop divergences for each component of the BFKL detectors, considering both the gluon source $\cO$ and the quark source $J$ separately.  
\begin{itemize}
    \item \textbf{Gluon source}\\
    The one-loop $\epsilon$ poles are
    \be
    \< \cD^{\BFKL}_{J_L,gg}(z) \>_{[\cO]_R(p)}^{\text{1-loop}} &= \frac{\alpha_s}{4\pi \epsilon}  \left(\gamma_{\BFKL}(J_L) + \frac{8 n_f T_F}{3}\right)\<\cD_{J_L}^{\BFKL}\>_{\cO(p)}^{\text{tree}}+O(\epsilon^0)\,,\\
    \< \cD^{\BFKL}_{J_L,qg}(z) \>_{[\cO]_R(p)}^{\text{1-loop}} &= -\frac{\alpha_s}{4\pi \epsilon}   \frac{4 n_f T_F}{3} \<\cD_{J_L}^{\BFKL}\>_{\cO(p)}^{\text{tree}}+O(\epsilon^0)\,,\\
    \< \cD^{\BFKL}_{J_L,qq}(z) \>_{[\cO]_R(p)}^{\text{1-loop}} &=O(\epsilon^0)\,.
    \ee
    The $J_L= -2+4\epsilon$ poles at 1-loop are 
    \be
    \< \cD^{\BFKL}_{J_L,gg}(z) \>_{[\cO]_R(p)}^{\text{1-loop}} &= \frac{\alpha_s \mu^{2\epsilon}}{4\pi} \frac{\cR_2(\epsilon)}{J_L+2-4\epsilon}\<\cD^{\DGLAP}_{J_L,g}(z) \>_{\cO(p)}^{\text{tree}}+O((J_L+2-4\epsilon)^0)\,,\\
    \< \cD^{\BFKL}_{J_L,qg}(z) \>_{[\cO]_R(p)}^{\text{1-loop}} &= O((J_L+2-4\epsilon)^0)\,,\\
    \< \cD^{\BFKL}_{J_L,qq}(z) \>_{[\cO]_R(p)}^{\text{1-loop}} &= \frac{\alpha_s \mu^{2\epsilon}}{4\pi} \frac{\cR_3(\epsilon)}{J_L+2-4\epsilon}\<\cD^{\DGLAP}_{J_L,g}(z) \>_{\cO(p)}^{\text{tree}}+O((J_L+2-4\epsilon)^0)\,,
    \ee
    where $\cR_2$ is given in \eqref{eq:R2_expr} and $\cR_3$ is
    \be\label{eq:R3_expr}
    \cR_3(\epsilon) = -2n_f T_F (C_F - C_A/2)\frac{2^{2+2 \epsilon} \pi \left(2-5 \epsilon + 4 \epsilon ^2 \right) (\mu/\tilde{\mu})^{2\epsilon}\Gamma (1-3\epsilon )}{(3-4 \epsilon ) \Gamma (2-4 \epsilon ) \Gamma (2-\epsilon )}\,.
    \ee
    \item \textbf{Quark source}\\
    The one-loop $\epsilon$ poles are
    \be 
    \< \cD^{\BFKL}_{J_L,gg}(z) \>_{[J]_R(p)}^{\text{1-loop}} &= 0\,,\\
    \< \cD^{\BFKL}_{J_L,qg}(z) \>_{[J]_R(p)}^{\text{1-loop}} &= \frac{\alpha_s \mu^{2\epsilon}}{4\pi (p^2)^{\epsilon}} \left(\frac{C_A}{\epsilon^2} +\frac{3 C_A+\gamma_\BFKL(J_L)}{2\epsilon}\right)\<\cD_{J_L}^{\BFKL}\>_{J(p)}^{\text{tree}}
    +O(\epsilon^0)\,\\
    \< \cD^{\BFKL}_{J_L,qq}(z) \>_{[J]_R(p)}^{\text{1-loop}} &=\frac{\alpha_s \mu^{2\epsilon}}{4\pi (p^2)^{\epsilon}} \left(-\frac{2C_A}{\epsilon^2} -\frac{3 C_A}{\epsilon}\right)\<\cD_{J_L}^{\BFKL}\>_{J(p)}^{\text{tree}}
    +O(\epsilon^0)\,
    \ee
    The $J_L= -2+4\epsilon$ poles at 1-loop are
    \be
    \< \cD^{\BFKL}_{J_L,gg}(z) \>_{[J]_R(p)}^{\text{1-loop}} &= 0\,,\\
    \< \cD^{\BFKL}_{J_L,qg}(z) \>_{[J]_R(p)}^{\text{1-loop}} &= \frac{\alpha_s \mu^{2\epsilon}}{4\pi}\frac{\cR_4(\epsilon)}{J_L+2-4\epsilon}\<\cD^{\DGLAP}_{J_L,q}(z) \>_{J(p)}^{\text{tree}}+O((J_L+2-4\epsilon)^0)\,,\\
    \< \cD^{\BFKL}_{J_L,qq}(z) \>_{[J]_R(p)}^{\text{1-loop}} &= O((J_L+2-4\epsilon)^0)\,,
    \ee
    where $\cR_4$ is
    \be\label{eq:R4_expr}
    \cR_4(\epsilon) =  -C_A C_F \frac{3 \pi   \left(\epsilon ^2-5 \epsilon +2\right) (\tilde{\mu}/\mu)^{2\epsilon}\Gamma (-3\epsilon )}{2^{-2 \epsilon }\Gamma (2-4 \epsilon ) \Gamma (1-\epsilon )}\,.
    \ee
\end{itemize}

We notice that the combination in \eqref{eq:QCD_BFKL_decomposition} ensures that the one-loop $\epsilon$-divergence is identical for both quark and gluon sources and reproduces the one-loop BFKL anomalous dimension $\gamma_\text{BFKL}(J_L)$
\be
\< \cD^{\BFKL}_{J_L}(z) \>_{[\cO]_R(p)}^{\text{1-loop}} &= \frac{\alpha_s}{4\pi }  \frac{\gamma_{\BFKL}(J_L)}{\epsilon} \<\cD_{J_L}^{\BFKL}\>_{\cO(p)}^{\text{tree}}+O(\epsilon^0)\,,\\
\< \cD^{\BFKL}_{J_L}(z) \>_{[J]_R(p)}^{\text{1-loop}} &= \frac{\alpha_s}{4\pi } \frac{\gamma_{\BFKL}(J_L)}{\epsilon}   \<\cD_{J_L}^{\BFKL}\>_{J(p)}^{\text{tree}}+O(\epsilon^0)\,.
\ee
This is particularly crucial for the matrix element with the quark source, as individual components can exhibit double poles in $\epsilon$ at one-loop order.

The 1-loop $J_L=-2+4\epsilon$ pole for the BFKL detector is proportional to the DGLAP detectors, and the source operators $\cO$ and $J$ project onto the coefficients of $\cD_{J_L,g}^{\DGLAP}$ and $\cD_{J_L,q}^{\DGLAP}$, respectively:
\be
\< \cD^{\BFKL}_{J_L}(z) \>_{[\cO]_R(p)}^{\text{1-loop}} &= \frac{\alpha_s \mu^{2\epsilon}}{4\pi} \frac{\cR_2(\epsilon)+\cR_3(\epsilon)}{J_L+2-4\epsilon}\<\cD^{\DGLAP}_{J_L,g}(z) \>_{\cO(p)}^{\text{tree}}+O((J_L+2-4\epsilon)^0)\,,\\
\< \cD^{\BFKL}_{J_L}(z) \>_{[J]_R(p)}^{\text{1-loop}} &= \frac{\alpha_s \mu^{2\epsilon}}{4\pi} \frac{2\cR_4(\epsilon)}{J_L+2-4\epsilon}\<\cD^{\DGLAP}_{J_L,q}(z) \>_{\cO(p)}^{\text{tree}}+O((J_L+2-4\epsilon)^0)\,.
\ee
As a consistency check, we can calculate the ratio of these two coefficients  in the $\epsilon\to 0$ limit
\be
\lim_{\epsilon\to 0} {\cR_2(\epsilon)+\cR_3(\epsilon) \over 2\cR_4(\epsilon)} = \frac{C_A}{C_F}\,,
\ee
which again reflects the degeneracy of $\cD^{\BFKL}_{J_L}$ and $C_A \cD^{\DGLAP}_{J_L,g}+C_F \cD^{\DGLAP}_{J_L,q}$ at their intersection. These formulas directly predict the poles associated to mixing between the shadow DGLAP and BFKL trajectories, upon taking the shadow transform of both sides and replacing $J_L\to 2-d-J_L$. For example,
\be
\< \cD^{\BFKL}_{J_L}(z) \>_{[\cO]_R(p)}^{\text{1-loop}} &= \frac{\alpha_s \mu^{2\epsilon}}{4\pi} \frac{\cR_2(\epsilon)+\cR_3(\epsilon)}{-J_L-2\epsilon}\<\tl\cD^{\DGLAP}_{J_L,g}(z) \>_{\cO(p)}^{\text{tree}}+O((-J_L-2\epsilon)^0)\,.
\ee

\subsection{Mixing between DGLAP detectors and their shadows}

In section~\ref{sec:shadowDGLAP}, we explained how the gluon DGLAP detector mixes with its shadow at $J_L=-1$, and the relation with the subleading soft theorem. The 1-loop $J_L=-1$ pole can be summarized as
\be\label{eq:Dg_JLminus1pole}
\<\cD^{\DGLAP}_{J_L,g}(z)\> = \frac{\a_s\mu^{2\e}}{4\pi}\frac{\cS_1(\e)}{J_L+1}\left[C_A \<\tl{\cD}^{\DGLAP}_{J_L,g}(z)\> + C_F \<\tl{\cD}^{\DGLAP}_{J_L,q}(z)\> \right]\Big|_{J_L=-1}+ O((J_L+1)^0)\,,
\ee
where $\cS_1(\e)$ is given in \eqref{eq:S1_expr}. For the specific matrix elements computed in this work, we have
\be
\<\cD^{\DGLAP}_{J_L,g}(z)\>_{[\cO]_R(p)}^{\text{1-loop}} &= \frac{\a_s\mu^{2\e}}{4\pi}\frac{C_A\cS_1(\e)}{J_L+1}\<\tl{\cD}^{\DGLAP}_{J_L,g}(z)\>_{\cO(p)}^{\text{tree}} + O((J_L+1)^0), \\
\<\cD^{\DGLAP}_{J_L,g}(z)\>_{[J]_R(p)}^{\text{1-loop}} &= \frac{\a_s\mu^{2\e}}{4\pi}\frac{C_F\cS_1(\e)}{J_L+1}\<\tl{\cD}^{\DGLAP}_{J_L,q}(z)\>_{J(p)}^{\text{tree}} + O((J_L+1)^0).
\ee

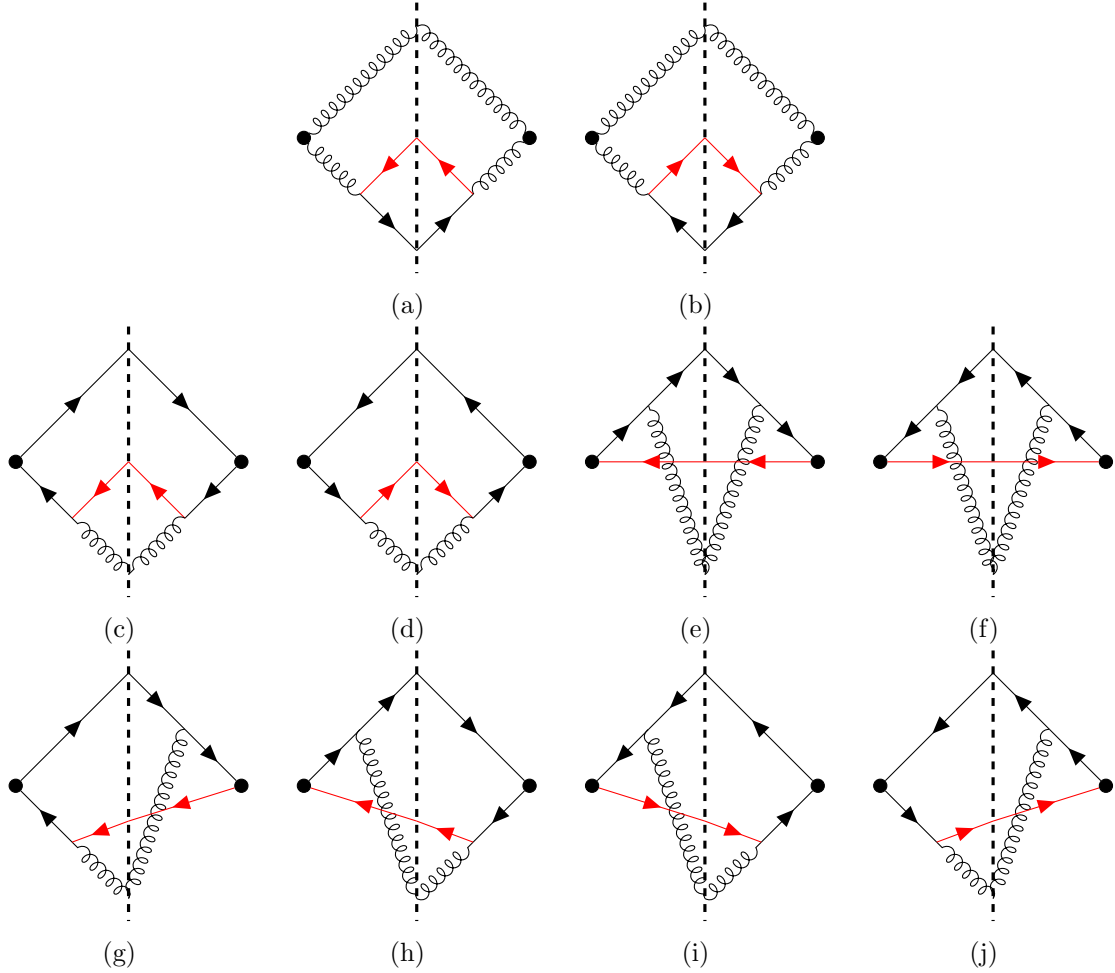
\begin{figure}[h]
    \centering
    \subfloat[]{\label{fig:Og_soft_q}
        \begin{tikzpicture}
            \node[draw, black, fill=black, circle, inner sep=0pt, minimum size=5pt] (a) at (0, 0){};
            \node[draw, black, fill=black, circle, inner sep=0pt, minimum size=5pt] (a2) at (-3, 0){};

            \begin{feynman}
                \vertex (b) at (-1.5, 1.5);
                \vertex (c) at (-1.5, -1.5);
                \vertex (d) at (-1.5, 0);
                \vertex (e) at (-0.75, -0.75);
                \vertex (e2) at (-2.25, -0.75);

                \diagram* {
                    (a) -- [gluon] (b),
                    (a2) -- [gluon] (b),
                    (a) -- [gluon] (e),
                    (a2) -- [gluon] (e2),
                    (c) -- [fermion] (e) -- [fermion,red] (d),
                    (d) -- [fermion,red] (e2) -- [fermion] (c)
                };
            \end{feynman}
            \draw[dashed,  line width=1.2pt] (-1.5, 1.8) -- (-1.5,-1.8);
        \end{tikzpicture}
    }\quad
    \subfloat[]{\label{fig:Og_soft_qb}
        \begin{tikzpicture}
            \node[draw, black, fill=black, circle, inner sep=0pt, minimum size=5pt] (a) at (0, 0){};
            \node[draw, black, fill=black, circle, inner sep=0pt, minimum size=5pt] (a2) at (-3, 0){};

            \begin{feynman}
                \vertex (b) at (-1.5, 1.5);
                \vertex (d) at (-1.5, -1.5);
                \vertex (c) at (-1.5, 0);
                \vertex (e) at (-0.75, -0.75);
                \vertex (e2) at (-2.25, -0.75);

                \diagram* {
                    (a) -- [gluon] (b),
                    (a2) -- [gluon] (b),
                    (a) -- [gluon] (e),
                    (a2) -- [gluon] (e2),
                    (c) -- [fermion,red] (e) -- [fermion] (d),
                    (d) -- [fermion] (e2) -- [fermion,red] (c)
                };
            \end{feynman}
            \draw[dashed,  line width=1.2pt] (-1.5, 1.8) -- (-1.5,-1.8);
        \end{tikzpicture}
    }\\
    \subfloat[]{\label{fig:J_soft_q1}
        \begin{tikzpicture}
            \node[draw, black, fill=black, circle, inner sep=0pt, minimum size=5pt] (a) at (0, 0){};
            \node[draw, black, fill=black, circle, inner sep=0pt, minimum size=5pt] (a2) at (-3, 0){};

            \begin{feynman}
                \vertex (b) at (-1.5, 1.5);
                \vertex (d) at (-1.5, -1.5);
                \vertex (c) at (-1.5, 0);
                \vertex (e) at (-0.75, -0.75);
                \vertex (e2) at (-2.25, -0.75);

                \diagram* {
                    (e2) -- [fermion] (a2) -- [fermion] (b) -- [fermion] (a) -- [fermion] (e),
                    (e) -- [fermion,red] (c),
                    (e) -- [gluon] (d),
                    (c) -- [fermion,red] (e2),
                    (d) -- [gluon] (e2)
                };
            \end{feynman}
            \draw[dashed,  line width=1.2pt] (-1.5, 1.8) -- (-1.5,-1.8);
        \end{tikzpicture}
    }\quad
    \subfloat[]{\label{fig:J_soft_qb1}
        \begin{tikzpicture}
            \node[draw, black, fill=black, circle, inner sep=0pt, minimum size=5pt] (a) at (0, 0){};
            \node[draw, black, fill=black, circle, inner sep=0pt, minimum size=5pt] (a2) at (-3, 0){};

            \begin{feynman}
                \vertex (b) at (-1.5, 1.5);
                \vertex (d) at (-1.5, -1.5);
                \vertex (c) at (-1.5, 0);
                \vertex (e) at (-0.75, -0.75);
                \vertex (e2) at (-2.25, -0.75);

                \diagram* {
                    (e) -- [fermion] (a) -- [fermion] (b) -- [fermion] (a2) -- [fermion] (e2),
                    (c) -- [fermion,red] (e),
                    (e) -- [gluon] (d),
                    (e2) -- [fermion,red] (c),
                    (d) -- [gluon] (e2)
                };
            \end{feynman}
            \draw[dashed,  line width=1.2pt] (-1.5, 1.8) -- (-1.5,-1.8);
        \end{tikzpicture}
    }\quad
    \subfloat[]{\label{fig:J_soft_q2}
        \begin{tikzpicture}
            \node[draw, black, fill=black, circle, inner sep=0pt, minimum size=5pt] (a) at (0, 0){};
            \node[draw, black, fill=black, circle, inner sep=0pt, minimum size=5pt] (a2) at (-3, 0){};

            \begin{feynman}
                \vertex (b) at (-1.5, 1.5);
                \vertex (d) at (-1.5, -1.5);
                \vertex (c) at (-1.5, 0);
                \vertex (e) at (-0.75, 0.75);
                \vertex (e2) at (-2.25, 0.75);

                \diagram* {
                     (a2) -- [fermion] (e2) -- [fermion] (b) -- [fermion] (e) -- [fermion] (a),
                    (e) -- [gluon] (d),
                    (d) -- [gluon] (e2),
                    (a) -- [fermion,red] (c) -- [fermion,red] (a2)
                };
            \end{feynman}
            \draw[dashed,  line width=1.2pt] (-1.5, 1.8) -- (-1.5,-1.8);
        \end{tikzpicture}
    }\quad
    \subfloat[]{\label{fig:J_soft_qb2}
        \begin{tikzpicture}
            \node[draw, black, fill=black, circle, inner sep=0pt, minimum size=5pt] (a) at (0, 0){};
            \node[draw, black, fill=black, circle, inner sep=0pt, minimum size=5pt] (a2) at (-3, 0){};

            \begin{feynman}
                \vertex (b) at (-1.5, 1.5);
                \vertex (d) at (-1.5, -1.5);
                \vertex (c) at (-1.5, 0);
                \vertex (e) at (-0.75, 0.75);
                \vertex (e2) at (-2.25, 0.75);

                \diagram* {
                     (a) -- [fermion] (e) -- [fermion] (b) -- [fermion] (e2) -- [fermion] (a2),
                    (e) -- [gluon] (d),
                    (d) -- [gluon] (e2),
                    (a2) -- [fermion,red] (c) -- [fermion,red] (a)
                };
            \end{feynman}
            \draw[dashed,  line width=1.2pt] (-1.5, 1.8) -- (-1.5,-1.8);
        \end{tikzpicture}
    }\\
    \subfloat[]{\label{fig:J_soft_q3}
        \begin{tikzpicture}
            \node[draw, black, fill=black, circle, inner sep=0pt, minimum size=5pt] (a) at (0, 0){};
            \node[draw, black, fill=black, circle, inner sep=0pt, minimum size=5pt] (a2) at (-3, 0){};

            \begin{feynman}
                \vertex (b) at (-1.5, 1.5);
                \vertex (d) at (-1.5, -1.5);
                \vertex (c) at (-1.5, -0.47);
                \vertex (e) at (-0.75, 0.75);
                \vertex (e2) at (-2.25, -0.75);

                \diagram* {
                    (e2) -- [fermion] (a2) -- [fermion] (b) -- [fermion] (e) -- [fermion] (a),
                    (e) -- [gluon] (d),
                    (d) -- [gluon] (e2),
                    (a) -- [fermion,red] (c) -- [fermion,red] (e2)
                };
            \end{feynman}
            \draw[dashed,  line width=1.2pt] (-1.5, 1.8) -- (-1.5,-1.8);
        \end{tikzpicture}
    }\quad
    \subfloat[]{\label{fig:J_soft_q4}
        \begin{tikzpicture}
            \node[draw, black, fill=black, circle, inner sep=0pt, minimum size=5pt] (a) at (0, 0){};
            \node[draw, black, fill=black, circle, inner sep=0pt, minimum size=5pt] (a2) at (3, 0){};

            \begin{feynman}
                \vertex (b) at (1.5, 1.5);
                \vertex (d) at (1.5, -1.5);
                \vertex (c) at (1.5, -0.47);
                \vertex (e) at (0.75, 0.75);
                \vertex (e2) at (2.25, -0.75);

                \diagram* {
                    (a) -- [fermion] (e) -- [fermion] (b) -- [fermion] (a2) -- [fermion] (e2),
                    (e) -- [gluon] (d),
                    (d) -- [gluon] (e2),
                    (e2) -- [fermion,red] (c) -- [fermion,red] (a)
                };
            \end{feynman}
            \draw[dashed,  line width=1.2pt] (1.5, 1.8) -- (1.5,-1.8);
        \end{tikzpicture}
    }\quad
    \subfloat[]{\label{fig:J_soft_qb3}
        \begin{tikzpicture}
            \node[draw, black, fill=black, circle, inner sep=0pt, minimum size=5pt] (a) at (0, 0){};
            \node[draw, black, fill=black, circle, inner sep=0pt, minimum size=5pt] (a2) at (3, 0){};

            \begin{feynman}
                \vertex (b) at (1.5, 1.5);
                \vertex (d) at (1.5, -1.5);
                \vertex (c) at (1.5, -0.47);
                \vertex (e) at (0.75, 0.75);
                \vertex (e2) at (2.25, -0.75);

                \diagram* {
                    (e2) -- [fermion] (a2) -- [fermion] (b) -- [fermion] (e) -- [fermion] (a),
                    (e) -- [gluon] (d),
                    (d) -- [gluon] (e2),
                    (a) -- [fermion,red] (c) -- [fermion,red] (e2)
                };
            \end{feynman}
            \draw[dashed,  line width=1.2pt] (1.5, 1.8) -- (1.5,-1.8);
        \end{tikzpicture}
    }\quad
    \subfloat[]{\label{fig:J_soft_qb4}
        \begin{tikzpicture}
            \node[draw, black, fill=black, circle, inner sep=0pt, minimum size=5pt] (a) at (0, 0){};
            \node[draw, black, fill=black, circle, inner sep=0pt, minimum size=5pt] (a2) at (-3, 0){};

            \begin{feynman}
                \vertex (b) at (-1.5, 1.5);
                \vertex (d) at (-1.5, -1.5);
                \vertex (c) at (-1.5, -0.47);
                \vertex (e) at (-0.75, 0.75);
                \vertex (e2) at (-2.25, -0.75);

                \diagram* {
                    (a) -- [fermion] (e) -- [fermion] (b) -- [fermion] (a2) -- [fermion] (e2),
                    (e) -- [gluon] (d),
                    (d) -- [gluon] (e2),
                    (e2) -- [fermion,red] (c) -- [fermion,red] (a)
                };
            \end{feynman}
            \draw[dashed,  line width=1.2pt] (-1.5, 1.8) -- (-1.5,-1.8);
        \end{tikzpicture}
    }

    \caption{Feynman diagrams for soft quark/anti-quark emission. Red lines represent the soft quark or anti-quark. (\ref{fig:Og_soft_q}) and (\ref{fig:Og_soft_qb}) are one-loop real emission of a soft quark and soft anti-quark from the gluon operator $\cO$, while (\ref{fig:J_soft_q1})-(\ref{fig:J_soft_qb4}) are matrix elements of the electromagnetic current $J$.}
    \label{fig:soft_quark_diagrams}
\end{figure}

The missing piece before resolving the mixing between DGLAP detectors and their shadows is the $J_L=-1$ pole of the quark DGLAP detector $\cD^{\DGLAP}_{J_L,q}$. This pole arises from the leading divergence in soft quark/anti-quark emission, as fermions exhibit less singular behavior in the soft limit. To the best of our knowledge, there is no universal theorem governing soft quark emission. Nevertheless, a diagrammatic approach can be employed to derive the quark counterpart of \eqref{eq:Dg_JLminus1pole}. At one loop, virtual diagrams for source operators $\cO$ and $J$ only have two particles in the final state and hence do not contribute a $J_L=-1$ pole. Figure~\ref{fig:soft_quark_diagrams} shows the one-loop real emission diagrams for matrix elements $\<\cD^{\DGLAP}_{J_L,q}\>_{\cO}$ and $\<\cD^{\DGLAP}_{J_L,q}\>_{J}$, in which red lines represent the quark/anti-quark being measured by the quark DGLAP detector. In the soft limit, diagrams (\ref{fig:Og_soft_q})-(\ref{fig:J_soft_qb1}) are more singular compared to (\ref{fig:J_soft_q2})-(\ref{fig:J_soft_qb4}) because there are two propagators that become nearly on-shell. In particular, diagrams (\ref{fig:Og_soft_q})-(\ref{fig:J_soft_qb1}) depict the physical picture by which $\cD^{\DGLAP}_{J_L,q}$ evolves into to the shadow detectors $\tl{\cD}^{\DGLAP}_{J_L,g}$ and $\tl{\cD}^{\DGLAP}_{J_L,q}$. 

\begin{figure}[ht]
    \centering
    \subfloat[]{\label{fig:softq_g_to_qq}
        \begin{tikzpicture}[scale=0.8]
            \node[label={[label distance=0.0cm]135:{\(\mu\)}}] (a) at (2, 0){};
            \node[label={[label distance=0.0cm]45:{\(\nu\)}}] (a2) at (-2, 0){};

            \begin{feynman}
                \vertex (b) at (1, 0);
                \vertex (c) at (-1, 0);
                \vertex[label=5:\({\color{red} p_s}\)] (d) at ( 0,1);
                \vertex[label=315:\(p_i\)] (e) at (0, 0);

                \diagram* {
                    (a) -- [gluon] (b),
                    (a2) -- [gluon] (c),
                    (c) -- [fermion] (e) -- [fermion] (b),
                    (b) -- [fermion,red] (d) -- [fermion,red] (c)
                };
            \end{feynman}
            \draw[dashed,  line width=1.2pt] (0, 1.5) -- (0,-1.0);
            \node[label={[label distance=0.0cm]0:{\(+\)}}] (a) at (1.7, 0){};
            \node[label={[label distance=0.0cm]135:{\(\mu\)}}] (a) at (6.4, 0){};
            \node[label={[label distance=0.0cm]45:{\(\nu\)}}] (a2) at (2.4, 0){};

            \begin{feynman}
                \vertex (b) at (5.4, 0);
                \vertex (c) at (3.4, 0);
                \vertex[label=5:\({\color{red} p_s}\)] (d) at (4.4, 1);
                \vertex[label=315:\(p_i\)] (e) at (4.4, 0);

                \diagram* {
                    (a) -- [gluon] (b),
                    (a2) -- [gluon] (c),
                    (b) -- [fermion] (e) -- [fermion] (c),
                    (c) -- [fermion,red] (d) -- [fermion,red] (b)
                };
            \end{feynman}
            \draw[dashed,  line width=1.2pt] (4.4, 1.5) -- (4.4,-1.0);
        \end{tikzpicture}
    }\quad
    \subfloat[]{\label{fig:softq_q_to_qg}
        \begin{tikzpicture}[scale=0.8]
            \node[label={[label distance=0.0cm]135:{\(\mu\)}}] (a) at (2, 0){};
            \node[label={[label distance=0.0cm]45:{\(\nu\)}}] (a2) at (-2, 0){};

            \begin{feynman}
                \vertex (b) at (1, 0);
                \vertex (c) at (-1, 0);
                \vertex[label=5:\({\color{red} p_s}\)] (d) at ( 0,1);
                \vertex[label=315:\(p_i\)] (e) at (0, 0);

                \diagram* {
                    (a) -- [fermion] (b),
                    (c) -- [fermion] (a2),
                    (c) -- [gluon] (e) -- [gluon] (b),
                    (b) -- [fermion,red] (d) -- [fermion,red] (c)
                };
            \end{feynman}
            \draw[dashed,  line width=1.2pt] (0, 1.5) -- (0,-1.0);
            \node[label={[label distance=0.0cm]0:{\(+\)}}] (a) at (1.7, 0){};
            \node[label={[label distance=0.0cm]135:{\(\mu\)}}] (a) at (6.4, 0){};
            \node[label={[label distance=0.0cm]45:{\(\nu\)}}] (a2) at (2.4, 0){};

            \begin{feynman}
                \vertex (b) at (5.4, 0);
                \vertex (c) at (3.4, 0);
                \vertex[label=5:\({\color{red} p_s}\)] (d) at (4.4, 1);
                \vertex[label=315:\(p_i\)] (e) at (4.4, 0);

                \diagram* {
                    (b) -- [fermion] (a),
                    (a2) -- [fermion] (c),
                    (b) -- [gluon] (e) -- [gluon] (c),
                    (c) -- [fermion,red] (d) -- [fermion,red] (b)
                };
            \end{feynman}
            \draw[dashed,  line width=1.2pt] (4.4, 1.5) -- (4.4,-1.0);
        \end{tikzpicture}
    }

    \caption{Diagrams for soft quark/antiquark emission from near onshell propagators.}
    \label{fig:soft_quark_splitting}
\end{figure}
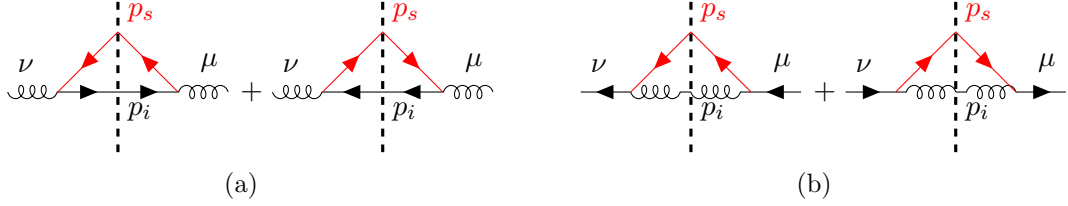

For diagram (\ref{fig:Og_soft_q}) and (\ref{fig:Og_soft_qb}), the underlying process is a nearly on-shell gluon splitting into a quark-antiquark pair, with momenta $p_s^\mu$ and $p_i^\mu$ (see figure \ref{fig:softq_g_to_qq}). In the soft limit $p_s^\mu\to 0$, we can approximate the numerator of the gluon propagator with $\sum_\lambda \varepsilon_{\lambda}^*(p_i)\otimes \varepsilon_{\lambda}(p_i)$. The corresponding soft emission kernel is
\be
S^{q\bar{q}}_{\mu\nu}(p_s,p_i) &= g^2 \tilde{\mu}^{2\epsilon}  (2n_f T_F)\frac{\mathrm{tr}[\slashed{p}_s \gamma^\rho \slashed{p}_i \gamma^\sigma]}{(2p_s\cdot p_i)^2}\left(\sum_{\lambda}\varepsilon_{\lambda \mu}^*(p_i)\varepsilon_{\lambda\rho}(p_i)\right) \left(\sum_{\lambda^\prime}\varepsilon_{\lambda^\prime \sigma}^*(p_i)\varepsilon_{\lambda^\prime\nu}(p_i)\right)\nn\\
&= g^2 \tilde{\mu}^{2\epsilon}   \frac{2n_f T_F}{p_s\cdot p_i} \sum_{\lambda} \varepsilon_{\lambda\mu}^*(p_i) \varepsilon_{\lambda\nu}(p_i)\,. \label{eq:softq_g_to_qq}
\ee
Similarly, we can obtain the soft quark/antiquark emission process in figure (\ref{fig:softq_q_to_qg})
\be
S^{qg}(p_s,p_i) = g^2 \tilde{\mu}^{2\epsilon} \frac{C_F}{2} \frac{d-2}{p_s \cdot p_i} \sum_{s^\prime} u_{s^\prime}(p_i)\otimes \bar{u}_{s^\prime}(p_i)\,,\label{eq:softq_q_to_qg}\\
S^{\bar{q}g}(p_s,p_i) = g^2 \tilde{\mu}^{2\epsilon} \frac{C_F}{2} \frac{d-2}{ p_s \cdot p_i} \sum_{s^\prime} v_{s^\prime}(p_i) \otimes \bar{v}_{s^\prime}(p_i)\,.\label{eq:softq_qb_to_qbg}
\ee

The universal kinematic factor $(p_s\cdot p_i)^{-1}$ appearing in different channels, shown in \eqref{eq:softq_g_to_qq}, \eqref{eq:softq_q_to_qg}, and \eqref{eq:softq_qb_to_qbg}, implies that the $J_L=-1$ pole in the one-loop matrix element $\langle \cD^{\DGLAP}_{J_L,q} \rangle$ is governed by the shadow transform of DGLAP operators. For example, we consider the $J_L=-1$ pole of $\<\cD^{\DGLAP}_{J_L,q}\>_{\cO}$ at which only real emission contributes
\be
&\<\cD^{\DGLAP}_{J_L,q}\>^{\text{1-loop}}_{[\cO]_R(p)} \nn\\
=& \int \frac{E^{-J_L}dE}{(2\pi)^{d-1}2E} \int \prod_{i=1}^{2} {d^{d}p_i \de^+(p_i^2) \over (2\pi)^{d-1}} (2\pi)^d \de^{(d)}(p-E z - p_1 - p_2)\left(|\cF^{\cO}_{q\bar{q}g}(E z, p_1, p_2)|^2 + (q\leftrightarrow \bar{q})\right)\nn\\
=& -\frac{g^2\tilde{\mu}^{2\epsilon}}{J_L+1}\frac{2 n_f T_F}{2^d \pi^{d-1}} \int d\mathrm{LIPS}_2 \frac{1}{z\cdot p_1} |\cF^{\cO}_{gg}(p_1,p_2)|^2 + \cdots\,.
\ee
In the same way, we can get the $J_L=-1$ pole of $\<\cD^{\DGLAP}_{J_L,q}\>_{J}$
\be
\<\cD^{\DGLAP}_{J_L,q}\>^{\text{1-loop}}_{[J]_R(p)} = -\frac{g^2\tilde{\mu}^{2\epsilon}}{J_L+1}\frac{d-2}{2^d \pi^{d-1}}\frac{C_F}{2} \int d\mathrm{LIPS}_2 \left(\frac{1}{z\cdot p_1}+\frac{1}{z\cdot p_2}\right) |\cF^{J}_{q\bar{q}}(p_1,p_2)|^2  + \cdots\,.
\ee
These results can be converted to the matrix elements of the shadow of DGLAP detectors
\be
\<\cD^{\DGLAP}_{J_L,q}(z)\>_{[\cO]_R(p)}^{\text{1-loop}} &= \frac{\a_s\mu^{2\e}}{4\pi}\p{\frac{-n_fT_F}{(1-\e)}}\frac{\cS_1(\e)}{J_L+1}\<\tl{\cD}^{\DGLAP}_{J_L,g}(z)\>_{\cO(p)}^{\text{tree}} + O((J_L+1)^0), \\
\<\cD^{\DGLAP}_{J_L,q}(z)\>_{[J]_R(p)}^{\text{1-loop}} &= \frac{\a_s\mu^{2\e}}{4\pi}\p{-\frac{C_F}{2}}\frac{\cS_1(\e)}{J_L+1}\<\tl{\cD}^{\DGLAP}_{J_L,q}(z)\>_{J(p)}^{\text{tree}} + O((J_L+1)^0).
\ee

\subsection{Recombination of trajectories in QCD}\label{sec:CFplot_QCD}

We now have all the necessary ingredients to compute the renormalized trajectories in QCD at one-loop. The method is essentially the same as in the pure YM case in section \ref{sec:CFplot_YM}, except that we now have to deal with five trajectories: gluon/quark DGLAP, gluon/quark shadow DGLAP, and the BFKL trajectory. The nondegenerate basis for the detectors can be written as
\be\label{eq:QCD_basis_withoutfake}
\mathbb{D}_{J_L} &= U_{\text{QCD}}\begin{pmatrix} \vspace{1mm} \mu^{J_L+2-2\e}\cD^{\DGLAP}_{J_L,g} \\  \vspace{1mm} \mu^{J_L+2-2\e}\cD^{\DGLAP}_{J_L,q} \\ \vspace{1mm} \mu^{-J_L}\tl{\cD}^{\DGLAP}_{J_L,g} \\ \vspace{1mm} \mu^{-J_L}\tl{\cD}^{\DGLAP}_{J_L,q} \\ \vspace{1mm} \cD^{\BFKL}_{J_L}  \end{pmatrix}, \nn \\
U_{\text{QCD}} &= \begin{pmatrix} 
1  & 0 & 1 & 0 & 0  \\ 
 0  & 1 & 0 & 1 & 0  \\ 
-\frac{1}{J_L+1-\e} & 0 &  \frac{1}{J_L+1-\e} & 0 & 0  \\
0 & -\frac{1}{J_L+1-\e} & 0 &  \frac{1}{J_L+1-\e} & 0   \\ 
\frac{C_A\frac{\pi^{1-\e}}{\G(1-\e)}}{(-2+2\e-J_L)(-2+2\e)} & \frac{C_F\frac{\pi^{1-\e}}{\G(1-\e)}}{(-2+2\e-J_L)(-2+2\e)} & \frac{C_A\frac{\pi^{1-\e}}{\G(1-\e)}}{J_L(-2+2\e)} & \frac{C_F\frac{\pi^{1-\e}}{\G(1-\e)}}{J_L(-2+2\e)} & \frac{1}{J_L(-2+2\e-J_L)} 
\end{pmatrix}.
\ee

Similar to the pure YM case, to find the renormalization matrix we consider the three trajectory intersections separately. We first construct a 3-by-3 matrix that removes all the divergences near the intersection between gluon/quark DGLAP and BFKL, and apply the spin shadow transform to find the matrix for the intersection between gluon/quark shadow DGLAP and BFKL. Then, we construct another 4-by-4 matrix for the intersection between gluon/quark DGLAP and their shadows. Finally, we perform the change of basis procedure explained in \ref{sec:CFplot_YM} to combine all three renormalization matrices in the nondegenerate basis \eqref{eq:QCD_basis_withoutfake}, and find the remaining matrix for the $1/\e$ divergences using \eqref{eq:Divmat_eps_div_eq}. With the full renormalization matrix, one can then use \eqref{eq:Dilatation_oneloop_eq} to obtain the dilatation matrix.

As explained in section \ref{sec:CFplot_YM}, we also need to introduce the ``fake trajectories" to remove unresolved $J_L$-divergences from the DGLAP and shadow DGLAP trajectory. In the QCD case, it turns out that we need four fake trajectories in total, and we give their definitions in appendix \ref{app:fake_trajectory}. The final renormalization matrix and the dilatation matrix including the fake trajectories are given in the ancillary file. See appendix \ref{app:anc} for more details. 

\begin{figure}
    \centering
    \begin{subfigure}{0.45\textwidth}
    \includegraphics[width = 7cm]{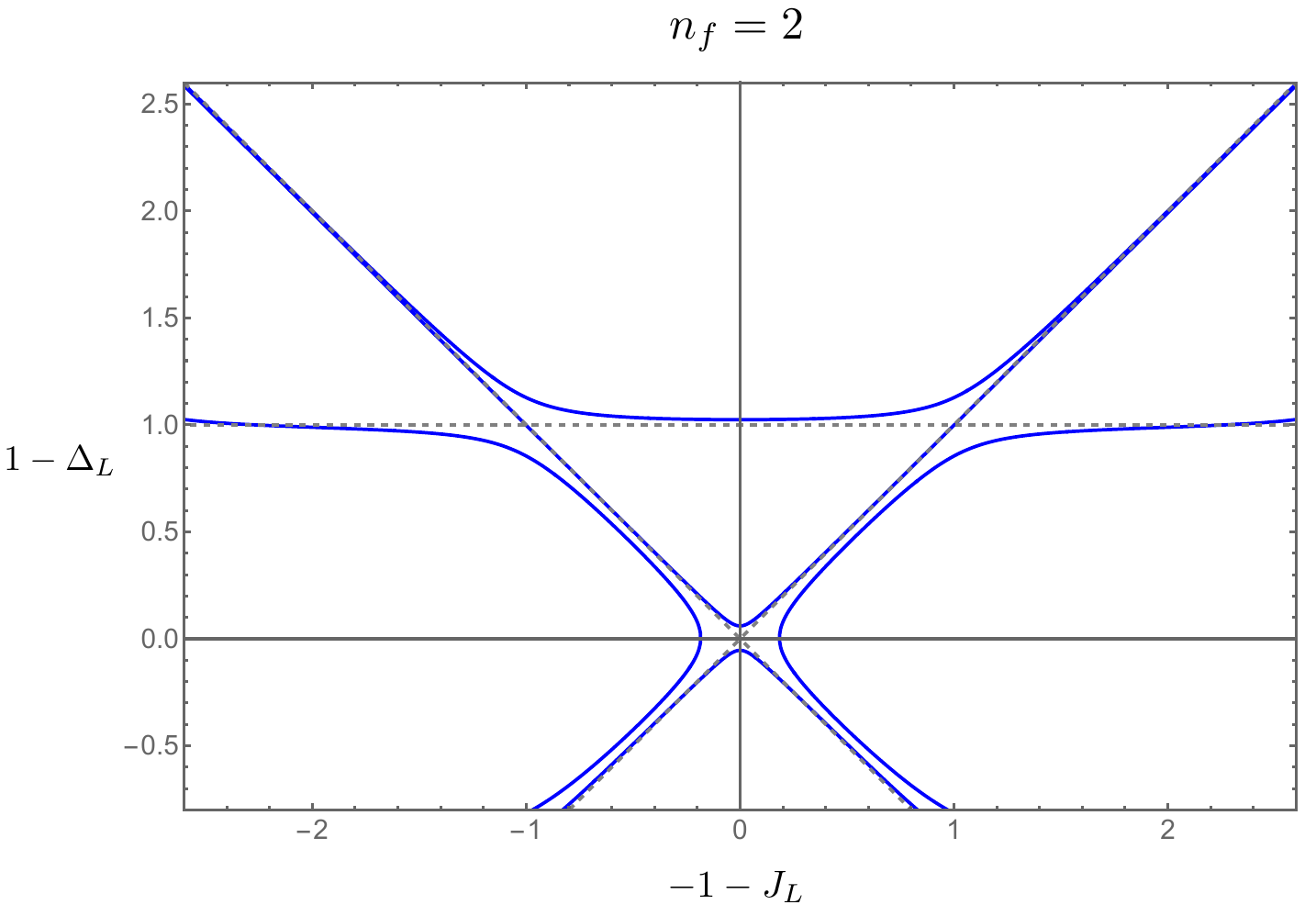}
    \end{subfigure}
    \begin{subfigure}{0.45\textwidth}
    \includegraphics[width = 7cm]{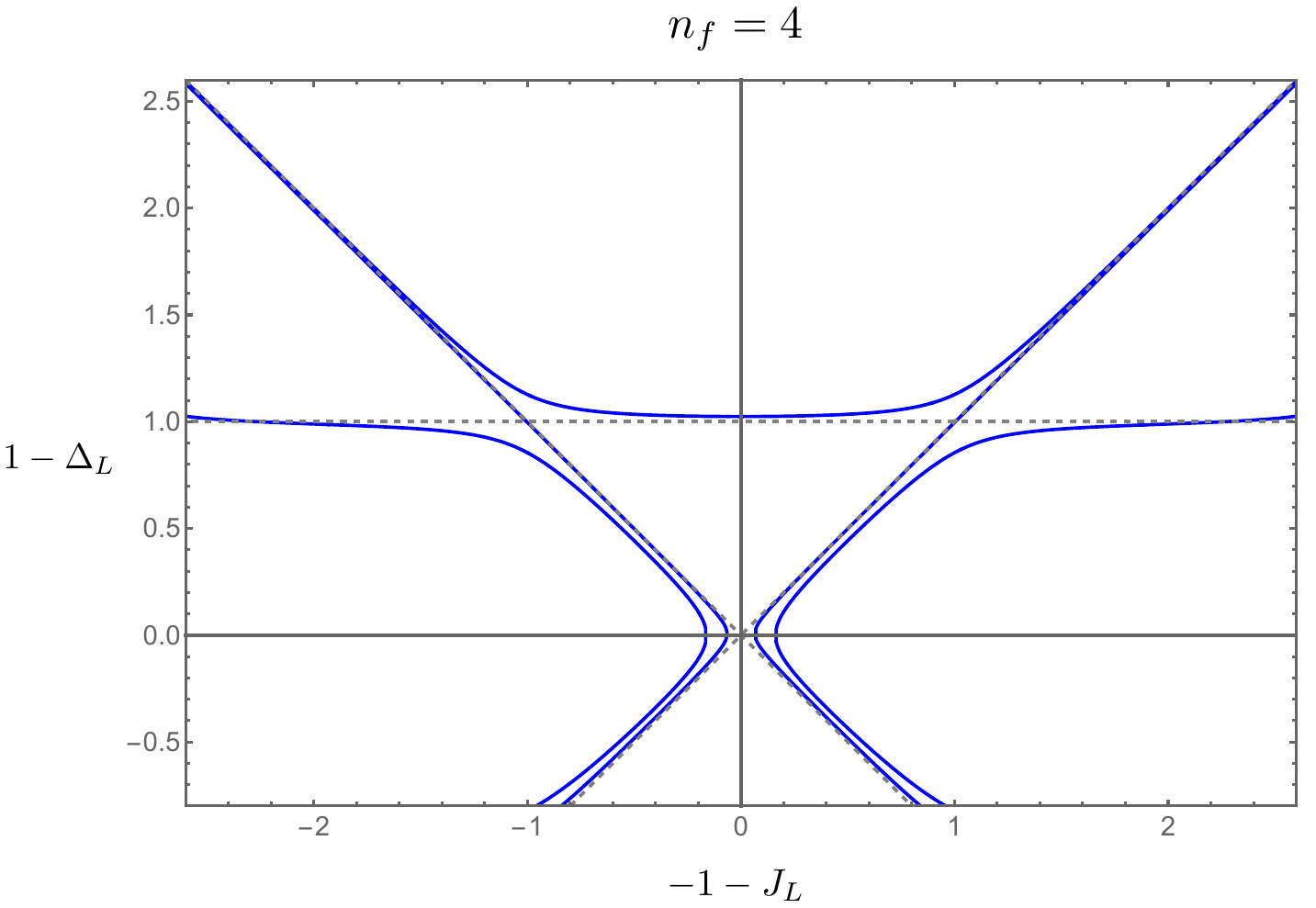}
    \end{subfigure}
    
    \vspace{1em}

    \begin{subfigure}[b]{0.5\textwidth}
    \centering
    \includegraphics[width=\textwidth]{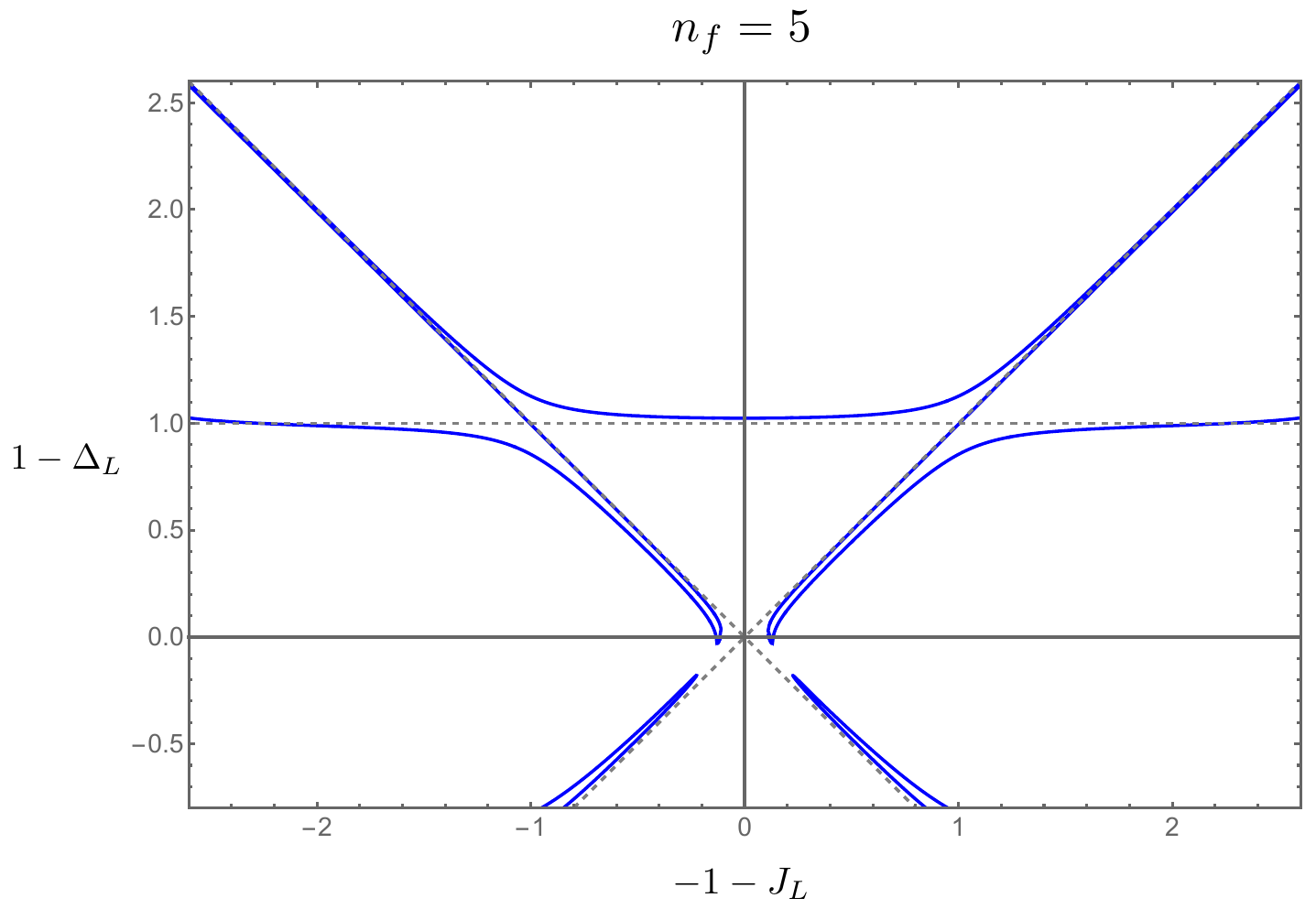}
    \end{subfigure}
    
    \caption{The Chew-Frautschi plot of QCD at 1-loop for $\a_s=0.01$ and $n_f=2,4,5$, where only the points with real $\De_L$ are plotted. Dashed gray lines are the trajectories in free theory.}
    \label{fig:CFplot_QCD}
\end{figure}

The characteristic equation of the dilatation matrix determines the value of $\De_L$ of the renormalized trajectories as a function of $J_L$. We give the plots of the renormalized trajectories at 1-loop in figure \ref{fig:CFplot_QCD} for $\a_s=0.01$ and various values of the number of quark flavors $n_f$.

We can analytically obtain various anomalous dimensions near the trajectory intersections. For example, for the DGLAP/BFKL intersection at $J_L=-2$, we find that the three solutions to $\De_L$ are given by
\be
\p{\De_L(J_L=-2)}_1&= \frac{4C_F n_f T_F}{3C_A \pi}\a_s + O(\a_s^{3/2}), \nn \\
\p{\De_L(J_L=-2)}_{2,3}&=\pm \sqrt{\frac{2C_A}{\pi}\a_s} + \frac{11C_A^2+4(C_A-2C_F) n_f T_F}{12C_A\pi} \a_s + O(\a_s^{3/2}).
\ee
The first solution does not have the $O(\a_s^{1/2})$ enhanced term, and agrees with the finite eigenvalue of the DGLAP anomalous dimension $\hat \g^{(0)}_{ij}$ at $J_L=-2$. In figure \ref{fig:CFplot_QCD}, this solution corresponds to the trajectory that is almost on top of the free theory trajectory near $J_L=-2$. On the other hand, the other two solutions with the $O(\a_s^{1/2})$ enhanced term are due to the recombination of the $C_F \cD^{\DGLAP}_{J_L,q} + C_A \cD^{\DGLAP}_{J_L,g}$ trajectory and the BFKL trajectory near the intersection.

Now let us consider the intersection between the DGLAP and shadow DGLAP trajectories at $J_L=-1$. As we can see in figure \ref{fig:CFplot_QCD}, the behavior of the recombination near this intersection has interesting dependence on $n_f$. We can understand this by looking at the residue of the DGLAP anomalous dimension at $J_L=-1$,
\be
\hat{\gamma}^{(0)}_{ij}(J_L) = \begin{pmatrix}
    \hat{\gamma}^{(0)}_{qq}(J_L) & \hat{\gamma}^{(0)}_{qg}(J_L)\\
    \hat{\gamma}^{(0)}_{gq}(J_L) & \hat{\gamma}^{(0)}_{gg}(J_L)
\end{pmatrix}_{ij} 
= \frac{2}{J_L+1} \begin{pmatrix}
    C_F & 2n_f T_F \\
    -2C_F & -2C_A 
\end{pmatrix}_{ij} + O((J_L+1)^0)\,.
\ee
For $n_f<3$, the above matrix has one positive and one negative eigenvalue, and the characteristic equation at $J_L=-1$ has two real and two complex solutions. On the other hand, for $n_f>3$ both eigenvalues are negative, and the characteristic equation has all complex solutions.\footnote{The case of $n_f=3$ is special. In this case, one of eigenvalues of the 1-loop DGLAP anomalous dimension becomes finite at $J_L=-1$. As a result, even after renormalizing the trajectories at 1-loop, one of the DGLAP trajectories would still intersect with its shadow at $J_L=-1$, which would naively suggest there is level crossing. However, both eigenvalues of the 2-loop DGLAP anomalous dimension are divergent at $J_L=-1$, and thus we expect the level crossing will still get resolved at higher loop order.}

It is also interesting to compute the values of $\De=1-J_L$ at $J=1-\De_L=0$, which determines the $x$-intercept of the renormalized trajectories on the plot in figure \ref{fig:CFplot_QCD}. We find four solutions given by
\be
\De(J=0) = 2\pm \sqrt{\frac{2C_A-C_F\pm\sqrt{(2C_A+C_F)^2-16C_Fn_f T_F}}{\pi}\a_s} + O(\a_s^{3/2}).
\ee
For $n_f=2$, two of the solutions become complex, and this explains why only two of the trajectories intersect the $x$-axis. For $n_f=4,5$ all four trajectories intersect the $x$-axis, and indeed we find four real solutions for $\De(J=0)$.

Finally, let us make some comments on a curious feature of the $n_f=5$ plot. We just saw that for $n_f>3$, the characteristic equation of the dilatation matrix has no real solutions near the DGLAP/shadow DGLAP intersection at $J_L=-1$. On the $n_f=5$ plot, we see that there is an additional region below the origin where the trajectories ``disappear" and there are no real solutions. It turns out that the behavior in this region is not due to recombination of trajectories -- the region should be considered far away from the trajectory intersection (in perturbation theory), and hence the trajectory is well-described by the DGLAP anomalous dimension. Indeed, we find that the matrix $\hat{\gamma}^{(0)}_{ij}(J_L=-1-J)$ for $n_f=5$ has complex eigenvalues for $J\in [-0.173,-0.020]$, which agrees with the range of this region.

\begin{figure}
    \centering
    \begin{subfigure}{0.45\textwidth}
    \includegraphics[width = 7cm]{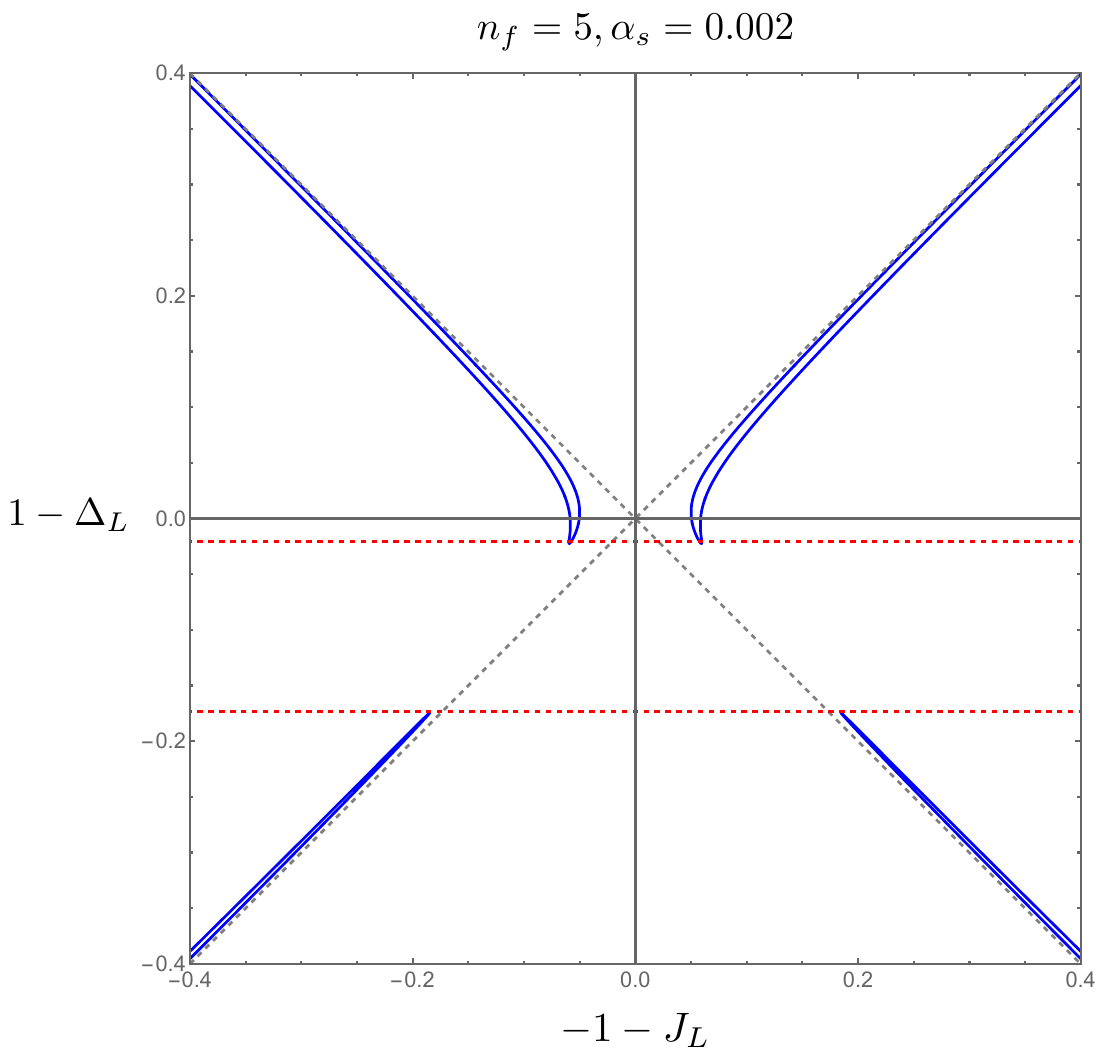}
    \end{subfigure}
    \begin{subfigure}{0.45\textwidth}
    \includegraphics[width = 7cm]{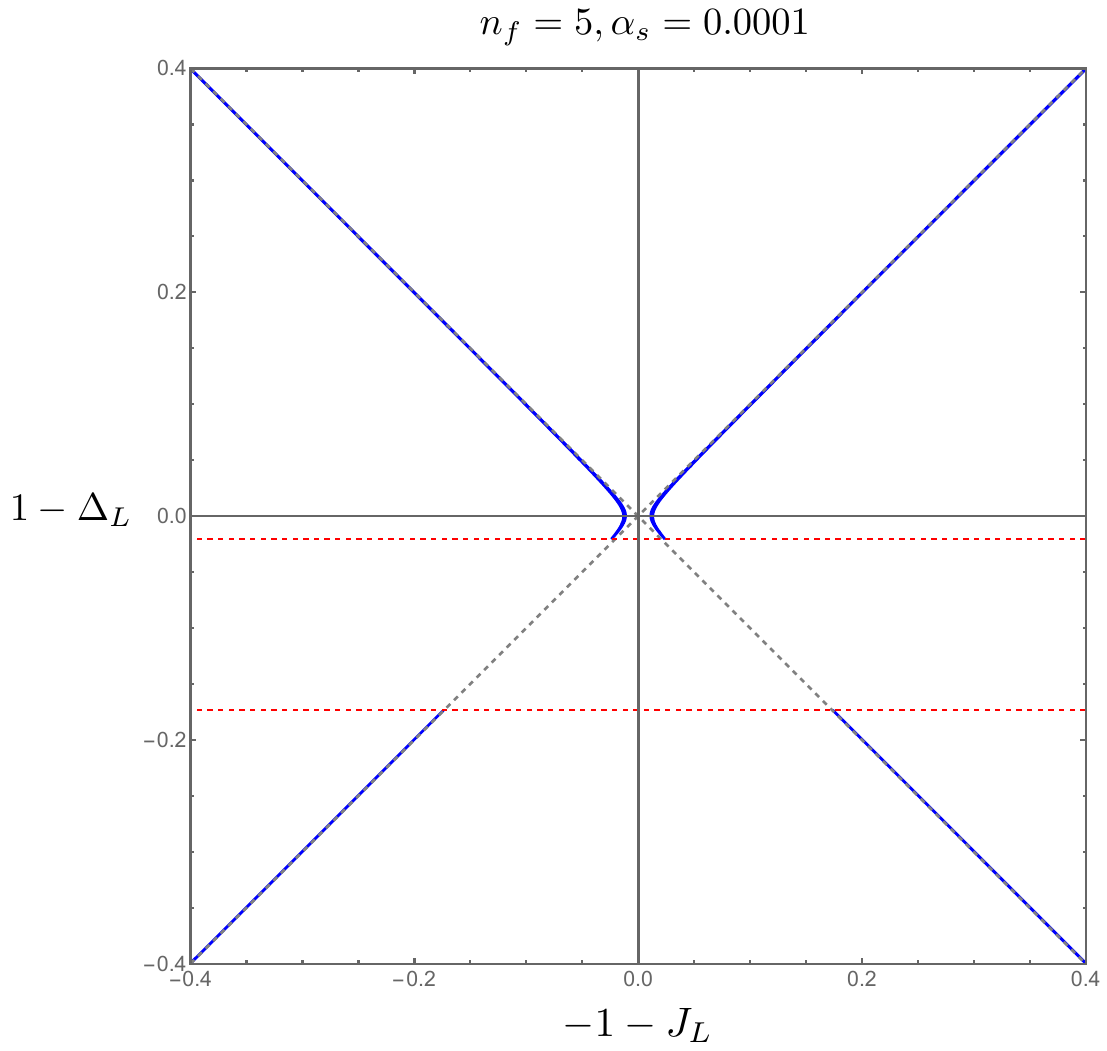}
    \end{subfigure}
    \caption{QCD Chew-Frautschi plot for $n_f=5$ and $\a_s=0.002, 0.0001$, where we zoom in on the region near the origin. The two dashed red lines denote the range $J\in [-0.173,-0.020]$ where the DGLAP anomalous dimension matrix has complex eigenvalues. As we decrease $\a_s$, the $O(\a_s^{1/2})$ splitting near the intersection decreases, but the region given by the dashed red lines remain the same.}
    \label{fig:CFplot_nf5_zoomin}
\end{figure}

In figure \ref{fig:CFplot_nf5_zoomin}, we plot the trajectories for $n_f=5$ and zoom in on the origin. To better illustrate features in perturbation theory, we choose the coupling to have smaller values $\a_s=0.002, 0.0001$. There are two regions where the trajectories disappear: one near the intersection point $J_L=-1$, and the other one given by the range $J\in [-0.173,-0.020]$ which is denoted by the two dashed red lines. As we change $\a_s$, the former region, which is due to recombination of trajectories, shrinks as $O(\a_s^{1/2})$. On the other hand, the latter region due to the behavior of the DGLAP anomalous dimension matrix remains unchanged.

\section{Applications to QCD phenomenology}

\subsection{Matching for detectors}
\label{sec:matching}

The first several sections of this work have been about EFT running for detectors. In this section, we discuss the idea of EFT {\it matching\/} for detectors, which is essential for phenomenological applications. In particular, we would like to understand matching across the QCD confinement threshold.

\subsubsection{Warmup: local operator matching}

As a warmup, let us first consider matching for local operators. Consider a ``microscopic" UV theory that flows at long distances to some IR effective theory. For simplicity, let us assume for the moment that the IR EFT is a CFT (not necessarily perturbative). The point-like measurements of this physical system are best described in terms of UV degrees of freedom. These are the local operators $\cO_i^\mathrm{UV}(x)$ of the microscopic theory.

We can insert $\cO_i^\mathrm{UV}$'s into the path integral and measure their correlation functions 
\be
\<\cO_1^\mathrm{UV}(x_1)\cdots \cO_n^\mathrm{UV}(x_n)\>.
\label{eq:uvcorrelator}
\ee
Assuming the microscopic theory is UV-complete, the correlation functions (\ref{eq:uvcorrelator}) make sense at arbitrary distance scales --- long or short. However, when the points $x_i$ are very far apart, the correlator should more efficiently be described within the IR EFT. The IR theory ``sees" the insertions $\cO_i^\mathrm{UV}(x)$ from far away, and the idea of local operator matching is that far-away point-like measurements should be expandable in a basis of point-like measurements $\cO^\mathrm{IR}_k(x)$ of the IR effective theory. This leads to the expansion
\be
\label{eq:localoperatormatching}
\cO^\mathrm{UV}_i(x) &= \sum_k c_{ik}\Lambda^{\De^\mathrm{UV}_i - \De^\mathrm{IR}_k} \cO^\mathrm{IR}_k(x).
\ee
Here, $\Lambda$ is the threshold scale between the UV and IR theories. The powers of $\Lambda$ are fixed by dimensional analysis in terms of the mass dimensions $\De_k^\mathrm{IR}$ of $\cO_k^\mathrm{IR}$. The $c_{ik}$ are numerical coefficients. The expansion (\ref{eq:localoperatormatching}) is suitable for computing long-distance correlators of $\cO^\mathrm{UV}_i(x)$ in an expansion in powers $1/(\Lambda |x_{ij}|)^{\De^\mathrm{IR}_k}$.

Generically, the expansion (\ref{eq:localoperatormatching}) is only constrained by symmetries. Any operator that can appear will appear. In particular the sum runs over an infinite set of IR local operators $\cO_k^\mathrm{IR}(x)$ of increasing scaling dimension $\De_k^\mathrm{IR}$.

As a concrete example, consider the case where the microscopic theory is the 3d Ising lattice model, described by spins $s_x\in \{\pm 1\}$ at lattice sites $x\in \Z^3$ with nearest-neighbor interactions. Let us tune this theory to the critical temperature, and let the IR EFT be the 3d Ising CFT. (Note that the UV theory is not even a continuum field theory, but local operator matching still works, as long as we interpret UV local operators as lattice operators.) An example of a UV local operator is the spin $s_x$ at a lattice site.  This is a $\Z_2$-odd measurement, so at long distances, it will be expandable in $\Z_2$-odd local operators of the 3d Ising CFT:
\be
\label{eq:isingexpansion}
s_x &= c_1 a^{\De_\sigma}\sigma(x) + c_2 a^{\De_\sigma+2} \ptl^2 \sigma(x) + a^{\De_\sigma+4}\p{c_3 \ptl^4 \sigma(x) + c_4 \sum_{\mu=1}^3 \ptl_\mu^4 \sigma(x)} + \dots
\nn\\
&\quad + c'_1 a^{\De_{\sigma'}} \sigma'(x) + \dots .
\ee
Here, $a=1/\Lambda$ is the lattice spacing, which sets the scale of the RG flow between the microscopic theory and the IR CFT.
The first term $\sigma(x)$ is the lowest-dimension $\Z_2$-odd scalar in the 3d Ising CFT. It appears together with an arbitrary series of derivatives that are invariant under the cubic symmetry of the lattice (but not necessarily the rotation symmetry of the IR EFT). There are also contributions from the next lowest dimension $\Z_2$-odd operator $\sigma'$, and so on. Inserting (\ref{eq:isingexpansion}) into a correlator allows us to compute long-distance correlators of lattice spins $s_x$ in an expansion in correlators of the IR CFT and the Wilson coefficients $c_1,c_2,$ etc.

\subsubsection{Detector matching is backwards!}

Now let us return to matching for detectors. Consider again an RG flow between a UV theory and an IR EFT. (We will now assume that both theories have exact Lorentz symmetry.) While local operators are point-like measurements, detectors are measurements at  infinity. Measurements at infinity are best described in terms of IR degrees of freedom. For example, we might have an ``$i$-particle detector"\footnote{We define creation and annihilation operators $a(p)$ and $a(p)^\dag$ so that they transform like Lorentz scalars. Their commutation relations are $[a(p),a(p')^\dag]=(2\pi)^{d-1} 2p^0 \de^{d-1}(\vec p- \vec p')$.}
\be
\cD^\mathrm{IR}_i(p) = a_i(p)^\dag a_i(p)
\ee
that detects particles of type $i$ moving with momentum $p$ out to infinity. Its 1-point matrix elements in a state $|\Psi\>$ are
\be
\<\cD^\mathrm{IR}_i(p)\>_\Psi &= \frac{1}{\sigma_\mathrm{tot}}\int d\sigma_{\Psi\to X} \sum_{a\in X\textrm{ with type $i$}} (2\pi)^{d-1} 2p^0 \de(\vec p - \vec p_a),
\ee
where the sum on the right-hand side runs over final state particles of type $i$.
 The very existence of such particles is built into the IR EFT, but may not be manifest from the UV description of the theory. 
 
 Because these IR detectors are well-defined, we can measure expectation values 
 \be
 \label{eq:correlatorofirdetectors}
 \<\Psi|\cD^\mathrm{IR}_1(p_1)\cdots \cD^\mathrm{IR}_n(p_n)|\Psi\>
 \ee
 in an arbitrary state $|\Psi\>$
regardless of its characteristics. For example, $|\Psi\>$ could be a high or low energy state with any number of particles. This is analogous to the well-definedness of correlators of UV local operators (\ref{eq:uvcorrelator}) at arbitrary positions.

However, if the state $|\Psi\>$ has very high energy, the UV degrees of freedom will ``see" the detectors $\cD^\mathrm{IR}_i(p_i)$ through the lens of the RG flow from the UV to the IR. The correlator (\ref{eq:correlatorofirdetectors}) should be more efficiently described in terms of measurements of UV detectors $\cD^\mathrm{UV}_k(z)$. This leads to the idea of expanding $\cD^\mathrm{IR}_i$ in terms of the $\cD^\mathrm{UV}_k$. While (\ref{eq:localoperatormatching}) is an expansion around a point $x$, matching of detectors is an expansion around infinity.

In doing this matching, we should take symmetries into account. It is natural to first classify detectors (both UV and IR) into irreps of the Lorentz group. When the IR particle is massless, Lorentz irreps with spin $J_L$ can be formed by integrating over energies with an appropriate weighting
\be
\label{eq:masslesslorentzdetector}
\cD^\mathrm{IR}_{J_L}(z) &\equiv \int \frac{dE}{E} E^{-J_L} \cD^\mathrm{IR}(p)|_{p=E z} \qquad (\textrm{massless particle}).
\ee
These are essentially DGLAP-like detectors for the IR effective theory.
For massive particles, Lorentz irreps can be formed by integrating against the unique (up to normalization) Lorentz-covariant kernel with homogeneity $J_L$ in $z$:\footnote{The  kernel $(2p\.z)^{J_L}$ can be thought of as the bulk to boundary propagator when $\R^{d-1,1}$ is viewed as the embedding space for the $d{-}2$ dimensional conformal group.}
\be
\label{eq:massivelorentzdetector}
\cD^\mathrm{IR}_{J_L}(z) &\equiv 
\frac{2\G(-J_L)m^{2-d-2J_L}}{\pi^{\frac{d-2}{2}} \G(\frac{2-d}{2}-J_L)} 
\int d^d p\, \de(p^2 - m^2) (2p\.z)^{J_L} \cD^\mathrm{IR}(p) \qquad (\textrm{massive particle}).
\ee
Here, we have chosen the factors out front so that (\ref{eq:massivelorentzdetector}) becomes (\ref{eq:masslesslorentzdetector}) in the limit $m\to 0$ when $J_L<\frac{2-d}{2}$. 

Because Lorentz symmetry is exact, an IR detector transforming as a Lorentz irrep with spin $J_L$ should have an expansion in UV detectors with the same $J_L$:
\be
\label{eq:detectorexpansion}
\cD^\mathrm{IR}_{J_L}(z) &= \sum_k C_k(J_L,\mu) \cD^\mathrm{UV}_{k,J_L}(z,\mu).
\ee
Here, we have allowed for the UV detectors to be defined in terms of some renormalization scale $\mu$. The combination (\ref{eq:detectorexpansion}) should be $\mu$-invariant, and this constrains the running of the coefficients $C_k(J_L,\mu)$ in terms of the anomalous dimensions of the $\cD_{k,J_L}^\mathrm{UV}$. From dimensional analysis, we may estimate the magnitude of the matching coefficients to be set by the threshold scale $\Lambda$ for the RG flow between the UV and the IR. As we will see later, this leads to an expansion of IR detector correlators (\ref{eq:correlatorofirdetectors}) in powers of $\Lambda/Q$, where $Q$ is the characteristic energy scale of the state $|\Psi\>$.

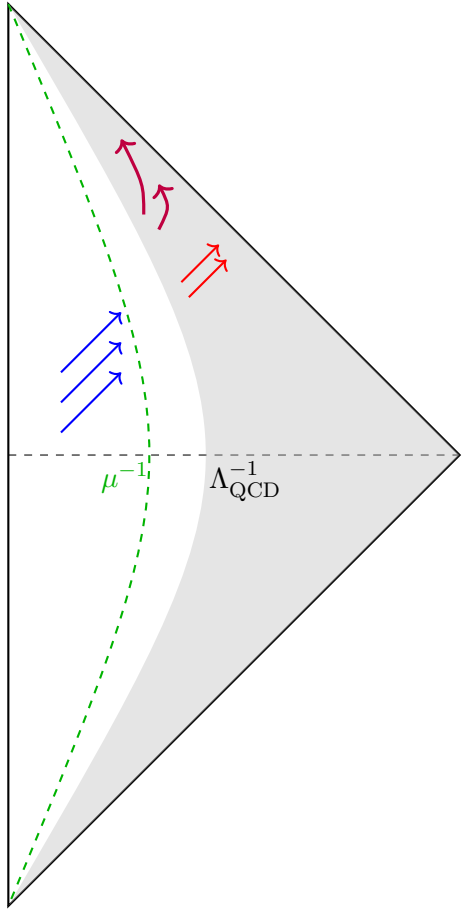
\begin{figure}[htbp]
\begin{center}
\begin{tikzpicture}

    \draw[thick] (0,6) -- (0,-6) -- (6,0) -- cycle;
    \draw[dashed] (0,0) -- (6,0);
    \draw[thick, dashed, green!70!black] (0,6) .. controls (2.5,0) .. (0,-6);
    \node[green!70!black, left] at (2.0,-0.3) {$\mu^{-1}$};
    \fill[lightgray, opacity=0.4] (0,6) .. controls (3.5,0) .. (0,-6) -- (6,0) -- cycle;
    \node[black] at (3.15,-0.35) {$\Lambda_{\text{QCD}}^{-1}$};
    \draw[->, thick, blue] (0.7,0.7) -- (1.5,1.5);
    \draw[->, thick, blue] (0.7,0.3) -- (1.5,1.1);
    \draw[->, thick, blue] (0.7,1.1) -- (1.5,1.9);
    \draw[->, thick, red] (2.3,2.3) -- (2.8,2.8);
    \draw[->, thick, red] (2.4,2.1) -- (2.9,2.6);
    \draw[->, line width = 1.2pt, purple] (1.8,3.2) .. controls (1.8, 3.6)..(1.5,4.2);
    \draw[->, line width = 1.2pt, purple] (2.0 ,3.0) .. controls (2.15, 3.3)..(2.0,3.6);
    
\end{tikzpicture}
\caption{Penrose diagram for matching IR detectors to UV detectors. In the UV region (white), degrees of freedom propagate towards null infinity (blue arrows). In perturbation theory, these degrees of freedom are measured by operators $\cD_{k,J_L}(z,\mu)$ associated with future null infinity, though they are defined with renormalization scale $\mu > \Lambda_\mathrm{QCD}$. In the IR region (gray), particles might propagate to null infinity (red arrows) or spatial infinity (crimson arrows).}
\label{fig:matching_Penrose}
\end{center}
\end{figure}

In more detail, we can think of the matching (\ref{eq:detectorexpansion}) in two steps. Let us imagine a hard process at the center of Minkowski space. We think of the region close to the origin as being described by the UV theory, while the region $|x|\gg \Lambda^{-1}$ is described by the IR theory (figure~\ref{fig:matching_Penrose}). A measurement of particles in the IR region maps to some kind of effective measurement in the UV region, and thus can be expanded in UV operators. However, we should then perform a second expansion of  the UV operators around future null infinity because Lorentz symmetry acts in a simpler way there. In particular, organizing operators into Lorentz irreps leads to an expansion in the $\cD^\mathrm{UV}_{k,J_L}(z,\mu)$. These detectors are still defined at a renormalization scale $\mu > \Lambda_\mathrm{QCD}$, but the formal location of the detectors is null infinity. The matrix elements of these UV detectors are computed in perturbation theory, which does not see the fact that excitations never actually propagate out to null infinity. (That information is entirely encoded in the matching coefficients $C_k(J_L,\mu)$.)

To summarize, for local operators it is natural to expand UV local operators in a series in IR local operators. However, for detectors, the natural matching is reversed: we expand IR detectors in UV detectors. In both cases, the expansions are constrained by symmetries. For detectors, Lorentz symmetry is a powerful tool, and it will allow us to pick out contributions from particular UV Regge trajectories in patterns of asymptotic particles.

\subsection{One-point event shape in collider physics}
\label{sec:comparingtodata}

The renormalized perturbative detectors that we have described in this paper cannot to be directly measured in collider experiments. However, they appear as intermediate objects via light-ray matching and thus play an important role in describing the perturbative evolution of collider observables. 

As an example, let us consider one of the simplest families of observables beyond the total cross-section: one-point event shapes at $e^+ e^-$ colliders 
\begin{equation}
    \< \mathbb{N}_{J_L} (z) \>_Q \equiv -\int d^d x\, e^{- i q\cdot x} \<\Omega|J_\mu(0) \mathbb{N}_{J_L}(z) J^\mu(x)|\Omega\>\,,
\end{equation}
where $Q$ is the energy component of  the momentum $q$ in the center-of-mass frame. Here, $\mathbb{N}_{J_L}$ is a hadron detector that measures all particles $i$ along the direction $\vec{z}$ with energy weighting $2-d-J_L$
\begin{equation}
    \mathbb{N}_{J_L}(z) = \sum_{i}\int {d^{d-1}\vec{p} \over (2\pi)^{d-1}2E} E^{2-d-J_L} \delta^{d-2}(\hat{p} - \hat{z}) a^\dagger_i(\vec{p}) a_i(\vec{p})\,.
\end{equation}
To avoid the complication of mass effects, we consider high-energy scattering so that the masses of hadrons are negligible. This correlation function can be reformulated in terms of a differential cross-section
\begin{equation}\label{eq:f_definition}
    f(J_L, Q) ={1 \over \sigma_{\text{tot}}} \sum_X \int d\sigma_{e^+e^-\to X}\, \sum_{a\in X}\left({E_a \over Q}\right)^{2-d-J_L} = {\Omega_{d-1} \over \sigma_{\text{tot}}}{\< \mathbb{N}_{J_L} (z) \>_Q \over Q^{2-d-J_L}}\,.
\end{equation}
Here,  $d\sigma_{e^+e^-\to X}$ is the differential cross-section for producing a hadronic final state $X$ in the $e^+e^-$ collision, 
$\sigma_{\text{tot}} = \sum_X \int d\sigma_{e^+e^-\to X} = -\int d^d x\, e^{- i q\cdot x} \<\Omega|J_\mu(0) J^\mu(x)|\Omega\>$ is the total cross-section, and $\Omega_{d-1}$ is the volume of the $(d-2)$-dimensional sphere.

In the high-energy limit, we propose that the hadronic detector $\mathbb{N}_{J_L}(z)$ matches onto perturbative partonic detectors as follows
\begin{equation} \label{eq:matching_h_to_p}
    \mathbb{N}_{J_L}(z) \simeq \sum_k C_k(J_L,\mu)\, [\cD_{J_L, k}]_R(z;\mu)\,,
\end{equation}
where $k$ runs over Regge trajectories at fixed $J_L$, $C_k$ is a matching coefficient that contains nonperturbative information about how partons evolves into hadrons. Meanwhile, $\mu$ is the renormalization scale at which our perturbative renormalized detectors $[\cD_{J_L, k}]_R$ are defined.

 When the detectors $[\cD_{J_L, k}]_R$ are leading twist DGLAP detectors, the matching coefficients $C_k$ are moments of the corresponding fragmentation functions, summed over all hadron flavors. In \eqref{eq:matching_h_to_p}, there is freedom to choose the factorization scale $\mu$. However, RG invariance requires that the $\mu$-dependence of the matching coefficients is determined by the RG equation for $[\cD_{J_L, k}]_R$. For example, suppose the detector RG equation is 
\begin{equation}\label{eq:detector_RG_schematic}
    \mu \frac{d}{d\mu} [\cD_{J_L, k}]_R(z;\mu) = \sum_{j}\gamma_{kj}(J_L;\mu) [\cD_{J_L, j}]_R(z;\mu)\,.
\end{equation}
The RG evolution of $C_k$ is then governed by the same anomalous dimension matrix:
\begin{equation}
    \mu \frac{d}{d\mu} C_k(J_L,\mu) = -\sum_j C_j(J_L,\mu) \gamma_{jk}(J_L;\mu)\,.
\end{equation}
Note that the $\mu$-dependence in $\gamma_{jk}$ only enters implicitly through $\alpha_s(\mu)$:
\begin{equation}
    \gamma_{jk}(J_L;\mu)\equiv \gamma_{jk}(J_L;\alpha_s(\mu))\,.
\end{equation} 

To determine which perturbative detectors dominate the matching equation \eqref{eq:matching_h_to_p} in the high-energy collisions, we employ dimensional analysis. In order to balance the classical dimension on both sides of Eq.~\eqref{eq:matching_h_to_p}, the coefficient $C_k$ should have non-vanishing classical dimension
\begin{equation}
    [C_k] = 2-d-J_L + \Delta_{L,k}^{(0)}\,,
\end{equation}
where $-\Delta_{L,k}^{(0)}$ is the classical dimension of the associated detector $\cD_{J_L, k}$.\footnote{As a reminder, we work in conventions where the mass dimension of a detector is {\it minus\/} $\Delta_L$, namely $[\cD]=-\Delta_L$. Furthermore, in perturbation theory, we have $\Delta_L = \Delta_L^{(0)} + \g_L(J_L)$, where $\Delta_L^{(0)}$ is the engineering dimension and $\g(J_L)$ is the anomalous dimension. These conventions are natural in CFT, where a detector transforms like a primary operator at infinity $\cO(\oo)$. The mass dimension of an operator at infinity is $[\cO(\oo)]=-\Delta_\cO$, as a consequence of the formula $\cO(\oo) = \lim_{x\to \oo} |x|^{2\Delta_\cO} \cO(x)$.} Since the typical energy scale for evolving from partons to hadrons is the  non-perturbative QCD scale $\Lambda_{\text{QCD}}$, we may estimate the magnitude of matching coefficients to be
\begin{equation}
    C_k \sim (\Lambda_{\text{QCD}})^{2-d-J_L + \Delta_{L,k}^{(0)}}\,.
\end{equation}
On the other hand, the matrix element of perturbative detectors typically scale as
\begin{equation}
    \frac{1}{\sigma_{\text{tot}}}\< [\cD_{J_L, k}]_R(z;\mu) \>_Q \sim Q^{-\Delta_{L,k}^{(0)}}\,.
\end{equation}
An important consequence of this dimensional analysis is that, in the high-energy limit where $Q\gg \Lambda_{\text{QCD}}$, detectors with lower dimension (larger $\Delta_L$) are suppressed in the perturbative matching. In other words, for a given fixed $J_L$, we expect that  the highest trajectory on the perturbative Chew-Frautschi plot generically dominates at large $Q$, unless the corresponding matching coefficient happens to be small.

Based on the matching proposal~\eqref{eq:matching_h_to_p}, we expect the observable $\< \mathbb{N}_{J_L} \>_Q$ to be sensitive to DGLAP/BFKL mixing when $J_L$ is near $-2$. To see how this observable can be related to perturbative trajectory, we consider the scale evolution of $\< \mathbb{N}_{J_L}\>_Q$
\begin{equation} \label{eq:scale_evo_1}
    Q\frac{d}{dQ}\<\mathbb{N}_{J_L}(z)\>_Q \simeq \sum_{k} C_k(J_L,Q)\, \left[Q\frac{d}{dQ}\<[\cD_{J_L, k}]_R(z;Q)\>_Q - \gamma_{k}(J_L, Q)  \<[\cD_{J_L, k}]_R(z;Q)\>_Q \right]\,,
\end{equation}
where we have chosen an operator basis such that the anomalous dimension matrix is diagonalized at the scale $Q$: $\gamma_{jk}(J_L;Q)= \gamma_k(J_L;Q) \delta_{jk}$.
In perturbation theory, $\<[\cD_{J_L, k}]_R(z;Q)\>_Q$ is only a non-trivial function of $\alpha_s(Q)$ 
\begin{equation}
    \<[\cD_{J_L, k}]_R(z;Q)\>_Q
    = Q^{-\Delta_{L,k}^{(0)}}(c_{k,0}(J_L) + c_{k,1}(J_L) \alpha_s(Q)+ c_{k,2}(J_L) \alpha_s(Q)^2+\cdots)\,.
\end{equation}
It is useful to re-organize Eq.~\eqref{eq:scale_evo_1} as
\begin{align} 
\label{eq:howwereorganize}
    Q\frac{d}{dQ}\<\mathbb{N}_{J_L}(z)\>_Q &\simeq \sum_{k} C_k(J_L,Q) \left[-\Delta_{L,k}(J_L;Q)  \<[\cD_{J_L, k}]_R(z;Q)\>_Q
    + Q\frac{d}{dQ}{\<[\cD_{J_L, k}]_R(z;Q)\>_Q \over Q^{-\Delta_{L,k}^{(0)}}} \right],
\end{align}
where $\Delta_{L,k}(J_L;\alpha_s(Q))=\Delta_{L,k}^{(0)} + \gamma_{k}(J_L;\alpha_s(Q))$ is the effective $\Delta_L$ of the detector at the scale $Q$. The second term in (\ref{eq:howwereorganize}) is relatively suppressed in the small coupling limit compared to the first term, due to smallness of the beta function:
\begin{equation}
    Q\frac{d}{dQ}{\<[\cD_{J_L, k}]_R(z;Q)\>_Q \over Q^{-\Delta_{L,k}^{(0)}}} = \beta(\alpha_s) \frac{\partial}{\partial \alpha_s} \left[{\<[\cD_{J_L, k}]_R(z;Q)\>_Q \over Q^{-\Delta_{L,k}^{(0)}}}\right]_{\alpha_s = \alpha_s(Q)}\,.
\end{equation}

As explained using dimensional analysis, we expect that the term with the lowest value of $\Delta_L$, which we call $\Delta_{L,\text{min}}(J_L;\alpha_s(Q))$, dominates the sum in the matching condition. Therefore, the approximate scale evolution of $\<\mathbb{N}_{J_L}(z)\>_Q$ is
\begin{equation}
     Q\frac{d}{dQ}\<\mathbb{N}_{J_L}(z)\>_Q
     \approx -\Delta_{L,\text{min}}(J_L;\alpha_s(Q))\, \<\mathbb{N}_{J_L}(z)\>_Q\,.
\end{equation}
Within a high-energy window $[Q_1,Q_2]$ where $\alpha_s(Q)$ does not change much, the approximate solution to this equation is
\begin{equation}
    \<\mathbb{N}_{J_L}(z)\>_Q \approx A_{J_L} Q^{-\Delta_{L,\text{min}}(J_L;\bar{\alpha}_s)}\,,
    \quad Q\in [Q_1, Q_2]\,,
\end{equation}
where $\bar{\alpha}_s$ is an effective average of the coupling constant in this window. Therefore, the function $f(J_L,Q)$ defined in Eq.~\eqref{eq:f_definition} has a similar power-law form with a modified exponent
\begin{equation}
    f(J_L,Q)\approx \tilde{A}_{J_L} Q^{-\Delta_{L,\text{min}}(J_L;\bar{\alpha}_s)-(2-d-J_L)}\,,
    \quad Q\in [Q_1, Q_2]\,.
\end{equation}

\subsection{Comparison with Monte Carlo simulation and CMS Open Data}
\label{sec:comparisontodata}

\begin{figure}
    \centering
    \includegraphics[width = 10cm]{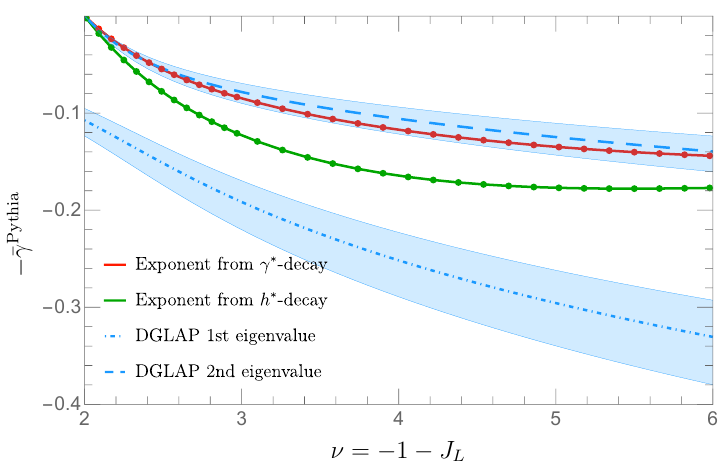}
    \caption{A plot of $-\bar{\gamma}_{\g^*/h^*}^\mathrm{Pythia}(\nu)$ as a function of $\nu=-1-J_L$. We compare them to the analytical results for one-loop DGLAP anomalous dimension eigenvalues in QCD (light blue). Due to quark/gluon mixing, $-\bar{\gamma}_{\g^*/h^*}^\mathrm{Pythia}(\nu)$ are located between these two eigenvalues.}
    \label{fig:data_DGLAP_comparison}
\end{figure}

To test the detector matching idea using the observable \eqref{eq:f_definition}, we generate virtual photon decay $(\gamma^* \to q\bar{q})$ and Higgs boson decay $(h^*\to gg)$ events using Pythia 8.3 and measure the corresponding $f_{\gamma^*}^{\text{Pythia}}(J_L,Q)$ and $f_{h^*}^{\text{Pythia}}(J_L,Q)$, in the energy range from $Q_1=250 \text{ GeV}$ to $Q_2=1 \text{ TeV}$ in windows of size $\Delta Q= 50 \text{ GeV}$. The energy weighting for each event is $\sum_a (E_a/Q)^{\nu-1}$, where $\nu=-1-J_L$. For $0<\nu<6$, we fit the $Q$-dependence with the power-law ansatz
\begin{equation}
    f_{\gamma^*}^{\text{Pythia}}(J_L,Q) = \tilde{A}^{\text{Pythia}}_{\gamma^*}(\nu) Q^{-\bar{\gamma}_{\gamma^*}^{\text{Pythia}}(\nu)}\,,\quad
    f_{h^*}^{\text{Pythia}}(J_L,Q) = \tilde{A}^{\text{Pythia}}_{h^*}(\nu) Q^{-\bar{\gamma}_{h^*}^{\text{Pythia}}(\nu)}\,.
\end{equation}
and extract the exponents $\bar{\gamma}_{\gamma^*}^{\text{Pythia}}(\nu)$ and $\bar{\gamma}_{h^*}^{\text{Pythia}}(\nu)$. Roughly speaking, these exponents play the role of anomalous dimensions that control the perturbative scale evolution. The interpretation of anomalous dimension is more clear when DGLAP physics dominates the evolution. This is shown in figure~\ref{fig:data_DGLAP_comparison}, in which we compared the leading-order DGLAP eigenvalues with $\bar{\gamma}_{\gamma^*}^{\text{Pythia}}(\nu)$ and $\bar{\gamma}_{h^*}^{\text{Pythia}}(\nu)$. Using two-loop running coupling effects, we vary the value of the strong coupling constant $\alpha_s$ from $\alpha_s(Q=125 $\text{ GeV}$)\approx 0.1127$ to $\alpha_s(Q=1 $\text{ TeV}$)\approx 0.0869$. Due to quark/gluon mixing, the extracted exponents are averages of the two DGLAP eigenvalues. We leave a more accurate prediction for these exponents to future work because it requires information about higher loop contributions as well as non-perturbative fragmentation.

\begin{figure}
    \centering
    \includegraphics[width = 10cm]{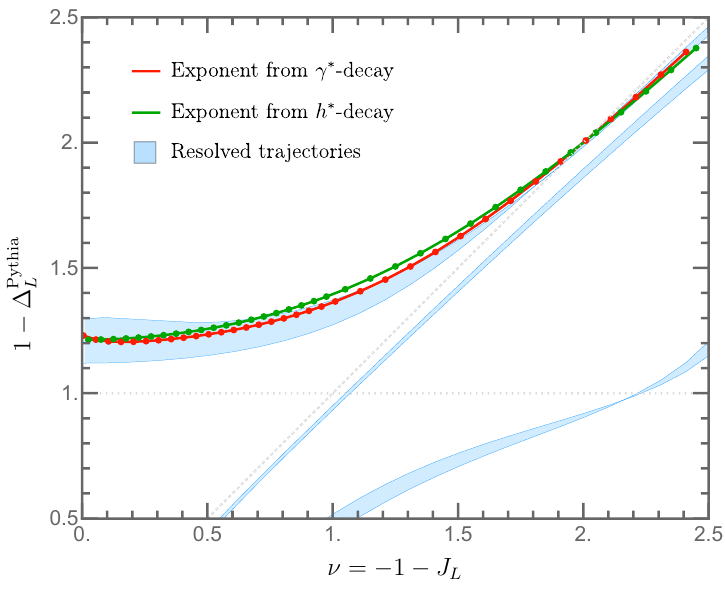}
    \caption{A plot of $1-\Delta_{L,\g^*/h^*}^\mathrm{Pythia}(\nu)$ as a function of $\nu=-1-J_L$. We compare them to the analytical results for $1-\Delta_L$ for the recombined DGLAP/BFKL trajectories in QCD (light blue).}
    \label{fig:data_intersection_comparison}
\end{figure}

The recombination of BFKL/DGLAP trajectories near the intersection region makes it ambiguous to use the concept of anomalous dimension. Instead, we will directly compare the dimension $-\Delta_L$ of the detector with the scaling exponents of the observable $\<\mathbb{N}_{J_L}(z)\>_Q$
\begin{equation}
    -\Delta_{L,\gamma^*}^{\text{Pythia}}(\nu) = -\bar{\gamma}_{\gamma^*}^{\text{Pythia}}(\nu) + \nu \,,\qquad
    -\Delta_{L,h^*}^{\text{Pythia}}(\nu) = -\bar{\gamma}_{h^*}^{\text{Pythia}}(\nu) + \nu\,.
\end{equation}
The comparison is given in figure~\ref{fig:data_intersection_comparison}, where we also choose to vary $\alpha_s$ from $125\text{ GeV}$ to $1\text{ TeV}$. Apart from scale variation, we also take into account uncertainty associated with whether to include the effect of recombination between DGLAP and shadow DGLAP in the Chew-Frautschi plot. The exponents from the event generator show very good agreement with the top trajectory near the intersection, providing strong evidence for our understanding of the recombination of DGLAP and BFKL trajectories, as well as our conceptual picture of matching for detectors.

\begin{figure}
    \centering
    \subfloat[]{\label{subfig:jet_data1}
    \includegraphics[width = 0.48\linewidth]{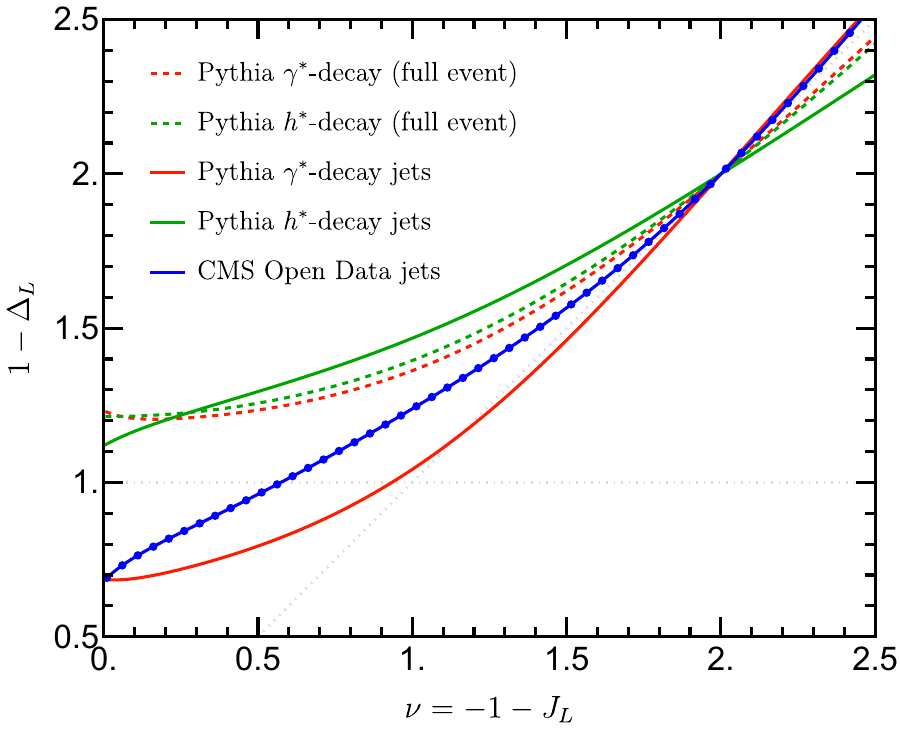}
    }\hfill
    \subfloat[]{\label{subfig:jet_data2}
    \includegraphics[width = 0.48\linewidth]{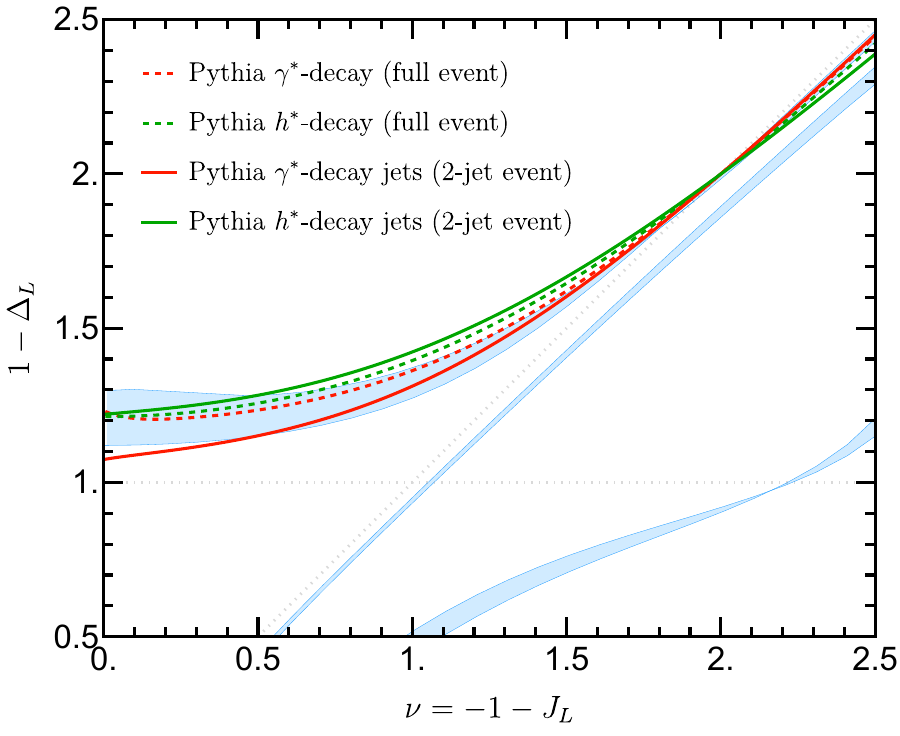}
    }
    \caption{Plots of $1-\Delta_{L}^\mathrm{jet}(\nu)$ as a function of $\nu=-1-J_L$. 
    In (\ref{subfig:jet_data1}), we show the $1-\Delta_{L}^\mathrm{jet}(\nu)$ for jet events from Pythia $\gamma^*/h^*$-decay simulation and CMS Open Data (solid lines). The Pythia jets are generated from $2\text{ TeV}$ decay events using anti-$k_T$ jet algorithm with jet radius chosen to be $0.4$. We compare them to $1-\Delta_{L,\g^*/h^*}^\mathrm{Pythia}$ (dashed lines) shown in figure~\ref{fig:data_intersection_comparison}. 
    In (\ref{subfig:jet_data2}), we compare $1-\Delta_{L}^\mathrm{jet}(\nu)$ from 2-jet decay events (solid lines) with $1-\Delta_{L,\g^*/h^*}^\mathrm{Pythia}$  in figure~\ref{fig:data_intersection_comparison} (dashed lines). We use 1 TeV to 2 TeV $\gamma^*/h^*$-decay events and the anti-$k_T$ jet algorithm with jet radius $0.4$, followed by selecting only 2-jet events. We also show analytical results for $1-\Delta_L$ for the recombined DGLAP/BFKL trajectories in QCD for comparison (light blue).}
    \label{fig:data_on_jets}
\end{figure}

However, it is not easy to access a wide range of high center-of-mass collision energies at $e^+ e^-$ colliders.\footnote{It would be interesting to analyze old LEP data, similar to what has been done recently for energy-energy correlation~\cite{Bossi:2025xsi}.} By contrast, a much wider spectrum of high-energy jets are available at the LHC, motivating us to generalize the observable \eqref{eq:f_definition} to jets
\be
 f_{\text{jet}}(J_L, E_{\text{jet}}) ={1 \over \sigma_{\text{tot}}^{\text{jet}}} \sum_X \int d\sigma^{\text{jet}}_X\, \sum_{a\in X}\left({E_a \over E_{\text{jet}}}\right)^{2-d-J_L} \,,
\ee
where $E_{\text{jet}}$ is the total energy of a jet. Following the same procedure, we can extract a scaling exponent by fitting the data
\be\label{eq:f_def_jet}
f_{\text{jet}}(J_L,E_{\text{jet}}) = \tilde{A}_{\text{jet}}(\nu) E_{\text{jet}}^{-\bar{\gamma}_{\text{jet}}(\nu)}\,,
\ee
and construct $-\Delta_{L}^{\text{jet}}(\nu) = -\bar{\gamma}_{\text{jet}}(\nu) + \nu$. In figure \ref{subfig:jet_data1}, we show the results extracted from Pythia jets (both $\gamma^*$- and $h^*$- decay) as well as CMS Open Data~\cite{CERNopendata,CMS2018policy,CMS2016_Jet_AOD}. The selected Pythia jets have energies in the range $500 \text{GeV} < E_{\text{jet}}< 1\text{ TeV}$. These jets were reconstructed using the anti-$k_T$ algorithm with radius parameter $R=0.4$. For CMS Open Data, we select the jets with transverse momenta in the range $375\, \text{GeV} <p_T< 1200\, \text{GeV}$ from the dataset~\cite{komiske_2019_3340205}.\footnote{This is based on a reprocessed dataset of jets culled from the CMS 2011A Open Data~\cite{CMS2016_Jet_AOD} and made public in the “MIT Open Data” format by~\cite{Komiske:2019jim,komiske_2019_3340205}.}

Compared to $1-\Delta_L^{\text{Pythia}}$ extracted from the full-event observable \eqref{eq:f_definition} in figure \ref{subfig:jet_data1}, we notice that $1-\Delta_L^{\text{jet}}$ from the jet observable \eqref{eq:f_def_jet} exhibits large deviations, reflecting the pitfalls of naively applying this framework to jets. The existence of a jet radius breaks Lorentz symmetry and can induce additional corrections in the scale evolution. However, we expect the effect of the jet algorithm only leads to a relatively mild modification to the spectrum relative to full-event observables.

To test this idea, we selected 2-jet events in our simulation, so that jets from $\gamma^*\to q\bar{q}$ and $h^*\to gg$ decay can be regarded as quark and gluon jets, respectively. The corresponding $1-\Delta_L^{\text{jet}}$ is given in figure \ref{subfig:jet_data2}, from which we can confirm our expectation that the jet algorithm alone does not qualitatively change the mixing spectrum. Instead, we expect the main source leading to the large deviation in figure \ref{subfig:jet_data1} is the quark/gluon fraction in the jet ensemble. For an ensemble of jets produced in a given collision process, this ratio exhibits energy dependence, changing the effective scaling exponent. However, this ratio is calculable for $pp$ collisions at the LHC using perturbative hard scattering matrix elements and parton distribution functions, see~\cite{Kaufmann:2015hma,Kang:2016mcy,Generet:2025vth} for example. Although the quark/gluon fraction is not determined in this work, we observe qualitatively that the CMS Open Data curve falls between the Pythia-simulated $\gamma^*$-decay and $h^*$-decay jets. This indirect comparison suggests that the CMS Open Data may contain signatures consistent with BFKL/DGLAP mixing. We leave a detailed study of LHC jets for the future.

\section{Discussion}
\label{sec:discussion}

In this work, we explored intrinsic detectors at $\scriplus$ in perturbative QCD, including the DGLAP (leading twist) and BFKL Regge trajectories. We studied their renormalization at one-loop, understanding how soft and collinear divergences lead to mixing among the DGLAP, shadow DGLAP, and BFKL trajectories at special locations on the Chew-Frautschi plot. We also hypothesized that general collider measurements at infinity, including quantities like hadron number flux, and hadron flux weighted by powers of energy, can be matched onto linear combinations of the detectors of perturbative QCD. Our hypothesis is supported by data, which shows that hadron number measurements (and their generalizations) scale with energy $Q$ in a way consistent with the perturbative anomalous dimensions of renormalized DGLAP/BFKL detectors at the appropriate values of $J_L$.

Although we found a consistent picture for DGLAP/BFKL mixing at one-loop, there are many open questions about the Chew-Frautschi plot of QCD. By our ``light-ray matching" hypothesis, these questions are relevant to understanding the scaling of general collider observables with energy. For example, Regge trajectories at higher $\Delta_L$ control subleading corrections to hadron number flux at large $Q$.

One important question is: how do we identify a ``complete" set of perturbative detectors in QCD in which any IR measurement can be expanded? For example, in this work we focused on BFKL operators built from a particular linear combination of quark and gluon color detectors $\cN^c(z) = \cN^c_g(z)+\cN^c_q(z)$ --- the combination that arises naturally in the leading soft gluon theorem. Other linear combinations lead to $1/\e^2$ divergences at one loop that do not cancel between real emission and virtual diagrams. What is the correct way to renormalize these more general detectors, and can they show up in matching? One possibility is that the $1/\e^2$ divergences signal mixing with additional detectors that we have not yet identified, similar to the role that $1/\e^2$ divergences play in renormalizing the Pomeron operator in the Wilson-Fisher theory \cite{Caron-Huot:2022eqs}. This question might also be related to the appearance in $\cN=4$ SYM of horizontal trajectories that receive $O(\sqrt {\alpha_s})$ anomalous dimensions for generic $J_L$ \cite{Ekhammar:2024neh}.\footnote{We thank Petr Kravchuk for these suggestions.} We have furthermore assumed without proof that the ``correct" detectors to study are those which are multiplicatively renormalized (as opposed to additively renormalized). This led us to focus on the color detectors $\cN^c(z)$, and not their energy-weighted generalizations $\cD_{J_L}^c(z)$. Do BFKL-type operators built out of energy-weighted color detectors have a role to play? Can they show up in matching computations as well?

In our calculation of DGLAP/BFKL mixing, the leading and subleading soft gluon theorems played an important role. Recombination of the DGLAP trajectory with other trajectories essentially expresses a tree-level soft gluon theorem for squared amplitudes/form factors. Because DGLAP anomalous dimensions possess additional poles at $J_L=-2+\mathbb{N}$, we expect additional recombinations at these locations, and these recombinations should be related to new (sub-)$^k$leading soft gluon theorems for squared-amplitudes/form factors. The simplest of them corresponds to $k=1$, which has been constrained by the LBK theorem to have opposite coefficients between the $J_L=-1$ pole and the $J_L = -2$ pole in both $\hat{\gamma}^{(0)}_{gg}(J_L)$ and $\hat{\gamma}^{(0)}_{gq}(J_L)$. This universality was first pointed out in the analysis of the classical part of the splitting functions~\cite{Dokshitzer:2006nm}. 
It would be very interesting to identify these new soft theorems, which may be controlled by matrix elements of detectors involving correlations between multiple hard particles in the final state. (Universal soft theorems in gravity should also imply interesting relationships between tree-level detectors in gravity \cite{Herrmann:2024yai}.) The recombination of detector trajectories provides a nice answer to the question: ``what happens to soft theorems at higher loop orders," at least for squared-amplitudes.

It is not immediately clear to us whether these lessons can be extended to amplitudes themselves (as opposed to their squares). Note that soft theorems for squared amplitudes might be much simpler than soft theorems for the amplitudes themselves. We saw an example of this for the mixing of the DGLAP trajectory with its shadow. The subleading soft gluon theorem for amplitudes involves the angular momentum operators $iJ_{\mu\nu}$. However, the $iJ_{\mu\nu}$ terms, which essentially contribute an imaginary part to the soft amplitude, disappeared after squaring, leaving only the shadow transform of a DGLAP detector. (This may be related to the analysis of subleading power corrections in Soft Collinear Effective Theory (SCET) and the LBK theorem \cite{Moult:2019mog}.) Can we find a detector observable that is instead sensitive to this imaginary part? It is also interesting to ask whether connections between soft theorems and asymptotic symmetries \cite{Strominger:2017zoo} can help illuminate detectors and their recombination.

The ``timelike" anomalous dimensions of DGLAP detectors are related by reciprocity to the ``spacelike" anomalous dimensions of local operators on the DGLAP trajectory. Now that we have understood timelike anomalous dimensions of BFKL operators both for generic $J_L$, and at the location where they mix with the DGLAP trajectory, can we use reciprocity to learn new things about BFKL physics in the Regge regime? Relatedly, resummation of splitting functions and hard partonic cross sections in the small-$x$ regime and match to DGLAP at moderate $x$ is important phenomenologically. Different approaches have been developed to address this problem in the past~\cite{Altarelli:2001ji,Ciafaloni:2003rd,White:2006yh}. It would be interesting to ask whether the BFKL/DGLAP recombination developed in this work can lead to new understanding of this important problem.

It is also natural to explore higher-twist Regge trajectories, along the lines of the studies \cite{Homrich:2022cfq,Klabbers:2023zdz,Homrich:2024nwc,Ekhammar:2024neh} for $\cN=4$ SYM and \cite{Henriksson:2023cnh} for the Wilson-Fisher theory, and also to explore Regge trajectories with odd signature or nonzero transverse spin \cite{Chang:2020qpj,Chang:2022ryc,Chen:2022jhb}. In addition, there exist an infinite number of tree-level BFKL-like trajectories at $\Delta_L=0$ built from higher products of color detectors $\cN^c(z)$. At higher loop orders, these undergo a complicated pattern of mixing with themselves \cite{Caron-Huot:2013fea}, and with $45^\circ$ Regge trajectories. It would be very interesting to untangle these threads, both at one loop and at higher order in perturbation theory.

Light-ray matching also provides an interesting window onto the mixing of horizontal trajectories: can we resolve different horizontal trajectories using precision studies of energy-weighted hadron flux at large $Q$? Can we explore nonlinear versions of BFKL evolution, such as the BK/JIMWLK equations \cite{Mueller:1994jq,Balitsky:1995ub,Kovchegov:1999yj,Jalilian-Marian:1996mkd,Jalilian-Marian:1997jhx,Iancu:2001ad}, and see evidence for them in collider data? In the context of soft gluon radiation, it is well-known that non-global logarithms~\cite{Dasgupta:2001sh} obey an evolution equation~\cite{Banfi:2002hw,Caron-Huot:2015bja,Becher:2015hka,Larkoski:2015zka,Banfi:2021owj,Becher:2021urs} that match the BK/JIMWLK equations under the mapping between the transverse plane and the celestial sphere~\cite{Hatta:2008st,Hatta:2009nd,Schwartz:2014wha}. It will be very interesting to explore it using the detector language.  Moving lower on the Chew-Frautschi plot, it would be interesting to identify and understand horizontal trajectories at larger $\Delta_L$. For example, do there exist subleading BFKL trajectories at $\Delta_L=1$ that complicate the mixing of DGLAP with its shadow at higher orders in perturbation theory?

Relatedly, let us highlight an important physical difference between large $Q$ in the context of detector physics, and the Regge limit. In both cases, the spectrum of light-ray operators plays an important role. In detector physics, the quantum number $\Delta_L$ controls the scaling of detector matrix elements with $Q$. In Regge physics, the quantum number $J$ controls matrix elements of light-ray operators at large boost. In a CFT, these are related by $\Delta_L=1-J$. However, in a non-CFT, there is a qualitative difference between large $Q$ and large boost. The Regge limit of large boost can be thought of as a large time limit in Rindler space, and the physics of chaos and the chaos bound applies \cite{Maldacena:2015waa}. There can be some initial transient growth of correlators in the Regge limit, but this growth eventually saturates (or decays away \cite{Caron-Huot:2020ouj}) at very large boost. In a CFT, saturation in the Regge limit is reflected in the requirement that the Regge intercept obey $J_0\leq 1$ nonperturbatively \cite{Caron-Huot:2017vep} (though this may be violated at fixed order in perturbation theory, for example at large $N$). The near-saturation regime is difficult to describe perturbatively, and there has been significant theoretical work on developing effective descriptions for the large boost regime in QCD, see \cite{Gelis:2010nm,Kovchegov:2012mbw} and references therein. By contrast, there is no saturation-type regime for detector matrix elements at very large $Q$ in QCD. Instead, at large $Q$ the running coupling $\alpha_s(Q)$ gets smaller and smaller, and our expressions for detector anomalous dimensions, including for example the $\sqrt{\alpha_s}$ effect (\ref{eq:crazysquareroot}), simply become more and more reliable. 

It would be nice to explore our light-ray matching hypothesis further both in perturbative examples and in real collider data.  Can we find evidence for subleading trajectories in collider data? It is also interesting to explore {\it complex\/} values of $J_L$ in collider data, as we will report on in upcoming work \cite{ComplexJL}. Furthermore, in matching to collider data, one should more systematically incorporate the effects of finite hadron masses by exploring expectation values of the massive detectors (\ref{eq:massivelorentzdetector}). It seems reasonable that detectors of massive particles (which live at future infinity) at fixed $J_L$  should match onto perturbative QCD detectors for massless partons  (which live at $\scriplus$) at the same $J_L$. Though this seems physically reasonable since masses can be treated perturbatively in the UV effective theory, it would be nice to verify this guess in a toy example that is entirely under perturbative control. Another question is how light-ray matching is related to the light-ray OPE \cite{Kologlu:2019mfz} and/or algebras of light-ray operators \cite{Cordova:2018ygx,Korchemsky:2021htm}.

\section*{Acknowledgements}
We especially thank Petr Kravchuk for initial collaboration on this work and many helpful insights.
We thank Arindam Bhattacharya, Ankita Budhraja, Cliff Cheung, Lance Dixon, Tom Hartman, Enrico Herrmann, Alexandre Homrich, Murat Kolo\u{g}lu, Kyle Lee, Yibei Li, Yue-Zhou Li, Juan Maldacena, Ian Moult, Julio Parra-Martinez, Andrea Puhm, Chia-Hsien Shen, Iain Stewart, Jesse Thaler, Raju Venugopalan,  Wouter Waalewijn, and Xiaoyuan Zhang for helpful discussions. We also thank the participants and organizers of the ``Energy Operators in Particle Physics, Quantum Field Theory and Gravity" workshop at the Simons Center for Geometry and Physics for stimulating discussions. HC is supported by the U.S. Department of Energy, Office of Science, Office of Nuclear Physics under grant Contract Number DE-SC0011090. DSD is supported in part by Simons Foundation grant 488657 (Simons Collaboration on the Nonperturbative Bootstrap) and the U.S. Department of Energy, Office of Science, Office of High Energy Physics, under Award Number DE-SC0011632. CHC is supported by a Kadanoff fellowship at the University of Chicago. HXZ is supported by National Science Foundation of China under contract No. 12425505.

\newpage

\appendix

\section{Equivalence of the two definitions of DGLAP detector}\label{app:twist-2}

In this appendix, we explain the equivalence of the two definitions of the DGLAP detector given by \eqref{eq:DGLAP_detector_g_def1} and \eqref{eq:DGLAP_detector_g_def} in free theory. We reproduce the definitions here for convenience,
\be
\cD^{\DGLAP}_{J_L,g}(z) &= \sum_{\l,c} \int_0^\oo \frac{E^{-J_L}dE}{(2\pi)^{d-1}2E}\left.\left[a^{\dag}_{\l,c}(p)a_{\l,c}(p)\right]\right|_{p=E z}\,, \label{eq:DGLAP_detector_g_def1_inAppendix} \\ 
\cD^{\DGLAP}_{J_L,g}(z) &= \frac{1}{C_{J_L}}\int d\a_1 d\a_2\ \p{(\a_1-\a_2+i\e)^{2\De_A+J_L}+(\a_2-\a_1+i\e)^{2\De_A+J_L}} \nn \\
&\qquad\qquad\qquad \times :F_a^{\nu}(\a_1,z)W^{ab}_{\mathrm{adj}}(\a_1,\a_2)F_{b\nu}(\a_2,z):. \label{eq:DGLAP_detector_g_def2_inAppendix}
\ee
In \eqref{eq:DGLAP_detector_g_def2_inAppendix} and the rest of this appendix we suppress the $(\bar z)$ label for brevity. To show the equivalence, we will start with the second definition \eqref{eq:DGLAP_detector_g_def2_inAppendix}, and show that it becomes the first definition \eqref{eq:DGLAP_detector_g_def1_inAppendix} after plugging in the mode expansion of the gauge field given by \eqref{eq:Amu_mode_expansion}. Throughout the derivation, we assume that we work in a gauge with nonsingular polarization $\varepsilon_\l(p)$ for all null $p$ (such as the lightcone gauge), so that that fields are ``sufficiently nice" at null infinity. See section \ref{sec:wilsonlinedetails} for more discussions on this.

Setting $x^\mu = Lz^\mu + \frac{\a}{4}\bar z^\mu$ in the mode expansion, we get
\be
A_{c}^\mu(Lz+\a \bar z/4) = \sum_{\l} \int_0^\oo \frac{d(\signplus \bar{z}\.p) }{(2\pi)^{d-1}2 (\signplus \bar{z}\. p)}\int d^{d-2}\vec{p}_\perp \left[ \varepsilon_{\lambda}^\mu(p)  a_{\l,c}(p) e^{\signminus i L {|\vec{p}_\perp|^2\over \bar{z}.p}} e^{\signminus i {\alpha \over 4}\bar{z}\.p} +\text{h.c.} \right],
\ee
where we have used the lightcone decomposition
\be
p^\mu = {z\. p \over 2} \bar{z}^\mu + {\bar{z}\. p \over 2} z^\mu + p_\perp^\mu\,,
\ee
and the on-shell condition $z\. p = |\vec{p}_\perp|^2/(\bar{z}\. p)$. In the $L\to \oo$ limit, the transverse  $\vec{p}_\perp$-integral can be done by stationary phase approximation
\be
\int d^{d-2} \vec p_\perp e^{\signminus i L |\vec{p}_\perp|^2/(\bar{z}.p)} f(\vec{p}_\perp) = {1\over L^{d-2\over 2}}  (\signminus i\pi \bar{z}\. p)^{d-2 \over 2} f(\vec{0} )+ O\p{\frac{1}{L^{\frac{d}{2}}}}.
\ee
In other words, the transverse momentum is localized at $\vec{p}_\perp = 0$ as hence the on-shell momentum should be proportional to $z^\mu$. Therefore, in the $L\to \oo$ limit we get
\be\label{eq:A_largeLlimit}
&A_{c}^\mu(Lz+\a \bar z/4) \nn \\
&=  \frac{1}{L^{\frac{d-2}{2}}} \sum_{\l} \int_0^\oo \frac{dE}{(2\pi)^{d-1}2E}\left[(-2\pi i E)^{\frac{d-2}{2}} \varepsilon_{\lambda}^\mu(p)  a_{\l,c}(p)e^{\signminus i\frac{\a E}{2}} +\text{h.c.} \right]\Big|_{p=Ez} + O\p{\frac{1}{L^{\frac{d}{2}}}},
\ee
where we have redefined the integration variable $\bar{z}\. p =\signplus 2E$.

We now consider the field strength $F$. Using the definition \eqref{eq:null_limit_def_special} and the above result, we can obtain the mode expansion of $F_\nu(\alpha,z)$,
\be\label{eq:Falphaz_mode_expansion}
&F^\nu_a(\a,z) \nn \\
&= \lim_{L \to \oo}\frac{L^{\frac{d-2}{2}}}{4}\bar z_{\mu} \p{\ptl^{\mu}A_a^{\nu}(Lz+\a \bar z/4) - \ptl^{\nu}A_a^{\mu}(Lz+\a \bar z/4) + gf_{abc}A_b^{\mu}(Lz+\a \bar z/4)A_c^{\nu}(Lz+\a \bar z/4)} \nn \\
&=\frac{1}{4}\sum_\l \int \frac{dE}{(2\pi)^{\frac{d}{2}}2E}\left[(-i)^{\frac{d}{2}}E^{\frac{d-2}{2}} \p{ p\.\bar z\varepsilon_{\lambda}^{\nu} - p^\nu\bar z\.\varepsilon_\l}  a_{\l,a}(p)e^{\signminus i\frac{\a E}{2}} +\text{h.c.} \right]\Big|_{p=Ez}\, ,
\ee
where the $gf_{abc}A_bA_c$ term vanishes because it scales as $\frac{1}{L^{\frac{d-2}{2}}}$. 

Let us plug \eqref{eq:Falphaz_mode_expansion} into the definition \eqref{eq:DGLAP_detector_g_def2_inAppendix}. Due to \eqref{eq:A_largeLlimit}, the Wilson line becomes $1+O(L^{\frac{2-d}{2}})$ in the $L\to \oo$ limit. So, the product $:F_a^\nu(\a_1,z)W^{ab}_{\mathrm{adj}}(\a_1,\a_2)F_{b\nu}(\a_2,z):$ can be written as
\be\label{eq:Falphaz_product_mode_expansion}
&:F_a^\nu(\a_1,z)W^{ab}_{\mathrm{adj}}(\a_1,\a_2)F_{b\nu}(\a_2,z): \nn \\
&= \frac{1}{16}\sum_{\l,\l'} \int \frac{d E dE'}{(2\pi)^d} : \bigg[(-i)^{\frac{d}{2}}E^{\frac{d-2}{2}}\varepsilon^\nu_{\l} a_{\l,a}(E z)e^{\signminus i\frac{\a_1 E}{2}} +\text{h.c.} \bigg] \nn \\
&\qquad\qquad\qquad\qquad \times \bigg[(-i)^{\frac{d}{2}} (E')^{\frac{d-2}{2}}\varepsilon_{\l' \nu}  a_{\l',b}(E' z)e^{\signminus i\frac{\a_2 E'}{2}} +\text{h.c.} \bigg] : \, .
\ee
Note that all the contributions from the $\bar{z}\.\e$ term in \eqref{eq:Falphaz_mode_expansion} vanish because $p\.\e=p^2=0$, and the result is now independent of the auxiliary null vector $\bar z$. 

Moreover, the kernel in \eqref{eq:DGLAP_detector_g_def2_inAppendix} only depends on $\a_1-\a_2$. Therefore, if we change the integration variables to $\a_0=\a_1-\a_2$ and $\a_2$, the only dependence on $\alpha_2$ will be the $e^{\pm i\frac{\a_2 E}{2}}$ and $e^{\pm i\frac{\a_2 E'}{2}}$ factors in \eqref{eq:Falphaz_product_mode_expansion}. After integrating over $\a_2$, they become delta functions $\de(\pm E \pm E')$. Since $E,E'$ are positive, only two of them are nonzero, and we have
\be\label{eq:DDGLAP_beforealpha0int}
\cD^{\DGLAP}_{J_L,g}(z) =&-\frac{1}{4C_{J_L}} \int d\a_0 \p{(\a_0+i\e)^{2\De_A+J_L} + (-\a_0+i\e)^{2\De_A+J_L}}e^{\signplus i\frac{\a_0 E}{2}} \nn \\
&\qquad \sum_{\l} \int \frac{d E}{(2\pi)^{d-1}} E^{d-2} : a^{\dag}_{\l,a}(E z)a_{\l,a}(E z):,
\ee
where we have used $\varepsilon_{\l}\.\varepsilon_{\l'}=-\de_{\l\l'}$.

To compute the $\a_0$-integral, we can close the contour to the upper half plane, and integrate the discontinuity of the branch cut from $0$ to $\oo$. This gives
\be
&\int d\a_0 \p{(\a_0+i\e)^{2\De_A+J_L} + (-\a_0+i\e)^{2\De_A+J_L}}e^{\signplus i\frac{\a_0 E}{2}} \nn \\
=& \int_0^{\oo} d\a_0 \p{e^{i\pi(2\De_A+J_L)}-e^{-i\pi(2\De_A+J_L)}}\a_0^{2\De_A+J_L}e^{\signplus i\frac{\a_0 E}{2}} \nn \\
=&-2^{2+2\De_A+J_L}e^{i\frac{\pi}{2}(2\De_A+J_L)}\sin(\pi(2\De_A+J_L))\G(2\De_A+J_L+1)E^{-J_L-1-2\De_A}.
\ee
Plugging this into \eqref{eq:DDGLAP_beforealpha0int}, we finally have
\be
\cD^{\DGLAP}_{J_L,g}(z) =&\frac{1}{C_{J_L}} (2^{2\De_A+J_L+1}e^{i\frac{\pi}{2}(2\De_A+J_L)}\sin(\pi(2\De_A+J_L))\G(2\De_A+J_L+1)) \nn \\
&\times \sum_{\l} \int \frac{E^{-J_L} d E}{(2\pi)^{d-1}2E}  : a^{\dag}_{\l,a}(E z)a_{\l,a}(E z):.
\ee
Thus, after choosing $C_{J_L}$ to be \eqref{eq:CJL_definition}, this becomes \eqref{eq:DGLAP_detector_g_def1_inAppendix}.

Now let us describe how this computation is modified in Feynman gauge. In Feynman gauge, we can start with the Wightman propagator for the gauge field
\be
\<0|A^\rho(x) A^\nu(q)|0\> &= 2\pi(- g^{\mu\rho}) \de^+(p^2) e^{-i p\.x},
\ee
from which we compute 
\be
\label{eq:fawightman}
\<0|F^{(\bar z)\rho}(\a,z) A^\nu(p)|0\> &= \lim_{L\to \oo} \frac{L^{\frac{d-2}{2}}}{4}
\p{2\pi i(\bar z\.p g^{\rho \nu} - p^\rho \bar z^\nu) \de^+(p^2) e^{-ip\.(L z + \a \bar z/4)}}.
\ee
Squaring and integrating over $\alpha$'s, we obtain the same functional form for the DGLAP vertex as before. However, the tensor structure is different. Instead of a polarization sum, we obtain a contraction of two copies of the tensor in (\ref{eq:fawightman}):
\be
-\frac{(\bar z\.p g^{\rho \nu} - p^\rho \bar z^\nu)(\bar z\.p \de^\mu_\rho - p_\rho \bar z^\mu)}{(p\. \bar z)^2}
&= - g^{\mu\nu} + \frac{p^\mu \bar z^\nu + p^\nu \bar z^\mu}{p\. \bar z},
\ee
where we used that $p$ is null.

\section{Computing matrix elements of BFKL detector}
In this appendix, we give more details regarding the calculations of the BFKL detector matrix element. First, in \ref{app:distributions} we derive several distributional identities related to the integration kernel in the definition of $\cD^{\BFKL}_{J_L,g}$. Then, in \ref{app:BFKL_1loop}, we provide more details of the one-loop matrix element computation. Finally, in \ref{app:poles_BFKL_JL1JL2}, we explain the connection between the leading soft theorem and part of the $\e$-pole of the one-loop matrix elements of $\cD^{\BFKL}_{J_L,g}$.

\subsection{Distributional identities}\label{app:distributions}
We first derive the two identities \eqref{eq:distribution_identity_JLminus2} and \eqref{eq:BFKL_int_JLpole_collinear_maintext} in the main text. Our main target is the following integral:
\be
\int D^{d-2}z_1 D^{d-2}z_2\ \p{\frac{2z_1\.z_2}{(2z_1\.z)(2z_2\.z)}}^{-\frac{J_L}{2}}f(z_1,z_2),
\ee
where $f(z_1,z_2)$ is the matrix element of $\cD^c_{2-d}(z_1)\cD^c_{2-d}(z_2)$, and thus it has $z_1,z_2$-homogeneities $2-d$. This integral can be viewed as the embedding space integral for $\R^{d-2}$. If we instead write it in terms of $\R^{d-2}$ coordinates, and choose $z$ to correspond to the origin of $\R^{d-2}$, we have
\be
&\int d^{d-2}\vec y_1d^{d-2}\vec y_2\ \p{\frac{\vec y_{12}^2}{\vec y_1^2 \vec y_2^2}}^{\frac{-J_L}{2}}f(\vec y_1,\vec y_2) \nn \\
=&\int d\O_{d-2}d\O_{d-3}\int_0^{\oo}dr_1 \int_0^{\oo}dr_2 \int_0^{\pi}d\th\ r_1^{d-3}r_2^{d-3} \sin^{d-4}\th \p{\frac{r_1^2+r_2^2-2r_1 r_2 \cos\th}{r_1^2 r_2^2}}^{\frac{-J_L}{2}} f(\vec y_1,\vec y_2).
\ee
Making the change of variables
\be\label{eq:r1r2_to_rhov}
r_1=\r v,\quad r_2=\r(1-v),
\ee
we can rewrite the integral as
\be\label{eq:BFKL_int_inrhov}
&\int d\O_{d-2}d\O_{d-3}\int_{-1}^{1}dx \int_0^{\oo}d\r \int_0^{1}dv\ v^{J_L+d-3}(1-v)^{J_L+d-3} \r^{J_L+2d-5}\nn \\
&\qquad\qquad\qquad\qquad \times (1-2v(1-v)(1+x))^{-\frac{J_L}{2}}(1-x^2)^{\frac{d-5}{2}} f(\vec y_1,\vec y_2),
\ee
where $x=\cos\th$.

Due to the distributional identity
\be
x^{-1+\e}\th(x) = \frac{1}{\e}\de(x) + \ldots,
\ee
the integral can have poles in $J_L$ coming from the $v$ and $\r$ dependence of the integrand. First, let us focus on the $v$-integral. There are two possible regions of the $v$-integral that can lead to poles in $J_L$. One is $v=0$, which corresponds to $r_1=0$, or $z_1=z$, and the other one is $v=1$, which corresponds to $z_2=z$. Note that $f(\vec y_1,\vec y_2)$ is independent of $z$, so we don't expect it to have any singularities near these two regions. In other words, the factors that will give the $J_L$-poles are $v^{J_L+d-3}$ and $(1-v)^{J_L+d-3}$. We see that both factors lead to poles at $J_L=2-d$. The result is given by
\be
\frac{1}{J_L+d-2}\vol(S^{d-3})\int d\O_{d-3} \int_{-1}^{1}dx \int_0^{\oo}d\r\ \r^{d-3} (1-x^2)^{\frac{d-5}{2}} f(0,\vec y_2) + (1\leftrightarrow 2).
\ee
This remaining integral is just a spin shadow transform $\int D^{d-2}z' f(z,z')$. (Note that $f$ has homogeneity $2-d$.) So, the pole of the integral at $J_L=2-d$ can be written as
\be
&\int D^{d-2}z_1 D^{d-2}z_2\ \p{\frac{2z_1\.z_2}{(2z_1\.z)(2z_2\.z)}}^{-\frac{J_L}{2}}f(z_1,z_2) \nn \\
&\sim \frac{\vol(S^{d-3})}{J_L+d-2} \p{\int D^{d-2}z' f(z,z')+ \int D^{d-2}z' f(z',z)},
\ee
which is \eqref{eq:distribution_identity_JLminus2}.

In the above analysis of the integral \eqref{eq:BFKL_int_inrhov}, we have focused on the dependence on $v$ that can lead to the $J_L$-pole at $J_L=2-d$. In fact, there are other poles in $J_L$ that come from the $\rho$-dependent terms in the integrand of \eqref{eq:BFKL_int_inrhov}. In particular, from \eqref{eq:r1r2_to_rhov} we see that the $\r\to 0$ limit corresponds to the $z_1,z_2\sim z$ limit, where both $z_1$ and $z_2$ approach $z$. In this limit, the matrix element $f(\vec y_1,\vec y_2)$ can have collinear divergences. As we explain in the main text, for the matrix elements we consider in this work, the collinear divergence is given by \eqref{eq:NN_collinear_behavior}, which is equivalent to
\be
f(\vec y_1,\vec y_2) \sim \frac{C_{\text{coll}}}{\vec y_{12}^2}.
\ee
This implies the integral \eqref{eq:BFKL_int_inrhov} can be written as
\be
&\int d\O_{d-2}d\O_{d-3}\int_{-1}^{1}dx \int_0^{\oo}d\r \int_0^{1}dv\ v^{J_L+d-3}(1-v)^{J_L+d-3} \r^{J_L+2d-7}\nn \\
&\qquad\qquad\qquad \times (1-2v(1-v)(1+x))^{-\frac{J_L}{2}-1}(1-x^2)^{\frac{d-5}{2}}\p{C_{\text{coll}} + O(\r)}.
\ee
Due to the $\r^{J_L+2d-7}$ factor, the integral has a pole at $J_L=6-2d$. After doing the remaining integrals over $x$ and $v$, we obtain
\be\label{eq:BFKL_int_JLpole_collinear}
&\int d\O_{d-2}d\O_{d-3}\int_{-1}^{1}dx \int_0^{\oo}d\r \int_0^{1}dv\ v^{J_L+d-3}(1-v)^{J_L+d-3} \r^{J_L+2d-7}\nn \\
&\qquad\qquad\qquad\qquad \times (1-2v(1-v)(1+x))^{-\frac{J_L}{2}-1}(1-x^2)^{\frac{d-5}{2}}\p{C_{\text{coll}} + O(\r)} \nn \\
&\sim \frac{\vol(S^{d-3})\vol(S^{d-4})}{J_L-6+2d}\frac{\pi^{\frac{3}{2}}\G(2-\tfrac{d}{2})\G(\tfrac{3d}{2}-5)}{\G(\tfrac{5-d}{2})\G(\tfrac{d}{2}-1)\G(d-3)}C_{\text{coll}}.
\ee
This gives \eqref{eq:BFKL_int_JLpole_collinear_maintext}.

\subsection{One-loop matrix elements}\label{app:BFKL_1loop}
Here, let us give more details of the calculation for the BFKL one-loop matrix element in section \ref{sec:BFKL_1loop}. In particular we will compute explicitly the function $F_{ggg}(\z)$ defined in \eqref{eq:Fggg_definition}, and also give a derivation for \eqref{eq:BFKL_1loop_real_eq1}.

To compute $F_{ggg}(\z)$, we can first use the momentum-conserving delta function to remove the $k_3$-integral in left-hand side of \eqref{eq:Fggg_definition}. It is also convenient to make a change of variable $E_i = \frac{p^2}{2p\.z_i}\a_i$ so that the maximum value of $\a_i$ is $1$ due to energy conservation. Then, \eqref{eq:Fggg_definition} becomes
\be\label{eq:Fggg_eq1}
&(p^2)^{2d-4}(2p\.z_1)^{2-d}(2p\.z_2)^{2-d}\int \frac{\a_1^{d-2}d\a_1}{(2\pi)^{d-1}2\a_1}\frac{\a_2^{d-2}d\a_2}{(2\pi)^{d-1}2\a_2}2\pi \de(p^2(1-\a_1-\a_2+\a_1\a_2 \z)) \nn \\
&\qquad\qquad\qquad\qquad\qquad \times \cI^{\prime\cO,\text{tree}}_{ggg}(\tfrac{p^2}{2p\.z_1}\a_1 z_1,\tfrac{p^2}{2p\.z_2}\a_2 z_2,p-\tfrac{p^2}{2p\.z_1}\a_1 z_1 - \tfrac{p^2}{2p\.z_2}\a_2 z_2) \nn \\
&=(p^2)^{2d-4}(2p\.z_1)^{2-d}(2p\.z_2)^{2-d} F_{ggg}(\z),
\ee
where $\z$ is the cross-ratio defined in \eqref{eq:zeta_def}.

From the color structure of the three-particle form factor, one can show that the integrand $\cI^{\prime\cO,\text{tree}}_{ggg}$ \eqref{eq:Iprime_ggg_def} satisfies
\be
\cI^{\prime\cO,\text{tree}}_{ggg}=-\frac{C_A}{2}\cI^{\cO,\text{tree}}_{ggg},
\ee
where the expression of $\cI^{\cO,\text{tree}}_{ggg}$ is the squared form factor given in \eqref{eq:form_factor_sq_O_to_ggg}. Then, one can check that $\cI^{\prime\cO,\text{tree}}_{ggg}$ takes the form $\cI^{\prime\cO,\text{tree}}_{ggg}= p^2 \frac{1}{\a_1^2\a_2^2 \z (1-\a_1\z)(1-\a_2\z)}\tl \cI^{\prime\cO,\text{tree}}_{ggg}(\a_1,\a_2,\z)$, where $\tl \cI^{\prime\cO,\text{tree}}_{ggg}$ is a polynomial in $\a_1,\a_2$, and $\z$.

Therefore, by \eqref{eq:Fggg_eq1}, $F_{ggg}(\z)$ can be written as
\be
F_{ggg}(\z) = \int \frac{\a_1^{d-2}d\a_1}{(2\pi)^{d-1}2\a_1}\frac{\a_2^{d-2}d\a_2}{(2\pi)^{d-1}2\a_2} \frac{2\pi \de(1-\a_1-\a_2+\a_1\a_2 \z)}{\a_1^2\a_2^2 \z (1-\a_1\z)(1-\a_2\z)}\tl \cI^{\prime\cO,\text{tree}}_{ggg}(\a_1,\a_2,\z).
\ee
Since $\tl \cI^{\prime\cO,\text{tree}}_{ggg}$ is a polynomial, the integral can be computed using the following identity:
\be
&\int_0^1 d\a_1 d\a_2\,  \a_1^{d-3}\a_2^{d-3} \frac{\de(1-\a_1-\a_2+\a_1\a_2 \z)}{\a_1^2\a_2^2 (1-\a_1\z)(1-\a_2\z)} \a_1^{n_1}\a_2^{n_2} \nn \\
&=\frac{\G(d-4+n_1)\G(d-4+n_2)}{(1-\z)\G(2d-8+n_1+n_2)}{}_2F_1(d-4+n_1,d-4+n_2, 2d-8+n_1+n_2,\z).
\ee
Then, we obtain
\be\label{eq:Fggg_expr}
F_{ggg}(\z) = &-\frac{C_A^2(N_c^2-1)g^2 \tilde{\mu}^{2\e}}{2^{2d}\pi^{2d-3}(1-\z) \z }\Big(4 \z ^4 \tl f_{d,d}(\z )-16 \z ^3 \tl f_{d-1,d}(\z )+8 (d-1) \z ^2 \tl f_{d-2,d}(\z )\nn \\
&\qquad +12 \z ^2 \tl f_{d-1,d-1}(\z )-8 (d-2) \z  \tl f_{d-3,d}(\z )-8 (d-1) \z  \tl f_{d-2,d-1}(\z )\nn \\
&\qquad +2 (d-2) \tl f_{d-4,d}(\z )+4 (d-2) \tl f_{d-3,d-1}(\z )+3 (d-2) \tl f_{d-2,d-2}(\z ) \Big),
\ee
where $\tl f_{a,b}(\z) =\frac{\G(a)\G(b)}{\G(a+b)}{}_2F_1(a,b,a+b,\z)$.

We now explain how to derive \eqref{eq:BFKL_1loop_real_eq1}. As argued below \eqref{eq:BFKL_1loop_real_eq01}, the matrix element must be proportional to $(p^2)^{d-2-\frac{J_L}{2}}(2p\.z)^{J_L}$. Therefore, by \eqref{eq:BFKL_1loop_real_eq01} we have
\be
&\< \cD^{\BFKL}_{J_L,g}(z)\>_{[\cO]_R(p)}^{\text{1-loop,R}} \nn \\
&= \frac{\G(J_L+d-2)}{\G(\tfrac{J_L+d-2}{2})}(p^2)^{2d-4}\int D^{d-2}z_1 D^{d-2}z_2 \frac{(2z_1\.z_2)^{-\frac{J_L}{2}}(2p\.z_1)^{2-d}(2p\.z_2)^{2-d}}{(2z\.z_1)^{-\frac{J_L}{2}}(2z\.z_2)^{-\frac{J_L}{2}}}F_{ggg}(\z) \nn \\
&=C^{\text{1-loop,R}}_{\BFKL}\times (p^2)^{d-2-\frac{J_L}{2}}(2p\.z)^{J_L}
\ee
To extract the coefficient, we can integrate both sides against $\int D^{d-2}z(2p\.z)^{2-d-J_L}$. The integrals over $z$ can be evaluated using the identities \cite{Simmons-Duffin:2012juh}
\be\label{eq:zint_identity0}
\int D^{d-2}z (2p\.z)^{2-d} = \frac{\pi^{\frac{d-2}{2}}\G(\tfrac{d-2}{2})}{\G(d-2)}(p^2)^{\frac{2-d}{2}},
\ee
and\footnote{To derive this identity, one can first use Feynman/Schwinger parameterization to bring the integrand to the form $(2y\.z)^{2-d}$, and then use \eqref{eq:zint_identity0}.}
\be
&\int D^{d-2}z \frac{1}{(2z\.z_1)^{\frac{J_{L1}-J_L-J_{L2}}{2}}(2z\.z_2)^{\frac{J_{L2}-J_L-J_{L1}}{2}}(2z\.p)^{J_L+d-2}} \nn \\
=&\frac{\pi^{\frac{d-2}{2}}\G(\tfrac{d-2-J_{L1}+J_L+J_{L2}}{2})\G(\tfrac{d-2-J_{L2}+J_L+J_{L1}}{2})}{\G(\tfrac{d-2}{2})\G(J_L+d-2)}\nn \\
&\times  \frac{(p^2)^{\frac{2-d}{2}-J_L}}{(2p\.z_1)^{\frac{J_{L1}-J_L-J_{L2}}{2}}(2p\.z_2)^{\frac{J_{L2}-J_L-J_{L1}}{2}}}  {}_2F_1(\tfrac{J_{L1}-J_L-J_{L2}}{2},\tfrac{J_{L2}-J_L-J_{L1}}{2},\tfrac{d-2}{2},1-\z).
\ee

The coefficient $C^{\text{1-loop,R}}_{\BFKL}$ is then given by
\be
C^{\text{1-loop,R}}_{\BFKL} &= \frac{\G(d-2)}{\G(\tfrac{d-2}{2})^2}(p^2)^{d-2}\nn \\
&\times \int D^{d-2}z_1 D^{d-2}z_2 (2p\.z_1)^{2-d}(2p\.z_2)^{2-d}\z^{-\frac{J_L}{2}}{}_2F_1(-\tfrac{J_L}{2},-\tfrac{J_L}{2},\tfrac{d-2}{2},1-\z)F_{ggg}(\z).
\ee
This can be further simplified to an integral over the cross-ratio $\z$. The remaining integrals give the volume factor $\vol(S^{d-2})\vol(S^{d-3})$, and the Jacobian is $2^{d-3}(\z(1-\z))^{\frac{d-4}{2}}$. We find
\be
C^{\text{1-loop,R}}_{\BFKL} &= \frac{\pi^{d-2}}{\G(\tfrac{d-2}{2})^2}\int_0^1 d\z\, (1-\z)^{\frac{d-4}{2}}\z^{\frac{d-4-J_L}{2}}{}_2F_1(-\tfrac{J_L}{2},-\tfrac{J_L}{2},\tfrac{d-2}{2},1-\z)F_{ggg}(\z),
\ee
which is equivalent to \eqref{eq:BFKL_1loop_real_eq1}.

\subsection{Poles in $J_{L1}, J_{L2}$}\label{app:poles_BFKL_JL1JL2}
In this section, we show that the $\e$-pole of $F_{ggg}(\z)$ (defined in \eqref{eq:Fggg_definition}) is related to the leading soft theorem. To do this, it is helpful to consider the more general horizontal trajectory detector $\cD_{J_{L1},J_{L2},J_L,g}$\eqref{eq:GeneralizedBFKL_definition} with generic $J_{L1}$ and $J_{L2}$. For an $(n+1)$-gluon state with form factor $\cF_{n+1}$, the matrix element of $\cD_{J_{L1},J_{L2},J_L,g}$ is given by
\be
&\<\cD_{J_{L1},J_{L2},J_L,g}(z)\>_{\cF_{n+1}}\nn \\
&=\frac{\G(d-2+J_L)}{\G(\tfrac{d-2+J_L+J_{L1}-J_{L2}}{2})\G(\tfrac{d-2+J_L-J_{L1}+J_{L2}}{2})} \int D^{d-2}z_1 D^{d-2}z_2 \<\cP_{-\tl J_{L1}}(z_1)\cP_{-\tl J_{L2}}(z_2)\cP_{-J_L}(z)\> \nn \\
&\times \frac{1}{(n-1)!}\int \frac{E_1^{-J_{L1}}dE_1}{(2\pi)^{d-1}2E_1}\frac{E_2^{-J_{L2}}dE_2}{(2\pi)^{d-1}2E_2}\left[\prod_{i=3}^{n+1}\frac{d^dk_i\de^{+}(k_i^2)}{(2\pi)^{d-1}}\right](2\pi)^{d}\de^{(d)}(p-\sum_i k_i)\nn \\
&\qquad\qquad\qquad \times\left. \<\cF_{n+1}|T_1^c T_2^c|\cF_{n+1}\>\right|_{\substack{k_1\to E_1 z_1 \\ k_2 \to E_2 z_2}},
\ee
This integral can have $J_{L1}$-poles due to the soft limit $E_1 \to 0$ from the $E_1$-integral, and similar $J_{L2}$-poles from the $E_2 \to 0$ limit. When setting $J_{L1}=J_{L2}=2-d$, they become poles in $\e$.

Let us focus on the $J_{L1}$-pole here, since the analysis for the $J_{L2}$-pole is identical. The leading soft theorem \eqref{eq:soft_theorem} implies
\be
&\<\cD_{J_{L1},J_{L2},J_L,g}(z)\>_{\cF_{n+1}}\nn \\
&=\frac{\G(d-2+J_L)}{\G(\tfrac{d-2+J_L+J_{L1}-J_{L2}}{2})\G(\tfrac{d-2+J_L-J_{L1}+J_{L2}}{2})} \int D^{d-2}z_1 D^{d-2}z_2 \<\cP_{-\tl J_{L1}}(z_1)\cP_{-\tl J_{L2}}(z_2)\cP_{-J_L}(z)\> \nn \\
&\times \frac{1}{(n-1)!}\int \frac{E_1^{-J_{L1}}dE_1}{(2\pi)^{d-1}2E_1}\frac{E_2^{-J_{L2}}dE_2}{(2\pi)^{d-1}2E_2}\left[\prod_{i=3}^{n+1}\frac{d^dk_i\de^{+}(k_i^2)}{(2\pi)^{d-1}}\right](2\pi)^{d}\de^{(d)}(p-\sum_i k_i)\nn \\
& \times \bigg(g^2\tilde{\mu}^{2\e}\frac{1}{E_1^2}\sum_{i,j=2}^{n+1}\frac{\e_1^* \.k_i \e_1 \.k_j}{z_1\.k_i z_1\.k_j}(if^{acb}) \left.\<\cF_{n}|T_i^a T_2^c T_j^b|\cF_{n}\>\right|_{k_2\to E_2 z_2}+O(E_1^{-1})\bigg).
\ee
Therefore, the leading soft behavior leads to a pole at $J_{L1}=-2$ given by
\be
&\<\cD_{J_{L1},J_{L2},J_L,g}(z)\>_{\cF_{n+1}}\nn \\
&\sim \frac{\G(d-2+J_L)}{\G(\tfrac{d-2+J_L+J_{L1}-J_{L2}}{2})\G(\tfrac{d-2+J_L-J_{L1}+J_{L2}}{2})} \int D^{d-2}z_1 D^{d-2}z_2 \<\cP_{-\tl J_{L1}}(z_1)\cP_{-\tl J_{L2}}(z_2)\cP_{-J_L}(z)\> \nn \\
&\times \frac{1}{J_{L1}+2}\frac{1}{2^d\pi^{d-1}}\frac{1}{(n-1)!}\int\frac{E_2^{-J_{L2}}dE_2}{(2\pi)^{d-1}2E_2}\left[\prod_{i=3}^{n+1}\frac{d^dk_i\de^{+}(k_i^2)}{(2\pi)^{d-1}}\right](2\pi)^{d}\de^{(d)}(p-\sum_i k_i)\nn \\
& \times g^2\tilde{\mu}^{2\e}\sum_{i,j=2}^{n+1}\frac{k_i\.k_j}{z_1\.k_i z_1\.k_j}(if^{acb}) \left.\<\cF_{n}|T_i^a T_2^c T_j^b|\cF_{n}\>\right|_{k_2\to E_2 z_2},
\ee
where we have also performed the polarization sum over $\e_1$ using \eqref{eq:Pi_lightconegauge}.\footnote{The contributions from the gauge-dependent term $\frac{z_1^\mu n^\nu+z_1^\nu n^\mu}{z_1\.n}$ vanish due to the gauge invariance condition $\sum_i T_i^a|\cF_n\>=0$.} One can show that in the sum over $i,j$, we get vanishing contribution from $i,j = 3,\ldots, n$ because of the contraction $if^{abc}T_i^a T_2^c T_j^b$. On the other hand, for $i=3,\ldots, n, j=2$ we have
\be
if^{acb}\<\cF_n|T_i^a T_2^c T_2^b|\cF_n\> = -\frac{C_A}{2}\<\cF_n|T_2^a T_i^a|\cF_n\>.
\ee
Moreover, since either $k_i$ or $k_j$ is $k_2=E_2 z_2$, the $z_1$-integral can be evaluated using
\be
&\int D^{d-2}z_1 \<\cP_{-(4-d)}(z_1)\cP_{-\tl J_{L2}}(z_2)\cP_{-J_L}(z)\> \frac{2z_2\.z_3}{(2z_1\.z_2)(2z_1\.z_3)}  \nn \\
&= \frac{\pi^{\frac{d-2}{2}}\G(\tfrac{d-4}{2})\G(\tfrac{2-d-J_L-J_{L2}}{2})\G(\tfrac{d+J_L+J_{L2}}{2})}{\G(\tfrac{-2-J_L-J_{L2}}{2})\G(\tfrac{-4+2d+J_L+J_{L2}}{2})}\<\cP_{-\tl J_{L2}}(z_2)\cP_{0}(z_3)\cP_{-J_L}(z)\>.
\ee

Combining everything and using \eqref{eq:numberdetector_measure_identity}, we find
\be
&\<\cD_{J_{L1},J_{L2},J_L,g}(z)\>_{\cF_{n+1}}\nn \\
&\sim -\frac{g^2\tilde{\mu}^{2\e}C_A }{J_{L1}+2}\frac{\G(d-2+J_L)\G(\tfrac{d-4}{2})\G(\tfrac{2-d-J_L-J_{L2}}{2})}{2^{d-1}\pi^{\frac{d}{2}}\G(\tfrac{d-2+J_L-2-J_{L2}}{2})\G(\tfrac{-2-J_L-J_{L2}}{2})\G(\tfrac{-4+2d+J_L+J_{L2}}{2})}  \nn \\
&\times \int D^{d-2}z_2 D^{d-2}z_3 \<\cP_{-\tl J_{L2}}(z_2)\cP_{0}(z_3)\cP_{-J_L}(z)\>\nn \\
& \times \frac{1}{(n-2)!}\int\frac{E_2^{-J_{L2}}dE_2}{(2\pi)^{d-1}2E_2}\frac{E_3^{d-2}dE_3}{(2\pi)^{d-1}2E_3}\left[\prod_{i=4}^{n+1}\frac{d^dk_i\de^{+}(k_i^2)}{(2\pi)^{d-1}}\right](2\pi)^{d}\de^{(d)}(p-\sum_i k_i)\nn \\
&\qquad\qquad\qquad \times \left.\<\cF_{n}|T_2^a T_3^a|\cF_{n}\>\right|_{\substack{k_2\to E_2 z_2 \\ k_3 \to E_3 z_3}}.
\ee
We can then recognize that the above expression is proportional to the matrix element $\<\cD_{2-d,J_{L2},J_L,g}(z)\>_{\cF_n}$. Therefore, the result can be written as
\be\label{eq:BFKL_JL1pole_result}
&\<\cD_{J_{L1},J_{L2},J_L,g}(z)\>_{\cF_{n+1}}\nn \\
&\sim -\frac{g^2\tilde{\mu}^{2\e}C_A }{J_{L1}+2}\frac{\G(\tfrac{d-4}{2})\G(\tfrac{2-d-J_L-J_{L2}}{2})\G(\tfrac{J_L-J_{L2}}{2})}{2^{d-1}\pi^{\frac{d}{2}}\G(\tfrac{d-4+J_L-J_{L2}}{2})\G(\tfrac{-2-J_L-J_{L2}}{2})}\<\cD_{2-d,J_{L2},J_L,g}(z)\>_{\cF_n}.
\ee

In summary, we have shown that the matrix element of $\cD_{J_{L1},J_{L2},J_L,g}(z)$ in an $(n+1)$-gluon state has a pole at $J_{L1}=-2$ proportional to the matrix element of $\cD_{2-d,J_{L2},J_L,g}(z)$ in an $n$-gluon state. In particular, if we set $J_{L1}=2-d$, then the pole at $J_{L1}=-2$ becomes a pole in $\e$ of $\cD_{2-d,J_{L2},J_L,g}(z)$ which is proportional to the matrix element of itself. This corresponds to the $\e$-pole of the $F_{ggg}(\z)$ function in section \ref{sec:BFKL_1loop}. This analysis also shows why $J_{L1}=J_{L2}=2-d$ is special. It is the only possible value of $J_{L1}, J_{L2}$ such that the operator can be multiplicatively renormalized. For other values of $J_{L1}, J_{L2}$, the divergence \eqref{eq:BFKL_JL1pole_result} will be additively renormalized. A similar phenomenon appears when studying horizontal trajectories of the Wilson-Fisher theory \cite{Caron-Huot:2022eqs}.

\section{The ``fake trajectory" trick} \label{app:fake_trajectory}
In this appendix, we give the tree-level and one-loop matrix elements of the ``fake trajectory" introduced in section \ref{sec:CFplot_YM} and \ref{sec:CFplot_QCD}. We stress again that these trajectories are not physical. The only purpose is to remove the $J_L$-divergences in the dilatation matrix which come from the mixing between DGLAP or shadow DGLAP with some trajectories we currently do not understand. Therefore, we will choose the matrix elements to be as simple as possible, so long as they suffice to remove the $J_L$-divergences.

\begin{figure}
    \centering
    \includegraphics[width = 10cm]{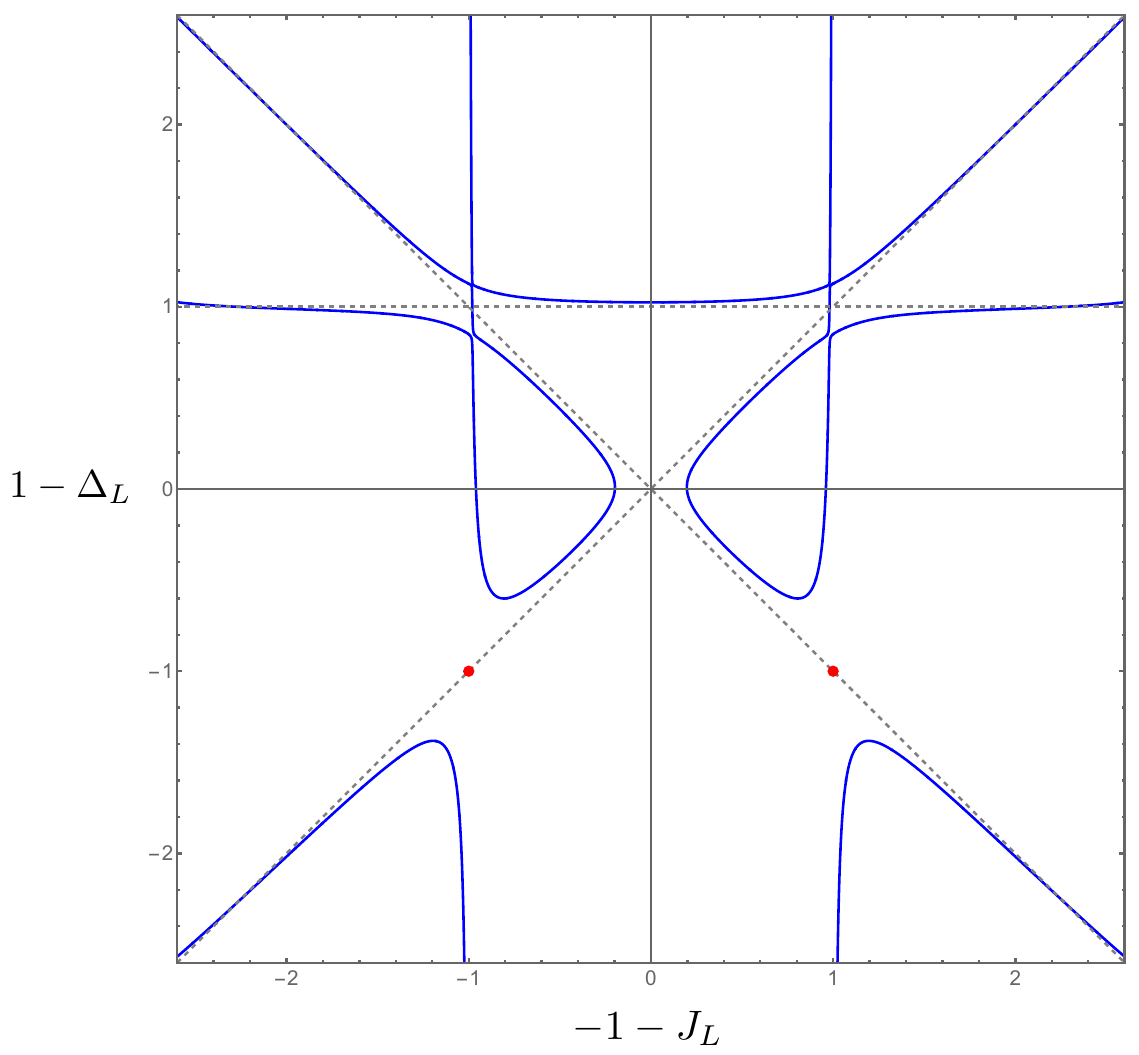}
    \caption{The Chew-Frautschi plot in pure YM at one-loop without introducing fake trajectories. Gray dashed lines are the trajectories in free theory. The red dots indicate the locations where the DGLAP and shadow DGLAP trajectories have unresolved $J_L$-poles, which infect the other parts of the plot.}
    \label{fig:CFplot_nofake}
\end{figure}

To better demonstrate the motivation for introducing the fake trajectories, we show in figure \ref{fig:CFplot_nofake} what the Chew-Frautschi plot would look like in pure YM at one-loop if we only considered the DGLAP, shadow DGLAP, and the BFKL trajectory. Even though one can still roughly see the trajectories being recombined near the intersections we are interested in, they are contaminated by additional divergences in $J_L$ of the DGLAP and shadow DGLAP trajectories, which come from the intersections marked by the red dots in the figure. We know that they must be resolved locally. In other words, to remove the divergences, we only need the behavior of the trajectories near each intersection. Conversely, the recombination at, for example, the DGLAP/BFKL intersection does not depend on how the red-dot intersections get resolved. Since our focus is on the regions away from the red dots, it's sufficient to find one possible way of resolving the red-dot intersections (even if it's not the physically correct one), so that the region we are interested in is not affected by them.

\subsection{Pure YM}
In the pure YM case in section \ref{sec:CFplot_YM}, we introduce a single fake trajectory given by a horizontal trajectory at $J=-1$. The bare detector $\cD^{\text{Fake}}_{J_L}$ has $\De_L=2$ for all $J_L$. The trajectory intersects with DGLAP at $J_L=4-d$ and with shadow DGLAP at $J_L= -2$. Since the two intersections are related by a spin shadow transform, we will focus on the intersection with DGLAP. 

The tree-level matrix element of $\cD^{\text{Fake}}_{J_L}$ in the gluon source is given by
\be
\<\cD^{\text{Fake}}_{J_L}\>^{\text{tree}}_{\cO(p)} =  \frac{d-2}{2^{d+1}\pi^{d-2}}(N_c^2-1)(2p\.z)^{J_L}(p^2)^{\frac{d-J_L-2}{2}}.
\ee
For generic $J_L$, the one-loop matrix element of $\cD^{\text{Fake}}_{J_L}$ has a pole in $\e$ given by
\be\label{eq:Fake_epspole}
\<\cD^{\text{Fake}}_{J_L}\>^{\text{1-loop}}_{[\cO]_R(p)} = \frac{\a_s}{4\pi}\frac{C_A}{\e}\p{-2\frac{A_{\text{Fake}}}{J_L(J_L+2)} + \g_{\text{Fake}}(J_L)}\<\cD^{\text{Fake}}_{J_L}\>^{\text{tree}}_{\cO(p)} + O(\e^0).
\ee

In order for the intersection to be resolved, the behavior of the fake trajectory and the DGLAP trajectory near the intersection should display the three effects described in section \ref{sec:structureofmixing}. In particular, we impose that the two bare detectors become identical at the tree-level intersection,
\be\label{eq:Fake_DGLAP_identity}
\cD^{\text{Fake}}_{4-d} = \cD^{\DGLAP}_{4-d,g}.
\ee
The pole of one-loop matrix element of the DGLAP detector at $J_L=0$ satisfies
\be\label{eq:DGLAP_JLpole_atzero}
\<\cD^{\DGLAP}_{J_L,g}\>^{\text{1-loop}}_{[\cO]_R(p)} = \frac{\a_s\tilde{\mu}^{2\e}}{4\pi}\frac{1}{J_L}\frac{C_A 2^{4\e}\pi^{\frac{1}{2}+\e}(4-13\e+14\e^2-5\e^3+2\e^4)}{\e(1-\e)\G(\tfrac{3}{2}-\e)}\<\cD^{\text{Fake}}_{J_L}\>^{\text{tree}}_{\cO(p)} + O(J_L^0).
\ee
On the other hand, the one-loop matrix element of $\cD^{\text{Fake}}_{J_L}$ has a pole at $J_L=4\e$ given by
\be\label{eq:Fake_JLpole}
\<\cD^{\text{Fake}}_{J_L}\>^{\text{1-loop}}_{[\cO]_R(p)} = \frac{\a_s\tilde{\mu}^{2\e}}{4\pi}\frac{C_A}{J_L-4\e}\p{-\frac{ A_{\text{Fake}}}{\e} + \cR_{\text{Fake}}(\e)}\<\cD^{\DGLAP}_{J_L,g}\>^{\text{tree}}_{\cO(p)} + O((J_L-4\e)^0).
\ee
Note that \eqref{eq:Fake_epspole} and \eqref{eq:Fake_JLpole} are chosen such that the double pole $\frac{1}{\e J_L}$ of the two equations agree with each other.

Finally, one also has to check that the one-loop matrix elements are consistent with the identity \eqref{eq:Fake_DGLAP_identity} and the $J_L$-pole of DGLAP \eqref{eq:DGLAP_JLpole_atzero}. As shown in \ref{sec:matrixelements_conditions}, this imposes nontrivial conditions on $A_{\text{Fake}}$, $\cR_{\text{Fake}}(\e)$, and $\g_{\text{Fake}}(J_L)$. We find that the simplest possible solution satisfying the conditions is given by
\be
A_{\text{Fake}} = 8,\quad \cR_{\text{Fake}}(\e) = \frac{2^{3-2\e}(5+3\g_E-3\log(4\pi))}{3(1-\e)\pi^{2\e}},\quad \g_{\text{Fake}}(J_L)=0.
\ee
This completely fixes the divergences of the one-loop matrix elements of $\cD^{\text{Fake}}_{J_L}$ near its intersection with the DGLAP trajectory.

\subsection{QCD}
The fake trajectories in the QCD case in section \ref{sec:CFplot_QCD} can be constructed in a similar way. However, due to the gluon/quark degeneracy, we need to introduce two horizontal trajectories at $J=-1$. Moreover, the DGLAP anomalous dimension matrix $\hat{\g}^{(0)}(J_L)$ in QCD has divergences at $J_L=1$ (which are absent in pure YM). This corresponds to $\De_L=3$ or $J=-2$, and thus we will also introduce two more fake horizontal trajectories at $J=-2$ to cancel the divergences. In total, we have four fake trajectories.

For the two fake horizontal trajectories at $J=-1$, their tree-level matrix elements in the gluon source and quark source are given by
\be
\<\cD^{\text{Fake}}_{J_L,i}\>^{\text{tree}}_{\cO(p)} &=  \frac{d-2}{2^{d+1}\pi^{d-2}}(N_c^2-1)(2p\.z)^{J_L}(p^2)^{\frac{d-J_L-2}{2}}, \nn \\
\<\cD^{\text{Fake}}_{J_L,i}\>^{\text{tree}}_{J(p)} &= \a_i \frac{d-2}{2^{d-3}\pi^{d-2}}N_c (2p\.z)^{J_L}(p^2)^{\frac{d-J_L-4}{2}},
\ee
where $i=1,2$. At the tree-level intersection $J_L=4-d$, they become identical to linear combinations of the DGLAP detectors:
\be\label{eq:Fake_identity_QCD}
\cD^{\text{Fake}}_{4-d,i} = \cD^{\DGLAP}_{4-d,g} + \a_i \cD^{\DGLAP}_{4-d,q}.
\ee

The $\e$-poles of the one-loop matrix elements are
\be
\<\cD^{\text{Fake}}_{J_L,i}\>^{\text{1-loop}}_{[\cO]_R(p)} = \frac{\a_s}{4\pi}\frac{C_A}{\e}\p{-2\frac{A_{\text{Fake},i}}{J_L(J_L+2)} + \g_{\text{Fake},i}(J_L)}\<\cD^{\text{Fake}}_{J_L,i}\>^{\text{tree}}_{\cO(p)} + O(\e^0), \nn \\
\<\cD^{\text{Fake}}_{J_L,i}\>^{\text{1-loop}}_{[J]_R(p)} = \frac{\a_s}{4\pi}\frac{C_A}{\e}\p{-2\frac{A_{\text{Fake},i}}{J_L(J_L+2)} + \g_{\text{Fake},i}(J_L)}\<\cD^{\text{Fake}}_{J_L,i}\>^{\text{tree}}_{J(p)} + O(\e^0).
\ee
In other words, the two fake trajectories do not mix with each other at generic $J_L$. At $J_L=4\e$,  the $J_L$-poles are given by
\be
\<\cD^{\text{Fake}}_{J_L,i}\>^{\text{1-loop}}_{[\cO]_R(p)} &=\frac{\a_s\tilde{\mu}^{2\e}}{4\pi}\frac{C_A}{J_L-4\e}\p{-\frac{ A_{\text{Fake},i}}{\e} + \cR^{(g)}_{\text{Fake},i}(\e)}\<\cD^{\DGLAP}_{J_L,g}\>^{\text{tree}}_{\cO(p)} + O((J_L-4\e)^0), \nn \\
\<\cD^{\text{Fake}}_{J_L,i}\>^{\text{1-loop}}_{[J]_R(p)} &=\frac{\a_s\tilde{\mu}^{2\e}}{4\pi}\frac{C_A}{J_L-4\e}\a_i\p{-\frac{ A_{\text{Fake},i}}{\e} + \cR^{(q)}_{\text{Fake},i}(\e)}\<\cD^{\DGLAP}_{J_L,q}\>^{\text{tree}}_{J(p)} + O((J_L-4\e)^0).
\ee
After imposing the identity \eqref{eq:Fake_identity_QCD} for the one-loop matrix elements, we find that for $i=1$ we can choose
\be\label{eq:Fake_QCD_solution}
A_{\text{Fake},1} &= \frac{4C_A+C_F+\sqrt{\De}}{C_A},\, \a_1=\frac{4C_A-C_F-\sqrt{\De}}{8n_f T_F},\, \g_{\text{Fake},1}(J_L)= 0, \nn \\
\cR^{(g)}_{\text{Fake},1}(\e) &= \frac{92C_A-3C_F-16n_f T_F -3\sqrt{\De}+6(4C_A+C_F+\sqrt{\De})(\g_E - \log(4\pi))}{3C_A2^{1+2\e}\pi^{2\e}(1-\e)}, \nn \\
\cR^{(q)}_{\text{Fake},1}(\e) &=\frac{4(4C_A-2C_F+\sqrt{\De})+(4C_A+C_F+\sqrt{\De})(\g_E - \log(4\pi))}{C_A(2\pi)^{2\e}(1-\e)},
\ee
where $\De=(4C_A-C_F)^2-16C_Fn_f T_F$. The matrix elements for $i=2$ can be obtained by replacing $\sqrt{\De} \to -\sqrt{\De}$.

Finally, we can follow the same procedure to construct the matrix elements of the two horizontal trajectories at $J=-2$, which will cancel the $J_L$-pole of DGLAP at $J_L=1$ and the $J_L$-pole of shadow DGLAP at $J_L=-3$. Let us denote them as $\cD^{\prime\text{Fake}}_{J_L,i}$, and simply give the final results. The tree-level matrix elements are
\be
\<\cD^{\prime\text{Fake}}_{J_L,i}\>^{\text{tree}}_{\cO(p)} &=  \frac{d-2}{2^{d+1}\pi^{d-2}}(N_c^2-1)(2p\.z)^{J_L}(p^2)^{\frac{d-J_L-3}{2}}, \nn \\
\<\cD^{\prime\text{Fake}}_{J_L,1}\>^{\text{tree}}_{J(p)} &= \frac{C_F(d-2)}{n_f T_F(2\pi)^{d-2}}N_c (2p\.z)^{J_L}(p^2)^{\frac{d-J_L-5}{2}},\quad \<\cD^{\prime\text{Fake}}_{J_L,2}\>^{\text{tree}}_{J(p)} =0.
\ee
At the tree-level intersection with DGLAP $J_L=5-d$, the bare operators $\cD^{\prime\text{Fake}}_{J_L,i}$ obey
\be
\cD^{\prime\text{Fake}}_{5-d,1} &= \cD^{\DGLAP}_{5-d,g} + \frac{C_F}{2n_f T_F}\cD^{\DGLAP}_{5-d,q}, \quad \cD^{\prime\text{Fake}}_{5-d,2} = \cD^{\DGLAP}_{5-d,g}.
\ee

The $\e$-poles of the one-loop matrix elements are
\be
\<\cD^{\prime\text{Fake}}_{J_L,1}\>^{\text{1-loop}}_{[\cO]_R(p)} = -\frac{\a_s}{4\pi}\frac{C_F}{\e}\frac{16}{(J_L+3)(J_L-1)}\<\cD^{\prime\text{Fake}}_{J_L,1}\>^{\text{tree}}_{\cO(p)} + O(\e^0), \quad \<\cD^{\prime\text{Fake}}_{J_L,2}\>^{\text{1-loop}}_{[\cO]_R(p)} =0, \nn \\
\<\cD^{\prime\text{Fake}}_{J_L,1}\>^{\text{1-loop}}_{[J]_R(p)} = -\frac{\a_s}{4\pi}\frac{C_F}{\e}\frac{16}{(J_L+3)(J_L-1)}\<\cD^{\prime\text{Fake}}_{J_L,1}\>^{\text{tree}}_{J(p)} + O(\e^0), \quad \<\cD^{\prime\text{Fake}}_{J_L,2}\>^{\text{1-loop}}_{[J]_R(p)} =0.
\ee
The one-loop matrix elements also have poles at $J_L=1+4\e$ which can be written as
\be
\<\cD^{\prime\text{Fake}}_{J_L,1}\>^{\text{1-loop}}_{[\cO]_R(p)} &=-\frac{\a_s\tilde{\mu}^{2\e}}{4\pi}\frac{\p{\frac{4C_F}{\e} + \frac{22C_A+4n_f T_F -6C_F(1+\g_E-\log(4\pi))}{3(1-\e)2^{-1+2\e}\pi^{2\e}}}}{J_L-1-4\e}\<\cD^{\DGLAP}_{J_L,g}\>^{\text{tree}}_{\cO(p)} + O((J_L-1-4\e)^0), \nn \\
\<\cD^{\prime\text{Fake}}_{J_L,2}\>^{\text{1-loop}}_{[\cO]_R(p)} &=-\frac{\a_s\tilde{\mu}^{2\e}}{4\pi}\frac{\frac{22C_A+4n_f T_F }{3(1-\e)2^{-1+2\e}\pi^{2\e}}}{J_L-1-4\e}\<\cD^{\DGLAP}_{J_L,g}\>^{\text{tree}}_{\cO(p)} + O((J_L-1-4\e)^0), \nn \\
\<\cD^{\prime\text{Fake}}_{J_L,1}\>^{\text{1-loop}}_{[J]_R(p)} &=-\frac{\a_s\tilde{\mu}^{2\e}}{4\pi}\frac{2C_F}{n_f T_F}\frac{\p{\frac{C_F}{\e} + \frac{2n_f T_F + C_F(1-\g_E+\log(4\pi))}{(1-\e)(2\pi)^{2\e}}}}{J_L-1-4\e}\<\cD^{\DGLAP}_{J_L,q}\>^{\text{tree}}_{J(p)} + O((J_L-1-4\e)^0), \nn \\
\<\cD^{\prime\text{Fake}}_{J_L,2}\>^{\text{1-loop}}_{[J]_R(p)} &=-\frac{\a_s\tilde{\mu}^{2\e}}{4\pi}\frac{\frac{4^{1-\e}C_F}{(1-\e)\pi^{2\e}}}{J_L-1-4\e}\<\cD^{\DGLAP}_{J_L,q}\>^{\text{tree}}_{J(p)} + O((J_L-1-4\e)^0).
\ee

\section{Details of the ancillary file}\label{app:anc}
The ancillary file contains four matrices: $\tt{DivMatPureYM}$, $\tt{DilatationPureYM}$, $\tt{DivMatQCD}$, and $\tt{DilatationQCD}$. Their precise definitions are as follows.

$\tt{DivMatPureYM}$ and $\tt{DilatationPureYM}$ are 4-by-4 matrices acting on the multiplet
\be\label{eq:regular_basis_pureYM_withFake}
\mathbb{D}_{J_L} &= U_{\text{YM-Fake}}\begin{pmatrix} \vspace{1mm} \mu^{J_L+2-2\e}\cD^{\DGLAP}_{J_L,g} \\ \vspace{1mm} \mu^{-J_L}\tl{\cD}^{\DGLAP}_{J_L,g} \\ \cD^{\BFKL}_{J_L,g} \\ \mu^2 \cD^{\text{Fake}}_{J_L}\end{pmatrix}, \nn \\
U_{\text{YM-Fake}}&=\begin{pmatrix} \vspace{1mm} 1 & 1 & 0 & 0 \\ \vspace{1mm} -\frac{1}{J_L+1-\e} & \frac{1}{J_L+1-\e} & 0 & 0 \\ \frac{C_A\frac{\pi^{1-\e}}{\G(1-\e)}}{(-2+2\e-J_L)(-2+2\e)} & \frac{C_A\frac{\pi^{1-\e}}{\G(1-\e)}}{J_L(-2+2\e)} & \frac{1}{J_L(-2+2\e-J_L)} &0 \\ \frac{1}{2(J_L-2\e)(1+\e)} & -\frac{1}{2(J_L+2)(1+\e)} & 0 & -\frac{1}{(J_L+2)(J_L-2\e)} \end{pmatrix}.
\ee
The renormalized detectors are given by acting $\p{1+\tt{DivMatPureYM}}^{-1}$ on the bare detectors,
\be
\left[\mathbb{D}_{J_L}\right]_R =\p{1+\text{\tt{DivMatPureYM}}}^{-1}\, \mathbb{D}_{J_L}.
\ee
Lastly, $\tt{DilatationPureYM}$ is the action of the dilatation operator on the renormalized detectors at $\e=0$:
\be
D\left[\mathbb{D}_{J_L}\right]_R =\mathscr{D} \left[\mathbb{D}_{J_L}\right]_R,\quad  \left.\mathscr{D}\right|_{\e\to 0}=\text{\tt{DilatationPureYM}}.
\ee

Similarly, $\tt{DivMatQCD}$ and $\tt{DilatationQCD}$ are 9-by-9 matrices acting on the following multiplet:
\be
&\mathbb{D}_{J_L} = U_{\text{QCD-Fake}}\begin{pmatrix} \vspace{1mm} \mu^{J_L+2-2\e}\cD^{\DGLAP}_{J_L,g} \\  \vspace{1mm} \mu^{J_L+2-2\e}\cD^{\DGLAP}_{J_L,q} \\ \vspace{1mm} \mu^{-J_L}\tl{\cD}^{\DGLAP}_{J_L,g} \\ \vspace{1mm} \mu^{-J_L}\tl{\cD}^{\DGLAP}_{J_L,q} \\ \vspace{1mm} \cD^{\BFKL}_{J_L} \\ \vspace{1mm} \mu^2 \cD^{\text{Fake}}_{J_L,1} \\ \vspace{1mm} \mu^2 \cD^{\text{Fake}}_{J_L,2} \\ \vspace{1mm} \mu^3 \cD^{\prime\text{Fake}}_{J_L,1} \\ \vspace{1mm} \mu^3 \cD^{\prime\text{Fake}}_{J_L,2} \end{pmatrix}, \nn \\
&U_{\text{QCD-Fake}}= \nn \\
&\scalebox{0.65}{$
\begin{pmatrix} 
1  & 0 & 1 & 0 & 0 & 0 & 0 & 0 & 0 \\ 
 0  & 1 & 0 & 1 & 0 & 0 & 0 & 0 & 0 \\ 
-\frac{1}{J_L+1-\e} & 0 &  \frac{1}{J_L+1-\e} & 0 & 0 & 0 & 0 & 0 & 0 \\
0 & -\frac{1}{J_L+1-\e} & 0 &  \frac{1}{J_L+1-\e} & 0 & 0 & 0 & 0 & 0  \\ 
\frac{C_A\frac{\pi^{1-\e}}{\G(1-\e)}}{(-2+2\e-J_L)(-2+2\e)} & \frac{C_F\frac{\pi^{1-\e}}{\G(1-\e)}}{(-2+2\e-J_L)(-2+2\e)} & \frac{C_A\frac{\pi^{1-\e}}{\G(1-\e)}}{J_L(-2+2\e)} & \frac{C_F\frac{\pi^{1-\e}}{\G(1-\e)}}{J_L(-2+2\e)} & \frac{1}{J_L(-2+2\e-J_L)} & 0 & 0 & 0 & 0 \\ 
-\frac{1}{(2+2\e)(-J_L+2\e)} & -\frac{\a_1}{(2+2\e)(-J_L+2\e)} & -\frac{1}{(J_L+2)(2+2\e)} & -\frac{\a_1}{(J_L+2)(2+2\e)} & 0 & \frac{1}{(J_L+2)(-J_L+2\e)} & 0 & 0 & 0 \\ 
-\frac{1}{(2+2\e)(-J_L+2\e)} & -\frac{\a_2}{(2+2\e)(-J_L+2\e)} & -\frac{1}{(J_L+2)(2+2\e)} & -\frac{\a_2}{(J_L+2)(2+2\e)} & 0  & 0 & \frac{1}{(J_L+2)(-J_L+2\e)}  & 0 & 0 \\ 
-\frac{1}{(4+2\e)(1-J_L+2\e)} & -\frac{C_F}{2n_f T_F(4+2\e)(1-J_L+2\e)} & -\frac{1}{(J_L+3)(4+2\e)} & -\frac{C_F}{2n_f T_F(J_L+3)(4+2\e)} & 0 & 0 & 0 & \frac{1}{(J_L+3)(1-J_L+2\e)} & 0 \\ 
-\frac{1}{(4+2\e)(1-J_L+2\e)} & 0 & -\frac{1}{(J_L+3)(4+2\e)} & 0 & 0 & 0 & 0  & 0 & \frac{1}{(J_L+3)(1-J_L+2\e)} 
\end{pmatrix}
$},
\ee
where the coefficients $\a_i$ are given by \eqref{eq:Fake_QCD_solution} and the description below.

\section{One-loop matrix elements in QCD}\label{app:QCD_1loop}

In this appendix, we collect the results of one-loop calculation of matrix elements used in this paper. The procedure of extracting the divergences are given in section \ref{sec:DGLAP_1loop}, \ref{sec:BFKL_1loop} and appendix \ref{app:BFKL_1loop}. 

\subsection{Matrix elements of DGLAP detectors}

In section \ref{sec:DGLAP_1loop}, we provide the calculation of the gluon DGLAP detector matrix element $\<\cD^{\DGLAP}_{J_L,g}\>_{\cO}$ at one loop. Here we list other cases.

\begin{figure}[ht]
    \centering
    
    \subfloat[]{%
        \begin{tikzpicture}
            \node[draw, black, fill=black, circle, inner sep=0pt, minimum size=5pt] (a) at (0, 0){};
            \node[draw, black, fill=black, circle, inner sep=0pt, minimum size=5pt] (a2) at (-3, 0){};

            \begin{feynman}
                \vertex (b) at (-1.5, 1.5);
                \vertex (c) at (-1.5, -1.5);
                \vertex (d) at (-1.5, 0);
                \vertex (e) at (-0.75, -0.75);
                \vertex (e2) at (-2.25, -0.75);

                \diagram* {
                    (a) -- [gluon] (b),
                    (a) -- [gluon] (e),
                    (a2) -- [gluon] (b),
                    (a2) -- [gluon] (e2),
                    (c) -- [fermion] (e) -- [fermion] (d) -- [fermion] (e2) -- [fermion] (c),
                };
            \end{feynman}
            \draw [dashed, line width=1.2pt] (-1.5, 1.8) -- (-1.5, -1.8);
            \node[draw, red, fill=red, circle, inner sep=0pt, minimum size=5pt] (d) at (-1.5, -1.5){};
            \node[right] at (-1.5,-1.5) {${\color{red} \cD^{\DGLAP}_{J_L,q}(z)}$};
        \end{tikzpicture}
    }\quad
    \subfloat[]{%
        \begin{tikzpicture}
            \node[draw, black, fill=black, circle, inner sep=0pt, minimum size=5pt] (a) at (0, 0){};
            \node[draw, black, fill=black, circle, inner sep=0pt, minimum size=5pt] (a2) at (-3, 0){};

            \begin{feynman}
                \vertex (b) at (-1.5, 1.5);
                \vertex (c) at (-1.5, -1.5);
                \vertex (d) at (-1.5, 0);
                \vertex (e) at (-0.75, -0.75);
                \vertex (e2) at (-2.25, -0.75);

                \diagram* {
                    (a) -- [gluon] (b),
                    (a) -- [gluon] (e),
                    (a2) -- [gluon] (b),
                    (a2) -- [gluon] (e2),
                    (c) -- [fermion] (e2) -- [fermion] (d) -- [fermion] (e) -- [fermion] (c),
                };
            \end{feynman}
            \draw [dashed, line width=1.2pt] (-1.5, 1.8) -- (-1.5, -1.8);
            \node[draw, red, fill=red, circle, inner sep=0pt, minimum size=5pt] (d) at (-1.5, -1.5){};
            \node[right] at (-1.5,-1.5) {${\color{red} \cD^{\DGLAP}_{J_L,q}(z)}$};
        \end{tikzpicture}
    }
    
    \caption{Feynman diagrams for one-loop matrix element $\<\cD^{\DGLAP}_{J_L,q}(z)\>_{\cO(p)}$. Red dot represents the DGLAP measurement that sums quark and antiquark contributions.}
    \label{fig:DGLAP_q_in_O}
\end{figure}
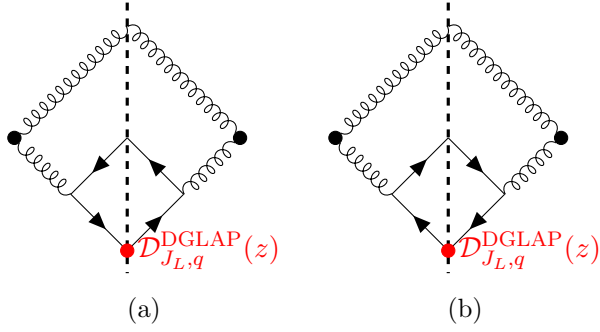

\begin{figure}[ht]
    \centering
    
    \subfloat[]{%
        \begin{tikzpicture}
            \node[draw, black, fill=black, circle, inner sep=0pt, minimum size=5pt] (a) at (0, 0){};
            \node[draw, black, fill=black, circle, inner sep=0pt, minimum size=5pt] (a2) at (-3, 0){};

            \begin{feynman}
                \vertex (b) at (-1.5, 1.5);
                \vertex (c) at (-1.5, -1.5);
                \vertex (d) at (-1.5, 0);
                \vertex (e) at (-0.75, -0.75);
                \vertex (e2) at (-2.25, -0.75);

                \diagram* {
                    (a) -- [fermion] (b) -- [fermion] (a2) -- [fermion] (e2) -- [fermion] (c) -- [fermion] (e) -- [fermion] (a),
                    (e) -- [gluon] (d) -- [gluon] (e2),
                };
            \end{feynman}
            \draw [dashed, line width=1.2pt] (-1.5, 1.8) -- (-1.5, -1.8);
            \node[draw, red, fill=red, circle, inner sep=0pt, minimum size=5pt] (d) at (-1.5, 0){};
            \node[above] at (-1.5,0) {${\color{red} \cD^{\DGLAP}_{J_L,g}(z)}$};
        \end{tikzpicture}
    }\quad
    \subfloat[]{%
        \begin{tikzpicture}
            \node[draw, black, fill=black, circle, inner sep=0pt, minimum size=5pt] (a) at (0, 0){};
            \node[draw, black, fill=black, circle, inner sep=0pt, minimum size=5pt] (a2) at (-3, 0){};

            \begin{feynman}
                \vertex (b) at (-1.5, 1.5);
                \vertex (c) at (-1.5, -1.5);
                \vertex (d) at (-1.5, 0);
                \vertex (e) at (-0.75, -0.75);
                \vertex (e2) at (-2.25, 0.75);

                \diagram* {
                    (a) -- [fermion] (b)-- [fermion] (e2) -- [fermion] (a2)  -- [fermion] (c) -- [fermion] (e) -- [fermion] (a),
                    (e) -- [gluon] (d) -- [gluon] (e2),
                };
            \end{feynman}
            \draw [dashed, line width=1.2pt] (-1.5, 1.8) -- (-1.5, -1.8);
            \node[draw, red, fill=red, circle, inner sep=0pt, minimum size=5pt] (d) at (-1.5, 0){};
            \node[above] at (-0.5,0) {${\color{red} \cD^{\DGLAP}_{J_L,g}(z)}$};
        \end{tikzpicture}
    }\quad
    \subfloat[]{%
        \begin{tikzpicture}
            \node[draw, black, fill=black, circle, inner sep=0pt, minimum size=5pt] (a) at (0, 0){};
            \node[draw, black, fill=black, circle, inner sep=0pt, minimum size=5pt] (a2) at (-3, 0){};

            \begin{feynman}
                \vertex (b) at (-1.5, 1.5);
                \vertex (c) at (-1.5, -1.5);
                \vertex (d) at (-1.5, 0);
                \vertex (e) at (-0.75, 0.75);
                \vertex (e2) at (-2.25, 0.75);

                \diagram* {
                    (a)-- [fermion] (e) -- [fermion] (b) -- [fermion] (e2) -- [fermion] (a2)  -- [fermion] (c)  -- [fermion] (a),
                    (e) -- [gluon] (d) -- [gluon] (e2),
                };
            \end{feynman}
            \draw [dashed, line width=1.2pt] (-1.5, 1.8) -- (-1.5, -1.8);
            \node[draw, red, fill=red, circle, inner sep=0pt, minimum size=5pt] (d) at (-1.5, 0){};
            \node[below] at (-1.5,0) {${\color{red} \cD^{\DGLAP}_{J_L,g}(z)}$};
        \end{tikzpicture}
    }\quad
    \subfloat[]{%
        \begin{tikzpicture}
            \node[draw, black, fill=black, circle, inner sep=0pt, minimum size=5pt] (a) at (0, 0){};
            \node[draw, black, fill=black, circle, inner sep=0pt, minimum size=5pt] (a2) at (-3, 0){};

            \begin{feynman}
                \vertex (b) at (-1.5, 1.5);
                \vertex (c) at (-1.5, -1.5);
                \vertex (d) at (-1.5, 0);
                \vertex (e) at (-0.75, 0.75);
                \vertex (e2) at (-2.25, -0.75);

                \diagram* {
                    (a)-- [fermion] (e) -- [fermion] (b)  -- [fermion] (a2) -- [fermion] (e2)  -- [fermion] (c)  -- [fermion] (a),
                    (e) -- [gluon] (d) -- [gluon] (e2),
                };
            \end{feynman}
            \draw [dashed, line width=1.2pt] (-1.5, 1.8) -- (-1.5, -1.8);
            \node[draw, red, fill=red, circle, inner sep=0pt, minimum size=5pt] (d) at (-1.5, 0){};
            \node[below] at (-0.5,0) {${\color{red} \cD^{\DGLAP}_{J_L,g}(z)}$};
        \end{tikzpicture}
    }
    
    \caption{Feynman diagrams for one-loop matrix element $\<\cD^{\DGLAP}_{J_L,g}(z)\>_{J(p)}$. Red dot represents the DGLAP measurement for gluon.}
    \label{fig:DGLAP_g_in_J}
\end{figure}
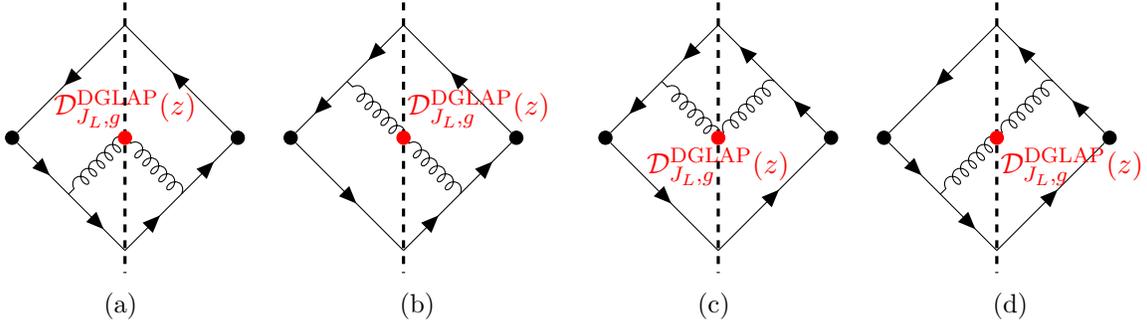

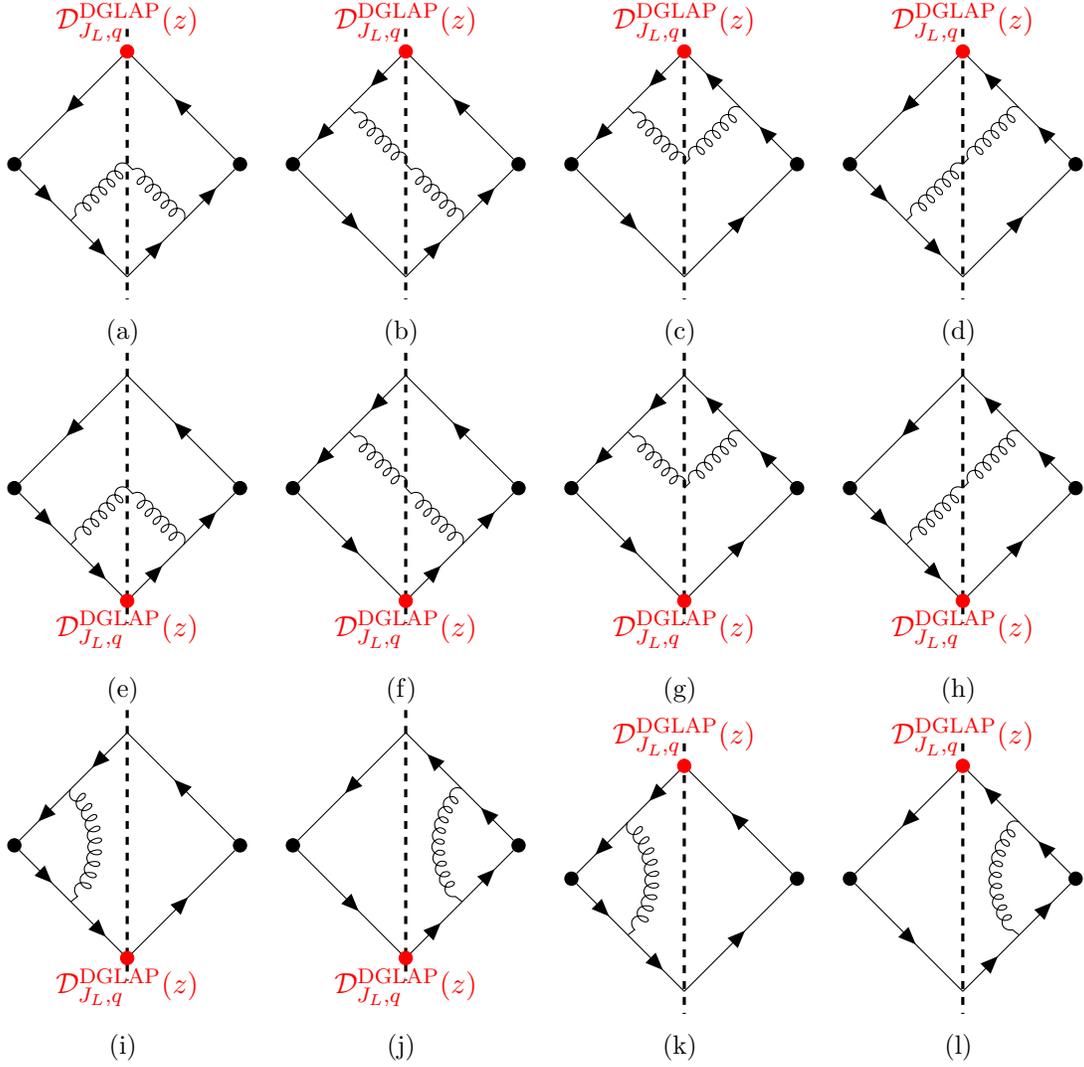
\begin{figure}[ht]
    \centering
    
    \subfloat[]{%
        \begin{tikzpicture}
            \node[draw, black, fill=black, circle, inner sep=0pt, minimum size=5pt] (a) at (0, 0){};
            \node[draw, black, fill=black, circle, inner sep=0pt, minimum size=5pt] (a2) at (-3, 0){};

            \begin{feynman}
                \vertex (b) at (-1.5, 1.5);
                \vertex (c) at (-1.5, -1.5);
                \vertex (d) at (-1.5, 0);
                \vertex (e) at (-0.75, -0.75);
                \vertex (e2) at (-2.25, -0.75);

                \diagram* {
                    (a) -- [fermion] (b) -- [fermion] (a2) -- [fermion] (e2) -- [fermion] (c) -- [fermion] (e) -- [fermion] (a),
                    (e) -- [gluon] (d) -- [gluon] (e2),
                };
            \end{feynman}
            \draw [dashed, line width=1.2pt] (-1.5, 1.8) -- (-1.5, -1.8);
            \node[draw, red, fill=red, circle, inner sep=0pt, minimum size=5pt] (d) at (-1.5, 1.5){};
            \node[above] at (-1.5,1.5) {${\color{red} \cD^{\DGLAP}_{J_L,q}(z)}$};
        \end{tikzpicture}
    }\quad
    \subfloat[]{%
        \begin{tikzpicture}
            \node[draw, black, fill=black, circle, inner sep=0pt, minimum size=5pt] (a) at (0, 0){};
            \node[draw, black, fill=black, circle, inner sep=0pt, minimum size=5pt] (a2) at (-3, 0){};

            \begin{feynman}
                \vertex (b) at (-1.5, 1.5);
                \vertex (c) at (-1.5, -1.5);
                \vertex (d) at (-1.5, 0);
                \vertex (e) at (-0.75, -0.75);
                \vertex (e2) at (-2.25, 0.75);

                \diagram* {
                    (a) -- [fermion] (b)-- [fermion] (e2) -- [fermion] (a2)  -- [fermion] (c) -- [fermion] (e) -- [fermion] (a),
                    (e) -- [gluon] (d) -- [gluon] (e2),
                };
            \end{feynman}
            \draw [dashed, line width=1.2pt] (-1.5, 1.8) -- (-1.5, -1.8);
            \node[draw, red, fill=red, circle, inner sep=0pt, minimum size=5pt] (d) at (-1.5, 1.5){};
            \node[above] at (-1.5,1.5) {${\color{red} \cD^{\DGLAP}_{J_L,q}(z)}$};
        \end{tikzpicture}
    }\quad
    \subfloat[]{%
        \begin{tikzpicture}
            \node[draw, black, fill=black, circle, inner sep=0pt, minimum size=5pt] (a) at (0, 0){};
            \node[draw, black, fill=black, circle, inner sep=0pt, minimum size=5pt] (a2) at (-3, 0){};

            \begin{feynman}
                \vertex (b) at (-1.5, 1.5);
                \vertex (c) at (-1.5, -1.5);
                \vertex (d) at (-1.5, 0);
                \vertex (e) at (-0.75, 0.75);
                \vertex (e2) at (-2.25, 0.75);

                \diagram* {
                    (a)-- [fermion] (e) -- [fermion] (b) -- [fermion] (e2) -- [fermion] (a2)  -- [fermion] (c)  -- [fermion] (a),
                    (e) -- [gluon] (d) -- [gluon] (e2),
                };
            \end{feynman}
            \draw [dashed, line width=1.2pt] (-1.5, 1.8) -- (-1.5, -1.8);
            \node[draw, red, fill=red, circle, inner sep=0pt, minimum size=5pt] (d) at (-1.5, 1.5){};
            \node[above] at (-1.5,1.5) {${\color{red} \cD^{\DGLAP}_{J_L,q}(z)}$};
        \end{tikzpicture}
    }\quad
    \subfloat[]{%
        \begin{tikzpicture}
            \node[draw, black, fill=black, circle, inner sep=0pt, minimum size=5pt] (a) at (0, 0){};
            \node[draw, black, fill=black, circle, inner sep=0pt, minimum size=5pt] (a2) at (-3, 0){};

            \begin{feynman}
                \vertex (b) at (-1.5, 1.5);
                \vertex (c) at (-1.5, -1.5);
                \vertex (d) at (-1.5, 0);
                \vertex (e) at (-0.75, 0.75);
                \vertex (e2) at (-2.25, -0.75);

                \diagram* {
                    (a)-- [fermion] (e) -- [fermion] (b)  -- [fermion] (a2) -- [fermion] (e2)  -- [fermion] (c)  -- [fermion] (a),
                    (e) -- [gluon] (d) -- [gluon] (e2),
                };
            \end{feynman}
            \draw [dashed, line width=1.2pt] (-1.5, 1.8) -- (-1.5, -1.8);
            \node[draw, red, fill=red, circle, inner sep=0pt, minimum size=5pt] (d) at (-1.5, 1.5){};
            \node[above] at (-1.5,1.5) {${\color{red} \cD^{\DGLAP}_{J_L,q}(z)}$};
        \end{tikzpicture}
    }\\
    \subfloat[]{%
        \begin{tikzpicture}
            \node[draw, black, fill=black, circle, inner sep=0pt, minimum size=5pt] (a) at (0, 0){};
            \node[draw, black, fill=black, circle, inner sep=0pt, minimum size=5pt] (a2) at (-3, 0){};

            \begin{feynman}
                \vertex (b) at (-1.5, 1.5);
                \vertex (c) at (-1.5, -1.5);
                \vertex (d) at (-1.5, 0);
                \vertex (e) at (-0.75, -0.75);
                \vertex (e2) at (-2.25, -0.75);

                \diagram* {
                    (a) -- [fermion] (b) -- [fermion] (a2) -- [fermion] (e2) -- [fermion] (c) -- [fermion] (e) -- [fermion] (a),
                    (e) -- [gluon] (d) -- [gluon] (e2),
                };
            \end{feynman}
            \draw [dashed, line width=1.2pt] (-1.5, 1.8) -- (-1.5, -1.8);
            \node[draw, red, fill=red, circle, inner sep=0pt, minimum size=5pt] (d) at (-1.5, -1.5){};
            \node[below] at (-1.5,-1.5) {${\color{red} \cD^{\DGLAP}_{J_L,q}(z)}$};
        \end{tikzpicture}
    }\quad
    \subfloat[]{%
        \begin{tikzpicture}
            \node[draw, black, fill=black, circle, inner sep=0pt, minimum size=5pt] (a) at (0, 0){};
            \node[draw, black, fill=black, circle, inner sep=0pt, minimum size=5pt] (a2) at (-3, 0){};

            \begin{feynman}
                \vertex (b) at (-1.5, 1.5);
                \vertex (c) at (-1.5, -1.5);
                \vertex (d) at (-1.5, 0);
                \vertex (e) at (-0.75, -0.75);
                \vertex (e2) at (-2.25, 0.75);

                \diagram* {
                    (a) -- [fermion] (b)-- [fermion] (e2) -- [fermion] (a2)  -- [fermion] (c) -- [fermion] (e) -- [fermion] (a),
                    (e) -- [gluon] (d) -- [gluon] (e2),
                };
            \end{feynman}
            \draw [dashed, line width=1.2pt] (-1.5, 1.8) -- (-1.5, -1.8);
            \node[draw, red, fill=red, circle, inner sep=0pt, minimum size=5pt] (d) at (-1.5, -1.5){};
            \node[below] at (-1.5,-1.5) {${\color{red} \cD^{\DGLAP}_{J_L,q}(z)}$};
        \end{tikzpicture}
    }\quad
    \subfloat[]{%
        \begin{tikzpicture}
            \node[draw, black, fill=black, circle, inner sep=0pt, minimum size=5pt] (a) at (0, 0){};
            \node[draw, black, fill=black, circle, inner sep=0pt, minimum size=5pt] (a2) at (-3, 0){};

            \begin{feynman}
                \vertex (b) at (-1.5, 1.5);
                \vertex (c) at (-1.5, -1.5);
                \vertex (d) at (-1.5, 0);
                \vertex (e) at (-0.75, 0.75);
                \vertex (e2) at (-2.25, 0.75);

                \diagram* {
                    (a)-- [fermion] (e) -- [fermion] (b) -- [fermion] (e2) -- [fermion] (a2)  -- [fermion] (c)  -- [fermion] (a),
                    (e) -- [gluon] (d) -- [gluon] (e2),
                };
            \end{feynman}
            \draw [dashed, line width=1.2pt] (-1.5, 1.8) -- (-1.5, -1.8);
            \node[draw, red, fill=red, circle, inner sep=0pt, minimum size=5pt] (d) at (-1.5, -1.5){};
            \node[below] at (-1.5,-1.5) {${\color{red} \cD^{\DGLAP}_{J_L,q}(z)}$};
        \end{tikzpicture}
    }\quad
    \subfloat[]{%
        \begin{tikzpicture}
            \node[draw, black, fill=black, circle, inner sep=0pt, minimum size=5pt] (a) at (0, 0){};
            \node[draw, black, fill=black, circle, inner sep=0pt, minimum size=5pt] (a2) at (-3, 0){};

            \begin{feynman}
                \vertex (b) at (-1.5, 1.5);
                \vertex (c) at (-1.5, -1.5);
                \vertex (d) at (-1.5, 0);
                \vertex (e) at (-0.75, 0.75);
                \vertex (e2) at (-2.25, -0.75);

                \diagram* {
                    (a)-- [fermion] (e) -- [fermion] (b)  -- [fermion] (a2) -- [fermion] (e2)  -- [fermion] (c)  -- [fermion] (a),
                    (e) -- [gluon] (d) -- [gluon] (e2),
                };
            \end{feynman}
            \draw [dashed, line width=1.2pt] (-1.5, 1.8) -- (-1.5, -1.8);
            \node[draw, red, fill=red, circle, inner sep=0pt, minimum size=5pt] (d) at (-1.5, -1.5){};
            \node[below] at (-1.5,-1.5) {${\color{red} \cD^{\DGLAP}_{J_L,q}(z)}$};
        \end{tikzpicture}
    }\\
    \subfloat[]{%
        \begin{tikzpicture}
            \node[draw, black, fill=black, circle, inner sep=0pt, minimum size=5pt] (a) at (0, 0){};
            \node[draw, black, fill=black, circle, inner sep=0pt, minimum size=5pt] (a2) at (-3, 0){};

            \begin{feynman}
                \vertex (b) at (-1.5, 1.5);
                \vertex (c) at (-1.5, -1.5);
                \vertex (d) at (-1.5, 0);
                \vertex (e) at (-2.25, 0.75);
                \vertex (e2) at (-2.25, -0.75);

                \diagram* {
                    (a) -- [fermion] (b) -- [fermion] (e)  -- [fermion] (a2) -- [fermion] (e2)  -- [fermion] (c)  -- [fermion] (a),
                    (e) -- [gluon, quarter left] (e2),
                };
            \end{feynman}
            \draw [dashed, line width=1.2pt] (-1.5, 1.8) -- (-1.5, -1.8);
            \node[draw, red, fill=red, circle, inner sep=0pt, minimum size=5pt] (d) at (-1.5, -1.5){};
            \node[below] at (-1.5,-1.5) {${\color{red} \cD^{\DGLAP}_{J_L,q}(z)}$};
        \end{tikzpicture}
    }\quad
    \subfloat[]{%
        \begin{tikzpicture}
            \node[draw, black, fill=black, circle, inner sep=0pt, minimum size=5pt] (a) at (0, 0){};
            \node[draw, black, fill=black, circle, inner sep=0pt, minimum size=5pt] (a2) at (-3, 0){};

            \begin{feynman}
                \vertex (b) at (-1.5, 1.5);
                \vertex (c) at (-1.5, -1.5);
                \vertex (d) at (-1.5, 0);
                \vertex (e) at (-0.75, 0.75);
                \vertex (e2) at (-0.75, -0.75);

                \diagram* {
                    (a) -- [fermion] (e)-- [fermion] (b)   -- [fermion] (a2)   -- [fermion] (c) -- [fermion] (e2)  -- [fermion] (a),
                    (e) -- [gluon, quarter right] (e2),
                };
            \end{feynman}
            \draw [dashed, line width=1.2pt] (-1.5, 1.8) -- (-1.5, -1.8);
            \node[draw, red, fill=red, circle, inner sep=0pt, minimum size=5pt] (d) at (-1.5, -1.5){};
            \node[below] at (-1.5,-1.5) {${\color{red} \cD^{\DGLAP}_{J_L,q}(z)}$};
        \end{tikzpicture}
    }\quad
    \subfloat[]{%
        \begin{tikzpicture}
            \node[draw, black, fill=black, circle, inner sep=0pt, minimum size=5pt] (a) at (0, 0){};
            \node[draw, black, fill=black, circle, inner sep=0pt, minimum size=5pt] (a2) at (-3, 0){};

            \begin{feynman}
                \vertex (b) at (-1.5, 1.5);
                \vertex (c) at (-1.5, -1.5);
                \vertex (d) at (-1.5, 0);
                \vertex (e) at (-2.25, 0.75);
                \vertex (e2) at (-2.25, -0.75);

                \diagram* {
                    (a) -- [fermion] (b) -- [fermion] (e)  -- [fermion] (a2) -- [fermion] (e2)  -- [fermion] (c)  -- [fermion] (a),
                    (e) -- [gluon, quarter left] (e2),
                };
            \end{feynman}
            \draw [dashed, line width=1.2pt] (-1.5, 1.8) -- (-1.5, -1.8);
            \node[draw, red, fill=red, circle, inner sep=0pt, minimum size=5pt] (d) at (-1.5, 1.5){};
            \node[above] at (-1.5,1.5) {${\color{red} \cD^{\DGLAP}_{J_L,q}(z)}$};
        \end{tikzpicture}
    }\quad
    \subfloat[]{%
        \begin{tikzpicture}
            \node[draw, black, fill=black, circle, inner sep=0pt, minimum size=5pt] (a) at (0, 0){};
            \node[draw, black, fill=black, circle, inner sep=0pt, minimum size=5pt] (a2) at (-3, 0){};

            \begin{feynman}
                \vertex (b) at (-1.5, 1.5);
                \vertex (c) at (-1.5, -1.5);
                \vertex (d) at (-1.5, 0);
                \vertex (e) at (-0.75, 0.75);
                \vertex (e2) at (-0.75, -0.75);

                \diagram* {
                    (a) -- [fermion] (e)-- [fermion] (b)   -- [fermion] (a2)   -- [fermion] (c) -- [fermion] (e2)  -- [fermion] (a),
                    (e) -- [gluon, quarter right] (e2),
                };
            \end{feynman}
            \draw [dashed, line width=1.2pt] (-1.5, 1.8) -- (-1.5, -1.8);
            \node[draw, red, fill=red, circle, inner sep=0pt, minimum size=5pt] (d) at (-1.5, 1.5){};
            \node[above] at (-1.5,1.5) {${\color{red} \cD^{\DGLAP}_{J_L,q}(z)}$};
        \end{tikzpicture}
    }
    
    \caption{Feynman diagrams for one-loop matrix element $\<\cD^{\DGLAP}_{J_L,q}(z)\>_{J(p)}$, consisting of both real emission and virtual diagrams. Red dot represents the DGLAP measurement for quark and antiquark.}
    \label{fig:DGLAP_q_in_J}
\end{figure}

\begin{itemize}
    \item \textbf{Quark DGLAP detector, Gluon Source}\\
    At one-loop, only real emission contributes to $\<\cD^{\DGLAP}_{J_L,q}\>^{\text{1-loop}}_{\cO}$. The Feynman diagrams are shown in figure~\ref{fig:DGLAP_q_in_O}, which contain virtual gluon splitting into $q \bar{q}$ pair. For generic $J_L$, this one-loop matrix element has $1/\epsilon$ pole due to collinear divergence
    \be
    \<\cD^{\DGLAP}_{J_L,q}(z)\>^{\text{1-loop}}_{[\cO]_R(p)} =
    \frac{\alpha_s}{4\pi \e} (2n_f T_F) \frac{2(J_L^2+J_L+2)}{(J_L-1)J_L(J_L+1)} \<\cD^{\DGLAP}_{J_L,g}(z)\>^{\text{tree}}_{[\cO]_R(p)} + O(\epsilon^0)\,.
    \ee
    $\<\cD^{\DGLAP}_{J_L,q}\>^{\text{1-loop}}_{\cO}$ does not have $J_L\sim -2$ pole because the soft limit of quarks is suppressed compared to gluon. Its leading pole is at $J_L=-1$
    \be
    \<\cD^{\DGLAP}_{J_L,q}(z)\>^{\text{1-loop}}_{[\cO]_R(p)} = -\frac{\alpha_s \tilde{\mu}^{2\epsilon}}{4\pi} \frac{(N_c^2-1) n_f T_F}{J_L+1} \frac{(p^2)^{2-\epsilon}}{z\cdot p} \frac{ \Gamma (2-\epsilon )}{4^{3 - 2 \epsilon} \pi ^{2 - 3 \epsilon}\epsilon ^2 \Gamma (-2 \epsilon )}\nn\\ 
    + O((J_L+1)^0)\,.
    \ee
    \item \textbf{Gluon DGLAP detector, Quark Source}\\
    Feynman diagrams in figure \ref{fig:DGLAP_g_in_J} give the non-vanishing contribution to $\<\cD^{\DGLAP}_{J_L,g}\>^{\text{1-loop}}_{J}$. For generic $J_L$, the $\epsilon$ pole of $\<\cD^{\DGLAP}_{J_L,g}(z)\>^{\text{1-loop}}_{[J]_R(p)}$ is 
    \be 
    \<\cD^{\DGLAP}_{J_L,g}(z)\>^{\text{1-loop}}_{[J]_R(p)} = \frac{\alpha_s}{4\pi \e} C_F \frac{2(J_L^2 + J_L + 2)}{J_L(J_L+1)(J_L+2)} \<\cD^{\DGLAP}_{J_L,q}(z)\>_{J(p)}^{\text{tree}} + O(\e^0)\,.
    \ee
    The $J_L$ poles at $-2$ and $-1$ are
    \be
    & \<\cD^{\DGLAP}_{J_L,g}(z)\>^{\text{1-loop}}_{[J]_R(p)} = \frac{\alpha_s \tilde{\mu}^{2\e}}{4\pi} \frac{C_F N_c}{J_L+2} \frac{(p^2)^{2-\e}}{(z\cdot p)^2} \frac{2^{4\epsilon }  (\epsilon -1) \Gamma (-\epsilon )}{\pi ^{2 - 3\epsilon }\Gamma (1-2\epsilon )} + O((J_L+2)^0)\,. \\
    & \<\cD^{\DGLAP}_{J_L,g}(z)\>^{\text{1-loop}}_{[J]_R(p)} = \frac{\alpha_s \tilde{\mu}^{2\e}}{4\pi} \frac{C_F N_c}{J_L+1} \frac{(p^2)^{1-\e}}{z\cdot p}  \frac{2^{4 \epsilon +1}  (\epsilon -1)^2 \Gamma (-\epsilon )}{\pi ^{2 - 3 \epsilon}\Gamma(1-2 \epsilon )}+ O((J_L+1)^0)\,.
    \ee
    \item \textbf{Quark DGLAP detector, Quark Source}\\
    The real emission and virtual correction diagrams for $\<\cD^{\DGLAP}_{J_L,q}\>^{\text{1-loop}}_{J}$ are shown in figure~\ref{fig:DGLAP_q_in_J}. The quark wavefunction counterterm cancels the composite operator counterterm for $J_\mu$. The one-loop $\e$-pole is 
    \be 
    \<\cD^{\DGLAP}_{J_L,q}(z)\>^{\text{1-loop}}_{[J]_R(p)} = \frac{\alpha_s}{4\pi \e} C_F \left[4(\psi(-J_L)+\gamma_E )- {2 \over J_L(J_L+1)} -3\right]
    \<\cD^{\DGLAP}_{J_L,q}(z)\>^{\text{tree}}_{J(p)}\nn\\
    + O(\e^0)\,.
    \ee
    The leading $J_L$ pole is at $J_L=-1$
    \be
    \<\cD^{\DGLAP}_{J_L,q}(z)\>^{\text{1-loop}}_{[J]_R(p)} = -\frac{\alpha_s \tilde{\mu}^{2\e}}{4\pi} \frac{C_F N_c}{J_L+1} \frac{(p^2)^{1-\e}}{z\cdot p}
    \frac{2^{4\epsilon } (\epsilon -1)^2 \Gamma (-\epsilon )}{\pi ^{2 - 3\epsilon }\Gamma(1-2 \epsilon )} + (O(J_L+1)^0)\,.
    \ee
\end{itemize}

\subsection{Matrix elements of BFKL detectors}

In this section, we list the Feynman diagrams for calculating the one-loop matrix elements of BFKL detector \eqref{eq:QCD_BFKL_decomposition}. The final expressions of one-loop divergences are listed in \ref{subsec:QCD-divergences}. We will not show the diagrams for one-loop wavefunction and composite operator renormalization counterterm. Their contributions are proportional to the corresponding tree-level matrix elements, where: for $\langle \cD_{J_L}^{\BFKL} \rangle_{\cO}$, the proportionality constant is given in \eqref{eq:delta_g}; for $\langle \cD_{J_L}^{\BFKL} \rangle_{J}$, the proportionality constant vanishes.

\subsubsection*{Gluon Source}

\begin{figure}[htbp]
    \centering
    \subfloat[]{
        \begin{tikzpicture}
            \node[draw, black, fill=black, circle, inner sep=0pt, minimum size=5pt] (a) at (0, 0){};

            \begin{feynman}
                \vertex (b) at (-1.5, 1.5);
                \vertex (c) at (-1.5, -1.5);
                \vertex (d) at (-{sqrt(2)*1.5}, 0);

                \diagram* {
                    (a) -- [gluon] (b),
                    (a) -- [gluon] (c),
                    (a) -- [gluon] (d),
                };
            \end{feynman}

            \pgfmathsetmacro{\toshift}{3}
            \node[scale=1.3] at ({0.45},0) {$+$};
            \node[scale=1.3] at ({\toshift+0.45},0) {$+$};
            \node[scale=1.3] at ({2*\toshift+0.45},0) {$+$};
            \node[draw, black, fill=black, circle, inner sep=0pt, minimum size=5pt] (a) at (\toshift, 0){};
            \begin{feynman}
                \vertex (b) at ({\toshift-1.5}, 1.5);
                \vertex (c) at ({\toshift-1.5}, -1.5);
                \vertex (d) at ({\toshift-sqrt(2)*1.5}, 0);
                \vertex (e) at ({\toshift-0.75}, 0.75);

                \diagram* {
                    (a) -- [gluon] (b),
                    (a) -- [gluon] (c),
                    (e) -- [gluon] (d),
                };
            \end{feynman}

            \node[draw, black, fill=black, circle, inner sep=0pt, minimum size=5pt] (a) at ({2*\toshift}, 0){};
            \begin{feynman}
                \vertex (b) at ({2*\toshift-1.5}, 1.5);
                \vertex (c) at ({2*\toshift-1.5}, -1.5);
                \vertex (d) at ({2*\toshift-sqrt(2)*1.5}, 0);
                \vertex (e) at ({2*\toshift-0.75}, -0.75);

                \diagram* {
                    (a) -- [gluon] (b),
                    (a) -- [gluon] (c),
                    (e) -- [gluon] (d),
                };
            \end{feynman}

            \node[draw, black, fill=black, circle, inner sep=0pt, minimum size=5pt] (a) at ({3*\toshift}, 0){};
            \begin{feynman}
                \vertex (b) at ({3*\toshift-1.5}, 1.5);
                \vertex (c) at ({3*\toshift-1.5}, -1.5);
                \vertex (d) at ({3*\toshift-sqrt(2)*1.5}, 0);
                \vertex (e) at ({3*\toshift-0.75}, -0.75);

                \diagram* {
                    (a) -- [gluon] (d),
                    (a) -- [gluon] (c),
                    (e) -- [gluon] (b),
                };
            \end{feynman}

            \pgfmathsetmacro{\toshift}{4}
            \node[scale=1.5] at ({-\toshift+1.2},0) {$=$};
            \begin{feynman}
                \vertex (a) at (-\toshift, 0);
                \vertex (b) at ({-\toshift-1.5}, 1.5);
                \vertex (c) at ({-\toshift-1.5}, -1.5);
                \vertex (d) at ({-\toshift-sqrt(2)*1.5}, 0);
                \vertex (e) at ({-\toshift-0.75}, 0.75);

                \diagram* {
                    (a) -- [gluon] (b),
                    (a) -- [gluon] (c),
                    (a) -- [gluon] (d),
                };
            \end{feynman}
            \node[draw, gray, fill=gray, circle, inner sep=0pt, minimum size=30pt] (a) at (-\toshift, 0){};
        \end{tikzpicture}
    }\\
    \subfloat[]{%
        \begin{tikzpicture}
            \node[draw, black, fill=black, circle, inner sep=0pt, minimum size=5pt] (a) at (0, 0){};

            \begin{feynman}
                \vertex (b) at (-1.5, 1.5);
                \vertex (c) at (-1.5, -1.5);
                \vertex (i1) at (-0.75, 0.75);
                \vertex (i2) at (-0.75, -0.75);

                \diagram* {
                    (a) -- [gluon] (b),
                    (a) -- [gluon] (c),
                    (i1) -- [gluon, half right, looseness=1.1] (i2),
                };
            \end{feynman}

            \pgfmathsetmacro{\toshift}{2.8}
            
            \node[draw, black, fill=black, circle, inner sep=0pt, minimum size=5pt] (a) at (\toshift, 0){};
            \node[left,scale=1.2] at ({\toshift-1.5},0) {$+\frac{1}{2}$};
            \begin{feynman}
                \vertex (b) at ({\toshift-1.5}, 1.5);
                \vertex (c) at ({\toshift-1.5}, -1.5);
                \vertex (e) at ({\toshift-0.9}, 0.9);

                \diagram* {
                    (a) -- [gluon] (b),
                    (a) -- [gluon] (c),
                    (a) -- [gluon, half left, looseness=1.1] (e),
                };
            \end{feynman}

            \node[draw, black, fill=black, circle, inner sep=0pt, minimum size=5pt] (a) at ({2*\toshift}, 0){};
            \node[left, scale=1.2] at ({2*\toshift-1.5},0) {$+\frac{1}{2}$};
            \begin{feynman}
                \vertex (b) at ({2*\toshift-1.5}, 1.5);
                \vertex (c) at ({2*\toshift-1.5}, -1.5);
                \vertex (e) at ({2*\toshift-0.9}, -0.9);

                \diagram* {
                    (a) -- [gluon] (b),
                    (a) -- [gluon] (c),
                    (a) -- [gluon, half right, looseness=1.1] (e),
                };
            \end{feynman}

            \node[draw, black, fill=black, circle, inner sep=0pt, minimum size=5pt] (a) at ({3*\toshift+1.5}, 0){};
            \node[left,scale=1.2] at ({3*\toshift-1.5},0) {$+\frac{1}{2}$};
            \begin{feynman}
                \vertex (b) at ({3*\toshift-1.5}, 1.5);
                \vertex (c) at ({3*\toshift-1.5}, -1.5);
                \vertex (d) at ({3*\toshift}, 0);

                \diagram* {
                    (a) -- [gluon, half right, looseness=1.1] (d),
                    (a) -- [gluon, half left, looseness=1.1] (d),
                    (d) -- [gluon] (b),
                    (d) -- [gluon] (c),
                };
            \end{feynman}

            \pgfmathsetmacro{\toshift}{3.5}
            \node[scale=1.5] at ({-\toshift+1.2},0) {$=$};
            \begin{feynman}
                \vertex (a) at (-\toshift, 0);
                \vertex (b) at ({-\toshift-1.5}, 1.5);
                \vertex (c) at ({-\toshift-1.5}, -1.5);
                \vertex (e) at ({-\toshift-0.75}, 0.75);

                \diagram* {
                    (a) -- [gluon] (b),
                    (a) -- [gluon] (c),
                };
            \end{feynman}
            \node[draw, gray, fill=gray, circle, inner sep=0pt, minimum size=30pt] (a) at (-\toshift, 0){};
            \node[draw, white, fill=white, circle, inner sep=0pt, minimum size=15pt] (a) at (-\toshift, 0){};
        \end{tikzpicture}
    }
    \caption{Diagrammatic representations for pure gluon form factors associated to the operator $\cO$.}
    \label{fig:O_amp_rep}
\end{figure}
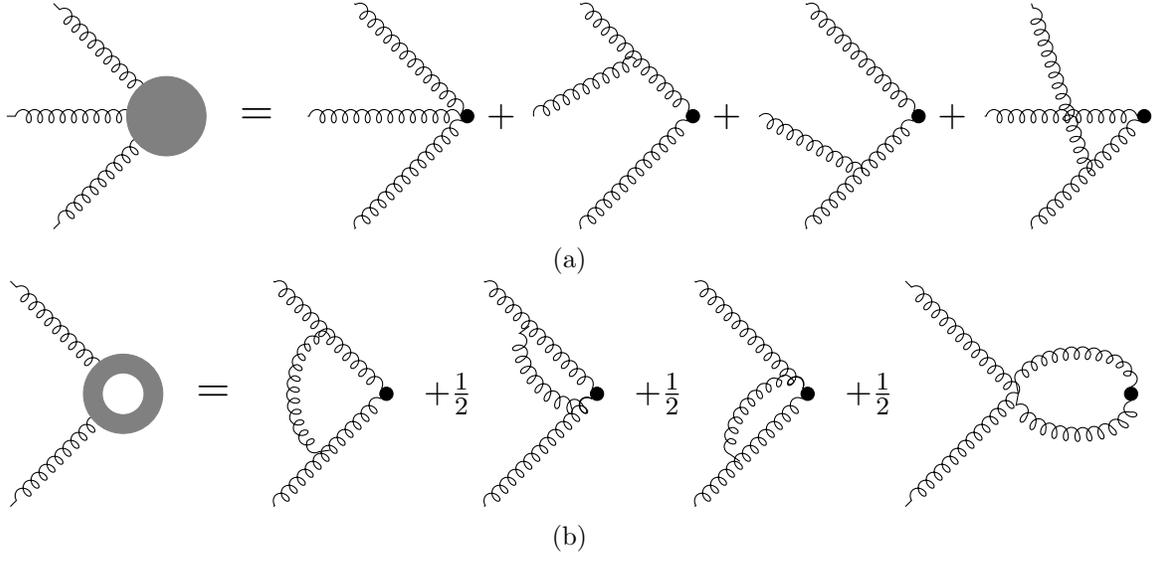

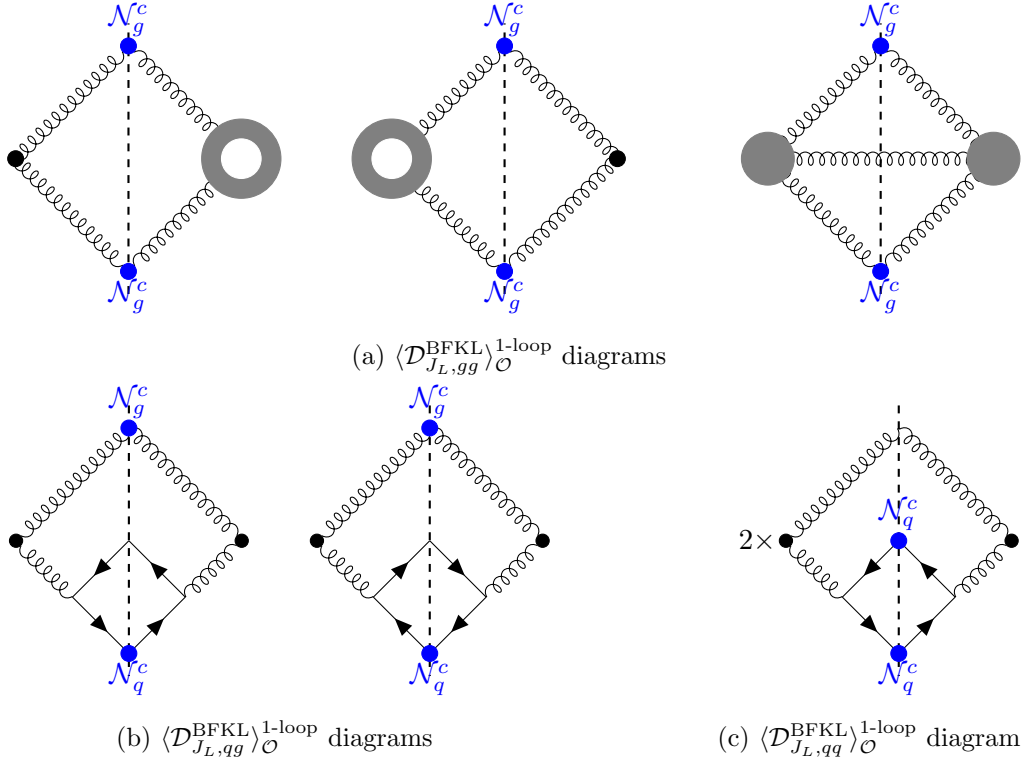
\begin{figure}[htbp]
    \centering
    \subfloat[$\<\cD^{\BFKL}_{J_L,gg}\>^{\text{1-loop}}_{\cO}$ diagrams]{\label{subfig:BFKL_O_gg}
        \begin{tikzpicture}
            \begin{feynman}
                \coordinate (O1) at (1.5,0);
                \coordinate (O2) at (-1.5,0);
                \coordinate (V1) at (0,1.5);
                \coordinate (V2) at (0,-1.5);
                
                \draw[dashed, thick] (0,-1.8) -- (0,1.8);
                
                \draw[gluon] (O1) -- (V1);
                \draw[gluon] (O1) -- (V2);
                \draw[gluon] (O2) -- (V1);
                \draw[gluon] (O2) -- (V2);
                
                \filldraw[blue] (V1) circle (3pt);
                \filldraw[blue] (V2) circle (3pt);
                \filldraw[black] (O1) circle (3pt);
                \filldraw[black] (O2) circle (3pt);
                \node[above] at (V1) {$\color{blue}\cN_g^{c}$};
                \node[below] at (V2) {$\color{blue}\cN_g^{c}$};
                \node[draw, gray, fill=gray, circle, inner sep=0pt, minimum size=30pt] (a) at (O1){};
                \node[draw, white, fill=white, circle, inner sep=0pt, minimum size=15pt] (a) at (O1){};

                \pgfmathsetmacro{\toshift}{5}
                \coordinate (O1) at ({\toshift+1.5},0);
                \coordinate (O2) at ({\toshift-1.5},0);
                \coordinate (V1) at ({\toshift},1.5);
                \coordinate (V2) at ({\toshift},-1.5);
                
                \draw[dashed, thick] ({\toshift},-1.8) -- ({\toshift},1.8);
                
                \draw[gluon] (O1) -- (V1);
                \draw[gluon] (O1) -- (V2);
                \draw[gluon] (O2) -- (V1);
                \draw[gluon] (O2) -- (V2);
                
                \filldraw[blue] (V1) circle (3pt);
                \filldraw[blue] (V2) circle (3pt);
                \filldraw[black] (O1) circle (3pt);
                \filldraw[black] (O2) circle (3pt);
                \node[above] at (V1) {$\color{blue}\cN_g^{c}$};
                \node[below] at (V2) {$\color{blue}\cN_g^{c}$};
                \node[draw, gray, fill=gray, circle, inner sep=0pt, minimum size=30pt] (a) at (O2){};
                \node[draw, white, fill=white, circle, inner sep=0pt, minimum size=15pt] (a) at (O2){};

                \coordinate (O1) at ({2*\toshift+1.5},0);
                \coordinate (O2) at ({2*\toshift-1.5},0);
                \coordinate (V1) at ({2*\toshift},1.5);
                \coordinate (V2) at ({2*\toshift},-1.5);
                
                \draw[dashed, thick] ({2*\toshift},-1.8) -- ({2*\toshift},1.8);
                
                \draw[gluon] (O1) -- (V1);
                \draw[gluon] (O1) -- (V2);
                \draw[gluon] (O2) -- (V1);
                \draw[gluon] (O2) -- (V2);
                \draw[gluon] (O1) -- (O2);
                
                \filldraw[blue] (V1) circle (3pt);
                \filldraw[blue] (V2) circle (3pt);
                \filldraw[black] (O1) circle (3pt);
                \filldraw[black] (O2) circle (3pt);
                \node[above] at (V1) {$\color{blue}\cN_g^{c}$};
                \node[below] at (V2) {$\color{blue}\cN_g^{c}$};
                \node[draw, gray, fill=gray, circle, inner sep=0pt, minimum size=20pt] (a) at (O1){};
                \node[draw, gray, fill=gray, circle, inner sep=0pt, minimum size=20pt] (a) at (O2){};
                
            \end{feynman}
        \end{tikzpicture}
    }\\
    \subfloat[$\<\cD^{\BFKL}_{J_L,qg}\>^{\text{1-loop}}_{\cO}$ diagrams]{\label{subfig:BFKL_O_qg}
    \begin{tikzpicture}
        \begin{feynman}

            \coordinate (O1) at (0,0);
            \coordinate (O2) at (-3,0);
            \coordinate (b) at (-1.5, 1.5);
            \coordinate (c) at (-1.5, -1.5);
            \coordinate (d) at (-1.5, 0);
            \coordinate (e) at (-0.75, -0.75);
            \coordinate (e2) at (-2.25, -0.75);
            \node[draw, black, fill=black, circle, inner sep=0pt, minimum size=5pt] (a) at (O1){};
            \node[draw, black, fill=black, circle, inner sep=0pt, minimum size=5pt] (a2) at (O2){};

            \diagram* {
                (a) -- [gluon] (b),
                (a) -- [gluon] (e),
                (a2) -- [gluon] (b),
                (a2) -- [gluon] (e2),
                (c) -- [fermion] (e) -- [fermion] (d) -- [fermion] (e2) -- [fermion] (c),
            };

            \draw [dashed, thick] (-1.5, 1.8) -- (-1.5, -1.8);
            \filldraw[blue] (b) circle (3pt);
            \filldraw[blue] (c) circle (3pt);
            \node[above] at (b) {$\color{blue}\cN_g^{c}$};
            \node[below] at (c) {$\color{blue}\cN_q^{c}$};

            \pgfmathsetmacro{\toshift}{4}
            \coordinate (O1) at (\toshift,0);
            \coordinate (O2) at ({\toshift-3},0);
            \coordinate (b) at ({\toshift-1.5}, 1.5);
            \coordinate (c) at ({\toshift-1.5}, -1.5);
            \coordinate (d) at ({\toshift-1.5}, 0);
            \coordinate (e) at ({\toshift-0.75}, -0.75);
            \coordinate (e2) at ({\toshift-2.25}, -0.75);
            \node[draw, black, fill=black, circle, inner sep=0pt, minimum size=5pt] (a) at (O1){};
            \node[draw, black, fill=black, circle, inner sep=0pt, minimum size=5pt] (a2) at (O2){};

            \diagram* {
                (a) -- [gluon] (b),
                (a) -- [gluon] (e),
                (a2) -- [gluon] (b),
                (a2) -- [gluon] (e2),
                (c) -- [fermion] (e2) -- [fermion] (d) -- [fermion] (e) -- [fermion] (c),
            };

            \draw [dashed, thick] ({\toshift-1.5}, 1.8) -- ({\toshift-1.5}, -1.8);
            \filldraw[blue] (b) circle (3pt);
            \filldraw[blue] (c) circle (3pt);
            \node[above] at (b) {$\color{blue}\cN_g^{c}$};
            \node[below] at (c) {$\color{blue}\cN_q^{c}$};
        \end{feynman}
    \end{tikzpicture}
    }\hspace{2cm}
    \subfloat[$\<\cD^{\BFKL}_{J_L,qq}\>^{\text{1-loop}}_{\cO}$ diagram]{\label{subfig:BFKL_O_qq}
    \begin{tikzpicture}
        \begin{feynman}

            \coordinate (O1) at (0,0);
            \coordinate (O2) at (-3,0);
            \coordinate (b) at (-1.5, 1.5);
            \coordinate (c) at (-1.5, -1.5);
            \coordinate (d) at (-1.5, 0);
            \coordinate (e) at (-0.75, -0.75);
            \coordinate (e2) at (-2.25, -0.75);
            \node[draw, black, fill=black, circle, inner sep=0pt, minimum size=5pt] (a) at (O1){};
            \node[draw, black, fill=black, circle, inner sep=0pt, minimum size=5pt] (a2) at (O2){};

            \diagram* {
                (a) -- [gluon] (b),
                (a) -- [gluon] (e),
                (a2) -- [gluon] (b),
                (a2) -- [gluon] (e2),
                (c) -- [fermion] (e) -- [fermion] (d) -- [fermion] (e2) -- [fermion] (c),
            };

            \draw [dashed, thick] (-1.5, 1.8) -- (-1.5, -1.8);
            \filldraw[blue] (d) circle (3pt);
            \filldraw[blue] (c) circle (3pt);
            \node[above] at (d) {$\color{blue}\cN_q^{c}$};
            \node[below] at (c) {$\color{blue}\cN_q^{c}$};
            \node[left] at (O2) {$2\times$};
        \end{feynman}
    \end{tikzpicture}
    }
    \caption{Feynman diagrams for the one-loop matrix element of the BFKL detector $\cD^{\BFKL}_{J_L}$ in the gluon source $\cO$. The components in the decomposition $\cD^{\BFKL}_{J_L} = \cD^{\BFKL}_{J_L,gg}+ 2\cD^{\BFKL}_{J_L,qg} + \cD^{\BFKL}_{J_L,qq}$ are listed in (\ref{subfig:BFKL_O_gg}), (\ref{subfig:BFKL_O_qg}) and (\ref{subfig:BFKL_O_qq}), respectively.}
    \label{fig:1loop_BFKL_O}
\end{figure}

As illustrated in figure~\ref{fig:O_amp_rep}, we employ diagrammatic representations for both the tree-level gluon emission and one-loop virtual correction form factors produced by the gluon source operator $\cO$. The one-loop Feynamn diagrams for $\<\cD^{\BFKL}_{J_L}\>_{\cO}$ are shown in figure~\ref{fig:1loop_BFKL_O}, where the $\cD^{\BFKL}_{J_L}$ is decomposed as $\cD^{\BFKL}_{J_L,gg}+ 2\cD^{\BFKL}_{J_L,qg} + \cD^{\BFKL}_{J_L,qq}$. Note that at one-loop order, the counterterms only contribute to $\<\cD^{\BFKL}_{J_L,gg}\>_{\cO}$ because there is no quark in the final state.

\subsubsection*{Quark Source}

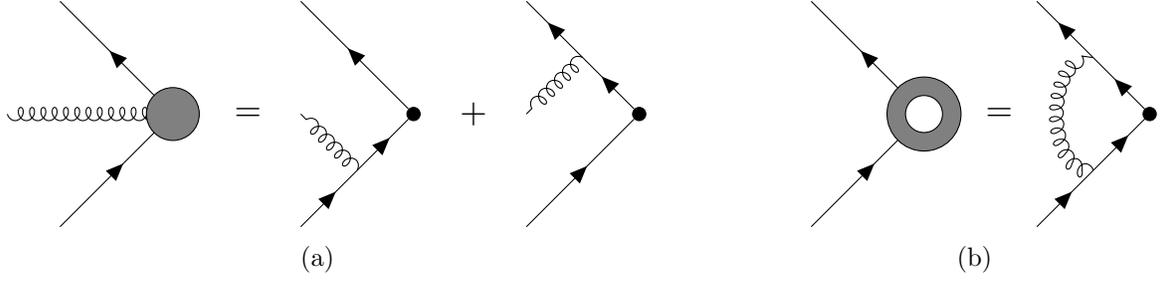
\begin{figure}[htbp]
    \centering
    \subfloat[]{
        \begin{tikzpicture}
            \begin{feynman}
                \node[draw, black, fill=black, circle, inner sep=0pt, minimum size=5pt] (a) at (0, 0){};
                \vertex (b) at (-1.5, 1.5);
                \vertex (c) at (-1.5, -1.5);
                \vertex (d) at (-1.5, 0);
                \vertex (e) at (-0.75, -0.75);

                \diagram* {
                    (c) --[fermion] (e) -- [fermion](a) -- [fermion] (b),
                    (e) -- [gluon] (d),
                };

                \pgfmathsetmacro{\toshift}{3}
                \node[scale=1.2] at (0.8,0) {$+$};
                \node[draw, black, fill=black, circle, inner sep=0pt, minimum size=5pt] (a) at (\toshift, 0){};
                \vertex (b) at ({\toshift-1.5}, 1.5);
                \vertex (c) at ({\toshift-1.5}, -1.5);
                \vertex (d) at ({\toshift-1.5}, 0);
                \vertex (e) at ({\toshift-0.75}, 0.75);

                \diagram* {
                    (c)  -- [fermion](a) --[fermion] (e) -- [fermion] (b),
                    (e) -- [gluon] (d),
                };

                \pgfmathsetmacro{\toshift}{-3.2}
                \node[scale=1.2] at ({\toshift+1},0) {$=$};
                \vertex (a) at (\toshift, 0);
                \vertex (b) at ({\toshift-1.5}, 1.5);
                \vertex (c) at ({\toshift-1.5}, -1.5);
                \vertex (d) at ({\toshift-2.2}, 0);

                \diagram* {
                    (c)  -- [fermion](a) -- [fermion] (b),
                    (d) -- [gluon] (a),
                };
                \draw[fill=gray] (a) circle (10pt);
            \end{feynman}
        \end{tikzpicture}
    }\hfill
    \subfloat[]{
        \begin{tikzpicture}
            \begin{feynman}
                \vertex (a) at (0, 0);
                \vertex (b) at (-1.5, 1.5);
                \vertex (c) at (-1.5, -1.5);
                \diagram* {
                    (c) -- [fermion](a) -- [fermion] (b),
                };
                \draw[fill=gray] (a) circle (14pt);
                \draw[fill=white] (a) circle (7pt);

                \node[scale=1.2] at (1,0) {$=$};
                \pgfmathsetmacro{\toshift}{3}
                \vertex (a) at (\toshift, 0);
                \vertex (b) at ({\toshift-1.5}, 1.5);
                \vertex (c) at ({\toshift-1.5}, -1.5);
                \vertex (d) at ({\toshift-0.75}, 0.75);
                \vertex (e) at ({\toshift-0.75}, -0.75);
                \diagram* {
                    (c) --[fermion] (e) -- [fermion](a) --[fermion](d) -- [fermion] (b),
                    (e) -- [gluon, half left, looseness=1.1] (d),
                };
                \node[draw, black, fill=black, circle, inner sep=0pt, minimum size=5pt] (a) at (\toshift, 0){};
            \end{feynman}
        \end{tikzpicture}
    }
    
    \caption{Diagrammatic representations for some form factors associated to $J_\mu$.}
    \label{fig:J_amp_rep}
\end{figure}

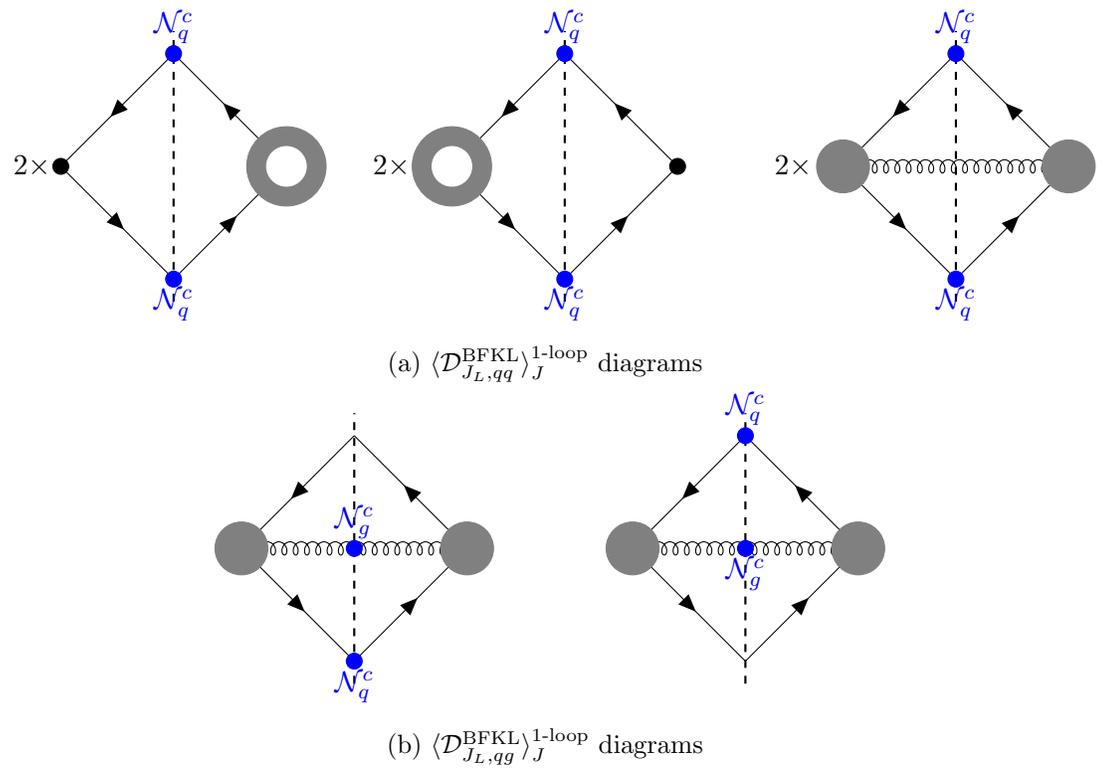
\begin{figure}[htbp]
    \centering
    \subfloat[$\<\cD^{\BFKL}_{J_L,qq}\>^{\text{1-loop}}_{J}$ diagrams]{\label{subfig:BFKL_J_qq}
        \begin{tikzpicture}
            \begin{feynman}
                \coordinate (O1) at (1.5,0);
                \coordinate (O2) at (-1.5,0);
                \coordinate (V1) at (0,1.5);
                \coordinate (V2) at (0,-1.5);
                \coordinate (V0) at (-2,0);
                
                \draw[dashed, thick] (0,-1.8) -- (0,1.8);
                
                \draw[fermion] (O1) -- (V1);
                \draw[fermion] (V2) -- (O1);
                \draw[fermion] (O2) -- (V2);
                \draw[fermion] (V1) -- (O2);
                
                \filldraw[blue] (V1) circle (3pt);
                \filldraw[blue] (V2) circle (3pt);
                \filldraw[black] (O1) circle (3pt);
                \filldraw[black] (O2) circle (3pt);
                \node[above] at (V1) {$\color{blue}\cN_q^{c}$};
                \node[below] at (V2) {$\color{blue}\cN_q^{c}$};
                \node[left] at (O2) {$2\times$};
                \node[draw, gray, fill=gray, circle, inner sep=0pt, minimum size=30pt] (a) at (O1){};
                \node[draw, white, fill=white, circle, inner sep=0pt, minimum size=15pt] (a) at (O1){};

                \pgfmathsetmacro{\toshift}{5.2}
                \coordinate (O1) at ({\toshift+1.5},0);
                \coordinate (O2) at ({\toshift-1.5},0);
                \coordinate (V1) at ({\toshift},1.5);
                \coordinate (V2) at ({\toshift},-1.5);
                \coordinate (V0) at ({\toshift-2},0);
                
                \draw[dashed, thick] ({\toshift},-1.8) -- ({\toshift},1.8);
                
               \draw[fermion] (O1) -- (V1);
               \draw[fermion] (V2) -- (O1);
               \draw[fermion] (O2) -- (V2);
               \draw[fermion] (V1) -- (O2);
                
                \filldraw[blue] (V1) circle (3pt);
                \filldraw[blue] (V2) circle (3pt);
                \filldraw[black] (O1) circle (3pt);
                \filldraw[black] (O2) circle (3pt);
                \node[above] at (V1) {$\color{blue}\cN_q^{c}$};
                \node[below] at (V2) {$\color{blue}\cN_q^{c}$};
                \node[left,xshift=-12pt] at (O2) {$2\times$};
                \node[draw, gray, fill=gray, circle, inner sep=0pt, minimum size=30pt] (a) at (O2){};
                \node[draw, white, fill=white, circle, inner sep=0pt, minimum size=15pt] (a) at (O2){};

                \coordinate (O1) at ({2*\toshift+1.5},0);
                \coordinate (O2) at ({2*\toshift-1.5},0);
                \coordinate (V1) at ({2*\toshift},1.5);
                \coordinate (V2) at ({2*\toshift},-1.5);
                \coordinate (V0) at ({2*\toshift-2},0);
                
                \draw[dashed, thick] ({2*\toshift},-1.8) -- ({2*\toshift},1.8);
                
                \draw[fermion] (O1) -- (V1);
                \draw[fermion] (V2) -- (O1);
                \draw[fermion] (O2) -- (V2);
                \draw[fermion] (V1) -- (O2);
                \draw[gluon] (O1) -- (O2);
                
                \filldraw[blue] (V1) circle (3pt);
                \filldraw[blue] (V2) circle (3pt);
                \filldraw[black] (O1) circle (3pt);
                \filldraw[black] (O2) circle (3pt);
                \node[above] at (V1) {$\color{blue}\cN_q^{c}$};
                \node[below] at (V2) {$\color{blue}\cN_q^{c}$};
                \node[left,xshift=-8pt] at (O2) {$2\times$};
                \node[draw, gray, fill=gray, circle, inner sep=0pt, minimum size=20pt] (a) at (O1){};
                \node[draw, gray, fill=gray, circle, inner sep=0pt, minimum size=20pt] (a) at (O2){};
                
            \end{feynman}
        \end{tikzpicture}
    }\\
    \subfloat[$\<\cD^{\BFKL}_{J_L,qg}\>^{\text{1-loop}}_{J}$ diagrams]{\label{subfig:BFKL_J_qg}
    \begin{tikzpicture}
        \begin{feynman}
            \pgfmathsetmacro{\toshift}{5.2}
            \coordinate (O1) at ({1.5},0);
            \coordinate (O2) at ({-1.5},0);
            \coordinate (V1) at (0,1.5);
            \coordinate (V2) at (0,-1.5);
            \coordinate (OO) at (0,0);
            
            \draw[dashed, thick] (0,-1.8) -- (0,1.8);
            
            \draw[fermion] (O1) -- (V1);
            \draw[fermion] (V2) -- (O1);
            \draw[fermion] (O2) -- (V2);
            \draw[fermion] (V1) -- (O2);
            \draw[gluon] (O1) -- (O2);
            
            \filldraw[blue] (OO) circle (3pt);
            \filldraw[blue] (V2) circle (3pt);
            \filldraw[black] (O1) circle (3pt);
            \filldraw[black] (O2) circle (3pt);
            \node[above] at (OO) {$\color{blue}\cN_g^{c}$};
            \node[below] at (V2) {$\color{blue}\cN_q^{c}$};
            \node[draw, gray, fill=gray, circle, inner sep=0pt, minimum size=20pt] (a) at (O1){};
            \node[draw, gray, fill=gray, circle, inner sep=0pt, minimum size=20pt] (a) at (O2){};

            \coordinate (O1) at ({\toshift+1.5},0);
            \coordinate (O2) at ({\toshift-1.5},0);
            \coordinate (V1) at ({\toshift},1.5);
            \coordinate (V2) at ({\toshift},-1.5);
            \coordinate (OO) at ({\toshift},0);
            
            \draw[dashed, thick] ({\toshift},-1.8) -- ({\toshift},1.8);
            
            \draw[fermion] (O1) -- (V1);
            \draw[fermion] (V2) -- (O1);
            \draw[fermion] (O2) -- (V2);
            \draw[fermion] (V1) -- (O2);
            \draw[gluon] (O1) -- (O2);
            
            \filldraw[blue] (OO) circle (3pt);
            \filldraw[blue] (V1) circle (3pt);
            \filldraw[black] (O1) circle (3pt);
            \filldraw[black] (O2) circle (3pt);
            \node[below] at (OO) {$\color{blue}\cN_g^{c}$};
            \node[above] at (V1) {$\color{blue}\cN_q^{c}$};
            \node[draw, gray, fill=gray, circle, inner sep=0pt, minimum size=20pt] (a) at (O1){};
            \node[draw, gray, fill=gray, circle, inner sep=0pt, minimum size=20pt] (a) at (O2){};
        \end{feynman}
    \end{tikzpicture}
    }
    \caption{Feynman diagrams for the one-loop matrix element of the BFKL detector $\cD^{\BFKL}_{J_L}$ in the quark source $J$. Since $\<\cD^{\BFKL}_{J_L,gg}\>_J^{\text{1-loop}}$ is zero, only diagrams for $\<\cD^{\BFKL}_{J_L,qq}\>_J^{\text{1-loop}}$ and $\<\cD^{\BFKL}_{J_L,qg}\>_J^{\text{1-loop}}$ are listed in (\ref{subfig:BFKL_J_qq}) and (\ref{subfig:BFKL_J_qg}) respectively.}
    \label{fig:1loop_BFKL_J}
\end{figure}

Similarly, we use a diagrammatic representation for relevant form factors of the operator $J_\mu$ (see figure~\ref{fig:J_amp_rep}). The Feynman diagrams for computing one-loop matrix element $\<\cD^{\BFKL}_{J_L}\>_J$ are shown in figure~\ref{fig:1loop_BFKL_J}. Unlike the gluon case, the matrix element $\<\cD^{\BFKL}_{J_L,gg}\>_J$ vanishes at this order because there is at most one gluon in the final state.

\clearpage 

\bibliographystyle{JHEP}
\bibliography{refs}

\end{document}